\documentclass[11pt,a4paper]{report}
\usepackage[utf8]{inputenc}
\usepackage[
backend=bibtex,
style=numeric,
sorting=nyt,
giveninits=true,
isbn=false
]{biblatex}
\AtEveryBibitem{\clearlist{language}}
\renewbibmacro{in:}{}
\usepackage[english]{babel}
\usepackage{amsmath}
\usepackage{amsfonts}
\usepackage{amssymb}
\usepackage{graphicx}
\usepackage{placeins}
\usepackage{subfiles}
\usepackage{amsmath,amssymb,amsthm}
\usepackage[utf8]{inputenc}
\usepackage[T1]{fontenc}
\usepackage{booktabs}
\usepackage{tocloft}
\usepackage[normalem]{ulem}
\usepackage{MnSymbol}
\usepackage{setspace}
\usepackage{float}
\usepackage{tikz}
\usepackage{url}
\usepackage{mathtools}
\usepackage{soul,color}
\usepackage{parskip}
\usepackage{fancyhdr}
\usepackage{subcaption}
\usepackage{caption}

\usepackage[numbib,nottoc]{tocbibind}
\usepackage{hyperref}
\usepackage{wrapfig}
\usepackage[left=2.4cm,top=2cm,bottom=2cm,right=2.4cm]{geometry}
\usetikzlibrary{arrows.meta}
\usetikzlibrary{shapes,arrows,backgrounds,fit,positioning}

\newcommand\Mycite[1]{%
	\citeauthor{#1}~[\citeyear{#1}]~\cite{#1} }

\usepackage{pdfpages}

\usepackage{titling}
\newcommand{\subtitle}[1]{
	\posttitle{
		\par\end{center}
	\begin{center}\large#1\end{center}
	\vskip0.5em}
	}
	
	\author{Erik Lørup}
	\title{Scaling of Structure and Dynamics in Molecular Liquids: Insights from Pressure Experiments and Molecular Dynamics}
	\subtitle{This thesis
		has been submitted to the PhD School of Science and Environment
		
		Main Supervisor: Kristine Niss,
		Co-Supervisor: Nick Bailey }
\begin{document}

\includepdf[page={1}]{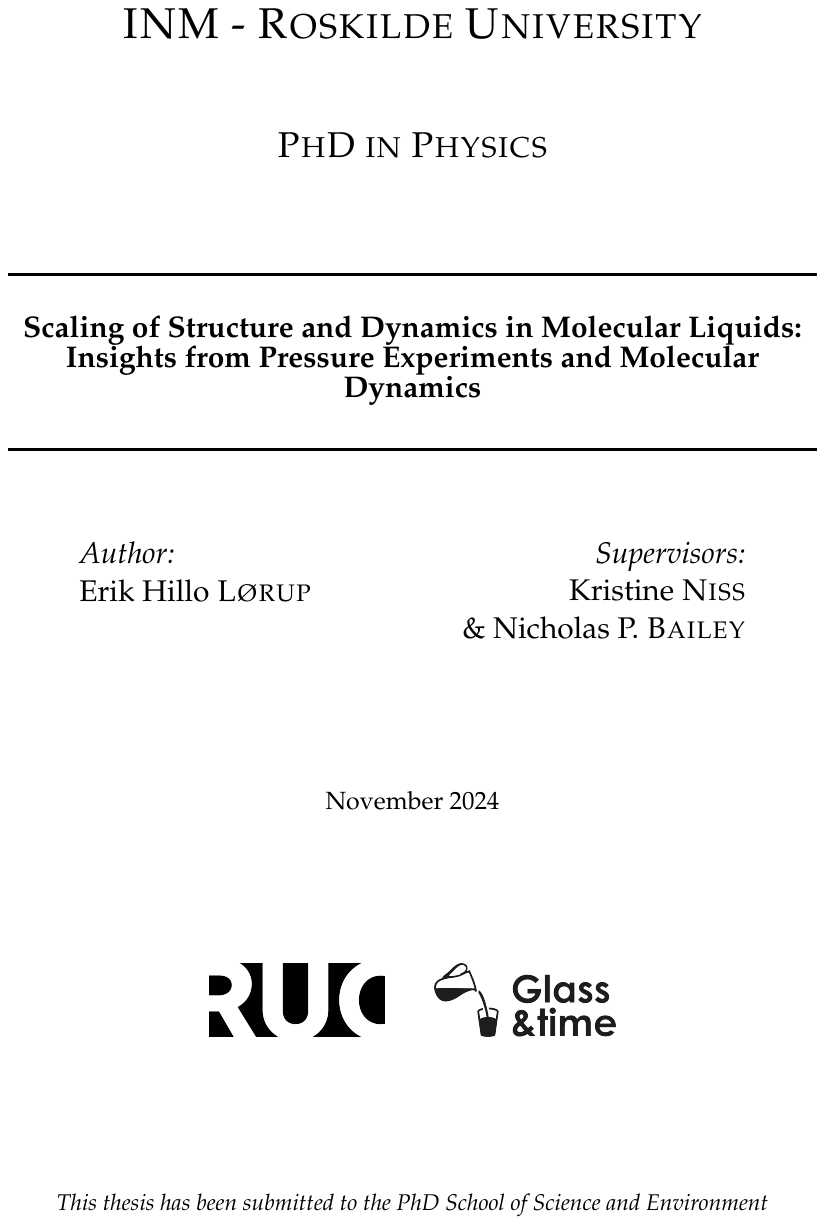}


\begin{abstract}

The overall goal of this thesis is to investigate the connection between the dynamics and structure of molecular glass formers, by testing different scaling laws for both. The inspiration for this work is the Isomorph theory \cite{IsomorphPaper1,IsomorphPaper2,IsomorphPaper3,IsomorphPaper4,IsomorphPaper5,Isomorph_v2,Dyre2014,Dyre2020} because it predicts a connection between structure and dynamics. The fundamental prediction of the Isomorph Theory is that there exist lines in the phase diagram where both structure and dynamics are invariant when presented in reduced units \cite{Dyre2014}. The prediction of constant dynamics has been tested and confirmed experimentally several times \cite{Dyre2014,Wase2018_nature,Wase2018,XiaoWence2015Itpf,}, but the structural prediction has never been confirmed experimentally. In this thesis, we have investigated the structural prediction for liquids where the prediction of constant dynamics has been shown experimentally. In chapter \ref{chapter:Cumene} we examine the van der Waals glass former cumene, which has previously been shown to obey the constant dynamics prediction \cite{Wase2018,Wase2018_nature,Ransom2017}. We present results of an experiment where we measure the first peak of $S(q)$ as a function of both pressure and temperature, and show that the measured structure is also invariant along lines of constant dynamics. In the same chapter, we also present results from MD simulations of cumene, which provide insight into the measured results and what role the intramolecular and intermolecular structure plays for $S(q)$. We conclude that cumene is the first liquid where one can experimentally show that both dynamics and structure are invariant along the same lines in the phase diagram.
In chapter \ref{chapter:Diamond} we extend our investigation of the fundamental prediction of the Isomorph Theory by testing several different glass shapes. We present results for the van der Waals bonded liquid DC704 and the hydrogen bonded liquid DPG, as well as preliminary results for the two van der Waals bonded liquids 5PPE and Squalane. We find that DC704 also has lines with constant dynamics and structure, while the prediction does not work for DPG. 
In chapter \ref{chapter:IDS} we test two new types of scaling of structure and dynamics, isochronal temperature scaling and isochronal density scaling. We examine literature data of the dynamics for cumene, DC704, and DPG, and the structural measurements presented in chapter \ref{chapter:Cumene} and \ref{chapter:Diamond}. We show the connection between density scaling, power-law density scaling, isochronal temperature scaling and isochronal density scaling. We discuss the consequences of isochronal temperature scaling and isochronal density scaling for fragility and how it can be used to examine the origin of breakdowns in power-law density scaling.
\end{abstract}
\renewcommand{\abstractname}{Resume}
\begin{abstract}
	
Det overordnede mål med denne afhandling er at undersøge sammenhængen mellem dynamikken og strukturen af molekylære glas-formere, ved at teste forskellige skaleringslove for begge. Inspiration for denne afhandling har været isomorf teorien\cite{IsomorphPaper1,IsomorphPaper2,IsomorphPaper3,IsomorphPaper4,IsomorphPaper5,Isomorph_v2,Dyre2014,Dyre2020} , fordi den forudsiger en sammenhæng mellem struktur og dynamik. Den grundlæggende forudsigelse i Isomorfteorien er, at der eksisterer linjer i fase-diagrammet, hvor både struktur og dynamik er invariant, når præsenteret i reducerede enheder \cite{Dyre2014}. Forudsigelsen om konstant dynamik er blevet testet og bekræftet eksperimentelt flere gange \cite{Dyre2014,Wase2018_nature,Wase2018,XiaoWence2015Itpf,}. I denne afhandling har vi undersøgt den strukturelle forudsigelse for væsker, hvor forudsigelsen om konstant dynamik er blevet vist eksperimentelt. I kapitel \ref{chapter:Cumene} undersøger vi van der Waals glas-formeren cumene, som tidligere har vist sig at overholde forudsigelsen om konstant dynamik \cite{Wase2018_nature,Wase2018,Ransom2017}. Vi præsenterer resultater fra et højtrykseksperiment, hvor vi måler den første peak af $S(q)$ som funktion af både tryk og temperatur, og viser at den målte struktur også er invariant langs linjer af konstant dynamik. I samme kapitel præsenterer vi også resultater fra MD-simuleringer af cumene, som giver indsigt i de målte resultater og hvad rolle den intramolekylære og intermolekylære struktur spiller for $S(q)$.  Vi konkluderer, at cumene er den første væske, hvor man eksperimentelt kan vise at både dynamik og struktur er invariant langs de samme linjer i fase-diagrammet.
I kapitel \ref{chapter:Diamond} udvider vi vores undersøgelse af den grundlæggende forudsigelse i Isomorfteorien ved at måle strukturen af flere forskellige glas-formere som funktion af tryk og temperatur. Vi præsenterer resultater for den van der Waals-bundne væske DC704 og den hydrogen-bundne væske DPG, samt foreløbige resultater for de to van der Waals bundne væsker 5PPE og Squalane. Vi finder at DC704 også har linjer med konstant dynamik og struktur, mens at forudsigelsen ikke er korrekt for DPG. 
I kapitel \ref{chapter:IDS} tester vi to nye typer skalering af struktur og dynamik, isochronal temperature scaling og isochronal density scaling. Vi undersøger litteratur data for dynamikken af cumene, DC704 og DPG, samt vores målte struktur data fra kapitel \ref{chapter:Cumene} og kapitel \ref{chapter:Diamond}. Vi viser sammenhængen mellem density scaling, power-law density scaling, isochronal temperature scaling og isochronal density scaling. Vi diskuterer hvilke konsekvenser isochronal temperature scaling og isochronal density scaling har for fragilitet, og hvordan det kan bruges til at undersøge grunden til at power-law density scaling bryder sammen.

\end{abstract}

\pagenumbering{Roman}
\chapter*{Acknowledgements}

As I am writing the acknowledgments to my thesis a few days before submission, I am sitting with a strange, ambiguous feeling. While I am happy to be nearing the completion of this thesis, I have been thoroughly enjoying my PhD studies at RUC, and I also feel sad that I will soon have to hand-in my thesis. Perhaps this could very well be some sort of Stockholm syndrome, but I currently have the feeling that it would have been a bigger punishment for Sisyphus to have been asked to leave the stone atop the hill in Tartarus and find something else to do. Especially when, from the top of that hill, there are a great view of many more steep hills it would be interesting to roll a boulder to the top of. The reason I have enjoyed the last three years is due to the many great people at IMFUFA and the Glass and Time group.  There are many people that I want to thank, both for making this thesis possible and for making the last three years enjoyable. A special thanks to Heine Larsen, both for helping with installing hopeless scientific software, but also for being 'the glue guy' of IMFUFA.

First and foremost, I want to thank my two supervisors, Kristine Niss and Nicholas Bailey. I am especially grateful to Kristine for approaching me with the opportunity of doing this project. During my bachelor's and master’s studies at Roskilde University, I never did a project about isomorph theory. I am glad to have the opportunity to mend that mistake. I want to thank Kristine for helping me understand the bigger picture of the work and introducing me to the world of experimental physics. I want to thank Nick for helping me keep an eye on the small details and for helping me navigate RUMD. I want to thank both my supervisors for giving me helpful "kicks in the butt" when needed, especially with the writing.  

Another important person in the creation of this thesis is Lorenzo Costigliola. The idea of testing isochronal density scaling and isochronal temperature scaling on data from molecular data originated from his progress report. In this thesis, Lorenzo is credited many times, and he had a large influence on this thesis.

I want to thank all the persons who have helped me with different beamtimes at large-scale facilities. This includes the local contracts at many beamtimes: Gabriel Cuello at ILL, Veijo Honkimaki and Ola Grendal at the ESRF, Oliver Alderman at ISIS, and Olga Shebanova and Thomas Zinn at DIAMOND. All of the crew joining me at the beamtimes also deserve to be thanked:  Kristine Niss, Bo Jakobsen, Henning Friis Poulsen, David Noriat, Christian Kjeldbjerg,  Søren Strandskov Sørensen and Calvin Carlson. A special thanks to Calvin Carlson for doing all the dirty work at almost all of my beamtimes.

I want to especially thank Oliver Alderman and the disordered materials group at ISIS Neutron and Muon source for hosting me for a four-month visit. Being at large-scale facilities is a truly unique experience and I am glad I got to experience this. The visit started a few projects that sadly did not make it into the thesis, but I still hope to finish them. I want to thank Dominic Fontes for succeeding in solving the crystalline structure of cumene. 

This project is partly funded by the ESS lighthouse SOLID. I would like to thank our SOLID collaborators, Morten Smedskjær and Faizal Jalaludeen, and the director of SOLID, Henning Friis Poulsen, for supporting and showing a genuine interest in this project.  

I want to also thank the RUCSAXS team, Dorthe Posselt, Tim Tejsner, Christian Kjeldbjerg, Bo Jacobsen and Jonathan M. Gow for helping with the measurements of different liquids as a function of temperature, shown in figure \ref{fig:liquid_examples}.  The Novo Nordisk Foundation is gratefully acknowledged for funding RUCSAXS – Roskilde University Interdisciplinary X-ray Scattering Hub – with grant NNF21OC0068491

I have used Trinka AI to proofread part of the thesis. Any corrections suggested by Trinka AI, have I decided either to implement or not. I want to send a special thank you to Ann Lauritzen, Calvin Carlson, Devan LaTray, Lorenzo Costigliola and Marc Dam for helping with the proofreading and corrections.


\clearpage
\tableofcontents

\clearpage

\pagenumbering{arabic}

 \chapter{Introduction}
 
\begin{figure}[H]
	\centering
	\includegraphics[width=0.7\textwidth]{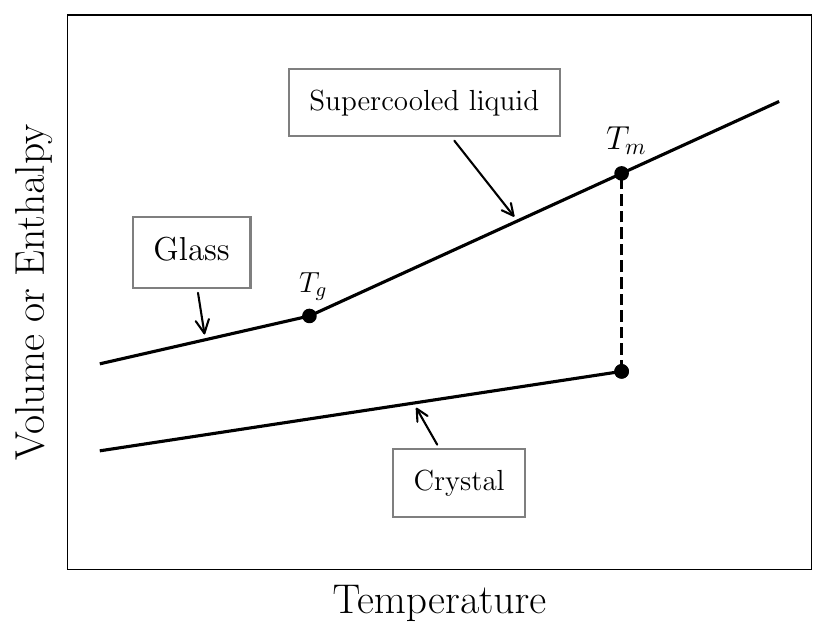}
	\caption{Illustration of the change in volume or enthalpy of a glass-former entering the glass phase by cooling. $T_g$ is the glass transition temperature, and $T_m$ is the melting temperature. When a liquid is cooled past its melting point it can either crystallize or become a supercooled liquid. If there is a phase transition into the crystal phase, there is an abrupt change in volume. If crystallization is avoided, the volume of the supercooled liquid will continue to decrease. When the liquid enters the glass phase, the liquid falls out of equilibrium. The molecules can no longer relax, so the volume changes less as a function of temperature compared to the liquid phase.}
	\label{fig:glass_transition_ex}
\end{figure}

When a liquid is subjected to a perturbation the liquid will relax to return to an equilibrium state \cite{Dyre2006C}. The time it takes the liquid to return to equilibrium is called the \textit{relaxation time}. The timescale of the relaxation time is noted by $\tau_\alpha$. When a liquid is cooled the relaxation time of the liquid increases. If the liquid is cooled fast enough to avoid crystallization the relaxation time will continue to increase. If cooling is continued, at some point the relaxation time of the liquid will be so long, the liquid can no longer reach equilibrium on an experimental time scale, and it falls out of equilibrium. This new phase is called a \textit{glass}, and the transition from a supercooled liquid to a glass, is called the \textit{glass transition}. The glass transition is characterized by the molecules freezing in; they can no longer adjust to the new temperature within an experimental time scale \cite{Hecksher2023}. The glass transition is dependent on the cooling rate, since the glass transition depends on the time scale of the cooling and the relaxation time. If it is possible to cool very fast the \textit{glass transition temperature}, $T_g$, is  higher compared to a slower cooling rate. A practical definition of the glass transition is $\tau_{\alpha} \approx 100 $ s. It is worth stressing that this relaxation time is the time scale of molecular displacement and/or re-orientations - the time scales of the bulk flow are much larger. An illustration of the change in volume of a glass-former entering the glass phase by cooling is shown in figure \ref{fig:glass_transition_ex}.
An example of how flow time scales are even longer is the famous pitch drop experiment, where a funnel full of bitumen slowly drips \cite{Pitchdrop}. In approximately 100 years a droplet has dropped nine times, and the world waits eagerly for the next one. Bitumen is at room temperature a super cooled liquid, that has an approximate relaxation time in the range of $10^{-4}$ s - $10^{-3}$ s \cite{Pitchdrop}. This is still many orders of magnitude in relaxation time from the glass transition. 

The relaxation time describes the time it takes for the liquid to return to equilibrium via molecular displacement and re-orientations.  Part of the explanation for the large change in relaxation time can be understood in terms of activation energy. If the dynamics are described by thermal fluctuation overcoming energy barriers, the relaxation time would be described by an Arrhenius law \cite{Dyre2006C}. However almost every glass-former deviates from the Arrhenius description of relaxation time and the relaxation time has a stronger temperature dependence, that can not alone be described by an Arrhenius law\cite{Dyre2006C,Angell1995}. The \textit{fragility} describes how much a liquid deviates from the Arrhenius description of relaxation time. These concepts are formally introduced in chapter \ref{chapter:dynamics}. The origin of this deviation from the Arrhenius description of relaxation time is unknown.   Recent studies have found an empirical connection between the fragility of different glass-formers and the structure, either in real space or reciprocal space \cite{Ryu2019,Ryu2020,Voylov2016,Wei2015}. The overall goal of this thesis is to investigate the connection between the dynamics and structure of molecular glass formers, by testing different scaling laws for both. Part of the motivation for examining this connection is the proposed structural origin for fragility. We will not examine this question directly, but we return to the question in the final discussion of the thesis.

\begin{figure}[h]
	\centering
	\begin{subfigure}[b] {0.495\textwidth}
		\includegraphics[width=0.95\textwidth]{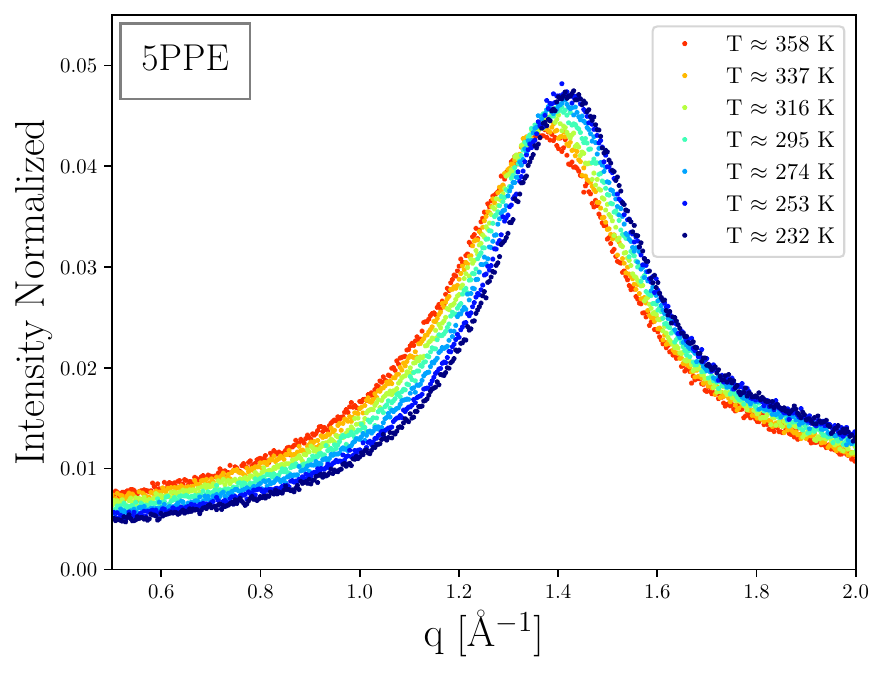}
		\caption{5PPE, $T_g = 245$ $^o$K \cite{Wase2018}}
		\label{fig:liquid_examples_a}
	\end{subfigure} \hfill
	\begin{subfigure}[b] {0.495\textwidth}
		\includegraphics[width=0.95\textwidth]{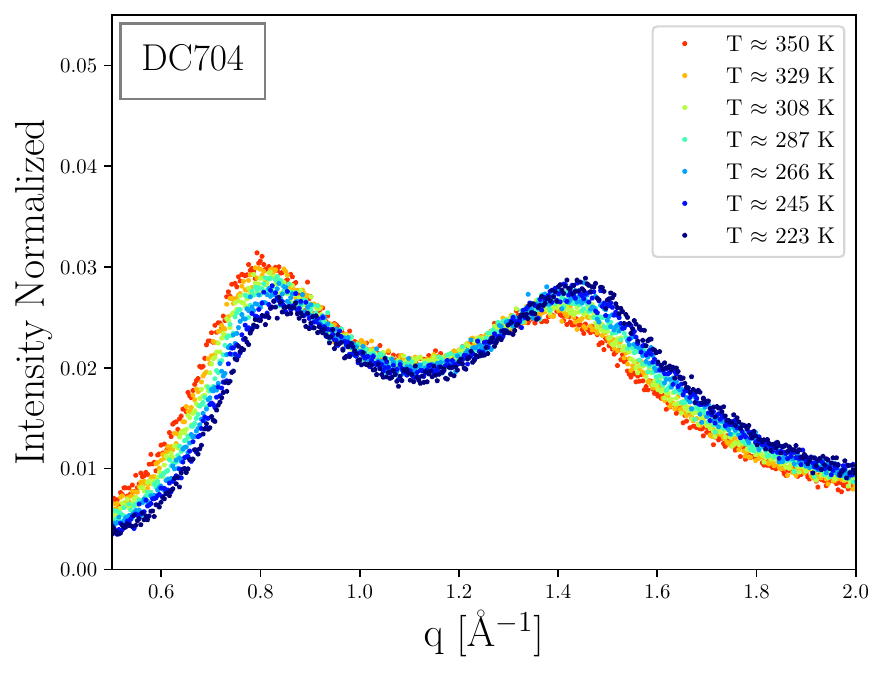}	
		\caption{DC704, $T_g = 210$ $^o$K \cite{HecksherTina2013Msog}}	
		\label{fig:liquid_examples_b}
	\end{subfigure}
	\begin{subfigure}[b] {0.495\textwidth}
		\includegraphics[width=0.95\textwidth]{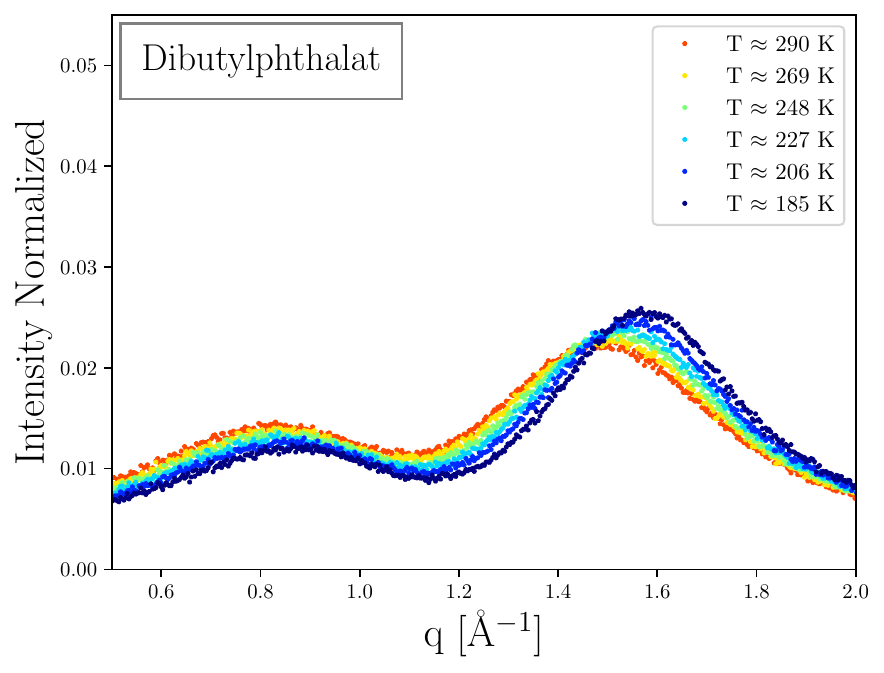}	
		\caption{Dibutylphthalat, $T_g = 178$ $^o$K \cite{Ruffer2008} }	
		\label{fig:liquid_examples_c}
	\end{subfigure}	
	\begin{subfigure}[b] {0.495\textwidth}
		\includegraphics[width=0.95\textwidth]{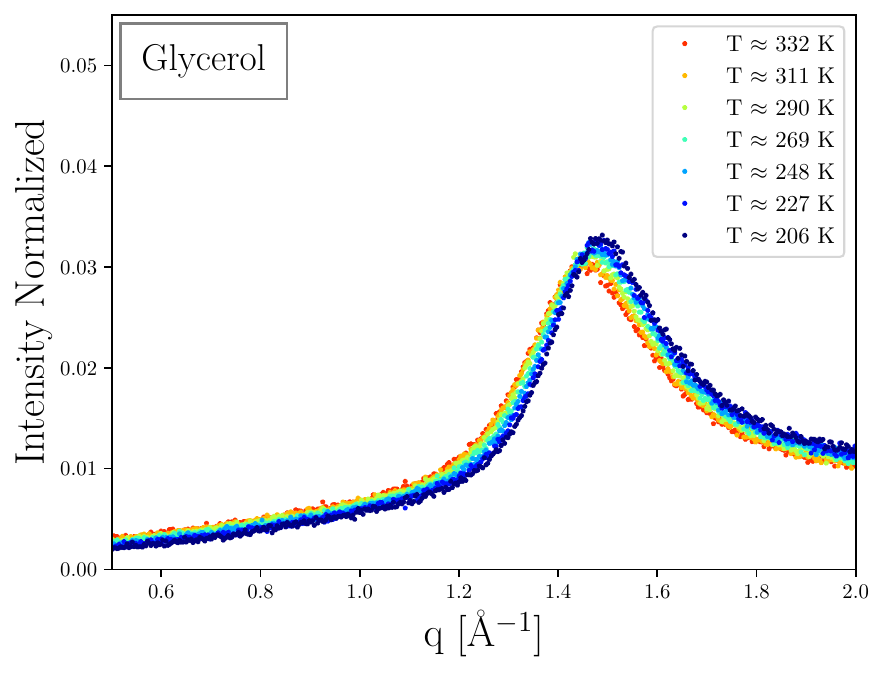}	
		\caption{Glycerol, $T_g = 188$ $^o$K \cite{Roland2005} }	
		\label{fig:liquid_examples_d}
	\end{subfigure}
	\begin{subfigure}[b] {0.495\textwidth}
		\includegraphics[width=0.95\textwidth]{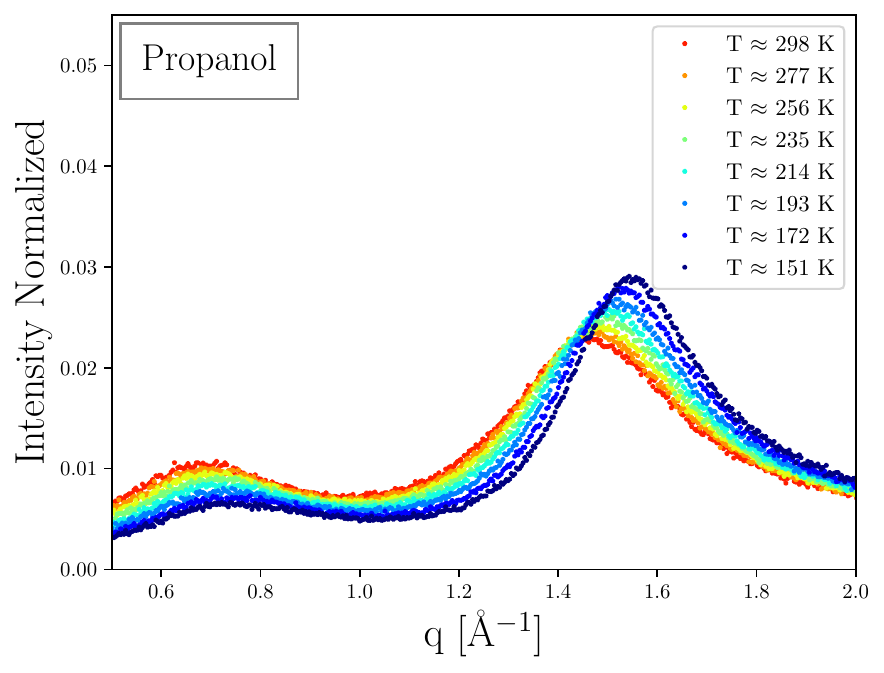}
		\caption{Propanol, $T_g = 98$ $^o$K \cite{Ramos2011} }
		\label{fig:liquid_examples_e}
	\end{subfigure} \hfill
	\begin{subfigure}[b] {0.495\textwidth}
		\includegraphics[width=0.95\textwidth]{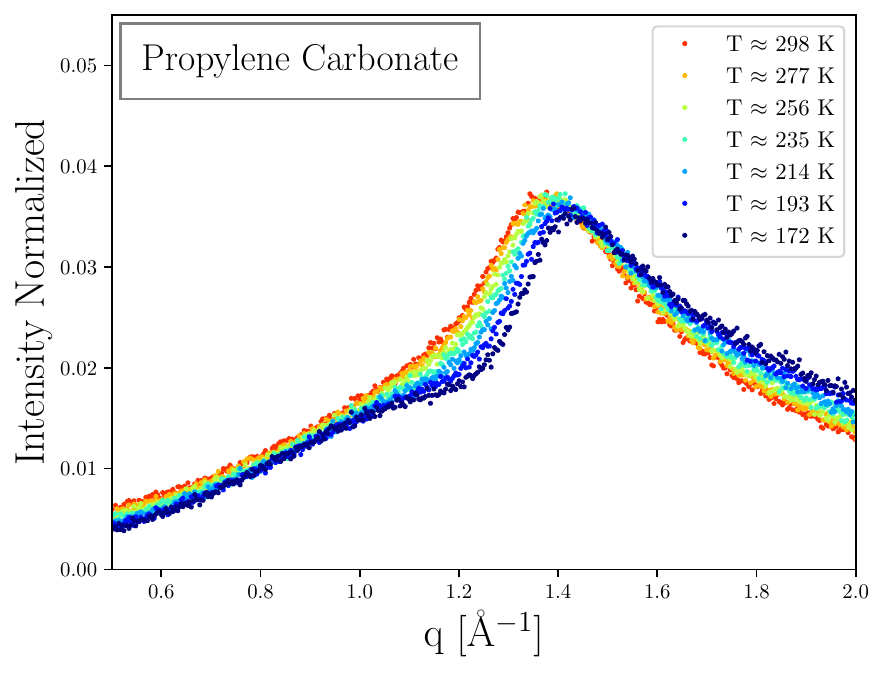}	
		\caption{Propylene carbonate, $T_g = 159$ $^o$K \cite{Casalini2005} }	
		\label{fig:liquid_examples_f}
	\end{subfigure}
	
	\caption{ Structural changes in the main structure peak of six molecular glass-formers measured by x-ray diffraction. The liquids are measured at RUCSAXS\small\textsuperscript{a}. The color scheme is relative to each liquid, but represents the temperature, where red is the highest temperature and blue is the lowest.   }
	\small\textsuperscript{a} Special thanks to RUCSAXS – Roskilde University Interdisciplinary X-ray Scattering Hub for helping with the measurements. 
	\label{fig:liquid_examples}
\end{figure}

Liquids and glasses are characterized by the lack of any long-range repeating structure. The structure of liquids can therefore seem like a counterintuitive notion.  Liquids and glasses contain short-range structures. This can originate from the fact that atoms cannot overlap and they repulse each other due to Pauli repulsion. This creates a preferred distance for nearest-neighbor atoms. Atoms can form molecules, and the  intramolecular structure of the molecule is also a structure in the liquid. Molecules also displace each other creating a preferred distance. Depending on the molecule, different intermolecular forces can also create structures in the liquid. Ionic liquids are a class of liquids, containing a positively charged cation and a negatively charged anion. The positive cation prefers to be closer to the negative anion. The introduction of positively and negatively charged molecules in the liquid can create longer range structure in the liquid \cite{Ionic_liquid_structure}. Hydrogen-bonded liquids are also known to have longer range structures \cite{Bolle2020,SidebottomDavid}. Experimentally the structure is normally measured by either x-ray or neutron diffraction.  In figure \ref{fig:liquid_examples}, the first structure peak, measured by x-ray diffraction for several molecular glass-forming liquids, is plotted as a function of temperature. How to measure structure experimentally is introduced formally in chapter \ref{chapter:Structure}. In this thesis we will focus on measures of two-particle structures such as the pair distribution function, $g(r)$, and the static structure factor, $S(q)$. More complex structural measures that try to capture the three dimensional structure of a liquid exists, but are beyond the scope of this thesis. To calculate measures such as the spatial density function \cite{Headen2019}, and four point correlation function \cite{Kob2020}, one need either simulations of the system or to conduct Reverse Monte Carlo simulations based on experimental data.

If there is a structural origin for the deviation from the Arrhenius description of relaxation time, as suggested by ref.\cite{Ryu2019,Ryu2020,Voylov2016,Wei2015}, then are structure and dynamics connected? The isomorph theory predicts the existence of \textit{isomorphs}, which are lines in the phase diagram where both structure and dynamics are invariant when presented in reduced units \cite{IsomorphPaper4,Dyre2014}. The underlying concepts will be introduced in greater detail in chapter \ref{chapter:dynamics}, but this is often called the fundamental prediction of isomorph theory. The prediction of invariant dynamics for molecular glass-formers has been shown to hold true in both experiments  \cite{GundermannDitte2011Ptde,Wase2018,Wase2018_nature,XiaoWence2015Itpf,} and in simulations \cite{Dyre2014,Ingebrigtsen2012,Veldhorst2014}. The structural prediction has been shown to be true in simulations \cite{Ingebrigtsen2012,Dyre2014,Stoppelman2022}, but there has been no systematic experimental study testing the structural prediction of isomorph theory. The reason for this is quite simple. If we want to show that the structure and the dynamics are invariant along an isomorph in the phase diagram, for the result to be interesting and nontrivial, we really have to show that the structure and dynamics change when moving away from that isomorph. When we compare the factor $\approx 10^{15}$ change in relaxation time, with the structural change shown in figure \ref{fig:liquid_examples}, it is understandable that previous studies have focused on the dynamical prediction rather than the structural prediction. 
 
One of the main goals of this thesis is to test the structural prediction of Isomorph theory. If the fundamental prediction of isomorph theory would be proven true, it would be of interest to many open areas of research. Mode-coupling theory is one of many areas in which this would be of interest. The reason for this is that mode-coupling theory tries to predict the relaxation dynamics of glass-forming materials directly from static, time-independent properties \cite{Janssen2018}. Another way of phrasing this is that the microscopic structure of a liquid predicts the dynamics of the liquid. If the structure predicts the dynamics of a liquid then it follows that both structure and dynamics should stay invariant along the same lines in phase diagram.

If there exists molecular liquids that obey the fundamental prediction of isomorph theory, then their structure and dynamics would follow in lockstep when moving around the phase diagram. The isomorph theory does not make any predictions about the change in structure or dynamics when moving between isomorphs, only that they are invariant along the same isomorph. If there is a structural connection between structure and fragility, molecular liquids that obey the isomorph theory is an obvious class of liquids to study. 

\section{Structure of the thesis}

This thesis contains two background chapters (chapters \ref{chapter:dynamics}, and \ref{chapter:Structure}), three results chapters (chapters \ref{chapter:Cumene}, \ref{chapter:Diamond}, and \ref{chapter:IDS}), and a concluding discussion (chapter \ref{chapter:discussion}). One of the overall goals of this thesis is to show the existence of \textit{pseudo-isomorphs}: Lines in the phase diagram where both the structure and dynamics are invariant when presented in reduced units.


Chapter \ref{chapter:dynamics} includes a general introduction to some key concepts relevant to the presented work. This includes the concepts of fragility and the scaling behavior of dynamics in liquids. The isomorph theory is also introduced along with how density scaling can be deduced from the isomorph theory. The new scaling concepts, \textit{isochronal density scaling} and \textit{isochronal temperature scaling}, are also introduced as well as how they can be derived from the isomorph theory.

In chapter \ref{chapter:Structure} a general introduction to structure in liquids is given. This includes the many relevant distribution functions, the relationship between them and how they can be measured experimentally. 

In chapter \ref{chapter:Cumene} we present the results of the first systematic test of the fundamental prediction of isomorph theory. We present high pressure structural measurements of cumene, a relatively good glass-former. The dynamical prediction has been shown to work for cumene to a high degree. The experimental data is supported by united atom molecular dynamics simulations of cumene. It is shown that for cumene both structure and dynamics are invariant along the same line in the phase diagram. This is the first experimental evidence of the existence of pseudo-isomorphs. 

In chapter \ref{chapter:Diamond} the search for pseudo-isomorphic liquids is extended to include several other glass-formers. We present measurements of the structure of the van der Waal liquid, DC704, and the hydrogen bonded liquid, DPG, as a function of pressure and temperature. Preliminary results for two other van der Waal liquids, 5PPE and squalane, are also presented. The van der Waal liquid, DC704, is shown to have pseudo-isomorphs. The hydrogen bonded liquid, DPG, is shown to not have lines of invariant structure and dynamics, in line with the predictions of isomorph theory\cite{Ingebrigtsen2012_simple_liquid}. 

In chapter \ref{chapter:IDS} we test two new scaling relations for the relaxation time: Isochronal density scaling and isochronal temperature scaling. We test the two new scaling relations on literature data of the dynamics for cumene, DC704, and DPG. The structural measurements presented in chapters \ref{chapter:Cumene} and \ref{chapter:Diamond} are reexamined with isochronal density scaling and isochronal temperature scaling. How isochronal density scaling and isochronal temperature scaling can be used to examine the breakdown of density scaling is also discussed.

In the final chapter of the thesis, chapter \ref{chapter:discussion}, the result of this thesis is discussed and further work is suggested. The consequences of the existence of pseudo-isomorphs is discussed in the context of fragility. The two new scaling relations, isochronal density scaling and isochronal temperature scaling, and they predictions for fragility is also discussed.
 

	\chapter{Dynamics and scaling of dynamics in glass-forming liquids.} \label{chapter:dynamics}
	
	In this section we will give a short introduction to a few of the central concepts needed to understand the work presented. The one of the overall goal of the thesis is to test for structural and dynamical invariance along the same lines in the phase diagram. This chapter will focus on scaling of dynamical properties in liquids, by introducing density scaling, power-law density scaling and in section \ref{sec:IDS_intro} we introduce the new concepts of isochronal density scaling and isochronal temperature scaling. The isomorph theory is introduced, along with the key concept of pseudo-isomorphs. The origin of the scaling behavior of dymamics is explained form the Isomorph theory
	
	\section{Fragility}
\begin{figure}[H]
\centering
\includegraphics[width=0.65\textwidth]{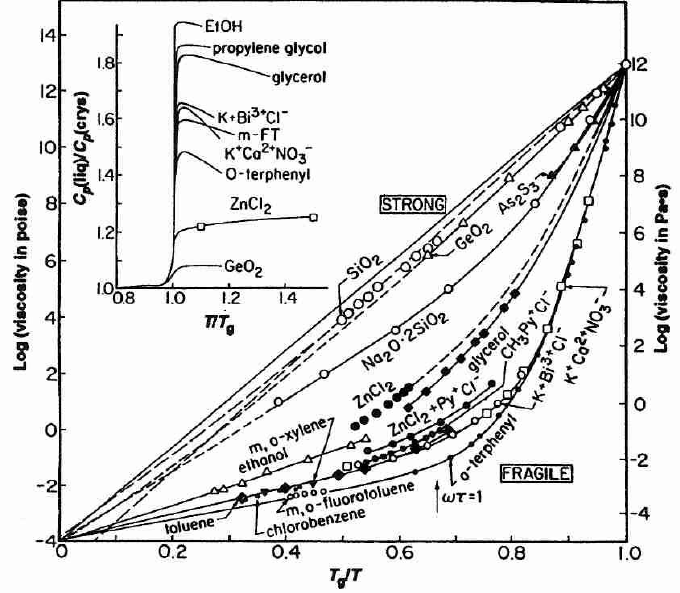}
\caption{Reprint of the Angell plot, originally published in ref. \cite{Angell1995}. The figure shows $T_g/T$ for several different glass formers on the x-axis. On the y-axis the viscosity is plotted on a $\log_{10}$ scale. An Arrhenius behavior of the relaxation time will look like a straight line in the Angell plot. Glass forming liquids with this behavior are classified as strong. Glass forming liquids that deviate from this, by showing non-linear, "super-Arrhenius" behavior are called fragile.}
\label{fig:Angell_Plot}
	\end{figure}

	In the introduction we described in words, the change in dynamics that happens for a liquid entering the glass. The relaxation time was defined as the time it takes the system to return to equilibrium. From entering the supercooled regime to the glass the relaxation time changes many orders of magnitude. In the introduction we describe how an Arrhenius description of the relaxation process, can partly explain the large change in relaxation time. However most glass-formers deviates from an Arrhenius behavior. A model to describe the relaxation process is as energy barriers to be overcome by thermal fluctuations \cite{Dyre2006C}. We let the dependent activation energy be state point dependent to account for the deviations from an Arrhenius behavior. In this picture the relaxation time or viscosity is given by
	
	\begin{equation}
\tau_{\alpha} (\rho,T) = \tau_0 \exp\left(\frac{\Delta E(\rho,T)}{k_B T}\right) \quad \quad \eta(\rho,T) = \eta_0 \exp\left( \frac{\Delta E(\rho,T)}{k_B T}\right)
	\end{equation}
	
	where $\Delta E(\rho,T)$ is the state point dependent activation energy. In figure \ref{fig:Angell_Plot} the iconic Angell plot is reprinted from ref. \cite{Angell1995} is reprinted. The figure shows $\log_{10}(\tau_\alpha)$ plotted against $T_g/T$. In the Angell plot an Arrhenius behavior of the relaxation time will be a straight line. An Arrhenius behavior corresponds to a state point-independent activation energy, $\Delta E(\rho,T) \approx \Delta E_0$. Liquids such as these are often called strong. Most of the glass formers shown in figure \ref{fig:Angell_Plot} deviate from the Arrhenius behavior. Liquids that deviate from the Arrhenius behavior are often called fragile. There are many ways to characterize the fragility of a liquid, the most common being the steepness index:

	\begin{equation} \label{eq:mp}
		m_P = \left.\frac{\partial \log_{10}(\tau_{\alpha})}{\partial T_g/T}\right|_{T=T_g , P = amb}
	\end{equation}

	 $m_P$ classifies the change in relaxation time as a function of temperature. Strong liquids that deviate very little from the Arrhenius behavior of the relaxation time, will have a smaller $m_P$-value, while fragile liquids will have a larger $m_P$.
	
\begin{figure}[H]
		\centering
		\includegraphics[trim = 30mm 80mm 15mm 80mm, clip,width=0.8\textwidth]{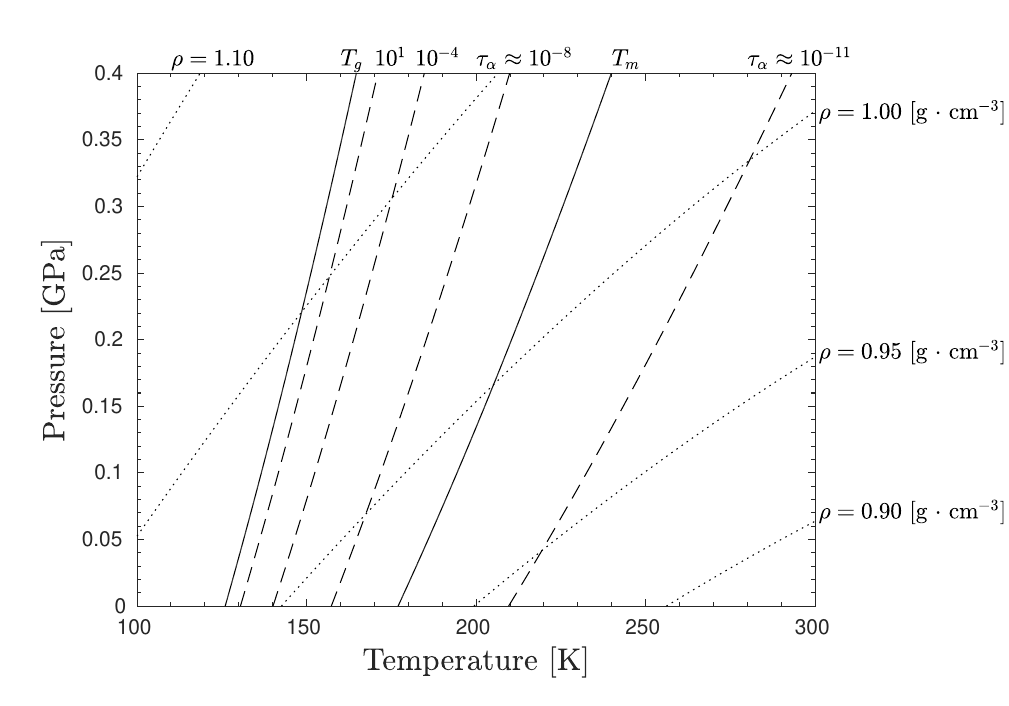}	
		\caption{The T-P phase diagram of cumene, with isochores and isochrones determined using the equation of state from \Mycite{Ransom2017}; further details are given in chapter \ref{chapter:Cumene}. The solid lines represent the glass transition and approximate melting line. The dotted lines are isochores, and the dashed lines are "isochrones", lines of constant relaxation time. In this thesis we often refer to the phase diagram, in order to know which isochrone a state point is placed. This means that we will often refer to the phase diagram, while we are only in the liquid phase. }
		\label{fig:phase_diagram_cumene_chatper_2}
	\end{figure}

The most common definition of fragility, defined in equation \ref{eq:mp}, measures the change in relaxation time as a function temperature along the ambient pressure isobar. It is clear that the relaxation time is state point dependent. On a similar note examining any phase diagram makes it clear that the change in relaxation time is dependent on more than temperature. In figure \ref{fig:phase_diagram_cumene_chatper_2} an example for an phase diagram is shown. In this thesis we will often refer to either the T-P phase diagram or the $\rho-T$ phase diagram, even though we are only looking at one phase, the liquid phase. While this is technically a zoom-in on the liquid phase of the phase diagram, this is often just called the phase diagram due to a lack of a better term.
	
	\subsubsection{Isochoric fragility}
	
	The definition of fragility from equation \ref{eq:mp} measures the change in relaxation time along the ambient pressure isobar. Experimentally this is the easiest to measure, since the only experimental variable we need to change is the temperature. However if one examines the phase diagram in figure \ref{fig:phase_diagram_cumene_chatper_2} the change in relaxation is clearly dependent more than just temperature. An obvious approach is to isolate the temperature dependence of the relaxation time, by examining the change in relaxation along an isochore. The isochoric fragility is defined as
	
	\begin{equation}
		m_\rho = \left.\frac{\text{d} \log_{10}(\tau_{\alpha})}{\text{d} T_g/T}\right|_{T=T_g, \rho = \rho_0}
	\end{equation}
	
	There is a direct relationship between $m_P$ and $m_\rho$, via the isobaric and the isochronic thermal expansivities \cite{Niss2007}
	
	\begin{equation}
 m_p = m_\rho \left(1 -\frac{\alpha_\rho}{\alpha_\tau}\right)
	\end{equation}
	
	where $\alpha_P = \left.-\frac{1}{\rho} \frac{\partial \rho}{\partial T} \right|_P$, is the isobaric expansivity, and $\alpha_\tau = \left.-\frac{1}{\rho} \frac{\partial \rho}{\partial T} \right|_\tau$ is the isochronic expansivity. \footnote{In section \ref{sec:isomorph_theory}, we define the important quantities $\gamma (\rho,T)$ in equation \ref{eq:gamma_def_2}. $\gamma(\rho,T)$ is related to isochronic expansivity by  $\gamma (\rho,T) = - \frac{1}{T \alpha_\tau}$. }. 
	\Mycite{Roland2005} found an empirical relation between  $m_P$ and $m_\rho$, based on data for around 20 small molecules and polymers
	
	\begin{equation}
m_P =(37 \pm 3) + (0.84 \pm 0.05) m_\rho
	\end{equation}
	
In general $m_\rho$ is smaller than the $m_P$ \cite{Roland2005}.
	
	\section{Density scaling}
	
	\textit{Density scaling} started as an empirical observation about the behavior of the relaxation time as function temperature and density. The temperature and density dependence of the relaxation time of a glass-former can be described as a function dependent on a single scaling variable \cite{Alba-Simionesco2002,Alba-Simionesco2004}:
	
	\begin{equation} \label{eq:density_scaling}
		\tau_\alpha(\rho,T) = f\left(\frac{e(\rho)}{T}\right) 
	\end{equation}
	
	where $e(\rho)$ is some function, only dependent on density. Throughout the thesis we will often refer to something being able to scale the data, or something collapsing the data. When we refer to data collapsing, we refer to several different measurements \textit{collapsing} into a single curve, dependent on only a single quantity, e.g. density scaling predicts that the relaxation time collapse into a single curve only dependent on $e(\rho)/T$.	This scaling relation is often not limited to the relaxation time, many "transport coefficients" like the viscosity and the diffusive coefficient can also be scaled. The shape of $e(\rho)$ have been controversial. Practically, $e(\rho)$ can often be assumed to be a power-law, which have been successful in scaling a number of molecular-glass formers, like van der Waals liquids \cite{Ransom2017,RolandCM2006Tsot}, Hydrogen-bonded liquids \cite{Casalini2004}, ionic liquids \cite{Lundin2021,Wase2020} and polymers\cite{Alba-Simionesco2004}. For a liquid where $e(\rho)$ is assumed to be a power-law, is in this thesis referred explicitly as \textit{power-law density scaling}. When we are referring to power-law density scaling, the scaling varible is referred as $\Gamma$:
	
	\begin{equation}\label{eq:powerlaw_density_scaling}
		\Gamma = \frac{\rho^\gamma}{T}
	\end{equation} 
	
\begin{figure}[H]
	\begin{subfigure}[t]{0.48\textwidth}
	\centering
	\includegraphics[width=0.95\textwidth]{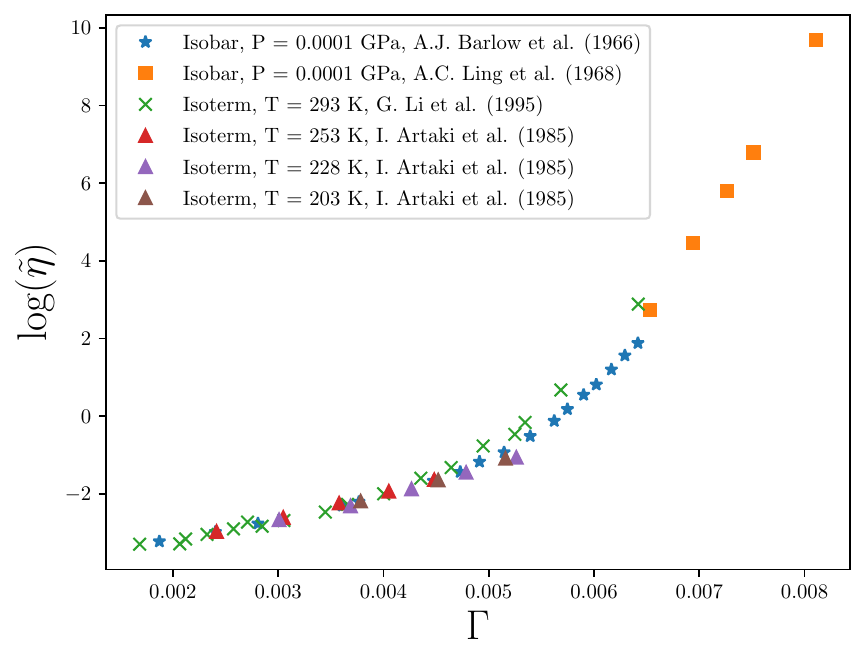}
	\caption{Cumene, $\gamma =4.8$}
	\label{fig:Cumene_dym_DS_chapter2}
		\end{subfigure}
			\begin{subfigure}[t]{0.48\textwidth}
		\centering
		\includegraphics[width=0.95\textwidth]{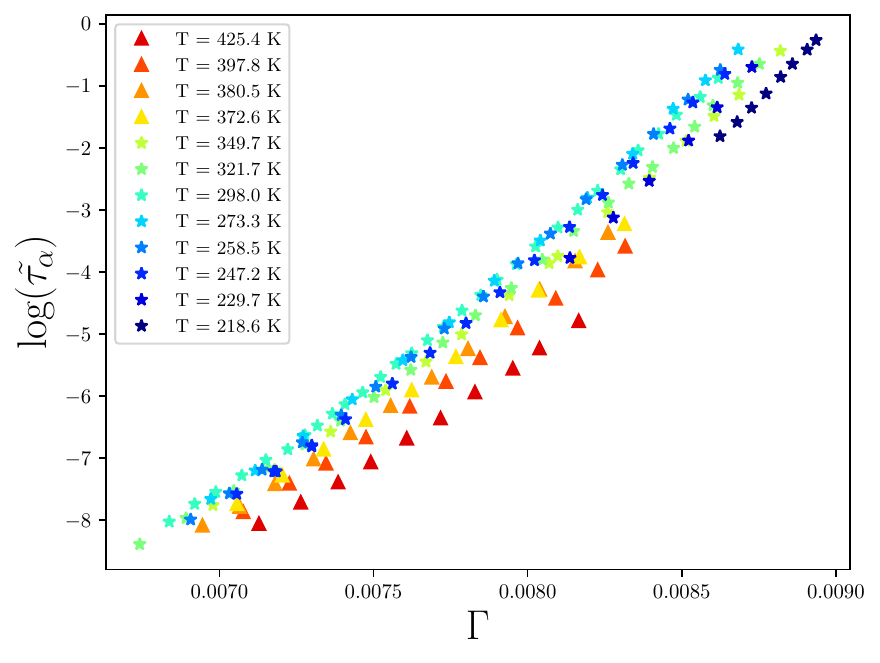}
		\caption{DC704, $\gamma=5.13$}
		\label{fig:Ransom_gamma_513_chapter2}
	\end{subfigure}
	
	\caption{Example of power-law density scaling and the break-down of power-law density scaling. Figure \ref{fig:Cumene_dym_DS_chapter2} is a recreation of figure 5 from \Mycite{Ransom2017}. For cumene literature measurement of the viscosity of cumene can be collapsed into a single curve, by power-law density scaling. The viscosity is plotted in reduced units, $\tilde{\eta}$ = $\eta \rho^{-\frac{2}{3}} T^{-\frac{1}{2}}$ and $\Gamma= \frac{\rho^\gamma}{T}$, where $\gamma=4.77$.
	Figure \ref{fig:Ransom_gamma_513_chapter2} shows a recreation of figure 3 from \Mycite{Ransom_DC704}. They demonstrated that power-law density scaling breakd down for DC704. The relaxation time is plotted in reduced units, $\tilde{\tau_\alpha} = \tau_\alpha \rho^{\frac{1}{3}} T^{-\frac{1}{2}}$, and $\Gamma= \frac{\rho^\gamma}{T}$, where $\gamma=5.13$. Both the data and the figure are discussed in further detail in chapter \ref{chapter:IDS} }
	\label{fig:Faxe_John}
\end{figure}

	Where $\gamma$ is a material specific constant. Reported values of $\gamma$ varies from 0.1 for Sorbitol \cite{Roland2005} to 8.5 for polychlorinated biphenyl with chlorine content 62 \% \cite{Voylov2016}. References \cite{Roland2005,Voylov2016} both contain an overview of reported $\gamma$. Power-law density scaling has been reported to break-down for several glass-formers, like van der Waals liquids such as, dibutyl phthalate \cite{Bøhling2012}, decahydroisoquinoline\cite{Bøhling2012} and DC704 \cite{Ransom_DC704}, as well for hydrogen bonded liquids, such as dipropylene glycol\cite{AdrjanowiczKarolina2017PNDo}. In figure \ref{fig:Faxe_John} an example from the literature of both power-law density scaling and the breakdown of power-law density scaling is shown.
	
The origin of this scaling behavior can be explained via isomorph theory. Both density scaling and power-law density scaling make assumptions about the shape of the isochrones in the $\rho,T$ phase diagram. Power-law density scaling assumes that the shape of the isochrone is a power-law. Density scaling does not make any assumptions about the functional form of the isochrone, expect that it is independent of temperature. In section \ref{sec:IDS_intro} the deviation of power-law density scaling and density scaling from isomorph theory will be shown explicitly.

\section{Isomorph theory} \label{sec:isomorph_theory}

The goal of this section is to give an introduction to the isomorph theory and its extension from model liquids into real-world liquids. The structure of the section will be a short introduction to the derivation of isomorph theory, ending with the fundamental prediction of isomorph theory. After this short introduction we will focus on the extension from model liquids into real-world liquids.  

The isomorph theory was first presented in a series of papers by the Glass and Time group at Roskilde University \cite{IsomorphPaper1,IsomorphPaper2,IsomorphPaper3,IsomorphPaper4,IsomorphPaper5}, and later in a revised form \cite{Isomorph_v2}. The overall idea with isomorph theory was not to create an all-encompassing theory to describe the liquid state. The idea is to describe a subclass of liquids with the characteristic that there are strong correlation between the potential energy and the virial. Any all-encompassing theory must also be able describe this simple class of liquids. This class of liquids are often called "R-simple" or "Roskilde-simple" liquids.  The structure of this section will be to introduce "Roskilde-simple" liquids and the fundamental prediction of isomorph theory.


As stated in the introduction Roskilde simple liquids are a class of liquids with correlation between the potential energy and the virial. To reach this definition we start by introduce the virial. The virial, W, quantifies deviation from the ideal gas, where the deviation are the contribution to the pressure coming from intermolecular interactions \cite{Dyre2014}:

\begin{equation}
	PV =Nk_bT +W
\end{equation}

The microscopic definition of the virial is given by

\begin{equation}
	W(\vec{R}) \equiv \frac{1}{3} \sum_{i}^{N} \vec{r_i}\vec{F_i} = - \frac{1}{3} \vec{R} \nabla U(\vec{R}) 
\end{equation}

Here $\vec{R}$ refers to the position of $N$ particles, $\vec{R} = \left(\vec{r_1}, \vec{r_2}, \cdots, \vec{r_N}\right)$. The virial and potential energy fluctuate in time at a given state point. The time dependent potential energy can be described as $U(t) = \left<U\right> + \Delta U \Rightarrow \Delta U = U(t) -\left<U\right>$, where $\left<U\right>$ is the time averaged potential energy at a given state point, and $\Delta U$ is the instantaneous potential energy fluctuation. The correlation coefficient between instantaneous fluctuation of  the virial and the potential energy is defined as:

\begin{equation}
	R \equiv \frac{\left<\Delta U \Delta W \right>}{\sqrt{\left<(\Delta U)^2 \right>}\sqrt{\left<(\Delta W)^2 \right>} }
\end{equation}

A liquid is defined as Roskilde simple if $R>0.9$. This definition is an arbitrary choice of cut-off, but can be used as an indicator of how strongly the potential energy and the virial correlate. \Mycite{Veldhorst2014} studied a system of flexible Lennard-Jones chains with 10 rigid segments. They found that $R<0.9$, but close to 0.9, and they found that many of the predictions of isomorph theory was true anyway.  For these R-simple systems with strong correlations between the potential energi and virial, it makes sense to describe the correlation between the potential energy and virial:
\begin{equation}\label{eq:Mads_P}
	\Delta U \approx \gamma \Delta W
\end{equation}

If the pair-wise particles interaction are governed by a inverse power-law potential, $U(r) \propto r^{-n}$, equation \ref{eq:Mads_P} is exact and $\gamma = \frac{n}{3}$. For many other model liquids governed by other potentials, the value of $\gamma$ is often state point dependent and defined as \cite{Dyre2020}:

\begin{equation}\label{eg:gamma}
	\gamma(\rho,T) \equiv \left.\frac{\partial \ln(T)}{\partial \ln(\rho)} \right|_{S_{ex}}
\end{equation} 

Where $S_{ex}$ referes to the excess entropy. The excess entropy is defined as the difference between the entropy at the current state point, and the ideal gas at the same state point:

\begin{equation}
	S_{ex} = S(\rho,T) - S_{id}(\rho,T)
\end{equation}

The excess entropy is zero for the ideal gas and negative in the liquid state since the disorder is larger in the gas. It can be shown that the definition of $\gamma$ in equation \ref{eg:gamma} can be calculated directly from slope of the potential energy and virial fluctuations  \cite{Dyre2020}:

\begin{equation}
\left.\frac{\partial \ln(T)}{\partial \ln(\rho)} \right|_{S_{ex}} = \frac{\left<\Delta U \Delta W\right>}{\left<(\Delta U)^2\right>} 
\end{equation}

The isomorph theory describes properties of the class of R-simple liquids.

\subsubsection{Predictions for R-simple systems}

Roskilde simple systems have \textit{hidden scale invariance}, meaning that for two configurations at a single state point, $\vec{R_a}$ and $\vec{R_b}$ there exists an uniform scaling conserving the ordering of the potential energies of the configurations:

\begin{equation}\label{eq:Hidden_scale}
	U\left(\vec{R_a}\right) < U\left(\vec{R_b}\right) \Rightarrow U\left(\lambda \vec{R_a}\right) < U\left(\lambda \vec{R_b}\right)
\end{equation}

 The microscopic excess entropy function can be defined as function of density and the potential energy of the configuration \cite{Dyre2020}.  

\begin{equation}
	S_{ex} (\vec{R}) \equiv S_{ex}\left(\rho, U(\vec{R})\right)
\end{equation}

The equation can be inverted to express the potential energy as a function of density and excess entropy:

\begin{equation}
	U(\vec{R})  = \left(\rho, S_{ex}(\vec{R})\right)
\end{equation}

By evaluating the microscopic $S_{ex}$ of two configurations at two different densities, it is possible to show, that for R-simple systems obeying hidden scale invariance (eq. \ref{eq:Hidden_scale}), the excess entropy is only dependent on the configuration in reduced units \cite{Isomorph_v2,Dyre2020}.

\begin{equation}
	S_{ex}(\vec{R}) = S_{ex}\left(\tilde{R}\right)
\end{equation}

Where $\tilde{R} = \rho^\frac{1}{3} \vec{R}$. The potential energy can be expressed as a function of density and the excess entropy of the configuration in reduced units

\begin{equation}\label{eq:pot_U}
	U(\vec{R})  = \left(\rho, S_{ex}(\tilde{R})\right)
\end{equation}

The reduced unit set used in isomorph theory derives from three fundamental quantities, length, energy and time \cite{Knudsen2024,IsomorphPaper4}. A quantity given in reduced unit is denoted by a tilde.

\begin{align*}
	\tilde{l} &= l / l_0 \quad \text{where} \quad l_0 =\rho^{-\frac{1}{3}}\\
	\tilde{E} &= E / E_0 \quad \text{where} \quad E_0 = k_bT \\
	\tilde{t} &= t / t_0 \quad \text{where} \quad t_0 = \rho^{-\frac{1}{3}} \left(\frac{m}{k_bT}\right)^{\frac{1}{2} }
\end{align*}

The reduced units that are used in this thesis, the assumption $k_b = m =1$ is made. The quantites that are presented this thesis, are the reduced length, reduced diffraction vector, $\tilde{q}$, reduced diffusion coefficient $\tilde{D}$ and reduced viscosity $\tilde{\eta}$. The relationship between the reduced units used later in the thesis and the experimental units is given by

\begin{align}
	\tilde{q} &= q \rho^{-\frac{1}{3}}\\
	\tilde{D} &= D \rho^{\frac{1}{3}} \left(\frac{1}{T}\right)^{\frac{1}{2}}\\
	\tilde{\eta} &= \eta \rho^{-\frac{2}{3}} \left(\frac{1}{T}\right)^{\frac{1}{2}}\\
	\tilde{\tau_\alpha} &= \tau_\alpha \rho^{\frac{1}{3}}\left(\frac{1}{T}\right)^{-\frac{1}{2} } \label{eq:dimless_tau}
\end{align}


\subsubsection{What are isomorphs?}

\textit{Configurational adiabats} are lines in the phase diagram of constant excess entropy. For R-simple systems \textit{isomorphs} are defined as the configurational adiabats. How do the forces behave along the configurational adiabats? In reduced coordinates the forces are given by:

\begin{equation}
	\tilde{F} = \vec{F} \frac{\rho^{-\frac{1}{3}}}{k_b T}
\end{equation}

The forces can also be derived from the potential energy, $\vec{F} =-\nabla U(\rho,S_{ex}(\tilde{R}))$: 

\begin{equation}
	\vec{F} = - \nabla U(\rho,S_{ex}(\vec{R})) = - \left. \frac{\partial U}{\partial S_{ex}} \right|_\rho \nabla S_{ex}(\tilde{R}) = -T \nabla S_{ex}(\tilde{R})
\end{equation}

Combining the two equations, the reduced forces become

\begin{equation}
	\tilde{F} = - \nabla S_{ex}(\tilde{R}) \frac{\rho^{-\frac{1}{3}}}{k_b}
\end{equation}

The reduced forces for R-simple system are given by a function of the reduced configuration. Two points along an configurational adiabat, and thereby also along an isomorph,  must have the same reduced force. From the fact that the reduced forces is invariant along isomorphs, it can be deduced that both that structure and dynamics are also invariant along isomorphs. This leads to the fundamental prediction of isomorph theory that states \textit{if two state points, ($\rho_1$,$T_1$) and ($\rho_2$,$T_2$) are isomorphic then they have same excess entropy, structure and dynamics in reduced units} \cite{Dyre2014}.

\subsection{Terminology of isomorph theory - Model liquids and real world liquids} \label{sec:thermology_of_isomorph_theory}

The framework of isomorph theory is developed on models of liquids. In theses models, liquids are often atomic liquids with no internal degree of freedom, and the interaction between particles is governed by a pair potential. The prediction of isomorph theory have been shown to work well for many molecular liquids \cite{Dyre2014,Wase2018,Wase2020,XiaoWence2015Itpf}. The goal of this section is to make a clear definition for the terminology used in this thesis, and show how there parallel between the terminology used for model systems and real-world liquids.

In computer simulations of model liquids, like the Lennard-Jones liquid, an isomorph is a line of invariant structure, dynamics and excess entropy. For model systems with bonds, if the bond is modeled with constraints, the strong correlation between virial and potential energy still holds true \cite{Veldhorst2014}. If the same bonds are modeled with a more realistic harmonic bond the correlations break down \cite{Olsen2016,Zahraa2023}. For these model liquids with harmonic bonds there still exist lines of constant structure and dynamics, but the excess entropy is not constant along these lines. These lines were named pseudo-isomorphs \cite{Olsen2016}, and could work as a extension from simple model liquids into more realistic models and real world liquids. A more detailed introduction to previous work on pseudo-isomorphs can be found in the introduction of chapter \ref{chapter:Cumene}. Pseudo-isomorph can be seen as a translation of isomorph into more realistic systems, where the requirement of constant excess entropy is relaxed.  

Another comparison between simple model liquids and real-world liquids that should be drawn is the comparison between \textit{excess entropy scaling} and \textit{isodynes}.  Excess entropy scaling is defined by Dyre as follows: "A liquid obeys excess-entropy scaling if its reduced dynamic properties at different temperatures and pressures are determined exclusively by $S_{ex}$" \cite{Dyre2018}. For R-simple systems isomorphs are configurational adiabats, so R-simple systems which have isomorphs also show excess entropy scaling. However there are cases where a system shows excess entropy scaling and not structural invariance along configurational adiabats. \Mycite{Knudsen2021} structured the Hansen and McDonald molten-salt model and found the dynamics of the model was invariant along lines of constant $S_{ex}$, however the structure was \textit{not} invarient along configurational adiabats. The molten-salt model can be seen as a model for ionic liquids. The same AUTHORS studied a more complex model of an ionic liquid \cite{Knudsen2024}. They found that for an united atom model of the ionic liquid Pyr14TFSI many dynamical quantities, like the mean squared displacement, the diffusion coefficient, viscosity, and the stress auto-correlation function all were invariant along the same lines in phase diagram. The structure along those lines was not invariant. They found that the model did not have isomorphs, as it had a R-values around 0. The model did not have pseudo-isomorphs, since the structure was not invariant.  The model did however have \textit{isodynes}, so lines in phase diagram along which several dynamical quantities are invariant \cite{Knudsen2024}. Measuring excess entropy experimentally is not easy, so by dropping the requirement of constrain of constant excess entropy, isodynes can be seen as the extension of excess entropy scaling from simple model liquids into real-world liquids.




\begin{table}[H]
	\centering
	\begin{tabular}{l|l|l}
		& Model liquids (with $S_{ex}$) & Real-world liquid \\ \hline
		Structure and dynamics & Isomorph                      & Pseudo-Isomorph  \\ \hline
		Dynamics               & Excess entropy scaling        & Isodyne         
	\end{tabular}
	\caption{The terminology of isomorph theory. Credit for this separation of the concepts should go to Lorenzo Costigliola and also partly inspired by ref. \cite{Knudsen2024}. }
\end{table}

\subsubsection{Identifying pseudo-isomorphs experimentally}

Along pseudo-isomorphs a number of dynamical and structural quantities are invariant when presented in reduced units. One of these invariant quantities are the reduced relaxation time, $\tilde{\tau}_{\alpha}$. The reduced relaxation time is defined in equation \ref{eq:dimless_tau} as:

\begin{equation*}
	\tilde{\tau_\alpha}= \tau_\alpha \rho^{\frac{1}{3}}\left(\frac{1}{T}\right)^{-\frac{1}{2} }
\end{equation*}

In a experimental setup we will change state points by varying temperature and pressure.  In an experiment the temperature could change by a factor 2, and the density by change by 15 \%. In comparison the relaxation time could change by 13 decades. The change in relaxation time is much larger that the corresponding change in temperature and density.  It is therefore fair to assume that along an pseudo-isomorph the relaxation time in experimental unit is approximately invariant also. One of the goals of this thesis is to test for the existence of pseudo-isomorphs. Experimentally we can therefore use the isochrone as our best candidate for the pseudo-isomorph or isodyne. If a liquid does not have pseudo-isomorph or isodynes, it would still have isochrones. The isochrone is the experimentalists best candidate for either isodynes or pseudo-isomorphs.

In equation \ref{eq:gamma_def} we introduced $\gamma(\rho,T)$ as the logarithmic derivative along an configurational adiabat. Instead of using lines of constant excess entropy to find isomorphs, in experiments we can use lines of constants relaxation time to find the pseudo-isomorphs. An extension of this definition of $\gamma$ \cite{Ingebrigtsen2012}:

\begin{equation}\label{eq:gamma_def}
	\gamma (\rho,T) \equiv  \left.\frac{\partial \ln\left(T\right)}{\partial \ln\left(\rho\right)}\right|_{\tau_{\alpha}}
\end{equation}

In the next section we will see how this definition of $\gamma$, can be used to derive power-law density scaling and density scaling, as well as the new concepts of isochronal density scaling and isochronal temperature scaling.

	\section[Isochronal Density Scaling]{Isochronal temperature scaling and Isochronal density scaling} \label{sec:IDS_intro}

In this section we will introduce the concepts of \textit{isochronal density scaling} and \textit{isochronal temperature scaling}, and the history of their derivation. The origin of this type of scaling stems from "freezing temperature scaling" and "freezing density scaling" that have been able to collapse many types of transport coefficients for model liquid systems  \cite{,Rosenfeld2000,Rosenfeld2001,KhrapakS.A.2024Fdso}. \Mycite{Rosenfeld2001} showed that, for a system where the interparticle interactions are driven by the Yukawa potential, the reduced viscosity and the reduced diffusion coefficient could be described by $\frac{T}{T_f(\rho)}$. \Mycite{Lorenzo2018} showed that the for the Lennard-Jones system, that the reduced viscosity also collapsed as function of $\frac{T}{T_f (\rho)}$, along with viscosity data of argon and methane. This scaling can be extended from the freezing line to any isomorph as the target isomorph \cite{Lorenzo_PHD,Lorenzo2019}, since the freezing line is an approximate isomorph \cite{Pedersen2016,Adrjanowicz2016}.  This generalization of freezing temperature scaling we denote as \textit{isochronal temperature scaling}. This can be expressed as follows: there exist some function, such that the relaxation time is a function of one varible, $T / T_{\text{target}}(\rho)$. 

\begin{equation}\label{eq:Isochronal_T_scaling}
	\tilde{\tau}(\rho,T) =	 f\left(\frac{T}{T_{\text{target}}(\rho)}\right)
\end{equation}

where $T_{\text{target}}(\rho)$ is the temperature along a target isochrone, at the same density. 

\Mycite{Khrapak2021} showed that for the Lennard-Jones system many of the transport coefficients, like the reduced self-diffusion and reduced shear viscosity, collapsed into a single curved when plotted against $\rho / \rho_F(T)$, where $\rho_F(T)$ is the freezing density for a given temperature.  Freezing density scaling have also collapsed transport coefficients in the Weeks-Chandler-Anderson system \cite{Khrapak2022} and transport coefficients of liquid nobel gasses \cite{KhrapakS.A2022FTaD,KhrapakS.A.2024Fdso}. 

\begin{equation}\label{eq:Isochronal_density_scaling}
	\tilde{\tau}(\rho,T) = f\left(\frac{\rho}{\rho_{\text{target}}(T)}\right)
\end{equation}

\subsection{Derivation of isochronal density/temperature scaling:}

In this section we will derive isochronal density and temperature scaling from the definition of the density scaling coefficient, along with power-law density scaling. The derivations presented in this section, is part of a work in progress with Lorenzo Costigliola, Kristine Niss and Nicholas Bailey. The results presented in chapter \ref{chapter:IDS}, where we test isochronal density scaling and isochronal temperature scaling on real-world liquid, the derivation presented should be credited to Lorenzo Costigliola.

We will see that by making assumption on the temperature and density dependence of $\gamma$, it is possible to derive both isochronal density and temperature scaling. The definition of the density scaling coefficient, $\gamma$, from equation \ref{eq:gamma_def} is the given as the logarithmic gradient of the isochrone in the $\left(\rho,T\right)$ phase diagram.

\begin{equation} \label{eq:gamma_def_2}
	\gamma(\rho,T) = \left. \frac{\partial \ln(T)}{\partial \ln(\rho)}\right|_{\tau_\alpha}
\end{equation}

The expression given in equation \ref{eq:gamma_def_2}, can be viewed as a differential equation, whose solution describes the shape of the isochrones in the phase diagram. If one assumes that $\gamma$ is a material specific constant, i.e. independent of $\rho$ and $T$, then for an initial position,$(\rho_{0},T_0)$ the solution the differential equation becomes:

\begin{align}
	\int_{T_0}^T d \ln T &= \gamma \int_{\rho_0}^\rho d \ln \rho \nonumber \\
	\ln\left(T\right) -\ln\left(T_0\right) &= \gamma\left(\ln\left(\rho\right) - \ln\left(\rho_0\right)\right) \nonumber\\
	\frac{T}{T_0} &= \left(\frac{\rho}{{\rho_0}}\right)^\gamma \nonumber \\
	\frac{{\rho_0}^\gamma}{T_0} &=\frac{\rho^\gamma}{T} 
\end{align}

This means that along an isochrone $\frac{\rho^\gamma}{T}$ is constant. If one denotes $\frac{{\rho_0}^\gamma}{T_0} =\Gamma$, then we have power-law density scaling as defined in equation \ref{eq:powerlaw_density_scaling}. If $\gamma$ is independent $\rho$ and $T$ the shape of the isochrone is a power-law. It can very simply be shown that systems with power-law density scaling also have isochronal temperature scaling and isochronal density scaling.

\subsubsection{Isochronal temperature scaling and Isochronal density scaling for systems with power-law density scaling}
For system with power-law density scaling the shape of the isochrone is a power-law. Along an isochrone the quantity $\Gamma$ is constant. For a single state point $(\rho_{0},T_0)$ one can predict the shape of the isochrone in the phase diagram:

\begin{align} \label{eq:IDS_describtion}
	\Gamma &= \frac{\rho_0^\gamma}{T_0} = \frac{\rho^\gamma}{T}  => \nonumber \\ 
	T(\rho)|_{IC}  &= T_0\left(\frac{\rho}{\rho_{0}}\right)^\gamma  \\
	\rho(T)|_{IC} &= \rho_{0} \left(\frac{T}{T_0}\right)^{\frac{1}{\gamma}}
\end{align}

The expression from equation \ref{eq:IDS_describtion} to describe the isochrones from a single state point is true for any state point along an isochrone. If we look at two isochoric state points along two different isochrones, $(\rho_{0},T_{1})$ and $(\rho_{0},T_{2})$ we can write:

\begin{align*}
	T(\rho)|_{IC1}  &= T_{1}\left(\frac{\rho}{\rho_{0}}\right)^\gamma \\
	T(\rho)|_{IC2} &= T_{2}\left(\frac{\rho}{\rho_{0}}\right)^\gamma
\end{align*}

where $IC1$ and $IC2$ refers to two different isochrones. Dividing the two equations we get:

\begin{equation}\label{eq:Isochronal_temperature_scaling}
	\frac{T(\rho)|_{IC1}}{T(\rho)|_{IC2}} = \frac{T_{1}}{T_{2}}
\end{equation}

The ratio between temperatures of two isochrones is constant at any density. If we keep one of the isochrones locked, and let us denote that as the "target isochrone". Then the value of $\frac{T}{T_\text{target}(\rho)}$ is unique for each isochrone. Another way of framing this is that the relaxation time can be described as a function of $\frac{T}{T_\text{target}(\rho)}$:

\begin{equation} 
	\tilde{\tau}(\rho,T) =	 f\left(\frac{T}{T_{\text{target}}(\rho)}\right)
\end{equation}

This is isochronal temperature scaling, as defined in equation \ref{eq:Isochronal_T_scaling}. The same trick can be used if we look at two state points along two different isochrones with the same temperature, $(\rho_{1},T_{0})$ and $(\rho_{2},T_{0})$:

\begin{equation}\label{eq:Isochronal_rho_scaling}
	\frac{\rho(T)|_{IC1}}{\rho(T)|_{IC2}} = \frac{\rho_{1}}{\rho_{2}}
\end{equation}

For systems with power-law density scaling the ratio between densities of two isochrones is constant at any temperature. If we keep one of the isochrones fixed, and let us denote that as the "target isochrone" then $\frac{\rho}{\rho_\text{target}(T)}$ is unique for each isochrone. We can write:

\begin{equation}
	\tilde{\tau}(\rho,T) = f\left(\frac{\rho}{\rho_{\text{target}}(T)}\right)
\end{equation}

\begin{figure}[H]
	\centering
	\includegraphics[width=0.65\textwidth]{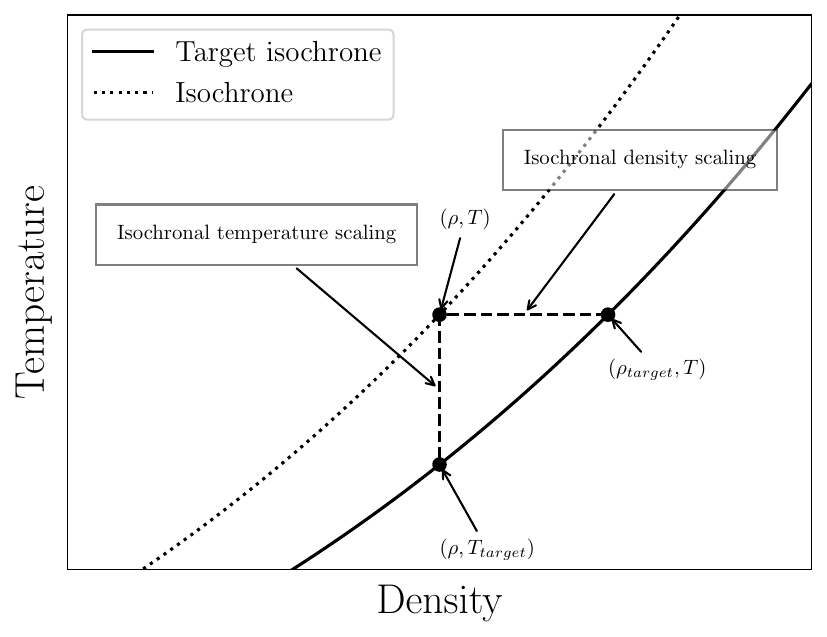}
	\caption{Illustration of isochronal temperature scaling and isochronal density scaling for two isochrones. The solid line is the target isochrone that is assumed to be known, and $(\rho,T)$ denote a given state point. The dotted line is the isochrone of the given state point.    }
	\label{fig:IDS_ITS_phase_ex}
\end{figure}

For both isochronal density scaling and isochronal temperature scaling to work, we only need to know the shape of the target isochrone. In figure \ref{fig:IDS_ITS_phase_ex} both types of isochronal scaling is illustrated. 

It is worth stressing that for systems with power-law density scaling, both isochronal density scaling and isochronal temperature scaling work. If we let $\gamma$ be dependent on either $\rho$ or $T$, we will see that either isochronal density scaling or isochronal temperature scaling is expected to work. If $\gamma$ is a function of both $\rho$ and T, then neither isochronal density scaling or isochronal temperature scaling is expected to work.

\subsubsection{$\gamma$ only dependent on $\rho$}

If we, instead of letting $\gamma$ be a material specific constant, let it depend on the density, then from equation \ref{eq:gamma_def} we can write up the differential equation.

\begin{align}
	\gamma(\rho) &= \left. \frac{\partial \ln(T)}{\partial \ln(\rho)}\right|_{\tau_\alpha} \Rightarrow \nonumber\\
	\int_{T_0}^T d \ln T &= \int_{\rho_0}^\rho \gamma(\rho) d \ln \rho \nonumber\\
	\ln(T)-\ln(T_0) &= \int_{0}^\rho \gamma(\rho) d \ln \rho -\int_{0}^{\rho_0} \gamma(\rho) d \ln \rho \nonumber \\	
	\ln(T)-\ln(T_0) &= \ln(e(\rho)) -\ln(e(\rho_0))\nonumber \\
	\frac{e(\rho_0)}{T_0} &= \frac{e\left(\rho\right)}{T}
\end{align}

Where the function $e(\rho)$ is  $\exp\left(\int \gamma(\rho) d \ln \rho\right)$. If one assumes that $\gamma$ is independent of the temperature then one can derive density scaling, as defined in equation \ref{eq:density_scaling}.  If we consider two state points along an isochore, $\left(\rho_{0},T_{1}\right)$ and $\left(\rho_{0},T_{2}\right)$ then the two isochrones can be described by: 

\begin{align}
	T(\rho)|_{IC1} &= T_{1}\frac{e(\rho)}{e\left(\rho_{0}\right)}  \\
	T(\rho)|_{IC2} &= T_{2}\frac{e(\rho)}{e\left(\rho_{0}\right)} 
\end{align}

The ratio between the two isochrones are given by:

\begin{equation}
	\frac{T(\rho)|_{IC1}}{T(\rho)|_{IC2}} = \frac{T_{1}}{T_{2}}
\end{equation}

This is same expression as in equation \ref{eq:Isochronal_temperature_scaling}. From here the argument to derive isochronal temperature scaling, as defined in equation \ref{eq:Isochronal_T_scaling}, is the same. If we let $\gamma$ be dependent on the density then we can derive isochronal temperature scaling and density scaling as defined in equation \ref{eq:density_scaling}. 




\subsubsection{$\gamma$ only depends on T}

If we consider the case where $\gamma$ is assumed independent of density, and only depends on the temperature.

\begin{align}
		\gamma(T) &= \left. \frac{\partial \ln(T)}{\partial \ln(\rho)}\right|_{\tau_\alpha} \Rightarrow \nonumber\\
\int_{\rho_0}^\rho d \ln \rho	 &=  \int_{T_0}^T \frac{1}{\gamma(T)} d \ln T \nonumber\\
\ln(\rho) - \ln\left(\rho_{0}\right)  &=  \ln\left(g(T)\right) - \ln\left(g(T_0)\right)  \nonumber\\
	\frac{g(T)}{g(T_0)} &= \frac{\rho}{\rho_0}  \label{eq:Jan_heinze}
\end{align}

Here $g(T)$ is defined as $\exp\left(\int_{0}^T \frac{1}{\gamma(T)}d \ln T\right)$.
From equation \ref{eq:Jan_heinze} it is possible to construct an expression like density scaling, $\frac{g\left(T\right)}{\rho}$. The procedure here is very similar to the case when assuming $\gamma$ only dependent on $\rho$. If we consider two state points along an isotherm, $\left(\rho_{1},T_{0}\right)$ and $\left(\rho_{2},T_{0}\right)$ then the two isochrones can be described by


\begin{align}
	\rho(T)|_{IC1} &= \rho_{1}\frac{g(T)}{g(T_0)}  \\
	\rho(T)|_{IC2} &= \rho_{2}\frac{g(T)}{g(T_0)}  
\end{align}

The ratio between the two isochrones are given by:

\begin{equation}
	\frac{\rho(T)|_{IC1}}{\rho(T)|_{IC2}} = \frac{\rho_{1}}{\rho_{2}}
\end{equation}

This is the same expression as in equation \ref{eq:Isochronal_rho_scaling}, and the argument for deriving isochronal density scaling is the same, but restated shortly. For two isochrones the ratio between densities along two isochrones at any given temperature is constant. If we keep one of the isochrones fixed, and let us denote that as the "target isochrone" then $\frac{\rho}{\rho_\text{target}(T)}$ is unique for each isochrone. We can write:

\begin{equation}
	\tilde{\tau}(\rho,T) = f\left(\frac{\rho}{\rho_{\text{target}}(T)}\right)
\end{equation}

\begin{figure}[H]
	\begin{subfigure}[t]{0.48\textwidth}
		\centering
		\includegraphics[width=0.99\textwidth]{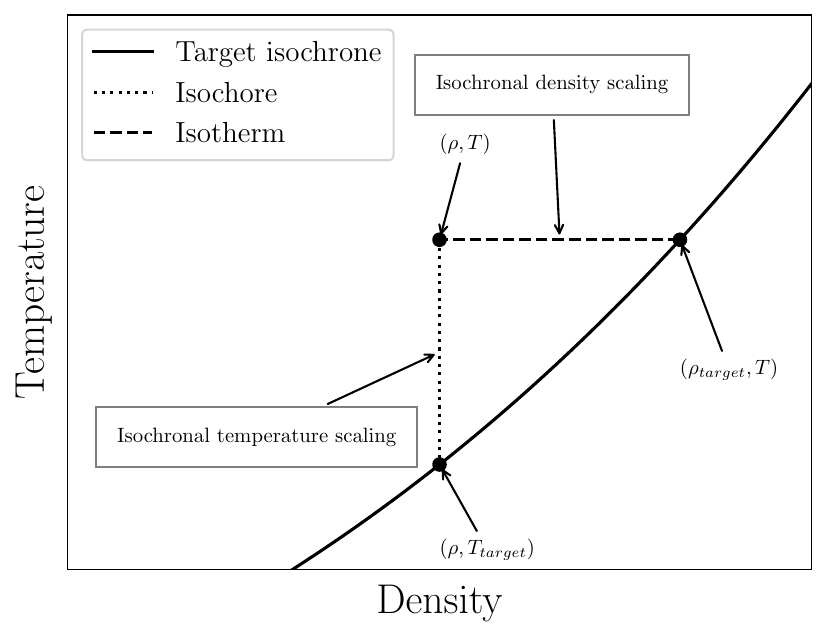}
		\caption{  $(\rho,T)$ phase diagram}
		\label{fig:IDS_ITS_phase_a}
	\end{subfigure}\hfill
	\begin{subfigure}[t]{0.48\textwidth}
		\centering
		\includegraphics[width=0.99\textwidth]{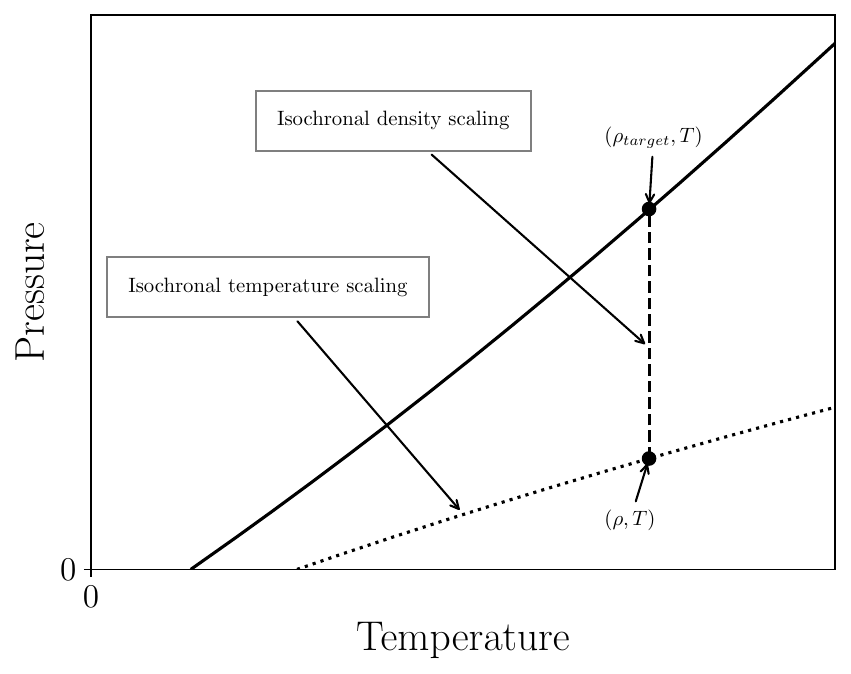}
		\caption{$(T,P)$ phase diagram}
		\label{fig:IDS_ITS_phase_b}
	\end{subfigure}
	\caption{Illustration of isochronal temperature scaling and isochronal density scaling in the $(\rho,T)$ phase diagram and in the $(T,P)$. This illustrates one of the strengths of isochronal density scaling. Practically, isochronal temperature scaling can run into the problem where $T_{target}(\rho)$ is at nonphysical negative pressures.  This is illustrated in figure \ref{fig:IDS_ITS_phase_b}. This is less of a problem in simulations of model liquids, but it can be if the target isochrone enters the gas-liquid coexistence part of the phase diagram \cite{Lorenzo2018}.}
	\label{fig:IDS_ITS_phase}
\end{figure}

This is the same expression as in equation \ref{eq:Isochronal_density_scaling}. The target isochrone used for the scaling, can really be any isochrone. In simulations of simple monoatomic systems the freezing line is relatively easy to identify. In experiments, an easily identifiable isochrone is the glass transition line.  In figure \ref{fig:IDS_ITS_phase} isochronal temperature scaling and isochronal density scaling is illustrated in the $(\rho,T)$-phase diagram and in the experimentalist's $(T,P)$-phase diagram. In practice, it is often easier to map to the target isochrone along the isotherm, than along the isochore. This is shown in figure \ref{fig:IDS_ITS_phase_b}, where $T_{target}(\rho)$ is at nonphysical negative pressures. This can also be an problem for isochronal density scaling, but it is less common, and can often be solved by choosing the target isochrone carefully


The generalized expression of isochronal temperature scaling should scale the relaxation time with any target isochrone. If that target isochrone is the glass-transition, then the relaxation times would be constant as a function of $T / T_g(\rho)$.  As a consequence of this the isochoric fragility is constant for liquids with density scaling. This is not a new result \cite{TarjusG2003Ddat}, but isochronal temperature scaling reproduces the same conclusion. Freezing-temperature scaling and freezing-density scaling can collapse many dynamical properties for model liquid system \cite{Lorenzo2018,Khrapak2021,KhrapakS.A.2024Fdso}. The generalized description, isochronal temperature/density scaling, has not yet been tested on experimental data. The result of such testing will be presented in chapter \ref{chapter:IDS}.

Isochronal density scaling assumes that $\gamma(\rho,T)$ is independent of the density, while isochronal temperature scaling assumes that $\gamma(\rho,T)$ is independent of the temperature. If $\gamma$ is a material specific constant, then both should be valid. The strength of isochronal density scaling and isochronal temperature scaling is that they give a natural way to probe the temperature and density dependency of $\gamma$. Power-law density scaling is the type of scaling most commonly found in the literature, and if power-law density scaling is valid, then $\gamma$ is a material specific constant, and both isochronal density scaling and isochronal temperature scaling should be valid. Power-law density scaling has been reported to breakdown for several glass-formers, for example the van der Waals liquids, dibutyl phthalate \cite{Bøhling2012}, decahydroisoquinoline\cite{Bøhling2012} and DC704 \cite{Ransom_DC704}, as well for hydrogen bonded liquids, such as dipropylene glycol\cite{AdrjanowiczKarolina2017PNDo}. The origin of the breakdown of power-law density scaling can be probed, by isochronal density scaling and isochronal temperature scaling. If the origin of the breakdown is the density dependence of $\gamma$, then isochronal density scaling should fail. The weak temperature dependence of $\gamma$ is the assumption of the original formulation of density scaling. If isochronal temperature scaling breaks down, then the temperature dependence of $\gamma$ plays an role in the breakdown of density scaling. If $\gamma$ is dependent on both temperature and density, then neither isochronal density scaling and isochronal temperature scaling should work. 

\chapter{Structure in Amorphous materials } \label{chapter:Structure}

 The goal of this chapter is to briefly introduce the concepts of structure in amorphous materials and  describe how to measure structure. The derivations presented in this chapter can be found in most textbooks on the subject. Most of the derivations of the different distribution functions follows "Underneath the Bragg peak" by \Mycite{UnderneathTheBraggPeaks}, with supplementary information from refs. \cite{AlsNielsen2011,Fischer2006,ThermalNeutronScattering, SidebottomDavid}.  The structural measures that are presented in this chapters are  two-particles structures, so often nearest neighbor distances between atoms and molecules. More complex structural measurement exist, but are not used in this thesis. In the end of the chapter software techniques to get microscopic information, such as Reverse Monte Carlo (RMC) and Emperical Potential Structural Refinement (EPSR) are introduced. The main focus of the chapter is the two particles correlation functions, such as the pair distribution function and static structure factor.
  
  Structure in amorphous materials seems like a counterintuitive notion since the defining characteristic of amorphous materials is their lack of repeating structures.  Amorphous materials are isotropic; thus, the three-dimensional position of the atoms can be described by a single non-vector variable, $r$ in real space or $q$ in inverse space. The structural measurements that describe amorphous materials are statistical averages of the structure. The structures described in this section are not the positions of a particle, but the probability of a particle at a given position. A common distinction for structural order is into, short-range, medium-range and long-range order. As mention earlier a characteristic of amorphous material is the lack of any long-range order. A contrast to this is crystalline material where the unit cell repeats though out the material. The short-range order is on the length scale of nearest-neighbor distances between structural unit, while the medium-range order embodies structures beyond the atomic level \cite{Ryu2020,Egami2020}. This distinction is commonly used for metallic and oxide glass-formers, since the structural unit are atoms. In a molecular glass-former there is a nearest neighbor distance between molecules, but the molecules themself also have a structure. The distinction into short-range and medium-range order is purely between structural units, but when the structural unit is a molecule the structural unit itself have a structure. The distinction into short-range and medium-range order comes muddier since the length scale between the intramolecular structure and the structural unit can overlap \cite{Ryu2020}.
  
The chapter first introduces the structure in real space and the pair distribution function, and then introduces how structure is measured via a scattering experiment and structure in reciprocal space. Then many of the common distribution functions are introduced, along with how to obtain microscopic structure from scattering data.

 \section{The pair distribution function.}
 The most intuitive structural measure is the pair distribution function, $g(r)$. The pair distribution describes the probability of the relative distance between atoms. Another way of phrasing this is, if one stands on a particle, then what is the probability of seeing a particle at a given $r$. It is defined as  

\begin{equation}\label{eq:gr}
	g(r) = \frac{1}{4\pi N \rho_0 r^2} \sum_i^N \sum_{j\neq i}^N \delta(r - r_{ij})
\end{equation}

where $r_{ij} = \left|r_i-r_j\right|$, so the inter-particle distance between two individual particles, $r_i$, $r_j$.  $N$ is the number of particles in the system and $\rho_{0}$ is the number density.  This definition of the pair distribution function is technically not continuous, but in any liquid, there are so many atoms that it can be considered continuous in practice. 
To provide motivation for this construction of $g(r)$, the $i\neq j$ sum over the Dirac-delta function is actually a local density. In three dimensions $\sum_j^N \delta(r - r_{ij})$ is the number of particles in a volume element, $dV = 4\pi r^2 dr$ from a single particle. The sum over $i$ averages the local density over every particles and the $\frac{1}{N} $ accounts for this averaging making it independent of sample size. The average number of particles in the volume element, $dV$, in the liquid would be $4 \pi r^2 dr \rho_{0}$. Thus, $g(r)$ is a measure of the local particle density relative to the global particle density. 

When $ r\rightarrow 0$, $g(r) \rightarrow 0$ the particles cannot overlap, so the probability of finding two particles at the same place is zero. While $r\rightarrow \infty$, $g(r) \rightarrow 1$, because at large r, there is no difference between local and global density. For real-world liquids, there is some range between $r=0$ and the nearest neighbor, where $g(r)$ is zero due to the Pauli repulsion between atoms. This creates a preferred distance for nearest-neighbor distances. For crystals the structure would reflect the repeating unit cell pattern, while for gasses there are not any preferred distance for nearest-neighbor distances. A very similar distribution function, to $g(r)$ is the \textit{pair density function}, $\rho(r)$

\begin{equation}
	\rho(r) = \rho_{0}g(r) = \frac{1}{4\pi N r^2} \sum_i^N \sum_{j\neq i}^N \delta(r - r_{ij})
\end{equation}

The behavior of the pair density function reflect the behavior of $g(r)$. The difference is that $\rho(r)$ reflect the actual number density of the sample, rather than only the relationship between the local particle density and the global particle density. This can also be seen in the limit of $\rho(r)$, since the limit when $r\rightarrow \infty$, $\rho(r) \rightarrow \rho_{0}$. In the literature, both are abbreviated rather confusingly as PDF \cite{UnderneathTheBraggPeaks}. The choice between using the pair distribution function and the pair density function often comes from whether the data are from simulations or experiments. The double sum in $g(r)$ can be easily calculated in a simulation, while $\rho(r)$ can be calculated in experiments where the density is not necessarily known.

From the definition in equation \ref{eq:gr}, it is clear that $g(r)$ can be separated into different partial pair distributions functions, with the only condition being that all the partial distribution function must sum to the total $g(r)$. For a molecular liquid, an obvious separation would be into an intra- and intermolecular $g(r)$. 

\begin{equation} \label{eq:gr_intra}
g(r) = g_{intra}(r) + g_{inter}(r)
\end{equation} 

\begin{figure}[H]
	\centering
	\includegraphics[width=0.75\textwidth]{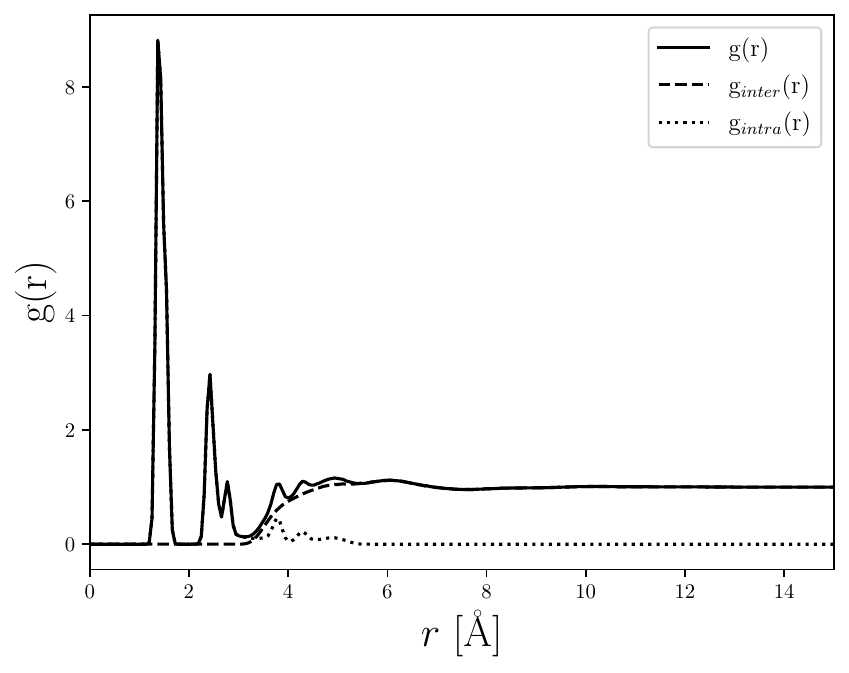}	
	\caption{Example of $g(r)$ for a united-atom model of cumene, that have been simulated as a part of the study in chapter \ref{chapter:Cumene}. The details of the simulations are described in section \ref{sec:Cumene_MD}. The solid black line is $g(r)$, the dashed line is $g_{inter}(r)$, while the dotted line is $g_{intra}(r)$. 
	 }
	\label{fig:g(r)}
\end{figure}

An example of this separation is shown in figure \ref{fig:g(r)}. The figure shows the $g(r)$, $ g_{intra}(r)$ and $ g_{inter}(r)$ for united atom model of cumene. The purpose of this figure is to illustrate $g(r)$ and its structures for a molecular liquid. The  model details are described in section \ref{sec:Cumene_MD}, but should not be necessary for understanding the figure. The first sharp peak can be attributed to the intramolecular bonds in the molecule. This is a well-defined length that repeats for every molecule. The molecule contains bonds with two different lengths, and the bonds vibrate, giving the width of the peaks. The second peak comes from the second nearest-neighbor distances, still within the molecule, but the peak intensity is smaller and the width is larger. The first intermolecular peak can be observed at approximately around $r = 6$ Å. Every atom repulse other atoms if they come too close, creating a "sweet spot" for the distances between atoms. The same is true for molecules, however, the effect is less pronounced.
	
Another logical separation of the pair distribution function is the division into different atomic species. This is presented in section \ref{sec:other_distribution_fuction}.

\section{Scattering}

In this section we will induce structure in inverse space, and how one can measure structure experimentally. Firstly, we will introduce scattering with a main focus on x-ray scattering. The derivation of $S(q)$ is independent of whether one uses x-ray or neutrons, but the terminology used is more consistent with x-ray scattering. The deviation of $S(q)$ follows ref. \cite{UnderneathTheBraggPeaks,Farrow2009}.

In a standard scattering experiment, a sample is hit by a beam of particles, typically consisting of either x-rays or neutrons, and the intensity of the scattered beam is measured as a function of the scattering angle, $2\theta$, and the wavelength of the beam. For x-rays the wavelength of the beam is monochromated making the energy of the beam well defined. For neutron scattering at a spallation source, the beam spans a wide range of wavelength and the wavelength of beam the is known from the time of flight of the neutrons from the source to the sample. This is possible because of the particle wave duality, making the velocity of the neutron related to the wavelength, $\lambda_{neutron} = \frac{h}{m \vec{v}}$.  The wavevector of the incoming beam is denoted, $\vec{k_0}$, and the wavevector of the scattered beam is denoted, $\vec{k_f}$. The diffraction vector, $\vec{q}$ is defined as

\begin{equation}
\vec{q} = \vec{k_0} - \vec{k_f}
\end{equation}

If the scattering is elastic, then the magnitude of the beam is conserved after scattering, $\left| \vec{k_0}\right| = \left| \vec{k_f}\right|$. For elastic scattering, the magnitude of the diffraction vector becomes:

\begin{equation}
	q = |\vec{q}| = \frac{4\pi\sin(\theta)}{\lambda}
\end{equation}

where $\lambda$ is the wavelength of the probing beam. The scattering process can separated into elastic and inelastic scattering, along with coherent and incoherent scattering, which are introduced in detail in section \ref{sec:coherent_scat}. In scattering experiments, structural information is often probed by elastic coherent scattering. In a scattering experiment, the scattered beam is measured by a detector that subtends the solid angle $\Delta \Omega$. Another essential scattering concept introduced is the \textit{differential cross-section}. The differential cross-section measures how well the sample scatters, independent of the intensity of the incoming beam and the number of atoms in the sample. Assuming that the sample is larger than the beam, the differental cross-section is defined as:

\begin{equation}
\frac{d \sigma}{d \Omega} = \frac{I_f}{I_0 N \Delta \Omega} =  \frac{\text{no. of x-ray photons/neutrons scattered per second into $\Delta \Omega$}}{I_0 N \Delta \Omega} 
\end{equation}

here, $I_0$ is the intensity of the incoming beam, $I_f$ is the measured number of photons/neutrons scattered into the detector per second.  $N$ is the number of particles in the sample that are illuminated by the beam and $\Delta \Omega$ is the solid angle that the detectors cover. In the following derivation of the structure factor, we focused on intensity, but other deviations prefer to use the differential cross-section. The differential cross-section and intensity are two closely related concepts. 

\subsection{Sample scattering amplitude}

A third related concept to the scattered intensity and the differential cross-section is the sample scattering amplitude. The sample scattering amplitude for a sample of $N$ atoms is given by
\begin{equation} \label{eq:scattering amplitude}
	\psi(\vec{q},t) =  \sum_{i}^{N} f_i(q) \exp(\text{i}\vec{q}\vec{r_i}(t))
\end{equation}

where $f_i(q)$ is the scattering length of the $i$'th atom, and $\vec{r_i}(t)$ referes to the position of the $i$'th atom. The reader should note that '$i$' is used as an index for the atoms, while 'i' is the imaginary unit. In this text '$i$' is exclusively used as subscript, while 'i' is mainly confined to the $\exp(...)$ term.  The relationship between the measured intensity and the sample scattering amplitude is given by

\begin{equation}
|\psi(\vec{q},t)|^2  = I(q)
\end{equation}

\subsubsection{Scattering lengths}

In this thesis, the scattering length of the $i$'th atom is denoted $f_i(q)$. This notation is more common for x-ray scattering than neutron scattering. In general, the scattering length depends on the interaction between the beam and the scattering centers. This interaction results in very different scattering lengths for x-ray and neutron scattering. For x-ray scattering, the wave interacts with the electron density of the atom, and the resulting scattering lengths are dependent on $q$. $f(q)$ is assumed to be a complex isotropic function. The limit of the x-ray scattering length is $f(0) = N_e$, where $N_e$ is the number of electrons in the atom. $f(q)$ for each atom type are normally taken from table values fitted on experimental data. The values used in this thesis are from ref.\cite{Waasmaier1995}. For neutron scattering, the scattering length  depends on the isotype of the atom, independent of the scattering vector, and a positive scattering length represents repulsive scattering potential \cite{Fischer2006}. The scattering lengths for neutrons are often denoted in literature by $b_i$. In this chapter we use $f_i(q)$ for the scattering length of the i'th atom, except in section \ref{sec:isotopic substitution} about the neutron specific technique isotopic substitution where $b_i$ is used. The experimental results presented in this thesis are from x-ray scattering, but the different scattering lengths of neutrons and x-rays make the techniques complimentary. In the following subsections, we will introduce incoherent and coherent scattering, and show how to derive $S(q)$ and the other distribution functions from the coherent scattering intensity. In a scattering experiment, the measured intensity has many different contributions: Coherent scattering, incoherent scattering, multiple scattering and scattering from the background. Isolating a coherent intensity from the total scattered intensity is not trivial.

\subsection{Coherent and incoherent scattering} \label{sec:coherent_scat}



 In a scattering experiment, if we have scattering from several scattering centers, then we have two imagined scenarios, assuming there are no contribution from multiple scattering and the background. One scenario where the phase difference between scattering center $i$, and scattering center $j$, are conserved. This is called \textit{coherent scattering}.  If the relative phase difference between the two scattering centers is lost, the scattering is called \textit{incoherent scattering}. If the scattering is incoherent, then there no interference between the two different scattered waves. The total incoherent intensity becomes the sum from each individual wave. For coherent scattering there is a conserved phase difference between the scattering centers, making the measured intensity the square of the sum of the wave amplitudes. Both are written explicitly later in the text.
 
 What kind of information can be obtained by the different types of scattering? If the relative location is conserved then the phase difference between the scattering centers also conserved. Most structural information is included in the coherent scattering. The information we are interested in is the structural information, e.g. the relative location between particles. To access this information it is not necessary to correlate the measurements in time. For the rest of this chapter, the time dependency of the scattering centers is suppressed.
 
 \begin{equation}
 	\psi(\vec{q}) =  \sum_{i}^{N} f_i (q)\exp (\text{i} \vec{q}\vec{r_i}) = \psi_i
 \end{equation}
 
 Information about the dynamics of the sample can also be access via scattering, however to access this information it is necessary to correlate the measurements in time. The incoherent scattering can be very useful to extract dynamical information about the same. However, the scope of this thesis is related to structural information; thus, we are only really interested in coherent scattering. However, understanding the time averaged incoherent scattering is important to understand the results of a scattering experiment.
 


The intensity of the scattered wave is given by the modulus squared of the sample scattering amplitude $\left|\psi(\vec{q})\right|^2$. For incoherent scattering, the phase difference is lost; thus, if the scattering is completely incoherent, the total intensity is the sum of the intensities from each atom:

\begin{equation}\label{eq:Zidane_1}
I_ {inc} = \sum_{i}^{N} \left|\psi_i(\vec{q})\right|^2 = \sum_{i}^{N} \psi_i^* \psi_i =\sum_{i}^{N} f^*_i(q)f_i(q) 
\end{equation}

This can be rewritten as a function of atom type as follows

\begin{equation}\label{eq:Zidane}
I_ {inc} = N \sum_{\alpha}^{} c_\alpha f_\alpha^*(q) f_\alpha(q) = N \left<f^2\right>
\end{equation}

where $c_\alpha$ is the concentration of the $\alpha$'th atom species, $c_\alpha = N_\alpha / N$, and $<f^2>$ is defined by the expression in equation \ref{eq:Zidane}. We can also define the sample-averaged scattering power, $\left<f\right> = \sum_{\alpha} c_\alpha f_\alpha $. It follows that:

\begin{equation}
\left<f\right>^2 = \sum_{\alpha} \sum_{\beta} c_\alpha c_\beta f_\alpha^*(q) f_\beta(q) = \frac{1}{N^2} \sum_{i}^{N} \sum_{j}^{N} f_j^*(q) f_i(q) 
\end{equation}

These definitions and expressions will be useful in a later chapter.

\subsubsection{Coherent scattering }

For coherent scattering the phase information is conserved. The full coherent scattering intensity is given by

\begin{equation}\label{eq:scattered_intensity} 
	I_{coh}(\vec{q}) =   \sum^N_{i} \sum_{j}^N f_i(q)f_j^*(q) \exp\left(\text{i} \vec{q} \cdot \vec{r_i}\right)  \exp\left(-\text{i} \vec{q} \vec{r_j} \right)  = \sum^N_{i} \sum_{j}^N f_i(q)f_j^*(q) \exp\left(\text{i} \vec{q} \cdot \vec{r_{ij}}\right) 
\end{equation}

where $\vec{r_{ij}} = \vec{r_i} -\vec{r_j}$. We can separate the double sum into two situations, one where the scattering comes from two different atoms, $i \neq j$ and one where the scattering comes from the same atom, $i =j$. If $i = j$ then $\vec{r_{ij}} = \vec{0}$:
\begin{equation}
	I_{coh}(\vec{q}) = \sum_{i} f_i^*(q)f_i(q) +  \sum^N_{i} \sum_{j\neq i}^N f_i(q)f_j(q) \exp\left(\text{i} \vec{q} \vec{r_{ij}}\right) 
\end{equation}

Using the result shown in equations \ref{eq:Zidane_1} and \ref{eq:Zidane} this can be rewritten as:

\begin{equation}
I_{coh}(\vec{q}) = N\left<f^2\right> + \sum^N_{i} \sum_{j\neq i}^N f_i(q)f_j(q) \exp\left(\text{i} \vec{q} \vec{r_{ij}}\right) 
\end{equation}

The static structure factor, $S(q)$ is often defined as follows \cite{UnderneathTheBraggPeaks,Farrow2009}
\begin{equation} \label{eq:sq_def}
	S(\vec{q}) = \frac{I_{coh}(\vec{q}) }{N \left<f\right>^2} -\frac{\left<f^2\right> -\left<f\right>^2 }{\left<f\right>^2}
\end{equation}

The term $\frac{\left<f^2\right> -\left<f\right>^2 }{\left<f\right>^2}$ is called Laue monotonic diffuse scattering term. The reason for this term is to the design the properties we want $S(q)$ to possess. First, we want  $S(q)$ to go to one, as $q \rightarrow \infty$. It will also prove practical later to normalize the intensity, with $\left<f\right>^2$ instead of $\left<f^2\right>$, then the Laue monotonic diffuse scattering term is needed to handle the cancellations of intensity when the scattering comes from a system with multiple scattering length. For a monoatomic system $\left<f\right>^2 = \left<f^2\right>$, so the monotonic diffuse scattering term becomes 0 in that case.

An alternative definition that is also used in literature is that the intensity is normalized by $N\left<f^2\right>$ instead of $N \left<f\right>^2$ \cite{UnderneathTheBraggPeaks}. Then the Laue monotonic diffuse scattering term is not needed.

If we plug equation \ref{eq:scattered_intensity} in the definition of the static structure factor, equation \ref{eq:sq_def} we obtain the following

\begin{align}
	S(\vec{q}) &= \frac{1}{N \left<f\right>^2 } \sum^N_{i=1} \sum_{j\neq i}^N f_i(q)f_j^*(q) \exp\left(\text{i} \vec{q} \cdot \vec{r_{ij}}\right) + \frac{1}{N \left<f\right>^2} \sum_{i}^{N} f_i(q)f_i^*(q)  -\frac{\left<f^2\right> -\left<f\right>^2 }{\left<f\right>^2} \nonumber \\
	S(\vec{q}) &= \frac{1}{N \left<f\right>^2 } \sum^N_{i=1} \sum_{j\neq i}^N f_i(q)f_j^*(q) \exp\left(\text{i} \vec{q} \cdot \vec{r_{ij}}\right) + \frac{N\left<f^2 \right>}{N \left<f\right>^2} -\frac{\left<f^2\right> -\left<f\right>^2 }{\left<f\right>^2} \nonumber \\
		S(\vec{q}) &=\frac{1}{N \left<f\right>^2 } \sum^N_{i=1} \sum_{j\neq i}^N f_i(q)f_j^*(q) \exp\left(\text{i} \vec{q} \cdot \vec{r_{ij}}\right) + 1 \nonumber \\
		 S(\vec{q}) -1 &= \frac{1}{N \left<f\right>^2 } \sum^N_{i=1} \sum_{j\neq i}^N f_i(q)f_j^*(q) \exp\left(\text{i} \vec{q} \cdot \vec{r_{ij}}\right) \label{eq:sq}
\end{align}

It is worth noting that the term $-1$ comes from the case, where $i=j$. 

\subsubsection{S(q) for isotropic samples}

If we assume that the sample is isotropic, as is the case for liquids, then it makes sense to calculate an orientation average for $S(\vec{q})$. If we place $\vec{q}$ along the z-axis, then the dot product between the diffraction vector and the vector between two different atoms becomes: 

\begin{equation}
	\vec{q} \cdot \vec{r_{ij}} = qr_{ij}\cos{\theta}
\end{equation}

where $\theta$ is the angle between $\vec{q}$ and $\vec{r_{ij}}$. If we calculate the orientational average, then both $r_{ij}$ and $\theta$ can take any value with equal probability. We calculate the angular averaged intensity of a pair of atoms in the sample:
\begin{equation}
	\overline{\exp \left({i\vec{q}\cdot\vec{r_{ij}}} \right)} = \frac{\int_0^{2\pi}\text{d}\phi \int_0^\pi \text{d} \theta\exp\left({\text{i}qr_{ij}\cos{\theta}} \right) r_{ij}^2 \sin{\theta}} {\int_0^{2\pi}\text{d}\phi \int_0^\pi \text{d} \theta r_{ij}^2\sin{\theta}}
\end{equation}

Lets first look at the denominator:

\begin{equation}
	\int_0^{2\pi} \int_0^\pi  r_{ij}^2\sin{\theta} d \theta d\phi = r_{ij}^2 \int_0^{2\pi} 2 d\phi =4 \pi r_{ij}^2
\end{equation}

The numerator is:

\begin{align}
	\int_0^{2\pi}d\phi \int_0^\pi \text{d}\theta \exp\left({\text{i}qr_{ij}\cos{\theta}} \right) r_{ij}^2 \sin{\theta}  &=\int_0^{2\pi} - \frac{r_{ij}^2}{\text{i}qr_{ij}} \left[ \exp\left(\text{i}qr_{ij}\cos(\theta)\right)\right]_0^\pi  d\phi \nonumber \\ 
	&= \int_0^{2\pi} - \frac{r_{ij}^2}{\text{i}qr_{ij}}  \left( -\exp(\text{i}qr_{ij}) + \exp(-\text{i}qr_{ij}) \right) d\phi  \nonumber\\
	&=  \frac{2\pi r_{ij}^2}{\text{i}qr_{ij}}  \left( \exp(\text{i}qr_{ij}) - \exp(-\text{i}qr_{ij}) \right) \nonumber \\
	& = \frac{4\pi r_{ij}^2 \sin(qr_{ij} )}{qr_{ij}}
\end{align}

Combining the denominator and the numerator we get:
\begin{equation}
	\overline{\exp \left({\text{i}\vec{q}\cdot\vec{r_{ij}}} \right)} = \frac{\sin(qr_{ij})}{qr_{ij}}
\end{equation}

Thus, the contributions from each atom pair to $S(q)$ is a sinc function with some weight from the scattering lengths, $f_i(q)f_j(q)$. Inserting into the expression of $S(q)$ from equation \ref{eq:sq} we get:

\begin{equation}\label{eq:sq_liquds}
	S(q) -1 = \frac{1}{N \left<f\right>^2 }\sum^N_ {i} \sum^N_{i\neq j} f_i(q)f_j(q) \frac{\sin(qr_{ij})}{qr_{ij}}
\end{equation}

\begin{equation}\label{eq:sq_liquds_v2}
	S(q) = \frac{1}{N \left<f\right>^2 } \sum^N_{i} \sum^N_{j}f_i(q)f_j(q) \frac{\sin(qr_{ij})}{qr_{ij}}
\end{equation}

Evaluating the high $q$ behavior of $S(q)$, it is clear that the right hand side of equation \ref{eq:sq_liquds} goes to 0 as $q \rightarrow \infty$, making $S(q) \rightarrow 1$ in the high q limit. The low q-limit can be shown to be related to the compressibility \cite{Fischer2006,ThermalNeutronScattering}:

\begin{equation}\label{eq:Rene_Henriksen}
	S(0) = \rho_m k_b T \chi_T 
\end{equation}

where $\chi_T$ is the isothermal compressibility, and $\rho_m$ is the number density of molecules in the sample. 

\subsection{Distribution function:} \label{sec:other_distribution_fuction}

This subsection introduces many other distributions function commonly used to describe structures from scattering experiments. They are introduced in less detail than the structure factor and the pair distribution function.

\subsubsection{The reduced structure function and the reduced pair distribution function}
Two distribution functions that are of interest to the experimentalist are the reduced structure function, $F(q)$, and the reduced pair distribution function, $G(r)$. 

\begin{equation}
	F(q) = q\left(S(q)-1\right)
\end{equation}

$G(r)$ is the Fourier transform of the reduced structure function.

\begin{equation} \label{eq:Yaro_Yaro}
	G(r) = \frac{2}{\pi} \int_0^\infty F(q)\sin(qr)\text{d}q = \frac{2}{\pi} \int_0^\infty q\left(S(q)-1\right)\sin(qr)\text{d}q = 4 \pi r \rho_0 [g(r) -1]
\end{equation}

where $\rho_0$ is the atomic number density of the sample and $g(r)$ is the pair distribution function defined in equation \ref{eq:gr}. This gives us a connection between our experimentally measured structure factor and the pair distribution function introduced in equation \ref{eq:gr}. The full deviation of this relationship is derived in ref. \cite{Farrow2009}.
Although $G(r)$ is less physically intuitive than $g(r)$, it can be calculated directly from experimental data without knowing the density of the sample. It can also be used to estimate the atomic number density of a sample, since  $g(r) \rightarrow 0$ when $r \rightarrow 0$, the low $r$ value of $G(r)$ goes to $-4\pi r \rho_0$. However as shown in figure \ref{fig:g(r)}, $g(r) = 0$ before $r=0$. The low $r$ behavior of $G(r)$ is negative going to 0, with a slope of $-4\pi r \rho_0$. The high $r$ behavior of $G(r)$ oscillates around 0, as $g(r) \rightarrow 1$ when $r \rightarrow \infty$.

The relationship in equation \ref{eq:Yaro_Yaro} can also be inverted, to calculate $S(q)$ from $g(r)$

\begin{equation}\label{eq:gr->sq}
S(q) -1 = \frac{1}{q} \int_{0}^{\infty} G(r)\sin(qr) \text{d}r =  \frac{4\pi \rho_0}{q} \int_{0}^{\infty} r\left(g(r)-1\right)\sin(qr) \text{d}r
\end{equation}

\subsubsection{The Radial distribution function:}
The radial distribution function is related to the pair distribution function:
\begin{equation}
	R(r) = 4\pi r^2 \rho_0 g(r) 
\end{equation}

The usefulness of the radial distribution function is that it gives the numbers of atoms in the annulus when integrated over an annulus. This is normally used to calculate the nearest neighbor coordination numbers in a glass.

\begin{equation}
	\int_{r_1}^{r_2} R(r) dr =\int_{r_1}^{r_2} 4\pi r^2 \rho_0 g(r) dr = \int_{r_1}^{r_2}\sum_i \sum_j \delta (r -r_{ij}) dr = N_c
\end{equation}

where $N_c$ is the number of atoms in the annulus. The radial distribution function has the same low $r$ limit as $g(r)$, as both of them go to 0, as $r \rightarrow 0$. Unlike $g(r)$, $R(r)$ grows with a factor of $r^2$ when $r$ increases, which makes it difficult to interpret at high $r$-values.

\subsection{Partial $S(q)$ and $g(r)$}

For molecular liquids or any system with more than one type of atom, it is interesting to analyze the structure of each type of atom in the sample instead of the combined structure of the whole system. The definition of $g(r)$ from equation \ref{eq:gr} is simply a sum of Dirac-delta function, with a common prefactor; massaging this sum we can separate it to atoms of different type. The same is true for the static structure factor $S(q)$. The most common approach is called Faber-Ziman formalism. Here the static structure factor between two types of atoms, $\alpha$ and $\beta$ is defined as

\begin{equation}
	S_{\alpha\beta}(q) =\frac{\left(\sum_\alpha c_\alpha f_\alpha(q) \right)^2´}{c_\alpha c_\beta f_\alpha(q) f_\beta(q)} S'_{\alpha \beta} (q) \quad \text{where} \quad S'_{\alpha \beta} (q)  =     \frac{1}{N \left<f\right>^2 } \sum^N_{i\in \alpha} \sum_{j \in \beta}^N f_i(q)f_j(q) \exp\left(i \vec{q} \vec{r_{ij}}\right)
\end{equation}

$S_{\alpha \beta}$ is the Faber-Ziman partial structure factor for atom pair $\alpha$,$\beta$. $S'_{\alpha \beta}(q)$ is simply the contribution to the total static structure factor from the scattering between atoms of type $\alpha$ and $\beta$. It is clear from the definition that

\begin{equation}
	S(q) = \sum_\alpha \sum_\beta S'_{\alpha \beta} (q)  \quad => \quad S(q) =  \sum_\alpha \sum_\beta \frac{c_\alpha c_\beta f_\alpha f_\beta}{\left(\sum_\alpha c_\alpha f_\alpha \right)^2} S_{\alpha \beta} (q)
\end{equation}

To provide some motivation for the scaling prefactor, they were chosen such that for all $S_{\alpha \beta}(q) \rightarrow 1$ as $q \rightarrow \infty$. It is also worth noting that $\left(\sum_\alpha c_\alpha f_\alpha \right)^2 =  \left< f \right>^2 $ and $ c_\alpha c_\beta f_\alpha f_\beta = \left<f_{\alpha \beta} \right>^2$. Just like in the case of the total structure factor the partial reduced pair distribution function can be defined:

\begin{align}
	G_{\alpha\beta}(r) &= \frac{2}{\pi} \int_0^\infty q \left(S_{\alpha\beta}(q) -1 \right) \sin(qr) dq\\
	G(r) &= \sum_\alpha \sum_\beta \frac{c_\alpha c_\beta f_\alpha f_\beta}{\left(\sum_\alpha c_\alpha f_\alpha \right)^2} G_{\alpha\beta} (r)
\end{align}

The partial pair distribution function would be:
\begin{equation}
	g_{\alpha \beta} (r) = \frac{1}{4 \pi N \rho_0 r^2} \frac{\left(\sum_\alpha c_\alpha f_\alpha \right)^2}{c_\alpha c_\beta f_\alpha f_\beta} \sum_{i\in \alpha} \sum_{j \in \beta } \delta (r - r_{ij})
\end{equation}

\begin{equation}
	g(r) = \sum_\alpha \sum_\beta \frac{c_\alpha c_\beta f_\alpha f_\beta}{\left(\sum_\alpha c_\alpha f_\alpha \right)^2} g_{\alpha\beta} (r)
\end{equation}

When calculated from simulation data the scattering lengths are normally ignored, but it is important when calculating the partial structure factors from scattering data or comparing with scattering data.

\subsection{Intra- and Intermolecular $S(q)$}

From the previous section, it should be clear that it is possible to separate contributions to $S(q)$ as desired. In this thesis we calculate the intra- and intermolecular contributions to the structure factor. Starting from equation \ref{eq:sq_liquds} the intermolecular contribution to $S(q)$ is given by:

\begin{equation} \label{eq:sq_seperation_structure}
	S(q) -1 = S_{intra} (q) + S_{inter} (q)
\end{equation}

The $S(q)-1$ is to emphasize that the self-scattering contributions are neither intra- or intermolecular.

\begin{equation}
S_{inter} (q) =  \frac{1}{N \left<f\right>^2 } \sum_{\lambda} \sum_{i \in \lambda} \sum_{j \notin \lambda} f_if_j \frac{\sin(qr_{ij})}{qr_{ij}}
\end{equation}

$\lambda$ denote which molecule the atoms are in. As mentioned earlier, we assume that the relative positions of the atoms in the liquid are independent of time. For the intramolecular structure the atoms vibrate, and we need to account for that. In the intramolecular structure, we can account for the vibration of the atoms relatively simply by the Debye–Waller term \cite{Fischer2006}. The position as a function of time, can be described as $r(t) = \left<\left< r\right>\right> + u(t)$ where $\left<\left< r\right>\right>$ denotes the time averaged position, and $u(t)$ denotes the displacement from the average position as a function of time. Including the Debye-Waller factor the intramolecular structure can be described by

\begin{equation}
	S_{intra} (q) =  \frac{1}{N \left<f\right>^2 } \sum_{\lambda} \sum_{i \in \lambda} \sum_{j \in \lambda, i\neq j} f_if_j \frac{\sin(qr_{ij})}{qr_{ij}} \exp \left(-\frac{1}{2} q^2 \left<\left< u^2\right>\right> \right)
\end{equation}

\begin{figure}[H]
	\begin{subfigure}[t]{0.48\textwidth}
		\centering
		\includegraphics[width=1\textwidth]{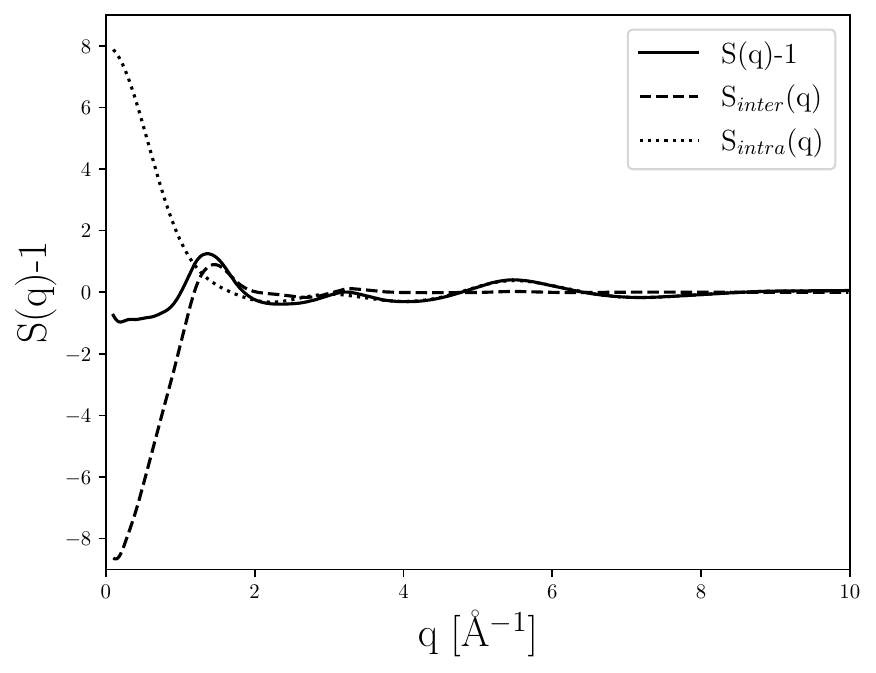}
			\caption{}
			\label{fig:Cumene_sq_example_v1_a}
	\end{subfigure}
	\begin{subfigure}[t]{0.48\textwidth}
	\centering
	\includegraphics[width=1\textwidth]{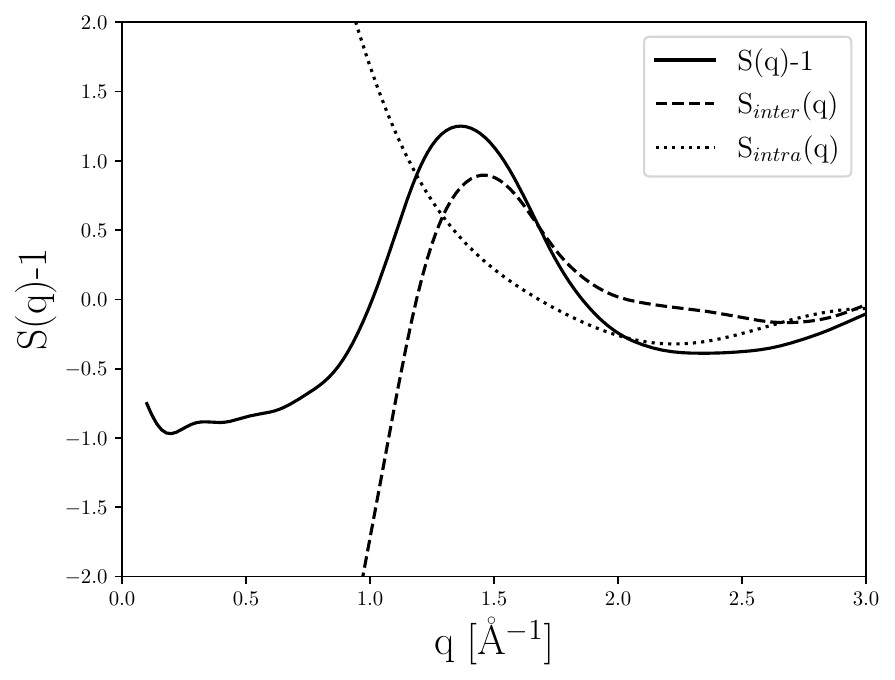}
	\caption{}
				\label{fig:Cumene_sq_example_v1_b}
\end{subfigure}
\caption{Example of $S(q)-1$ for a united-atom model of cumene, details are described in section \ref{sec:Cumene_MD}. Subfigure \ref{fig:Cumene_sq_example_v1_a} shows $S(q)-1$ in a larger q-range range, while subfigure \ref{fig:Cumene_sq_example_v1_b} focuses on the first peak. The solid black line is $S(q)-1$, the dashed line is $S_{inter}(q)$, while the dotted line is $S_{intra}(q)$. The $S(q)-1$ is calculated by Fourier-transforming $g(r) -1$, see equation \ref{eq:gr->sq}.}
\label{fig:Cumene_sq_example_v1}
\end{figure}

In figure \ref{fig:Cumene_sq_example_v1} an example of the static structure factor separated into intramolecular and intermolecular contributions based on a united-atom model of cumene. They are calculated by Fourier transforming the calculated $g(r)$. The behavior of $S_{inter}(q)$ and $S_{intra}(q)$ when $q \rightarrow 0$ can be explained by examining the analytical expression of $S_{intra}(q)$ and $S_{inter}(q)$. The limit of $S(q)$ when $q \rightarrow 0$ is finite and stated in equation \ref{eq:Rene_Henriksen}. The limit of $\frac{\sin(qr_{ij})}{qr_{ij}}$ when $q \rightarrow 0$ is 1. For this model, the intramolecular distance is smaller than the intermolecular distance. $r_{ij}^{(intra)} < r_{ij}^{(inter)} $  This causes $\frac{\sin(qr_{ij})}{qr_{ij}} \rightarrow 1$ faster for intramolecular distances than for intermolecular distances. For this model there are 9 atoms all with the same scattering length. The limit of the intramolecular contribution is

\begin{equation}\label{eq:sq_intra_limit}
\lim_{q\rightarrow 0} S_{intra} (q) = \frac{1}{N \left<f\right>^2 } \sum_{\lambda}^{N/9} \sum_{i \in \lambda}^9 \sum_{j \in \lambda, i\neq j}^8 f_if_j = \frac{8N\left<f\right>^2 }{N\left<f\right>^2 } = 8
\end{equation}

The limit of $S_{inter}$ becomes:

\begin{align} 
\lim_{q\rightarrow 0}\left( S(q) -1\right) - &= \lim_{q\rightarrow 0} S_{intra}(q) + \lim_{q\rightarrow 0} S_{inter}(q)  \Rightarrow \nonumber\\
\lim_{q\rightarrow 0} S_{inter}(q)&=\lim_{q\rightarrow 0}\left( S(q) -1\right) - \lim_{q\rightarrow 0} S_{intra}(q)  \nonumber \\
 \lim_{q\rightarrow 0} S_{inter}(q) &= \left(\rho_m k_b T \chi_T -1\right) -8
\end{align}

\subsection{Isotopic substitution} \label{sec:isotopic substitution}

Isotopic substitution is an experimental technique for calculating partial structure factors for a sample. The idea is itself quite simple, but powerful. For neutron scattering, the scattering lengths of an atom are different from isotope to isotope, but the scattering lengths are known. This technique is unique for neutron scattering and works best for a system with a large difference in the to scattering lengths. Since the technique is specific to neutron scattering, we change the notation of the scattering lengths to $b_i$ instead of $f_i(q)$. For a system with two atom types like carbon and hydrogen, substituting hydrogen with deuterium it is possible to calculate the partial structure factors with enough measurements. Hydrogen and deuterium are obvious isotopes to substitute, since the large difference in scattering length $b_H = -3.741$ fm, $b_D = 6.674$ fm.
There is an assumption that the structure of the molecule does not change with the substitution of the isotope, and only the scattering length changes. If a system is too complex to calculate the partial structure factors, substituting isotopes can exaggerate or hide the behavior of different atoms providing useful structural information for the sample.

Let us examine a simple organic liquid consisting only of carbon and hydrogen. To avoid confusion the notation have changed a bit, here $S_{C-H}$ denotes the Faber-Ziman carbon-hydrogen partial structure factor, while $S_{CH}$ denotes the measured total structure factor for a sample consisting only of carbon and hydrogen. In a neutron scattering experiment the measured static structure factor would be a sum of the partial static structure factors:

\begin{equation} \label{eq:S_CH}
	S_{CH}(q) = \frac{1}{(c_Hb_H +c_Cb_C )^2}\left(c_H^2 b_H^2 S_{H-H} (q)+c_C^2 b_C^2 S_{C-C} (q) +2 c_C c_H b_C b_H S_{C-H} (q)\right)
\end{equation}

If hydrogen is substituted with deuterium, then we would measure a different structure factor:

\begin{equation}\label{eq:S_CD}
	S_{CD}(q) = \frac{1}{(c_Db_D +c_Cb_C )^2} \left( c_D^2b_D^2 S_{H-H} (q)+c_C^2 b_C^2 S_{C-C} (q) +2c_C c_D b_C b_D S_{C-H} (q)\right)
\end{equation}

It is clear that $c_D = c_H$ since we have only changed isotopes not the amount of atoms in sample. To clean the expressions a little we note $\hat{S}_{\alpha\beta}(q) = \left(\sum_{\alpha} c_\alpha b_\alpha\right)^2 S_{\alpha\beta}(q)$.

The two structure factors have the $C-C$ partial structure factor term in common. If we subtract the two we cancel that term out:

\begin{equation}\label{eq:isotopic_substition_1}
	\hat{S}_{CD}(q) - \hat{S}_{CH}(q) = c_H^2\left(b_D^2 -  b_H^2 \right)S_{H-H} (q)+2 c_C c_H b_C\left( b_D -  b_H\right) S_{C-H} (q)
\end{equation}

If we wanted to exclude an other contribution, eg. $S_{C-H}$ we could subtract a weight contribution:

\begin{equation}\label{eq:isotopic_substition_2}
	\hat{S}_{CD}(q) - \frac{b_D}{b_H} \hat{S}_{CH}(q) = c_D^2b_D(b_D - b_H) S_{H-H}(q) + c_C^2 b_C^2 \left(1-\frac{b_D}{b_H}\right) S_{C-C}
\end{equation}

With two measurements it is possible to eliminate contributions from one partial structure factor, and we can choose which one to eliminate. We cannot determine all the partial structure factors as we are stuck with two equations and three unknowns. It is possible to create a third equation by measuring on a mixture of the deuterated sample and non-deuterated sample. The measured static structure factor would be:

\begin{equation}\label{eq:S_CHD}
	 \hat{S}_{CHD} (q) = c_C^2 b_C^2 S_{C-C} (q) +  c_H^2 b_{HD}^2  S_{H-H}(q) + 2  c_C c_H b_C b_{HD} S_{C-H} (q)
\end{equation}

where $b_{HD}$ is $x b_H + (1-x) b_D$ and $x$ is the fraction of nondeuterated sample in the mixture. Combining equations \ref{eq:S_CH},\ref{eq:S_CD} and \ref{eq:S_CHD} it is possible to isolate the partial static structure factors from the measurement. The results are presented here:

\begin{align}
	S_{H-H}(q)  &= \frac{x \hat{S}_{CH}(q) +(1-x)\hat{S}_{CD}(q) - \hat{S}_{HDC}(q)}{c_H^2 [x b_H^2 +(1-x)b_D^2 -b_{HD}^2 ]} \nonumber \\
	S_{C-H}(q)  &= \frac{\hat{S}_{CH}(q) - \hat{S}_{CD}(q) - (c_H^2b_H^2 - c_D^2b_D^2)S_{H-H}(q) }{2c_C c_H b_C (b_H-b_D)} \nonumber \\
	S_{C-C}(q)  &= \frac{\hat{S}_{CH}(q) - 2c_Cc_Hb_C b_H S_{C-H}(q) -c_H^2 c_H^2 S_{H-H} (q)}{c_C^2 b_C^2}
\end{align}


The above equations even though they seem complex, only depend on three measurements and a number of constants known from literature.







\section{Microscopic structure from scattering data}

So far in this chapter, it has been shown how starting from coherent scattering can define and calculate the static structure factor, different real-space correlation functions and even partial structure factors of a sample. However, each measure is a statistical average of the structure throughout the sample containing no microscopic knowledge of the measured sample.  One approach to obtain microscopic data is to model the experimental data by creating a model that would have the same measured structure factor in a scattering experiment. Reverse Monte Carlo (RMC), was historically the first successful approach to solving this inverse problem by creating a structural model from scattering data. It was first used on a simple disordered system, when was used to determine the disordered structure of liquid argon \cite{McGreevy1988}. The method described shortly; For some initial position, then a random particle does a random move, and then calculating the new $g(r)$. If the new $g(r)$ is at better fit to the measured data, then move is accepted if it creates a worse fit, it is accepted with a probability based on the experimental error. This algorithm is iterated until the goodness-of-fit reaches an equilibrium. 

The technique has evolved into different branches depending on the systems they wish to study. When the  complexity of the system increases, there can arise limitations on possible moves an atom can make. For a molecular liquid, a move that significantly change the bond lengths could provide a good fit to the experimental data, but the move is nonphysical. For each technique, the main idea of fitting a model to the measured structural data remains the same.

\subsubsection{Emperical Potential Structure Refinement}

Empirical Potential Structure Refinement (EPSR) is software technique created to fit a structural model to measured scattering data, with a particular focus on molecular liquids and glasses in mind \cite{Terban2022}. Practically, the interaction between particles is driven by three potentials: an intramolecular potential, an intermolecular potential, and an empirical potential that is refined to the diffraction data \cite{Soper2017}. The intra- and intermolecular potentials combined are known as the reference potential.
The intramolecular potential is modeled by a harmonic potential that describes each intermolecular atom interaction

\begin{equation}
	U_{intra} = C \sum_{\lambda} \sum_{\alpha \beta >\alpha} \frac{\left(r_{\alpha_\lambda\beta_\lambda} - d_{\alpha\beta}\right)^2}{2 w_{\alpha\beta}^2}
\end{equation}

where $r_{\alpha_i\beta_i}$ is the separation of the atoms $\alpha, \beta$ in molecule $\lambda$, $d_{\alpha\beta}$ is the average distance between atoms $\alpha,\beta$. $w_{\alpha\beta} = \frac{d_{\alpha\beta}}{\mu_{\alpha\beta}}$ and $\mu$ is the reduced mass of the atom pair $\alpha\beta$ in atomic mass units, $\mu_{\alpha \beta} = \frac{M_\alpha M_\beta}{M_\alpha + M_\beta}$. $C$ is a temperature-independent constant determined by comparing the simulated structure factors with the diffraction data at high $Q$. The intermolecular reference potential is often defined by a 12-6 Lennard-Jones potential with Coulomb charges if necessary :

\begin{equation}
	U_{\alpha\beta}(r_{\alpha_\lambda\beta_{\lambda'}}) = 4 \epsilon_{\alpha\beta} \left[ \left(\frac{\sigma_{\alpha\beta}}{r_{\alpha_\lambda\beta_{\lambda'}}}\right)^{12} - \left(\frac{\sigma_{\alpha\beta}}{r_{\alpha_\lambda\beta_{\lambda'}}}\right)^6 \right]  + \frac{q_\alpha q_\beta}{4\pi\epsilon_0r_{\alpha_\lambda\beta_{\lambda'}}}
\end{equation}

where $\lambda$ and $\lambda'$ denote distinct molecules. 

The goal of the empirical potential is to minimize the difference between the measured partial structure factors and the simulated partial structure factors. EPSR has different tricks for estimating the partial structure factors if full information is not known \cite{YoungsTristan2019}, but to describe the mechanism of EPSR, we assume that the partial structure factors are known. The potential is often chosen as the sum of $h$ Poisson functions:

\begin{align}
	U_{ij}^{ep} &= kT \sum_{n=1}^h C_kf_n(r) \\
	f_n(r) &= \frac{1}{4 \pi \sigma_r^3(n+2)!} \left( \frac{r}{\sigma_r} \right)^n \exp\left(- \frac{r}{\sigma_r} \right)
\end{align}

The motivation behind that potential is that it has an exact solution to its Fourier transform, so when fitting to the measurements, it can be done directly. In the q-space $C_k$' are fitted to minimize the difference between the measured and simulated partial structure factors. 

In EPSR, the movement of the atoms is done by Monte Carlo simulation, where the acceptance of a trail move is done to minimize the potential energy of the system. The total potential energy that the algorithm minimizes is the sum of the three potentials. Possible moves depend on the system, but can include, the rotation, or translation of the whole molecules, individual atom movement within the molecule, and rotation of side chains of the molecule.

With EPSR, RMC or any structural refinement technique, the results cannot give you the exact position of each molecule, only a solution that would give the same result in a scattering experiment. The validity of the simulation results can be strengthen by have several structure factors to model against. This would be done by using isotopic substitution and supplementing with x-ray scattering data.

\section{Correlations between structure and fragility} \label{sec:fragility}

In the literature there have been several empirical connection between the fragility of a glass-former and some kind of structural measurement. In this section we will introduce a few correlations

\Mycite{Voylov2016} correlated the change in the full width half maximum (FWHM) of the first peak in S(q) with the fragility of several different glass-forming liquids including metallic glassformers, oxidglasses, polymers and molecular glassformers. The authors correlate the relative change between the FWHM in the liquid and the glass, with fragility:

\begin{equation}
	\frac{FWHM_{liquid} - FWHM_{glass}}{FWHM_{liquid}} = \frac{\Delta FWHM}{den}
\end{equation}

For a molecular glass-former it is possible to seperate S(q) into intramolecular and intermolecular contributions, shown in equation \ref{eq:sq_seperation_structure}. Any change to the structure when moving in the phase diagrams comes almost exclusively from intermolecular changes in the liquid. If the intramolecular contribution is assumed to be invariant, then subtracting $S(q)$ at two different state points would quantify an intermolecular change.

\begin{figure}[h]
	\centering
\includegraphics[width=0.75\textwidth]{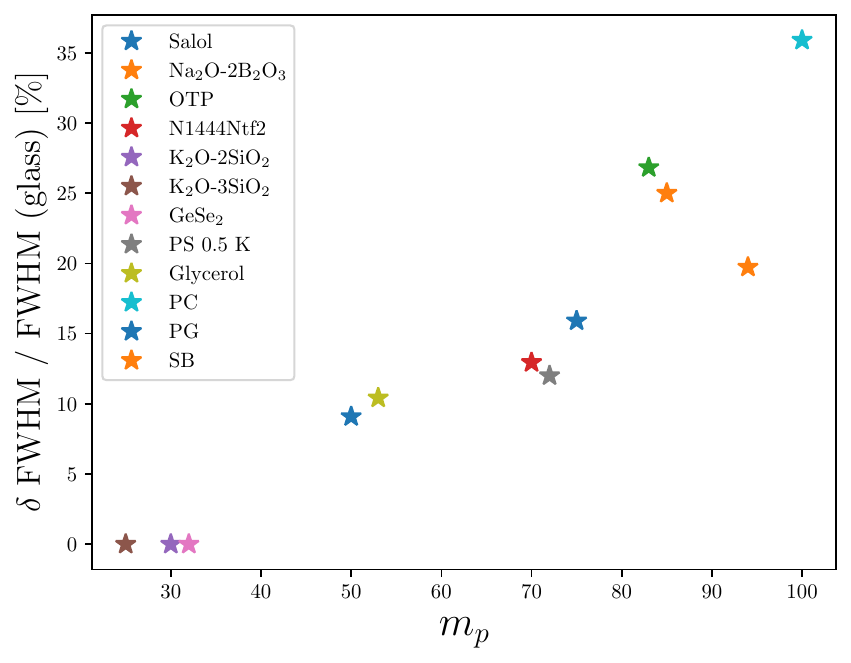}
\caption{Recreation of figure 3a from \Mycite{Voylov2016}. N1444Ntf2 is a room temperature ionic liquid, OTP  is orthoterphenyl, SB is sucrose benzoate, PC is propylene carbonate, PS is polystyrene, and PG is propylene glycol.  }
\label{fig:Voylov_fragility}
\end{figure}

	\Mycite{Ryu2020} correlate the medium-range order and the fragility for several different metallic glass-formers, using either experimental data or simulations. They find a correlation between fragility and the structural measure: 
\begin{equation}
	\frac{\xi_s(T_g)}{a}
\end{equation}

 where $\xi_s(T_g)$ is the structural correlation lengths evaluated at $T_g$ and $a$ is the average nearest-neighbor distance. A brief explanation of the origin for this correlation will be presented, but the details of the origin of the  are given in refs. \cite{Ryu2019,Ryu2020}. The motivation for this correlation, starts with the Ornstein-Zernike prediction of the behavior of the reduced pair distribution function, $G(r)$:

\begin{equation}\label{eq:G(r)_Ryu}
	G(r) = G_0(r) \exp\left(-\frac{r}{\xi_s}\right)
\end{equation}
where $G_0(r)$ the reduced PDF of the ideal glass state. The Fourier transform of a decaying exponential is a Lorentzian distribution. The Fourier transform of equation \ref{eq:G(r)_Ryu} becomes the convolution of an Lorentzian and the ideal glass structure factor $S_0(q)$. For many  metallic glass-formers the first  diffraction peak can be fitted with a Lorentzian function. This makes it possible to calculate $\xi_s$ and $a$ directly from fitting a Lorentzian to the first peak.

\begin{equation}
	\frac{\xi_s(T_g)}{a} = \frac{4}{3\pi} \frac{q_{max}(T_g)}{FWHM(T_g)}
\end{equation}

\begin{figure}[H]
	
	\centering
\includegraphics[width=0.75\textwidth]{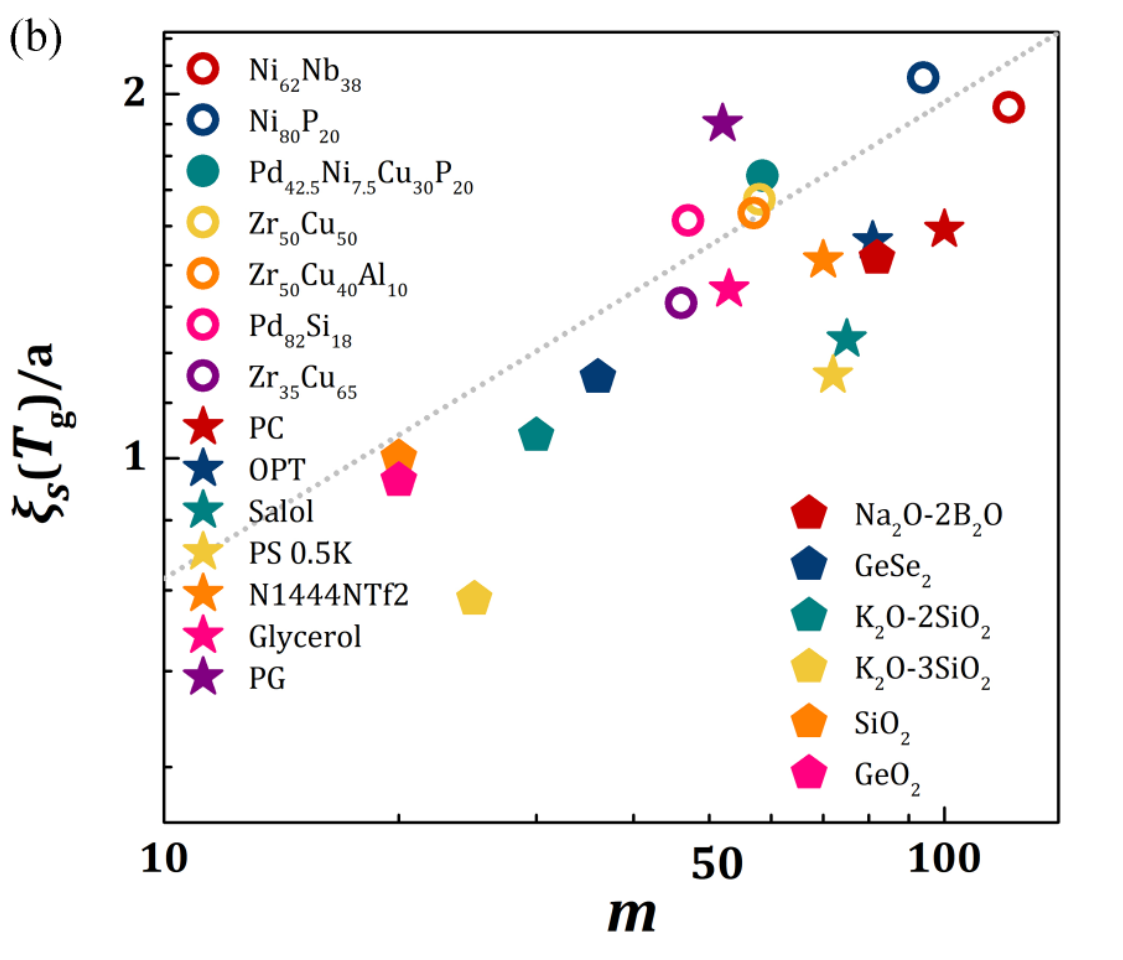}
\caption{Reprint of figure 1b from \Mycite{Ryu2020}. The round symbols are metallic glass formers and the open symbols are from simulations. The stars are data from organic glass-formers and the pentagons are network forming glasses. The data from organic glass-formers and the network-forming glasses are the same as in figure \ref{fig:Voylov_fragility}.}
\label{fig:Ryu_fragility}
\end{figure}

Where $q_{max}$ is q-value of the peak position of the first peak. In figure \ref{fig:Ryu_fragility} the proposed correlation between the structure correlation length and the fragility is shown \cite{Ryu2020}, for many different types of glass formers. 

Both \Mycite{Voylov2016} and \Mycite{Ryu2020} argue that they structural measure can be seen as describing the medium-range order. The medium-range order for metallic glass-formers are easier to interpret since the structural unit is an atom, while for molecular glass-formers the structural unit is a molecule making the medium range order less visible \cite{Ryu2020}.


\chapter[Pseudo-Isomorphs, part I]{Experimental Evidence of Pseudo-Isomorphs, part I} \label{chapter:Cumene}
 This chapter presents the results of the first systematic experimental tests of the existence of pseudo-isomorphs and the first experimental evidence of the existence of pseudo-isomorphs. For a liquid to have pseudo-isomorphs, there must exist lines in the phase diagram where, along these lines, both the structure and dynamics are invariant when presented in reduced units. The overall goal of this study is to examine variation of the structure of a molecular glass former, Cumene, along isochrones and compare it the variation along with isotherms, isobars, and isochores. In the literature, previous studies have focused on model liquids, and there are no experimental works testing the hypothesis. A few experimental studies have performed experiments where the existence of pseudo-isomorphs are not the main interest of the study, but still the results are relevant for our work. The following is a short introduction to the state of the field of research, first presenting results of simulation studies and then experimental studies.
		
\subsubsection{Simulations}

 \Mycite{Veldhorst2014} simulated a system of flexible Lennard-Jones chains with 10 rigid segments. The Lennard-Jones chain model is a course-grained model with the goal of mimicking the behavior of glassy polymer melts. In the model each Lennard-Jones particle may correspond to several monomers. The model has bonds and can therefore mimic a real-world molecular liquid. \Mycite{Veldhorst2014}  showed that although the model was not R-simple, by the simple (but arbitrary) criterion $R <0.9$, many of the predictions of isomorph theory were true for the system. They found that both the dynamics and the intermolecular structure collapsed in reduced units. The reduced unit for length $\tilde{r} = r \rho^{\frac{1}{3}}$, did not scale the intramolecular bonds. The term "Psuedo-isomorph" was coined by \Mycite{Olsen2016}, in a paper in which the same system was studied except with harmonic bonds instead of the rigid bonds used in \Mycite{Veldhorst2014}. \Mycite{Olsen2016} showed that the intramolecular degrees of freedom from harmonic bonds cause the breakdown of the system's otherwise strong virial potential energy correlations, but that many of the predictions of isomorph theory were true for these Pseudo-Isomorphs. \Mycite{ZahraaPHD} showed the existence of pseudo-isomorphs in many different models with internal degrees of freedom, including the symmetric dumbbell model, asymmetric dumbbell model and also Lewis-Wahnström OTP model  \cite{Zahraa2023}. Pseudo-isomorphs have been shown to exist for many model liquids and for more realistic models of liquids. \Mycite{Stoppelman2022} examined an all-atom propylene carbonate model using molecular dynamics simulations. They showed that the radial distribution function of the centers of mass for the molecules appeared to collapse along pseudo-isomorphic lines in the phase diagram. $g(r)$ of the center of mass for the molecules only measures the intermolecular structure, so the effects of the intramolecular bond are excluded.
\subsubsection{Experimental:}

One of the first experiments to attempt to examine the structure along an isochrone was by \Mycite{Tolle2001}, who used neutron scattering to measure the structure of orthoterphenyl (OTP) along isotherms, isobars, isochors, and isochrones. The original goal of Tolle's study was not to test the predictions of isomorph theory, but to test mode-coupling theory. The isomorph theory had not been developed yet. The results of his study are important to examine in the context of this study. In the \Mycite{Tolle2001} study, the isochore and isochrone had a shared state point, and the pressure change away from the state point was at most 55 MPa. The results were presented in experimental units, but the conclusions of Tölle was: "While along isobars and isotherms the static structure factor S(q) evolves continuously, it is nearly identical along an isochore and an isochrone"\cite{Tolle2001}. The results of \Mycite{Tolle2001} show that the structure of OTP change as a function of temperature and pressure, but with his measurements he is not able to see a difference in the structure between isochores and isochrones. Since the structure does change along isotherms and isobars, the only way for the isochores and isochrones to have the same structure is, if the isochore and isochrones are identical. This is not the case, but in a limited experimental range, it can be an approximation. The main reason Tolle is unable to separated the isochore and isochrone is likely the resolution of the measurements. The close the isochore and isochrone are in the phase diagram, the better resolution would be needed to separate the two measurements. The results of \Mycite{Tolle2001} shows that if we want to prove the existence of pseudo-isomorphs, a large experimental pressure range is needed to be able to differentiate between the isochores and isochrones. We want to prove that the structure is invariant along isochrones. To prove the structure along an isochrone is invariant, one really has to show that the structure change when not following the isochrone. So in order to prove the existence of pseudo-isomorphs, it is not enough to only show invariance along isochrones: we also have to show that the structure is not invariant along isochores.

\Mycite{Wase2020} measured the structure and dynamics of the ionic liquid "Pyr14TFSI" and tested density scaling on the liquid. In S(q), the ionic liquid has two peaks, a "charge peak" around 0.8-0.9 Å$^{-1}$, and a "structure peak" around 1.3-1.4Å$^{-1}$ \cite{Lundin2021}. They found that when plotted in reduced units, the structure peak was invariant along lines of isoconductivity, while the charge peak was not. Using density scaling, they found that the peak position of the structure peak collapses along lines of constant $\Gamma$, while the charge peak does not. \Mycite{Knudsen2024} examined a united-atom model of Pyr14TFSI and performed MD simulations of the same system \Mycite{Wase2020} measured experimentally. \Mycite{Knudsen2024} were able to show the same result of collapse of the structure peak and not the charge peak, using a similar change in density as \Mycite{Wase2020}. \Mycite{Knudsen2024} continued to compare the structure along isochrones and found that with a greater density difference, the structure peak also stopped collapsing. In \Mycite{PeterPHD} the collapse was also examined in real space with some of the partial pair distribution functions, and the structure did not appear to collapse along isochrones. If one combine the work of \Mycite{Wase2020} and \Mycite{Knudsen2024}, they have indirectly tested Pyr14TFSI for pseudo-isomorphs and found that it does not have pseudo-isomorphs. In the terminology of isomorph theory Pyr14TFSI have isodynes.
		
The existence of pseudo-isomorphs has been demonstrated in model liquids, but has only been indirectly tested for a single real-world liquid. For the existence of pseudo-isomorphs there has been no systematic test, and the few circumstantial tests that have been done have not provided experimental evidence for the existence of pseudo-isomorphs. The goal of this chapter is to present the results of one of the first systematic experimental tests for the existence of pseudo-isomorphs in real-world liquids. The sample we tested was a small molecular glass-former, cumene. The sample is introduced in detail in section \ref{sec:Cumene_sample}.  For cumene, the dynamical prediction of isomorph theory holds true in a large pressure \cite{Ransom2017} and temperature range \cite{Wase2018_nature,Wase2018,}. The candidates for pseudo-isomorphs are lines in the phase diagram with constant relaxation time, i.e. isochrones. Cumene has been shown to exhibit power-law density scaling, where the relaxation time can be described by a single variable, $\Gamma = \frac{\rho^\gamma}{T}$, where $\gamma = 4.8$ \cite{Ransom2017}. To find candidates for pseudo-isomorphs, we find lines of constant $\Gamma$ in phase diagram and measure the structure along these. 
		
The structure of this chapter is first an introduction of the sample, the experimental setup, and the process of subtracting the background for all measurements. Then, we compare the measured structure along isotherms, isobars, isochores, and isochrones. First, we simply plot the peaks next to each other, then we fit the peaks and compare the behaviors. In section \ref{sec:Cumene_MD} we use united-atom MD simulations of cumene to gain a microscopic understanding of the behavior of the structure along pseudo-isomorphs. 

\section{The sample: Cumene} \label{sec:Cumene_sample}

In this section we will give a general introduction to the sample cumene, and the equation of state (EoS) used to calculate the density.
In Appendix \ref{app:EOS} a general introduction to the Tait equation, is given. A equation of state is used to calculate the density in this chapter and chapter \ref{chapter:Diamond}, and the origin of the different expressions are also given.

\subsection{Cumene}
		
\begin{figure}[H]
			\centering
			\includegraphics[width=0.3\textwidth]{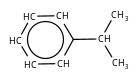}
			\caption{Cumene}
			\label{fig:cumene}
		\end{figure}

Cumene, C$_9$H$_{12}$, is a relatively small van der Waals-bonded molecular liquid that forms a glass relatively easily. The chemical structure of cumene is sketched in figure \ref{fig:cumene}. Cumene is a solvent and its main use in industry is for the production of acetone and phenol, and in nature it is often found in crude oil. For most people, it is not a chemical that affects their everyday life, but its claim to fame is that it is really an excellent model glass-forming liquid to study. It is a relatively small molecule with a simple intramolecular structure. For smaller hydrocarbons like benzene or toluene, it is more difficult to avoid crystallization when cooling. Cumene only consists of hydrogen and carbon, making it an excellent candidate for isotopic substitution, however this is beyond the scope of this study. The relatively small intramolecular structure would hopefully make it easier to separate the intra- and intermolecular contributions to the structure. Cumene tends to crystallize when heating in the supercooled regime and has a melting temperature of $T_m = 177 $ $^o$K. Cumene has a reported glass transition temperature of approximately 126-127 K at ambient pressure \cite{HansenHenrietteW.2017Cbfm}\cite{Wase2018}. The ambient pressure fragility of cumene is $m_p \approx 70$ \cite{Ransom2017,Wase2018}. The pressure dependence of the glass transition temperature was measured in a pressure range from ambient pressure to 4.55 GPa\cite{Ransom2017}. Measurements of the viscosity of Cumene have been conducted in temperatures from 300 to 130 K and at pressures up to 0.4 MPa by different authors \cite{BarlowA.J.1966Vbos}\cite{ARTAKI1985PEOT}\cite{LINGAC1968VOSO} \cite{LiG1995Patd}. The measurement of the viscosity  is discussed in detail in Chapter \ref{chapter:IDS}, where isochronal density scaling is tested.
		
\Mycite{Ransom2017} tested power-law density scaling in a pressure range from ambient pressures up to 4.55 GPa and found that with $\gamma =4.8$ literature data on the viscosity collapsed into a single curve. This is also discussed in detail in chapter \ref{chapter:IDS}, and  in figure \ref{fig:Cumene_dym_DS} the density scaling collapse is reproduced. The equation of state (EoS) used in this study was originally published in \Mycite{Ransom2017} as a part of the previously mentioned density scaling study. \Mycite{Ransom2017} extended the range of the equation of state originally published by \Mycite{Cibulka1999}. To extend the equation of state for Cumene, \Mycite{Ransom2017} use an EoS of toulene also based on work from \Mycite{Cibulka1999}. The toulene equation of state was fitted in a larger temperature range 179-583 $^o$K. Toulene and cumene have similar chemical structures. The equation of state have previously been used to show power-law density scaling of the dynamics of cumene in larger temperature and pressure range used in this experiment \cite{Ransom2017,Wase2018}. It was used in a pressure range from ambient pressure up to over 4.55 GPa and a temperature range from 130-303 K \cite{Ransom2017,Wase2018}. The temperature range used in this experiment is  $\approx$ 160 -290 K, and the highest pressure was 0.36 GPa, thus, well within the ranges previously used to fit the EoS. 

\subsubsection{Equation of state for Cumene}

The equation of state presented in \Mycite{Ransom2017} is fitted to the form shown in equation \ref{eq:EoS_v2}. In the fit, they let $C(T)$ be a temperature-dependent parameter. The equation of state is presented here:

		
\begin{equation}\label{eq:EoS_cumene}
\rho(T,P) = \frac{\rho_0(T)}{1 -C(T)\ln\left(\frac{B(T) +P}{B(T)+ P_0}\right)}
\end{equation}
		
where $P_0 = 0.0001 $ GPa, $B(T)$ and $C(T)$ are temperature-dependent parameters, and $\rho_0(T)$ is the ambient pressure density. The temperature-dependent parameter $B(T)$ is given by a fourth-order polynomial:
		
\begin{equation}\label{eq:BT}
B(T) = b_0 + b_1 \frac{T-T_0}{100}+ b_2 \left(\frac{T-T_0}{100}\right)^2 + b_3 \left(\frac{T-T_0}{100}\right)^3 + b_4 \left(\frac{T-T_0}{100}\right)^4
\end{equation}
		
where $b_0 = 0.111102$ GPa, $b_1 = -0.08954 $ GPa K$^{-1}$, $b_2 = 0.0226 $ GPa K$^{-2}$, $b_3 = -0.0034 $ GPa K$^{-3}$, $b_4 = -0.00028 $ GPa K$^{-4}$, and $T_0 = 298.15$ K.  $C(T)$ is given by
		
\begin{equation}
C(T) = 0.09373 -0.8 \left[(0.005004 \text{ K}^{-1})\frac{T-T_0}{100}\right]
\end{equation}
		
where $T_0$ is the same as in equation \ref{eq:BT}. For the temperature evolution of the ambient pressure density we use:
		
\begin{equation}
\rho_0(T) = \rho_0(T=273.2 \text{C})*(1-\alpha(T -273.2)) 
\end{equation}
		
where $\rho_{0}(T=273.2$ $ ^o$ $C) = 0.885$ g cm$^{-3}$, and $\alpha =0.000995 $ K$^{-1}$ \cite{KristinePHD}. This linear fit is based on measurements from \Mycite{BarlowA.J.1966Vbos}. The equation of state uses a fit on experimental data, and while imperfect, it can give accurate information about the density at a given state point. It is also worth stressing that this equation of state has been used successfully to find isochrones in the phase diagram \cite{Ransom2017,HansenHenrietteW.2017Cbfm,Wase2018}. Therefore even if the equation of state has some error, lines of constant $\Gamma = \frac{\rho^{4.8}}{T}$ have been shown to have invariant dynamics.  
		
\begin{figure}[H]
\centering
\includegraphics[trim = 30mm 80mm 15mm 80mm, clip,width=0.8\textwidth]{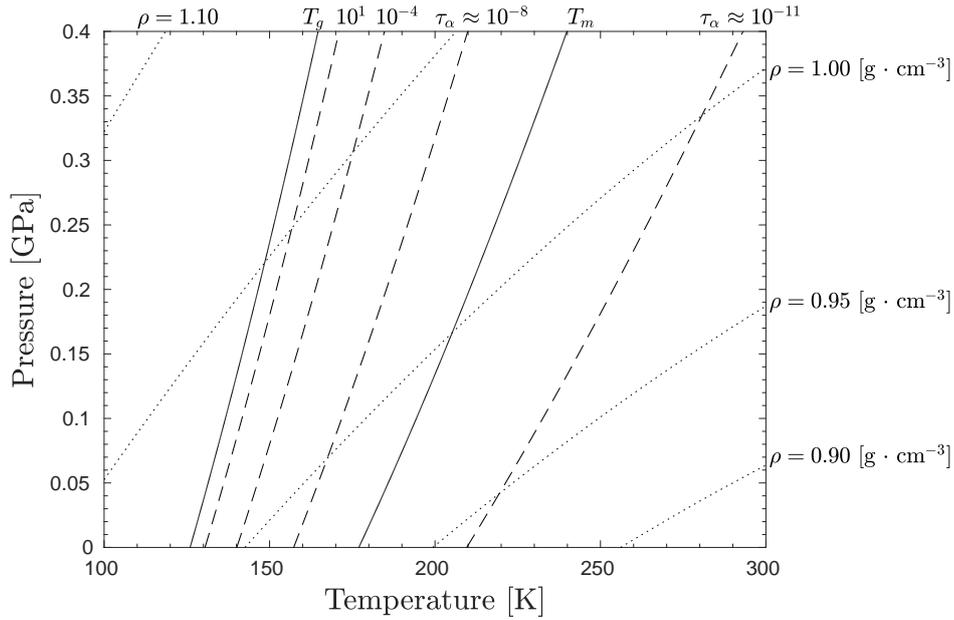}	
\caption{The P-T phase diagram of cumene, with isochores and isochrones determined using the equation of state from \Mycite{Ransom2017}. The solid lines represent the glass transition and approximate melting line, which are calculated as lines of constant $\Gamma$. The dotted lines are isochores, and the dashed lines are isochrones calculated as lines of constant $\Gamma$. The relaxation time at each isochrone is known at a single state point along each isochrone from ref. \cite{KristinePHD}. The relaxation time at each of the state points was measured at ambient pressure.}
\label{fig:phase_diagram_cumene}
\end{figure}

The phase diagram calculated from the equation of state is shown in figure \ref{fig:phase_diagram_cumene}. The solid lines represent the glass transition and melting line, extrapolated as lines of constant $\Gamma$. The melting line is approximately an isochrone \cite{Adrjanowicz2016} \cite{Pedersen2016}, while the glass transition is often experimentally defined by $\tau_{\alpha} = 100$ s. $\Gamma$ along the glass transition and melting line should be constant. The dashed lines are approximate isochrones that are calculated as lines of constant $\Gamma$. The dotted lines are isochores. The experiment covered a substantial region of the phase diagram so that we could compare the structures along isotherms, isobars, isochores, and isochrones.
		
\section{Experimental setup}
		
\begin{figure}[H]
\centering
\includegraphics[width=0.8\textwidth]{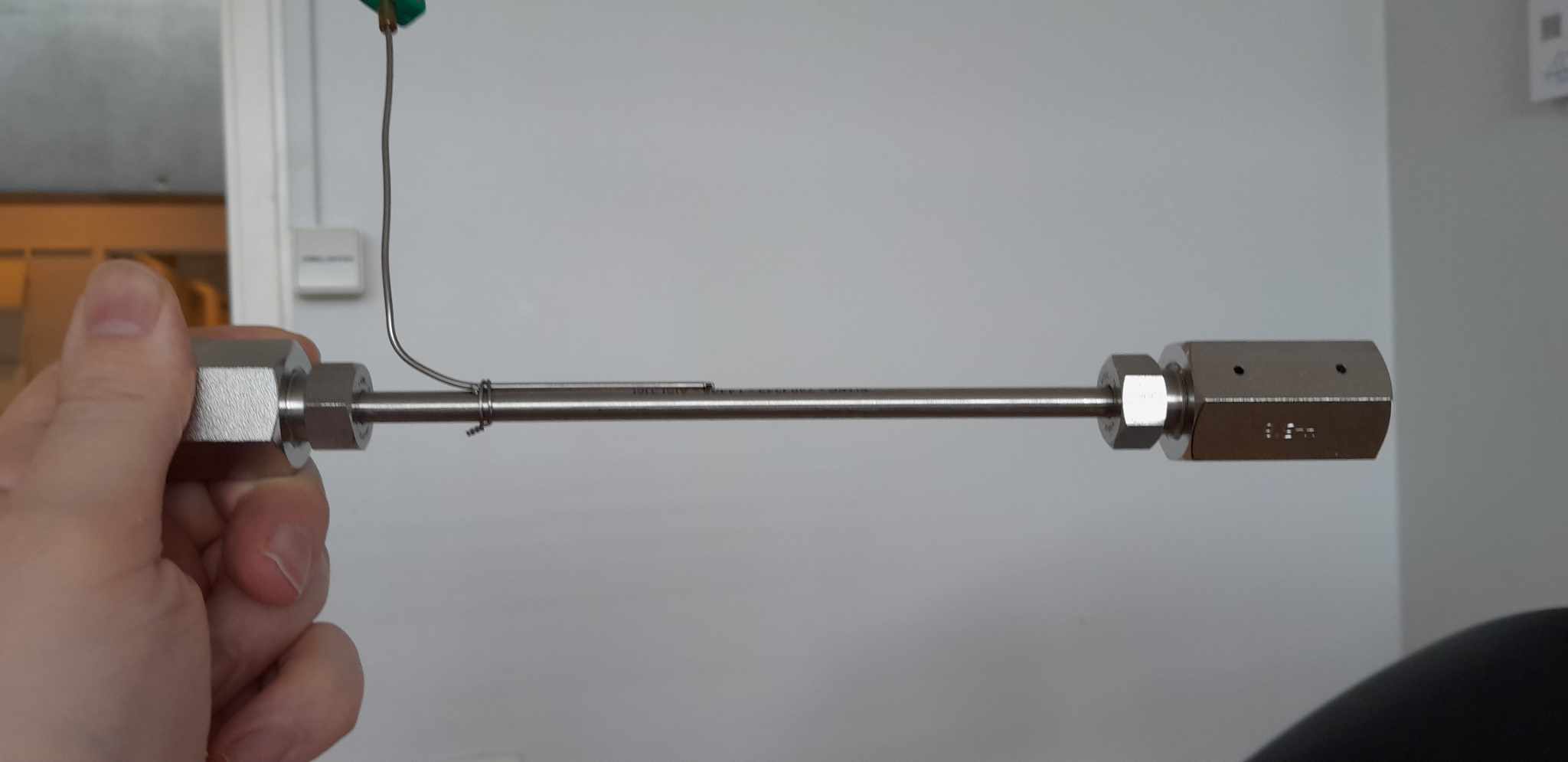}
\caption{The stainless steel nipple used as a pressure cell in the experiment (the author's fingers for size reference). The thermocouple was welded on the cell and used for temperature control during the experiment.  }
\label{fig:cell_1}
\end{figure}

The experiment was performed at the European Synchrotron Radiation Facility (ESRF) at the ID31 beamline. The cell used in  the experiment was a stainless steel nipple purchased from Sitec. The outer diameter of the cell is 6.35 mm (1/4 inch), and the inner diameter is 2.4 mm.  A picture of the cell can be seen in figure \ref{fig:cell_1}. For the experiment the energy was 76.921 keV. The high energy x-ray was needed to power through the steel walls of the cell and measure the sample. The detector used in the experiment was a Pilatus3 X CdTe 2M detector, and the sample-to-detector distance for the experiment was 417.741 mm.

\begin{figure}[H]
\centering
\includegraphics[trim = 20mm 90mm 20mm 50mm, clip,width=0.65\textwidth]{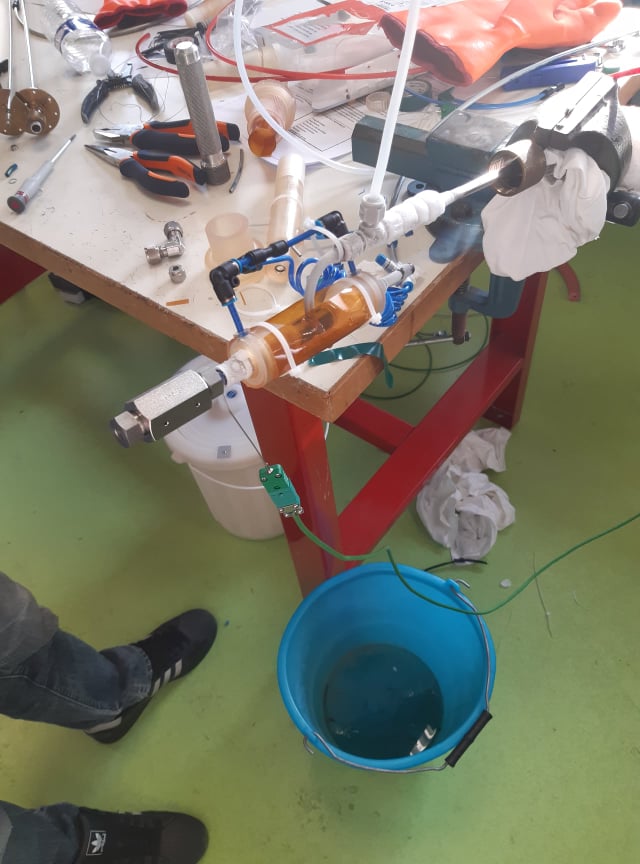}
\caption{The experimental setup as put in the beam. The pressure setup is not connected, but it is attached on the left side where the plug is. The pressure was applied directionally, and the sample was used as a pressure liquid. The temperature control is attached, with the cold flow coming from the cold tube, while the blue tubing controls the heat flow.   }
\label{fig:cell_2}
\end{figure}

In figure \ref{fig:cell_2} the experimental setup is shown with the cooling setup attached. The temperature was controlled by a constant cold flow and a varying heat flow. The cold flow of nitrogen gas was coming from a liquid nitrogen dewar. The cold flow is generated by natural evaporation of the liquid nitrogen. The cold flow could be slightly regulated by opening the valve.  The blue tubing in figure \ref{fig:cell_2} is heat flow. The heat flow was nitrogen gas at ambient temperature, the flow speed of ambient temperature nitrogen could be controlled to regulate the temperature. The pressure cell is shielded by capton foil to prevent ice condensation on the steel nipple. In figure \ref{fig:cell_2} the pressure setup is not attached to the cell. The pressure is applied though the left side of the cell, in figure \ref{fig:cell_2}. The pressure is applied through a flexible tube/capillary, and the pressure liquid is the sample itself. When the pressure is applied this way, if the liquid is at a very viscous state it can behave like a solid, giving uneven pressure distributions. When measuring along isobars, pressure was applied at high temperature and then cooled. When measuring along isotherms, the state points are chosen so that the sample is fluid enough to avoid this issue. 
		
The measuring protocol was along isotherms to measure in jumps of 10 $^o$K. After reaching the target temperature, we waited 10 minutes at the target temperature before measuring. This gives plenty of time for the liquid to reach equilibrium, given that the glass transition has an approximate relaxation time of 100s. The 10 minute wait was chosen to ensure that the measured temperature and the sample temperature had a chance to reach thermal equilibrium. The count time at each state point was 60s, and at each state point, we performed five measurements. To measure the isobars the pump was driven manually, meaning we had to enter the experimental hutch to change the pressure. The time taken to evacuate the hutch and prepare to measure was assumed sufficient for the sample to reach the target pressure. A DOI number was created for the data of the experiment, and the raw data will be made public via ref. \cite{ESRFdata} by the end of 2024.
		
\section{Results:} \label{sec:Cumene_exp_results}
		
\begin{figure}[H]
			\centering
			\includegraphics[trim = 30mm 80mm 15mm 80mm, clip,width=1.0\textwidth]{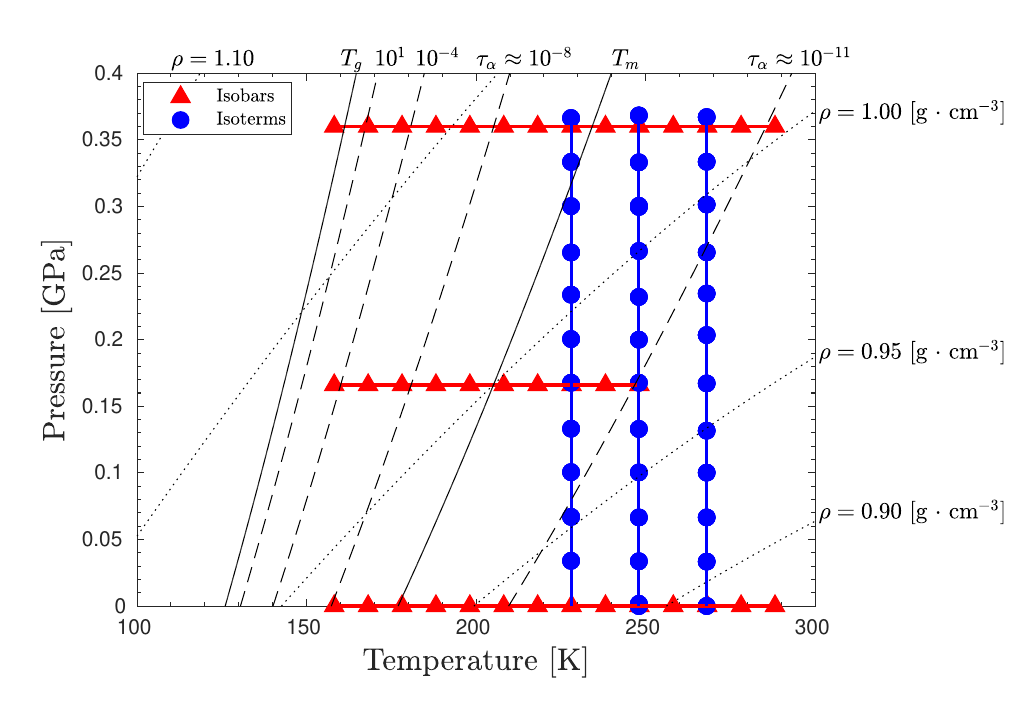}
			\caption{The state points measured in the experiment. We mapped the phase diagram by measuring along three isotherms and three isobars. The solid lines represent the glass transition and the approximate melting line, which are calculated as lines of constant $\Gamma$. The dotted lines represent isochores, and the dashed lines are isochrones calculated as lines of constant $\Gamma$.}
			\label{fig:phase_diagram_measurements_cumene}
		\end{figure}

In the experiment, we measured along three isotherms at 228 $^o$K, 248 $^o$K, and 268 K$^o$, and along three isobars at ambient pressure, 0.166 GPa, and 0.360 GPa. We wish to be able investigate enough of the phase diagram, to be able to compare the structure along isotherms, isobars, isochores and isochrones. The measured state points are shown in Figure \ref{fig:phase_diagram_measurements_cumene}.
		
		\subsection{Subtracting the background}
		
\begin{figure}[H]
			\begin{subfigure}[b] {0.495\textwidth}
				\includegraphics[trim = 30mm 80mm 40mm 80mm, clip=true,width=0.99\textwidth]{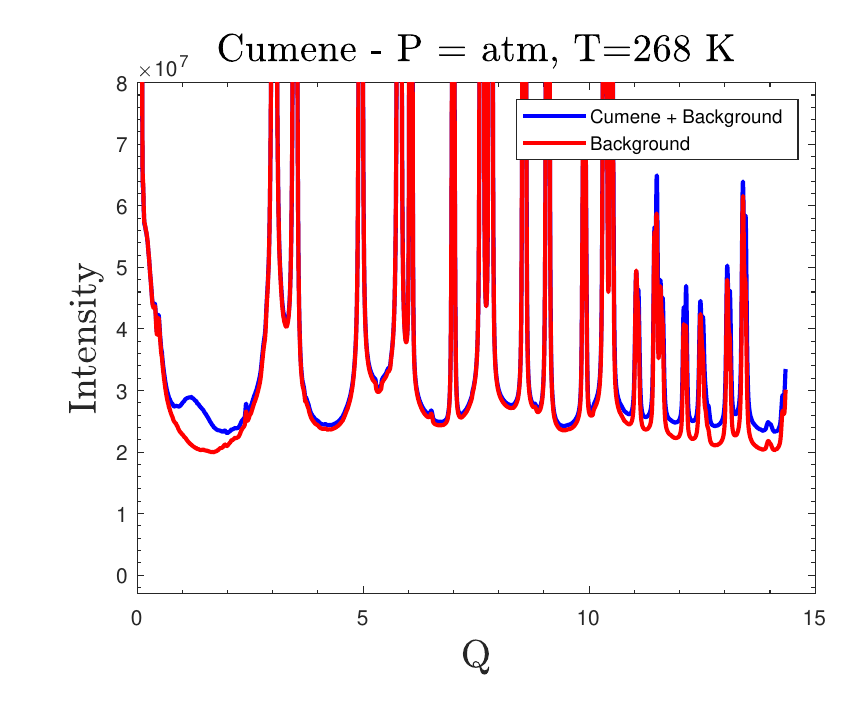}
				\caption{The empty cell in red and the cumene signal in the cell in blue for the total measured signal.}
				\label{fig:background_subtracktion_1_a}
			\end{subfigure}\hfill
			\begin{subfigure}[b] {0.495\textwidth}
				\includegraphics[trim = 30mm 80mm 40mm 80mm, clip=true,width=0.99\textwidth]{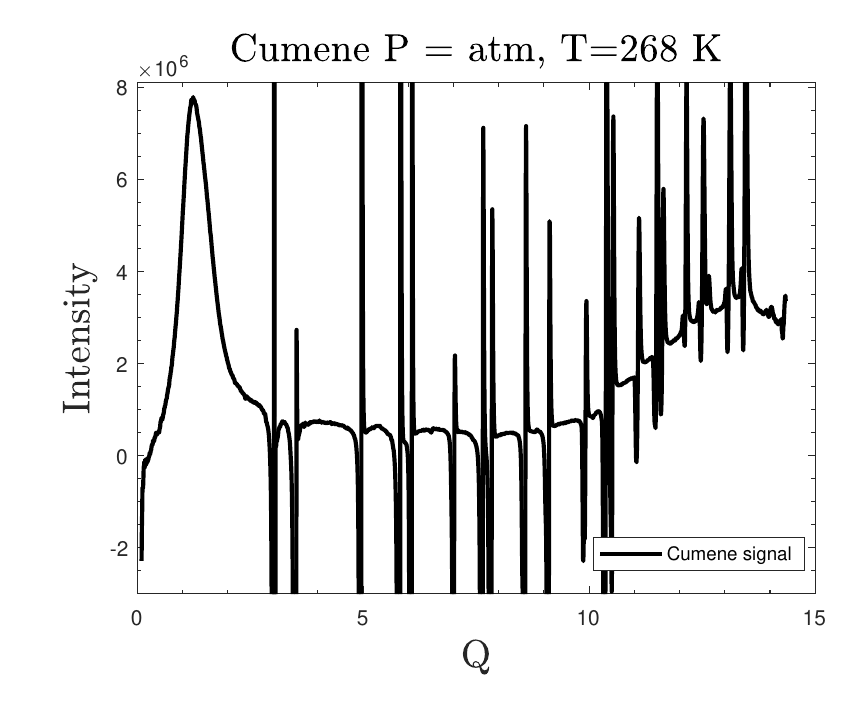}	
				\caption{The cumene signal calculated by subtracting the empty cell from the measurement.}	
				\label{fig:background_subtracktion_1_b}
			\end{subfigure}
			\caption{In subfigure \ref{fig:background_subtracktion_1_a} the background is plotted in red, and the background plus cumene signal is plotted in blue, for a state point at ambient pressure and T = 268 $^o$K. The isolated cumene signal is plotted in subfigure \ref{fig:background_subtracktion_1_b}. After q $\approx$ 3 Å$^{-1}$ the measured signal drowns in Bragg peaks from the cell, however it is possible to isolate the signal from the first peak.}
			\label{fig:background_subtracktion_1}
		\end{figure}

		To subtract a background from the measured data, we measured a temperature dependent background in the temperature range from 298-158 $^o$K, in jumps of 10 $^o$K. The measured background was the empty cell. It was not possible to measure a pressure dependent background. Due to the high pressures involved, it would not be safe to use a gas to apply the pressure. If another liquid was used to apply the pressure, it would most likely have an overlapping liquid signal that would influence the cumene signal. For each state point the background subtracted was that for the temperature closest to the measurement. The count time for each state point was 60 s. In figure \ref{fig:background_subtracktion_1} examples of the measured signal are shown. In subfigure \ref{fig:background_subtracktion_1_a}  the background and background + cumene signal are plotted on top of each other.  In subfigure \ref{fig:background_subtracktion_1_b} the background subtracted cumene signal is shown. After q $\approx 3 $ Å$^{-1}$ the cumene signal drowns in the remnants of the Bragg peaks from the cell, which are not possible to remove completely. In the very low q-range, the cumene signal becomes negative, but there is a window where we can measure the first structure peak of cumene. 
		
\begin{figure}[H]
			\begin{subfigure}[b] {0.495\textwidth}
				\includegraphics[trim = 30mm 80mm 40mm 80mm, clip=true,width=0.95\textwidth]{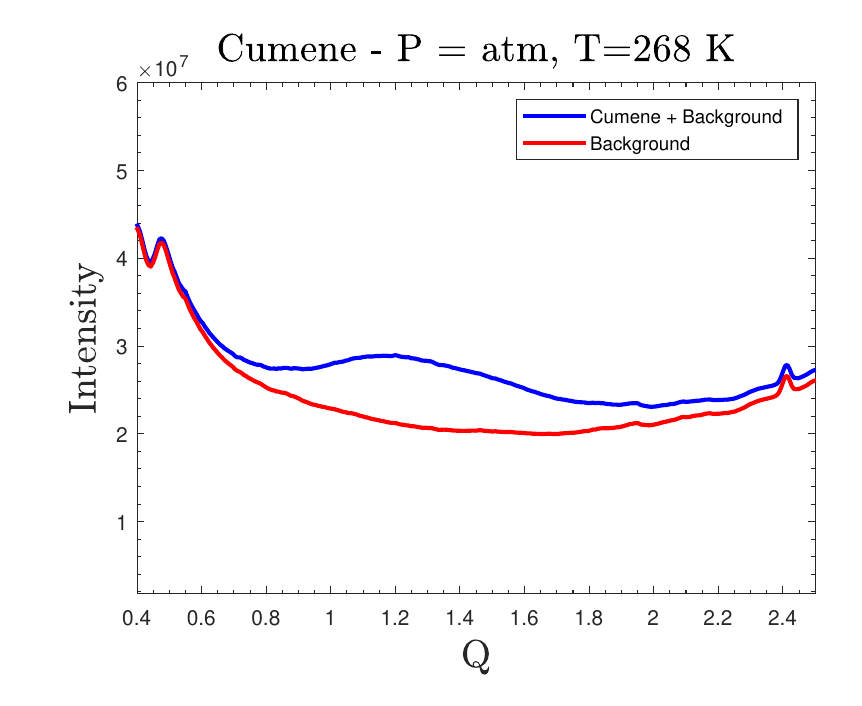}
				\caption{}
				\label{fig:background_subtracktion_2_a}
			\end{subfigure} \hfill
			\begin{subfigure}[b] {0.495\textwidth}
				\includegraphics[trim = 30mm 80mm 40mm 80mm, clip=true,width=0.95\textwidth]{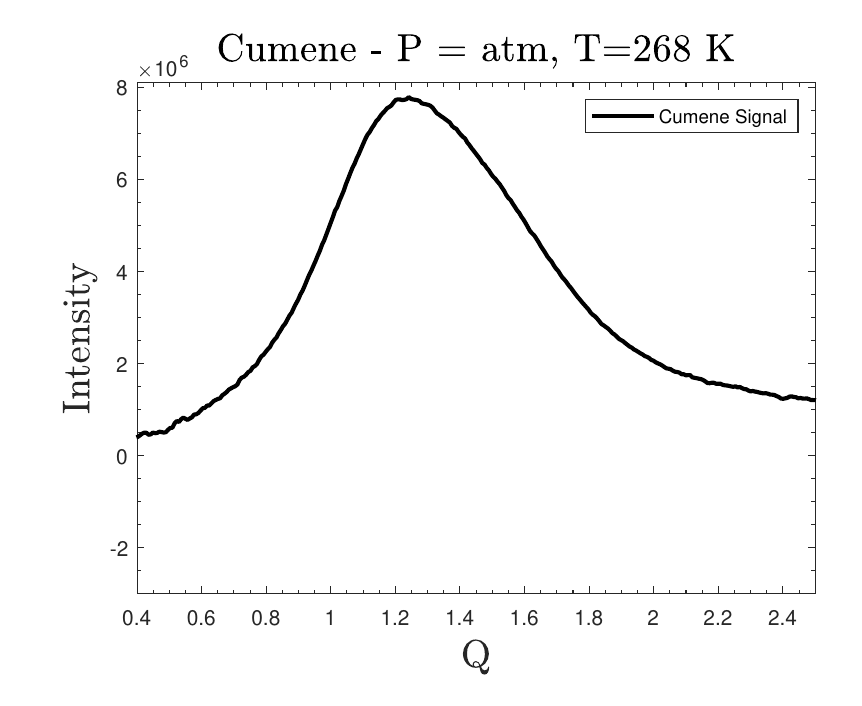}	
				\caption{}
				\label{fig:background_subtracktion_2_b}
			\end{subfigure}
			\caption{In \ref{fig:background_subtracktion_2_a} the empty cell signal is plotted in red and the cumene + cell signal in blue for the 1st structure peak. In \ref{fig:background_subtracktion_2_b},  the isolated cumene signal for the main structure peak.  }
			\label{fig:background_subtracktion_2}
		\end{figure}

\begin{figure}[H]
			\begin{subfigure}[b] {0.495\textwidth}
				\includegraphics[trim = 30mm 80mm 40mm 80mm, clip=true,width=0.95\textwidth]{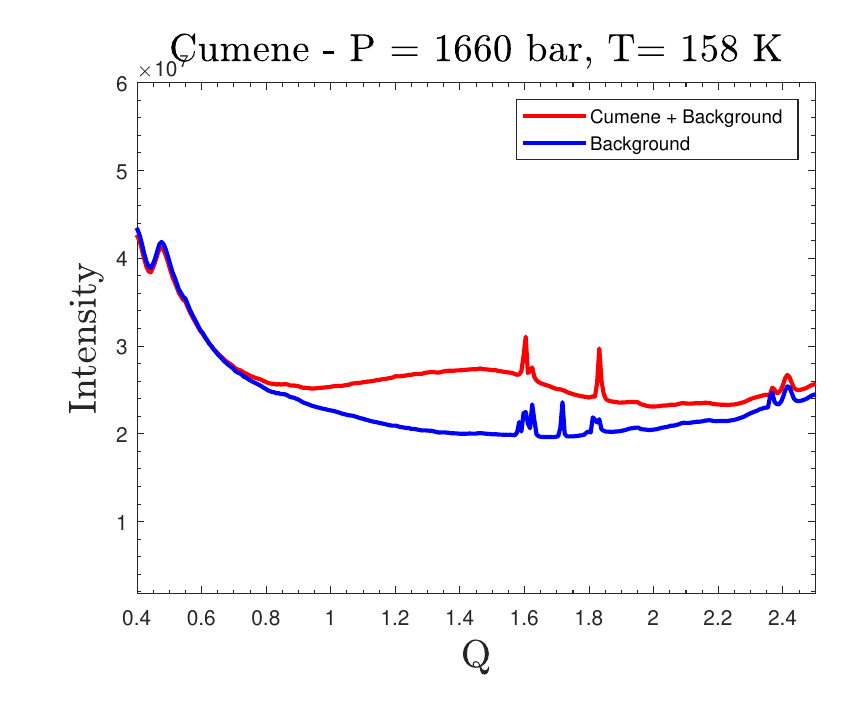}
				\caption{}
				\label{fig:background_subtracktion_3_a}
			\end{subfigure}
			\begin{subfigure}[b] {0.495\textwidth}
				\includegraphics[trim = 30mm 80mm 40mm 80mm, clip=true,width=0.95\textwidth]{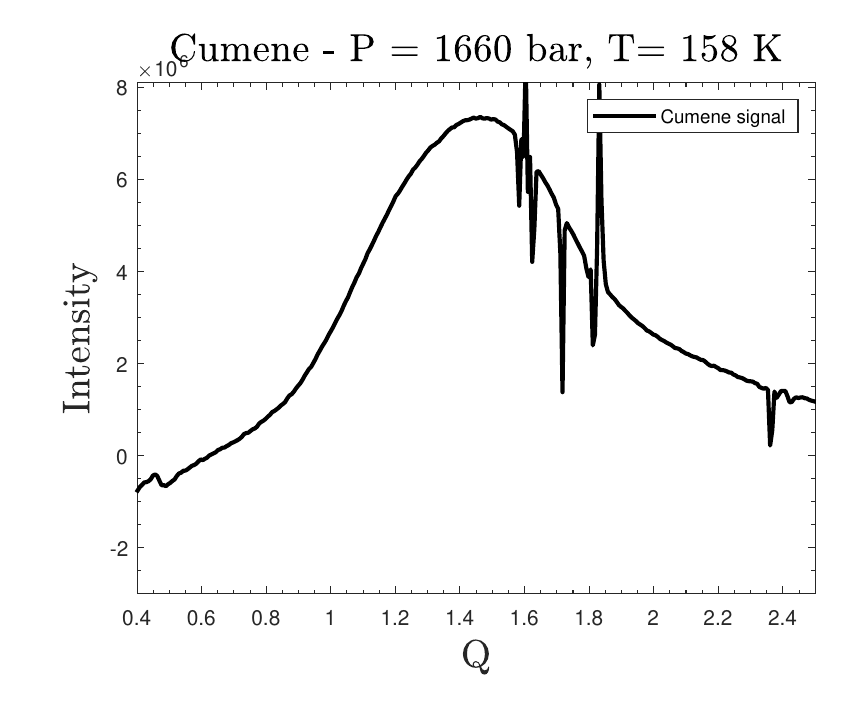}	
				\caption{}
				\label{fig:background_subtracktion_3_b}
			\end{subfigure}
			\caption{In subfigure \ref{fig:background_subtracktion_3_a} the empty cell (blue) and the cell + cumene signal (red) are plotted at P = 166 MPa, T =158 $^o$K. The effect of ice condensating on the cell can be seen. Ice can be seen in both the background and cumene signals. In subfigure \ref{fig:background_subtracktion_3_b} the cumene signal for the first peak calculated by subtracting the empty cell from the measurement. The effect of ice condensation on the cell can be seen on the isolated signal. The dips in the signal are when the ice signal is stronger in the measurement of the empty cell than the measurement of the background +  cumene signal.}
			\label{fig:background_subtracktion_3}
		\end{figure}

		In subfigure \ref{fig:background_subtracktion_2_a} and \ref{fig:background_subtracktion_2_b} the main structure peak of the cumene signal is shown. We were able to isolate the main structure peak of cumene at higher temperature. At lower temperatures, ice condenses on the cell both in the background and in the sample measurements. This can be seen by the characteristic triple peaks of hexagonal ice around 1.6 Å$^{-1}$\cite{IceCrystalPattern}. This caused problems on the right side of the peak, where the positive and negative Bragg peaks formed. The positive signal arises when we subtract too little ice signal, and the negative signal when we subtract too much ice. This is shown on subfigure \ref{fig:background_subtracktion_3_a}  and \ref{fig:background_subtracktion_3_b}. When cooled the ice evolved continuously and at different distances to the detector resulting in differences in the ice signal at each measurement. The problem with ice was only an issue for the isobaric measurements.

		
		\subsection{The temperature and pressure dependent behavior of the peak.}
		
\begin{figure}[H]
			\centering
			\includegraphics[trim = 30mm 80mm 40mm 80mm, clip=true,width=0.65\textwidth]{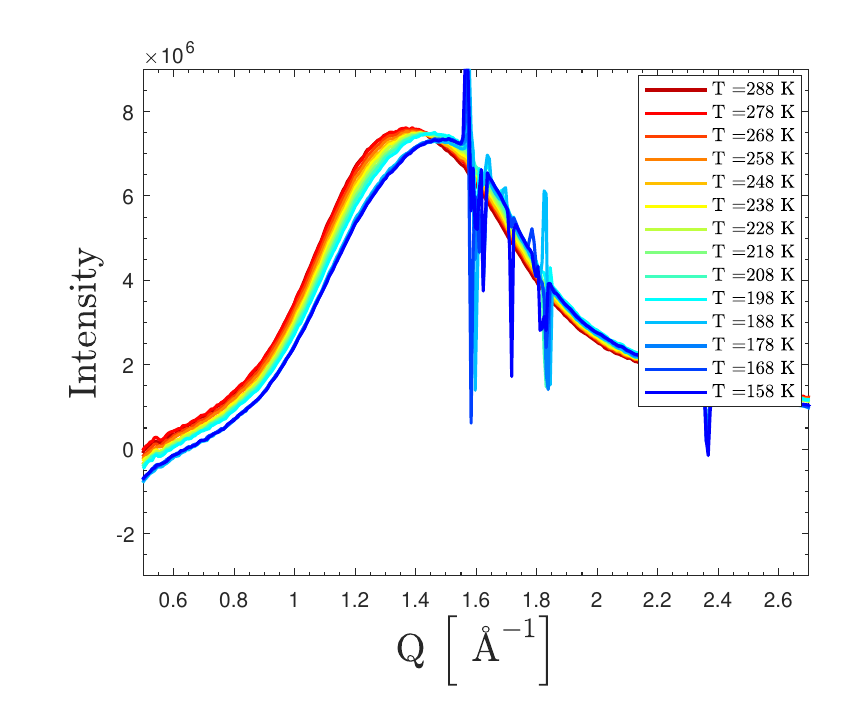}
			\caption{The evolution of the main structure peak along the 360 MPa isobar. Bragg peaks around 1.5-1.8 Å$^{-1}$ are from ice condensation on the steel tube \cite{IceCrystalPattern}. When the temperature was decreased, the peak moved to higher q values. The color scheme going from red to blue shows the temperature going from warmer to colder.  }
			\label{fig:cumene_3600_isobar}
			
		\end{figure}

\begin{figure}[H]
			\centering
			\includegraphics[trim = 30mm 80mm 40mm 80mm, clip=true,width=0.65\textwidth]{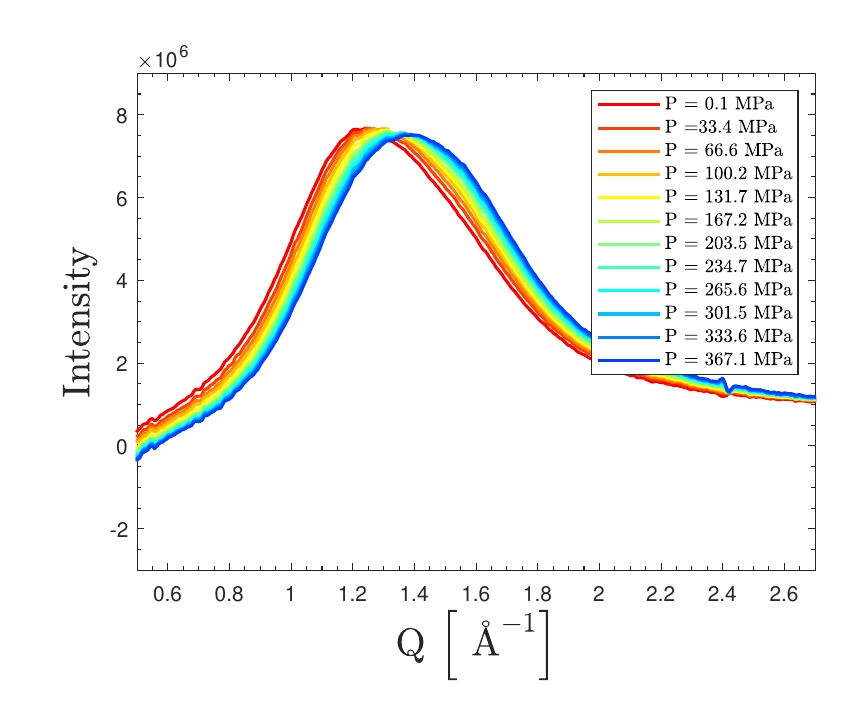}
			\caption{The pressure dependence of the structure along the 268 K isotherm. The color scheme changes from red to blue and shows the pressure going from ambient pressure to 360 MPa. When the pressure was increased, the peak moved to higher q values.  }
			\label{fig:cumene_268_isoterm}
		\end{figure}

		The temperature and pressure dependent behavior of the isolated cumene peak is examined in the following section. In this section, all the isolated cumene signals that are analyzed in a later section are presented.  In figure \ref{fig:cumene_3600_isobar} the main peak along the 360 MPa isobar is plotted. From the phase diagram in figure \ref{fig:phase_diagram_measurements_cumene} it can be seen that cumene should enter the glass along this isobar. The ice condensation is clearly visible on all the isobaric measurements. In figure \ref{fig:cumene_268_isoterm}, the pressure dependence of the main peak is shown. In general, the effect of increasing pressure seems similar to the effect of decreasing temperature, as the peak moves to higher q values, indicating that the molecules are moving closer together.

\begin{figure}[H]
			\begin{subfigure}[b] {0.495\textwidth}
				\centering
				\includegraphics[trim = 30mm 80mm 40mm 80mm, clip=true,width=0.99\textwidth]{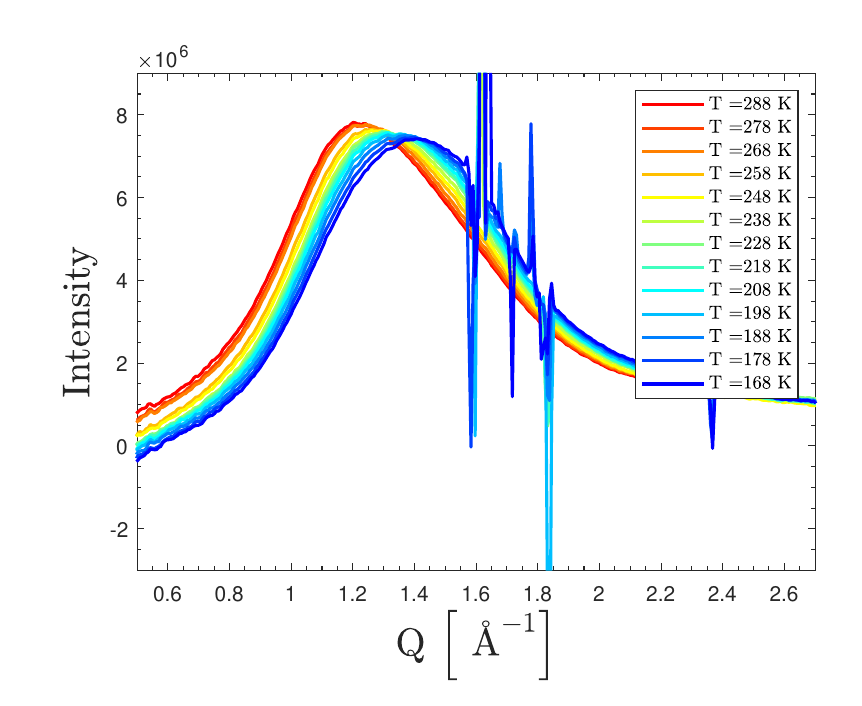}
				\caption{Isobar 0.1 MPa}
				\label{fig:cumene_atm_isobar}
			\end{subfigure}
			\begin{subfigure}[b] {0.495\textwidth}
				\centering
				\includegraphics[trim = 30mm 80mm 40mm 80mm, clip=true,width=0.99\textwidth]{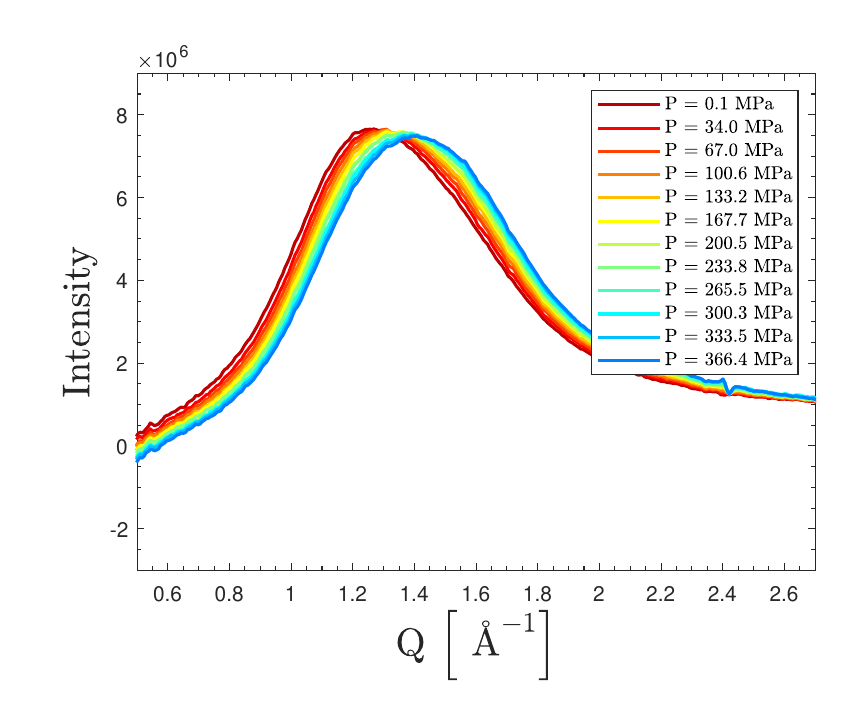}
				\caption{Isotherm 248 $^o$K }
				\label{fig:cumene_248_isoterm}
			\end{subfigure}
			\begin{subfigure}[b] {0.495\textwidth}
				\centering
				\includegraphics[trim = 30mm 80mm 40mm 80mm, clip=true,width=0.99\textwidth]{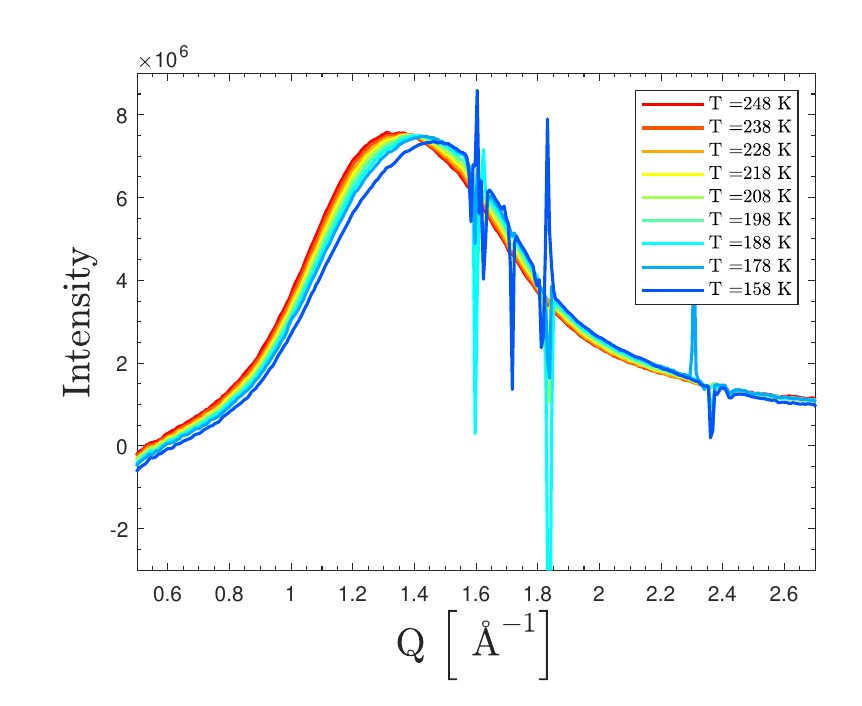}
				\caption{Isobar 166 MPa}
				\label{fig:cumene_1660_isobar}
			\end{subfigure}
			\begin{subfigure}[b] {0.495\textwidth}
				\centering
				\includegraphics[trim = 30mm 80mm 40mm 80mm, clip=true,width=0.99\textwidth]{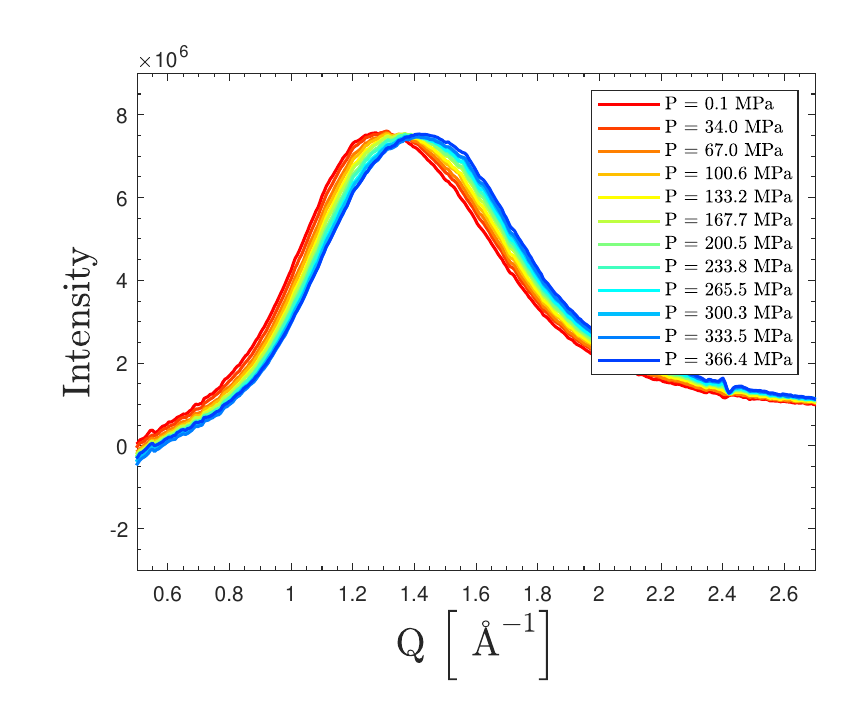}
				\caption{Isotherm 228 $^o$K }
				\label{fig:cumene_228_isoterm}
			\end{subfigure}
			\caption{The structure along the P = 0.1 MPa and P = 166 MPa isobars and T = 248 K and T = 228 K isotherms. The color scheme changes from red to blue and indicates going from low to high pressures for the isotherms, and high to low temperatures for the isobars. The structures along the isobars behave similarly to each other when compared to figure \ref{fig:cumene_3600_isobar}. The pressure dependence of the peak is similar for all three isotherms.}
		\end{figure}

		The peak of cumene seems to have some asymmetry at high temperatures and low pressures, and with increasing pressures and decreasing temperatures the peak becomes more symmetric as the peak position increases to higher q-values. The asymmetry can be seen in figures \ref{fig:cumene_3600_isobar} and \ref{fig:cumene_268_isoterm}, but also in the figures showing the background subtraction, see figure \ref{fig:background_subtracktion_2_b} and \ref{fig:background_subtracktion_3_b}. The structures along the other two isotherms are shown in figure \ref{fig:cumene_248_isoterm} and \ref{fig:cumene_228_isoterm}, and the structures along the other two isobars are shown in figure \ref{fig:cumene_atm_isobar} and  \ref{fig:cumene_1660_isobar}. In general, the structures along the isotherm and isobars behave similarly, as decreasing temperature and increasing pressure have a similar effect on the structure. 
		
		For the figures of both the isobars and isotherms, the peak is plotted in the q-range from 0.5-2.7 Å$^{-1}$. It is worth noting that around q $\approx$ 0.5-0.7 Å$^{-1}$, the background subtraction is not perfect as the intensity becomes negative. An example of this can be seen in figure \ref{fig:cumene_3600_isobar}, but it is present in all figures in this section. The background was measured only as a function of temperature, so pressure dependent changes to the background were not captured. Along the isotherms, the subtracted background was the same for all measurements, but the pressure effect on the background can be seen. However, a clear cumene peak could be observed for all measurements.
		
		\subsection{Structure along isochrones} \label{sec:cumene_exp_isochrone}
		
\begin{figure}[H]
			\centering
			\begin{subfigure}[b] {0.49\textwidth}
				\includegraphics[trim = 30mm 80mm 40mm 80mm, clip=true,width=0.99\textwidth]{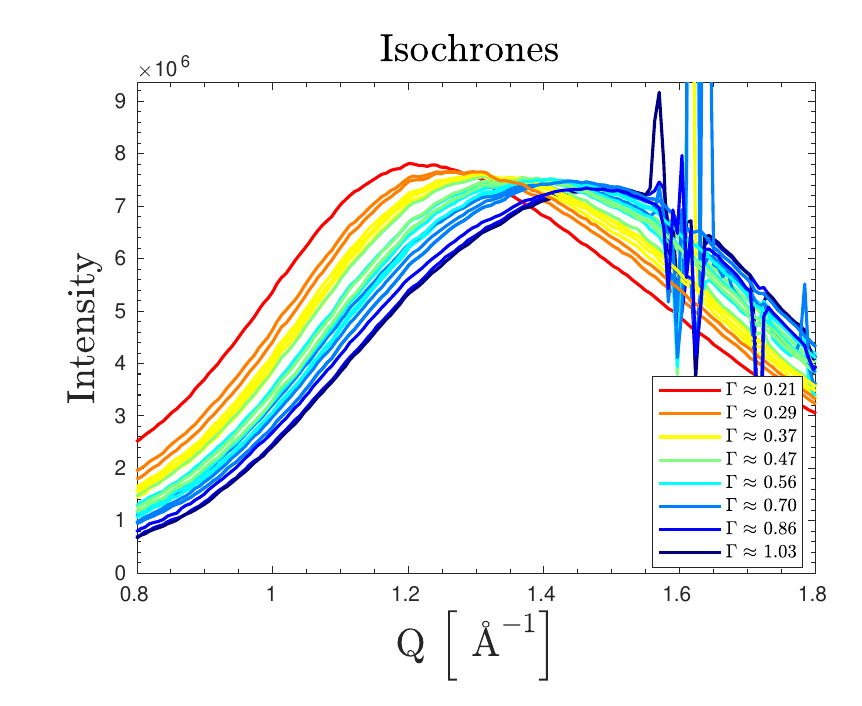}
				\caption{The measured first peak along different lines of constant $\Gamma$ in experimental units. $\Gamma$ is normalized to be 1 at $T_g$}
				\label{fig:rainbow_isochrones_a}
			\end{subfigure}\hfill
			\begin{subfigure}[b] {0.49\textwidth}
				\includegraphics[trim = 30mm 80mm 40mm 80mm, clip=true,width=0.99\textwidth]{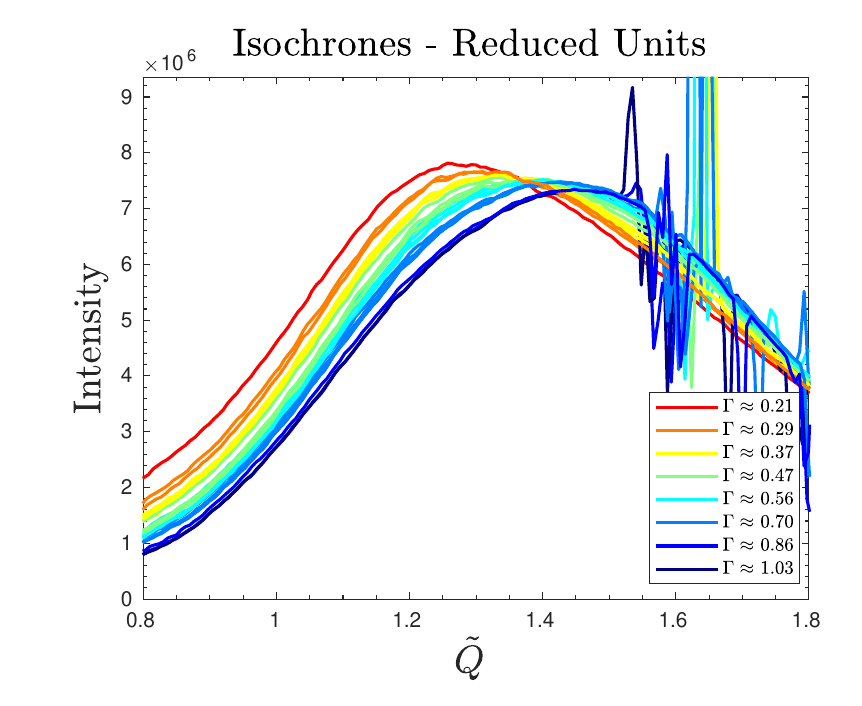}	
				\caption{The measured first peak along different isochrones in reduced units. $\Gamma$ is normalized to be 1 at $T_g$}	
				\label{fig:rainbow_isochrones_b}
			\end{subfigure}
			\begin{subfigure}[b] {0.70\textwidth}
				\centering
				\includegraphics[trim = 30mm 80mm 15mm 80mm, clip=true,width=0.99\textwidth]{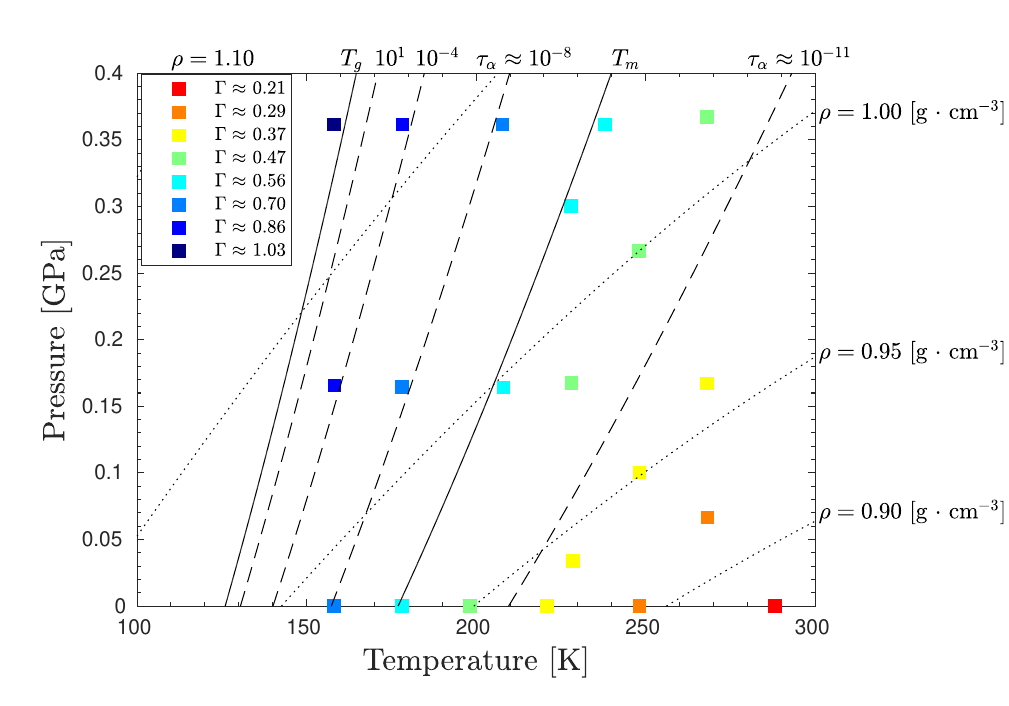}
				\caption{The state points plotted in subfigure (a) and (b). $\Gamma$ is normalized to be 1 at $T_g$.}	
				\label{fig:rainbow_isochrones_c}
			\end{subfigure}
			\caption{The main structure peak measured along many different isochrones presented in experimental units, shown in subfigure \ref{fig:rainbow_isochrones_a}, and in reduced units, $\tilde{q} =q \rho^{-\frac{1}{3}}$, shown in subfigure \ref{fig:rainbow_isochrones_b}. There are two main point we wish to show with this figure. Firstly, when the structure along the isochrones is presented in reduced units, the first structure peaks collapse upon each other. Secondly, the measured structure changes when moving from isochrone to isochrone.  $\Gamma$ is normalized to be 1.00 at $T_g$. In subfigure \ref{fig:rainbow_isochrones_c} the positions of the plotted state points are shown.  }
			\label{fig:rainbow_isochrones}
		\end{figure}

		The hypothesis we aimed to test is that cumene has pseudo-isomorphs. The relaxation time along a pseudo-isomorph is also invariant, so the experimental candidate for a pseudo-isomorph is an isochrone. The first and simplest approach to test for pseudo-isomorphs is to plot the structure in reduced units along an isochrone. In figure \ref{fig:rainbow_isochrones} the evolution of the 1st structure peak of cumene is shown for several state points along different isochrones. Here, the color represents different values of  $\Gamma$-value, and thereby different candidates for pseudo-isomorphs. In Figure \ref{fig:rainbow_isochrones_a} measurements at 20 different state points on 8 different isochrones are shown. In figure \ref{fig:rainbow_isochrones_b} the same measurements of the first structure peak are plotted in reduced units. The state point positions in the phase diagram are shown in figure \ref{fig:rainbow_isochrones_c}. The measurements along an isochrone collapse quite well on top of each other, and the peak shape and position of the scattered intensities differ when moving from isochrone to isochrone.  
		
\begin{figure}[H]
			\centering
			\begin{subfigure}[t]{0.54\textwidth}
				\includegraphics[trim = 30mm 80mm 15mm 80mm, clip=true,width=0.99\textwidth]{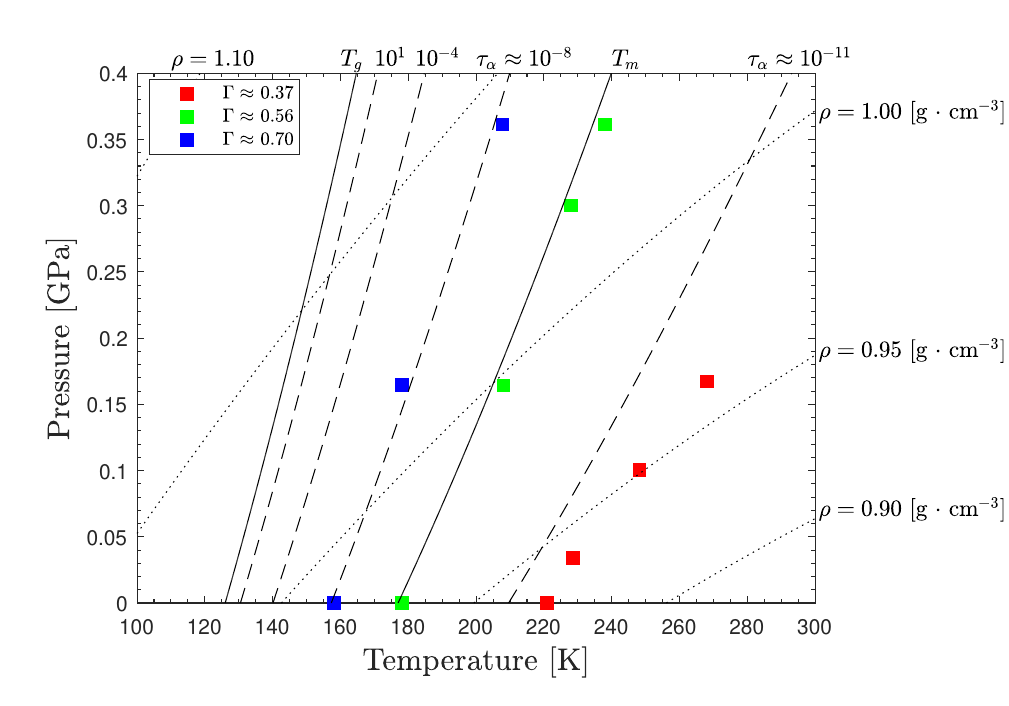}
				\caption{The state points plotted in figure \ref{fig:rainbow_zoom_b}}
				\label{fig:rainbow_zoom_a}
			\end{subfigure}
			\begin{subfigure}[t]{0.45\textwidth}
				\includegraphics[trim = 30mm 80mm 40mm 80mm, clip=true,width=0.99\textwidth]{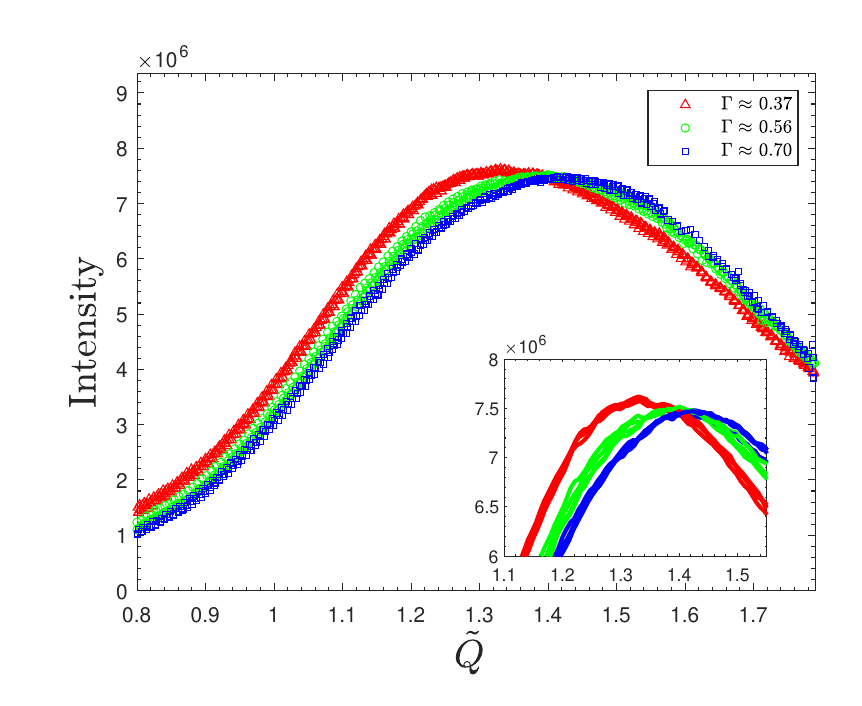}
				\caption{The main structure peak of Cumene plotted in reduced units, $\tilde{q} =q \rho^{-\frac{1}{3}}$}
				\label{fig:rainbow_zoom_b}	
			\end{subfigure}
			\caption{The structure along the three isochrones are presented in reduced units. In subfigure \ref{fig:rainbow_zoom_a} the position in the phase diagram of the 11 state points plotted in subfigure \ref{fig:rainbow_zoom_b}. The 11 measurements collapse into three distinct peaks, dependent only on which isochrone the state point is on. The signals from ice crystallizing on the cell have been removed. }
			\label{fig:rainbow_zoom}
		\end{figure}

		In Figure \ref{fig:rainbow_zoom} we have focused on three of the isochrones along which we have the most state points. In subfigure \ref{fig:rainbow_zoom_a} the position of the 11 state points in the phase diagram are shown, while in subfigure \ref{fig:rainbow_zoom_b} the main structure peak is shown in reduced units. The first structure peak collapsed quite well on top of each other when presented in reduced unit.
		
		The two main points that readers should take from the two figures, \ref{fig:rainbow_isochrones} and \ref{fig:rainbow_zoom} is firstly, the first structure peak is invariant along an isochrone when presented in reduced units. Second, the shape and position of the first structure peak visibly change when moving from isochrone to isochrone.
		
		In this subsection, we have demonstrated that the structure collapses nicely along an isochrone and that the structure is different between isochrones. It is also worth noting that the whole peak along an isochrone collapses, not just part of the peak. This is somewhat obscured by the ice peaks on the right side of the peak, but the parts of the peak that we can compare collapse. To investigate the general behavior of the peak, we fit the peak in the next section.
		
		\subsection{Fitting the peak} \label{sec:Split_Voigt}
		
		The experimental measurement of the first structure peak was fitted with a split pseudo-Voigt distribution. The pseudo-Voigt is a weighted sum of a Lorenzian and a Gaussian distribution. The split pseudo-Voigt are two pseudo-Voigt distributions, a lower and a higher profile, forced to meet at the center position of the peak.  We do not have a model to fit against, so the reason for choosing this distribution is not to reflect anything physical. It is just very good for in describing the peak. The six parameters we use to fit:
		
		\begin{align}
			p(1) &= \text{Peak value}\\
			p(2) &= \text{Center position}\\
			p(3) &= \text{Lower FWHM}\\
			p(4) &= \text{Lower profile shape factor}\\
			p(5) &= \text{Higher FWHM}\\
			p(6) &= \text{Higher profile shape factor}
		\end{align}
		
		The fitting function is a combined function with lower and higher parts based on the center position: 
		\begin{align}
			F(x) = F_l(x) \quad \text{if} \quad x < p(2) \nonumber\\
			F(x) = F_u(x) \quad \text{if} \quad x \geq p(2) \nonumber
		\end{align}
		
		Where
		\begin{equation}
			F_l(x) = p(1) \left(p(4) \frac{1}{1+\left(\frac{2(x-p(2))}{p(3)}\right)^2} + \left(1 - p(4)\right) \exp\left(-\frac{1}{2}\left(\frac{\left(x -p(2)\right)2 \sqrt{2 \log\left(2\right)}}{p(3)}\right)^2\right)\right)
		\end{equation}
		
		\begin{equation}
			F_u(x) = p(1) \left(p(6) \frac{1}{1 +\left(\frac{2(x-p(2))}{p(5)}\right)^2} + \left(1 - p(6)\right) \exp\left(-\frac{1}{2}\left(\frac{\left(x -p(2)\right)2 \sqrt{2 \log\left(2\right)}}{p(5)}\right)^2\right)\right)
		\end{equation}

		From the fitting, we used three measurements to describe the peak for each state point, the peak position,$q_{max}$, the full width half maximum of the peak, $FWHM$, and the "skewness" of the peak. The relationship between the lower FWHM and higher FWHM:
		
		\begin{align*}
			q_{max} &= p(2)\\ 
			\text{FWHM} &= \frac{p(3) +p(5)}{2}\\
			\text{Skewness} &= \frac{p(3)}{p(5)}
		\end{align*}
		
		The skewness measure is one if the peak is perfectly symmetric. If the skewness is less than unity the peak is skewed to the left, while skewness is greater than unity the peak is skewed to the right. When the experimental data was fitted, the data points affected by the ice condensation on the cell were ignored. The peaks were fitted in the q-range of $q \in [1,1.7]$ Å$^{-1}$. 
		
		\subsection{Comparing the structure along isochores and isochrones.} \label{sec:Cumene_rho_vs_Gamma}
		
In this section, we compare the behavior of the peak along isochore and isochrones. Firstly we reflect over the result of the previous section and results from literature. When moving around in the phase diagram, the effect of changing the density can be seen in the scattered intensity. The reduced units used in the study are chosen to scale out the effect of density. In figure \ref{fig:rainbow_isochrones} the effect of scaling out the density contributions can be seen. Density has an effect on the first peak, but when presented in reduced units, there are still clear structural changes. The origin of these structural changes is something other than density changes. \Mycite{Tolle2001} measured the structure of a van der waals liquid, OTP, using neutron scattering along an isochrone and isochore. The structure was measured in a pressure range of up to 90 MPa and temperatures between 290 and 339 $^o$K, and the paper showed that the measured intensities are invariant, in experimental units, along both isochrone and isochore. \Mycite{Tolle2001} also showed that the structure does change along isotherms and isobars. The structure does change, so the structure cannot be invariant along both isochrones and isochores, unless they are the same line. The reason \Mycite{Tolle2001} is not able to separate isochrones and isochores are likely due to a combination of the limited experimental range and the resolution of neutron scattering at the time of the experiment.

As mentioned in the introduction, it is not enough to show invariance along isochrones; we must also show that the structure changes along isotherms, isobars and isochores. We have already shown that the structure is invariant along isochrone, see figure \ref{fig:rainbow_isochrones}, and the structure changes when moving between isochrones, figure \ref{fig:rainbow_zoom}. We still need to compare the structure along isochores and isochrones.
		
\Mycite{Chen2021} studied the temperature-dependence of the structure of the three hydrogen-bonding glass formers, glycerol, xylitol and D-sorbitol. They found that the position of the main peak was close to invariant when presented in reduced units. This is a different behavior than what we see for cumene, as shown in figure \ref{fig:rainbow_isochrones}. \Mycite{Alba-Simionesco2024} plotted the structure along an isochore for the van der Waals liquid toluene and the polymer polubutadiene, and wrote about them: "[The plots] \textit{illustrate how the main peak of S(Q) and the short range order remains constant at constant density whatever is the temperature and the changes of the relaxation time or the effective activation energy; this is verified for van der Waals molecular liquids and polymers"}. The quote from \Mycite{Alba-Simionesco2024} refers to the whole peak shape and contradicts the results from Section \ref{sec:cumene_exp_isochrone}. The behavior of the structure along isochores and isochrones is important for revealing the existence of pseudo-isomorphs.
		
\begin{figure}[H]
			\centering
			\begin{subfigure}[b]{0.495\textwidth}
				\includegraphics[trim = 30mm 80mm 40mm 80mm, clip=true,width=0.99\textwidth]{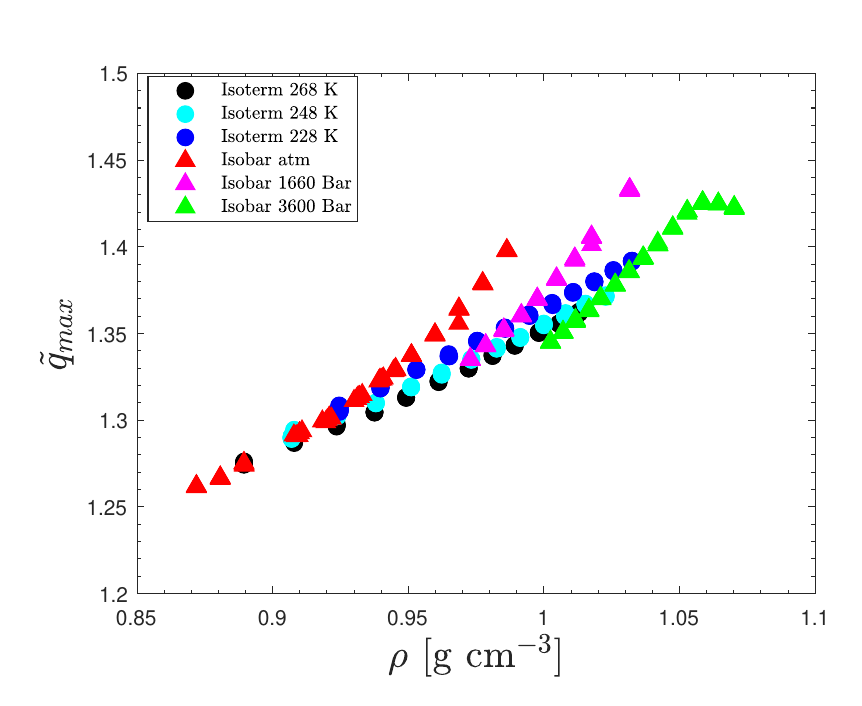}
				\caption{$\tilde{q}_{max}$ as a function of $\rho$}	
				\label{fig:fit_density_Gamma_compare_a}
			\end{subfigure}
			\begin{subfigure}[b]{0.495\textwidth}
				\includegraphics[trim = 30mm 80mm 40mm 80mm, clip=true,width=0.99\textwidth]{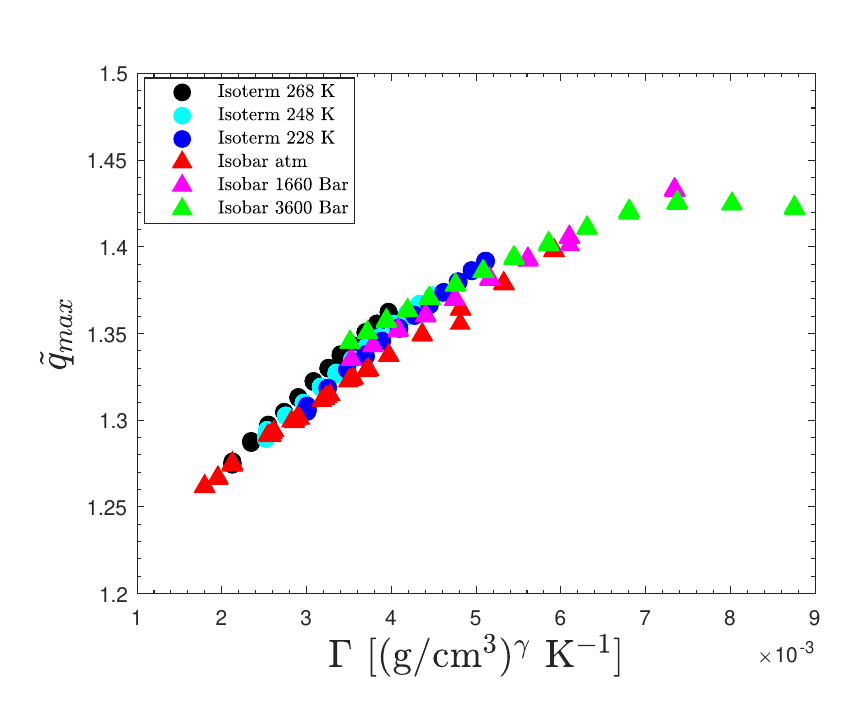}
				\caption{$\tilde{q}_{max}$ as a function of $\Gamma$}	
				\label{fig:fit_density_Gamma_compare_b}	
			\end{subfigure}
			
			\begin{subfigure}[b]{0.495\textwidth}
				\includegraphics[trim = 30mm 80mm 40mm 80mm, clip=true,width=0.99\textwidth]{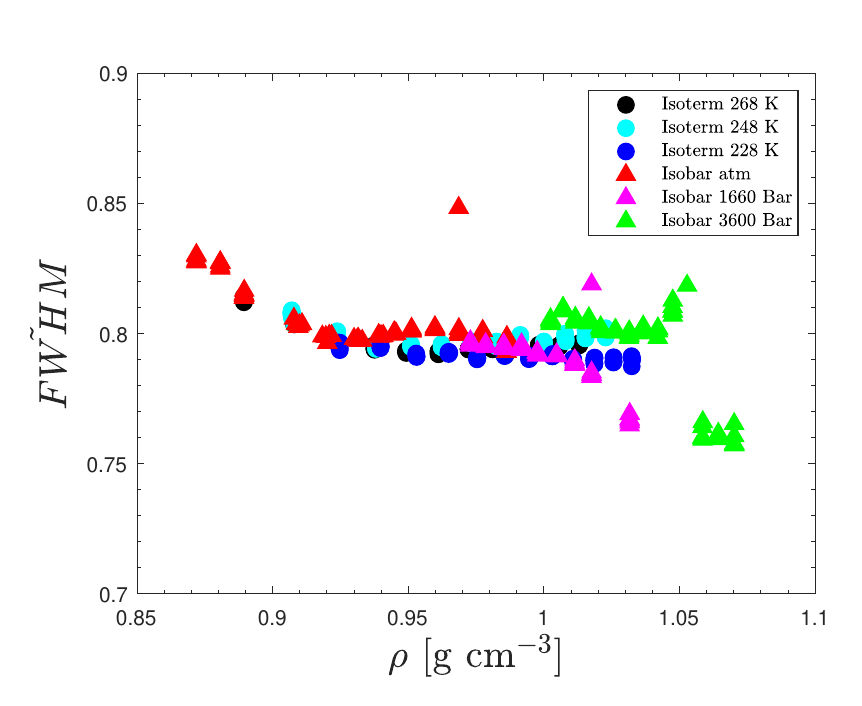}	
				\caption{FWHM as a function of $\rho$}	
				\label{fig:fit_density_Gamma_compare_c}
			\end{subfigure}
			\begin{subfigure}[b]{0.495\textwidth}
				\includegraphics[trim = 30mm 80mm 40mm 80mm, clip=true,width=0.99\textwidth]{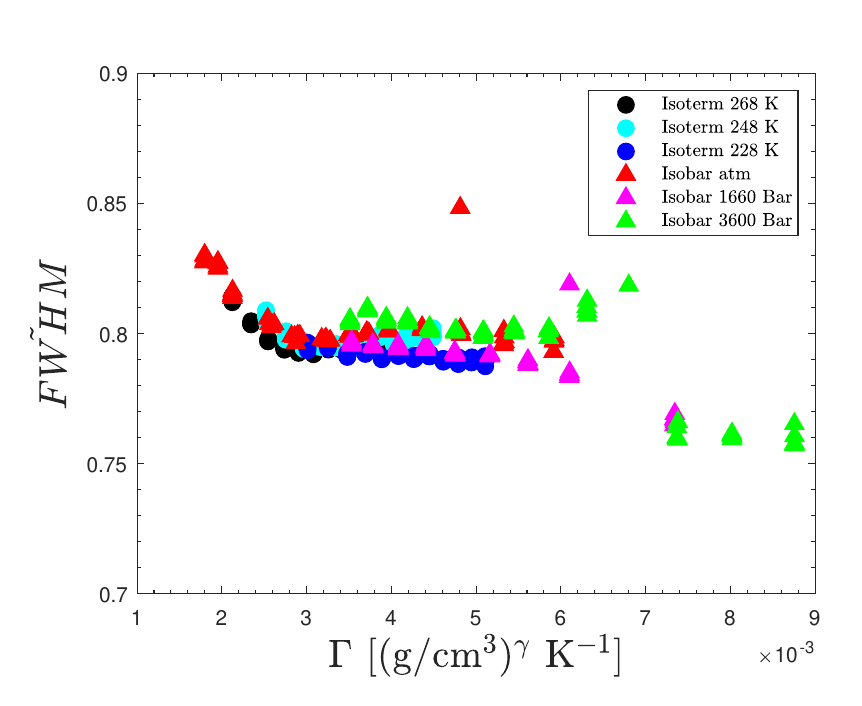}	
				\caption{FWHM as a function of $\Gamma$}
				\label{fig:fit_density_Gamma_compare_d}
			\end{subfigure}
			\begin{subfigure}[b]{0.495\textwidth}
				\includegraphics[trim = 30mm 80mm 40mm 80mm, clip=true,width=0.99\textwidth]{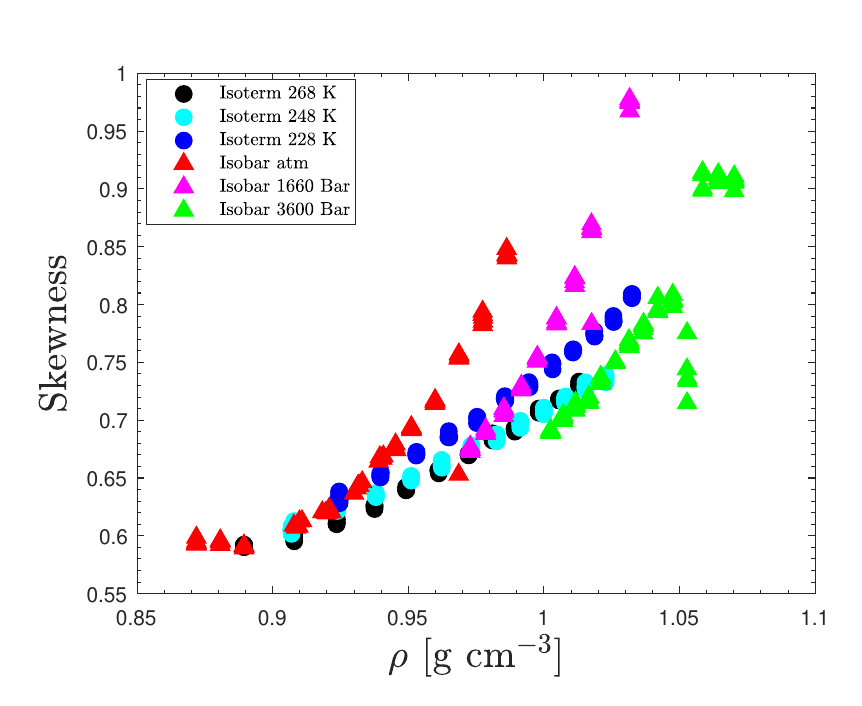}		
				\caption{Skewness of the peak as a function of $\rho$}
				\label{fig:fit_density_Gamma_compare_e}
			\end{subfigure}
			\begin{subfigure}[b]{0.495\textwidth}
				\includegraphics[trim = 30mm 80mm 40mm 80mm, clip=true,width=0.99\textwidth]{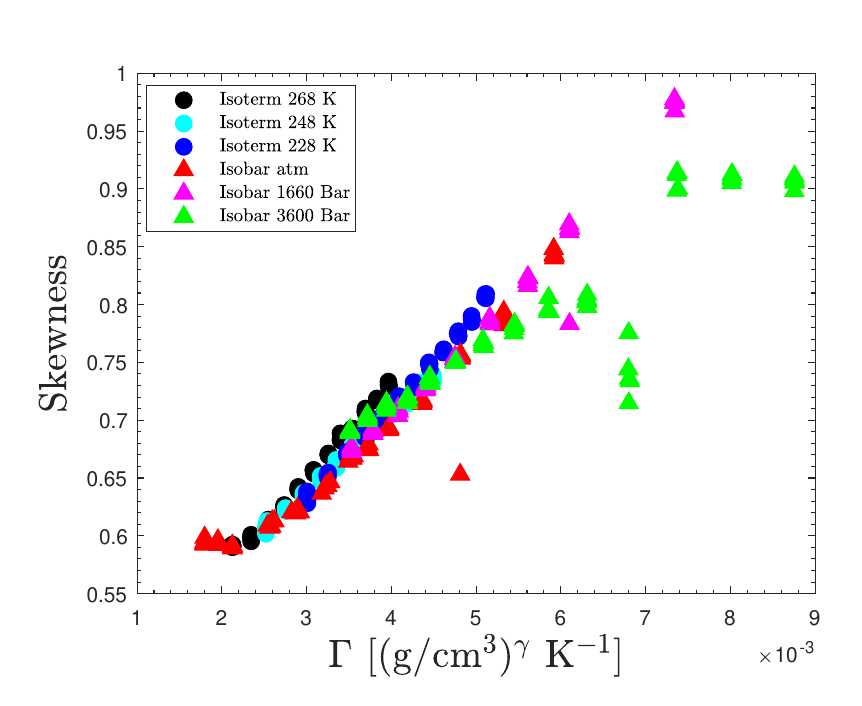}		
				\caption{Skewness of the peak as a function of $\Gamma$}
				\label{fig:fit_density_Gamma_compare_f}
			\end{subfigure}
			\caption{The evolution of the main structure peak of cumene, as a function of $\rho$ and $\Gamma$. The peak position, $\tilde{q}_{max}$, and the skewness of the peak appear to collapse better  as function $\Gamma$ then $\rho$. The FWHM did not change much for cumene, so it collapses regardless of what it is plotted against.   	}
			\label{fig:fit_density_Gamma_compare}
		\end{figure}

In figure \ref{fig:fit_density_Gamma_compare}, the three characteristics of the peak; the peak position, FWHM, and the skewness are compared as functions of $\rho$ and $\Gamma$. The peak position and skewness seem to collapse quite well as functions of $\Gamma$, while when plotted against the density, there is a clear difference between the state points. For cumene, the FWHM changes very little; thus, it can collapse as a function of most things, including $\rho$ and $\Gamma$. For the peak position and skewness the collapse along $\Gamma$ is good, but not perfect. It is however better than along the isochores. At each state point we measured 5 times. Due to the ice signal on the right side of the peak fitting the low temperature measurements gives some scatter on each measurement series. This can be seen especially on the FWHM and skewness measurements.

		\clearpage

Figure \ref{fig:fit_density_Gamma_compare} shows that for the van der Waal liquid, cumene, the structure collapses along an isochrone, and it is not only the peak position, but also the whole shape of the peak that is invariant along isochrones. It is also clear that the structure clearly changed along an isochore. \Mycite{Alba-Simionesco2024} plotted the structure of a very similar liquid, toluene, along an isochore and showed that the structure is invariant there. How should these two seemingly contradictory results be understood? Are the properties of cumene unique or can the two results be combined?
		
The first observation to answer this question is that the prediction of isomorph theory states that the structure should be invariant when presented in \textit{reduced units}. The isomorph reduced units for $q$ is $\tilde{q} = q \rho^{- \frac{1}{3}}$. When comparing structures along an isochore, one is de facto measuring in isomorph-reduced units, because $\rho$ is the same. We have seen in figure \ref{fig:rainbow_isochrones} that this has a visible effect on the position of the peak. The measured structure will change simply from the density changes. Along an isochore the structure will not change due to density changes, and when plotting in reduced units the structure will also not change due to density changes. It also gives some credit to the quote of \Mycite{Alba-Simionesco2024}. As a first approximation, density changes govern most of the structural changes measured in experimental units.  

A second observation is that the biggest difference between comparing q$_{max}$ and skewness as a function of $\rho$ and $\Gamma$ in figure \ref{fig:fit_density_Gamma_compare} comes when comparing low temperature isobaric measurements with high pressure isothermic measurements as a function of $\Gamma$. These points collapse as a function of $\Gamma$ but not $\rho$. If we compare these points in the phase diagram in figure \ref{fig:phase_diagram_measurements_cumene} these point are the measurements that are "furthest" away from each other in the phase diagram. If one tries to imagine the evolution of an isochore and an isochrone from the same state point, then if one only moves a short distance away from the state point, it is difficult to separate the isochore and isochrone. Starting from the low temperature isobaric measurements, the isochore and isochrone can diverge the most from each other. \Mycite{Tolle2001} found no changes in the structure for OTP along both an isochrone and an isochore, this is likely due to the limited distance in the phase diagram between the isochrone and the isochore and resolution of the measurements.

We have shown that there are structural changes when moving from isochrone to isochrone. When we measure the structure in experimental units as we move around in the phase diagram, we are measuring density changes and the changes from going from pseudo-isomorph to pseudo-isomorph. The size of the structural change from each contribution is likely different from liquid to liquid. For the hydrogen-bonded liquids studied in \Mycite{Chen2021} the non-density changes are small because the structure barely changes in reduced units. For cumene they are clearly visible.

\section{MD simulations of cumene} \label{sec:Cumene_MD}
		
The experimental results presented in section \ref{sec:Cumene_exp_results} showed a very good, but not perfect collapse of the structure along isochrones. Experimentally, we could only measure the first peak, and some of the measurements also had ice forming on the cell. The ice signal was overlapping with the right hand side of the cumene peak. The descriptors of the first peak - the peak position, FWHM, and skewness of the peak - are almost invariant along isochrones.  There are unanswered questions about behavior beyond the range of the measurements. The measurements collapsed, but not perfectly. Molecular dynamics simulations can provide a microscopic understanding of the structure of cumene. Previous  molecular dynamics simulations of cumene have focused on mixtures containing cumene. This includes the solubility of asphaltene in cumene\cite{TirjooAmin2019Mdso},  the super critical behavior of benzene, propene, and cumene mixtures \cite{YangXiaofeng2009MDSo}, and a cumene and 1,2,4-trimethylbenzene mixture confined in NaY zeolite \cite{VaranasiSrinivasaRao2015Sado}. The last two studies used a united-atom model (UA-model), in which the hydrogen atoms on a carbon atom are combined into a single atom, while the study of the solubility of asphaltene in cumene\cite{TirjooAmin2019Mdso} used an all-atom model (AA-model), in which the carbons and hydrogens are simulated as separate atoms. 
		
In molecular dynamics simulations, the vibrations of small hydrogens are often a computational limiting factor. The experimental results from section \ref{sec:Cumene_exp_results} are measured with x-ray scattering, where the scattering length of hydrogen is very small compared to the scattering length of carbon. The vibrations of small hydrogen atoms are often computationally expensive, and since we wish to compare the model with x-ray data, choosing a UA-model is a fair trade-off. The united atom model from refs. \cite{YangXiaofeng2009MDSo} and \cite{VaranasiSrinivasaRao2015Sado}, both use the TraPPE-UA potential \cite{TRAppE2000}. This potential is the starting point for constructing our MD model. 
		
		\subsection{United atom cumene model}
		
\begin{figure}[h]
			
			\begin{subfigure}[b]{0.495\textwidth}
				\includegraphics[width=0.99\textwidth]{Chapters/Figures/Cumene.jpg}
				\caption{Cumene}		
			\end{subfigure}
			\begin{subfigure}[b]{0.495\textwidth}
				\includegraphics[width=0.99\textwidth]{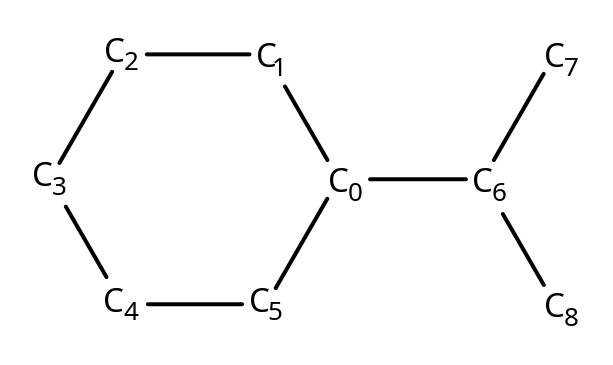}		
				\caption{UA model of Cumene}
			\end{subfigure}
			\caption{Cumene and the UA model of Cumene.}
			\label{fig:MD_model}
		\end{figure}

The model of cumene consists of nine CH$_x$ atoms, with four different types of non-bonded interactions. $C_0$ is a model of the ring carbon where the propyl group sits, $C_{1-5}$ are the pseudoatom of the ring CH. $C_6$ models the CH in the propyl group, and $C_7$ and $C_8$ are models of the CH$_3$ parts of the propyl group. The model with the position of each pseudoatom is shown in Figure \ref{fig:MD_model}. Both the bonded and non-bonded interactions were taken from the TraPPE-UA potentials \cite{TRAppE2000} with some modifications to the dihedral interaction. The system was simulated using Roskilde University Molecular Dynamics (RUMD) software which uses GPUs \cite{RUMD_paper}.
		
		The potential for non-bonded interactions is the 12-6 Lennard-Jones potential. The TraPPE potential allows for particles with charge; however, in the cumene model, each pseudoatom is charge-neutral:
		
		\begin{equation}
			U_{NB}(r_{ij})=4\epsilon_{ij} \left[ \left( \frac{\sigma_{ij}}{r_{ij}}\right)^{12} -\left( \frac{\sigma_{ij}}{r_{ij}}\right)^6 \right]
		\end{equation}
		
		\begin{table}[h]
			\centering
			\begin{tabular}{l|lll}
				& $\frac{\epsilon_{ij}}{k_b}$ [K] & $\sigma_{ij}$ [Å]  & $q$ [e]\\ \hline
				$C_0$ & 50.5       & 3.695 & 0                \\
				$C_{1-5}$, & 21.0       & 3.88 & 0                 \\
				$C_{6}$ & 10.0       & 4.65      & 0            \\
				$C_{7-8}$ & 98         & 3.75       & 0          
			\end{tabular}
			\caption{The nonbonded interactions for the UA-Cumene.}
		\end{table}

		\subsubsection{Bonds:}
		
		The TraPPE potential uses fixed bond lengths, whereas the in RUMD bonds are implemented by a harmonic potential:
		
		\begin{equation}
			U_{\text{bond}}( \Vec{r}) = \frac{1}{2} \sum_{bonds} k_s^{(i)} \left( r_{ij} - l_b^{(i)} \right)^2
		\end{equation}
		
		where $k_s^{(i)}$ is the spring constant of the $i$'th bond type, and $l_b^{(i)}$ the bond length of the $i$'th bond type. Fixed bond lengths are implemented in RUMD by ensuring the bonds are  stiff:
		
		\begin{table}[h]
			\centering
			\begin{tabular}{l|lll}
				bond    & $l_b$ [Å] & RUMD $\frac{k_s}{k_b}$ [K] & TraPPE\\ \hline
				$C_{0} - C_{1}  $ & 1.40 & 100000 &fixed              \\
				$C_{1} - C_{2}  $ & 1.40 & 100000 &fixed              \\
				$C_{2} - C_{3}  $ & 1.40 & 100000 &fixed              \\
				$C_{3} - C_{4}  $ & 1.40 & 100000 &fixed              \\
				$C_{4} - C_{5}  $ & 1.40 & 100000 &fixed              \\
				$C_{5} - C_{0}  $ & 1.40 & 100000 &fixed              \\
				
				$C_{0} - C_{6} $ & 1.54 & 100000 & fixed              \\
				$C_{6} - C_{7} $ & 1.54 & 100000 & fixed              \\
				$C_{6} - C_{8} $ & 1.54 & 100000 & fixed              \\
				
			\end{tabular}
			\caption{The bond lengths of the UA-Cumene. TraPPE uses fixed bond lengths.}
		\end{table}

		\subsubsection{Bond angle potential:}
		
		The angle potential in TraPPE is given by 
		\begin{equation}
			U_{\text{bend}} (\theta) = \frac{k_\theta}{2} \left(\theta - \theta_0 \right)^2    
		\end{equation}
		
		The TraPPE potential also uses rigid angles and this is implemented by making $k_\theta$ large. The values used in the simulations are listed in Table \ref{tab:angle_potential}.
		
		
		
		\begin{table}[h!]
			\centering
			\begin{tabular}{l|lll}
				& $\theta_0$ [deg] &RUMD $\frac{k_\theta}{k_b} $[K]& TraPPE $\frac{k_\theta}{k_b} $ [K]\\ \hline
				C$_{0}$ - C$_{1}$ -  C$_{2}$ & 120 & 100000&  Fixed             \\
				C$_{1}$ - C$_{2}$ -  C$_{3}$ & 120 & 100000&  Fixed              \\
				C$_{2}$ - C$_{3}$ -  C$_{4}$ & 120 & 100000&  Fixed              \\
				C$_{3}$ - C$_{4}$ -  C$_{5}$ & 120 & 100000&  Fixed             \\
				C$_{4}$ - C$_{4}$ -  C$_{0}$ & 120 & 100000&  Fixed              \\
				C$_{5}$ - C$_{0}$ -  C$_{1}$ & 120 & 100000&  Fixed          \\
				C$_{5}$ - C$_{0}$ -  C$_{6}$ & 120 & 100000&  Fixed              \\
				C$_{1}$ - C$_{0}$ -  C$_{6}$ & 120 & 100000&  Fixed              \\
				C$_{0}$ - C$_{6}$ -  C$_{7}$ & 112 & 62500  & 62500       \\
				C$_{0}$ - C$_{6}$ -  C$_{8}$ & 112 & 62500  & 62500     \\
				C$_{7}$ - C$_{6}$ -  C$_{8}$ & 112 & 62500  & 62500    \\
			\end{tabular}
			\caption{The parameters for the angle potential used for the UA-model, simulated using RUMD. The TRAppE potential uses fixed angles, which is implemented by making $\frac{k_\theta}{k_b}$ sufficient large. }
			\label{tab:angle_potential}
		\end{table}
		
		\FloatBarrier
		
		\subsubsection{Dihedral angle potential}
		
		The dihedral angle potential used in TraPPE potentials is a cosine function dependent on two parameters:
		
		\begin{equation}
			U_{torsion} (\phi) = e_0 \left[1 - cos\left(2 \phi + e_1 \right) \right]
		\end{equation}
		
		where the dihedral angle $\phi_{ijkl}$ is the angle between the planes containing atoms, $i$,$j$ and $k$ and the planes containing atoms $j$,$k$, and $l$. The \textit{cis} conformations correspond to $\phi_{ijkl} = 0^ {\text{o}}$, while \textit{trans} conformation corresponds to $\phi_{ijkl} = 180^ {\text{o}}$, \cite{TraPPE_web}. Clockwise rotation is defined as positive, and counterclockwise rotation is defined as negative. 
		
$\cos(\phi_{ijkl})$ is given by 
		
		\begin{equation}
			\cos{\left(\phi_{ijkl}\right)}  = \frac{\left(\Vec{v}_{ij} \times \Vec{v}_{jk} \right) \cdot\left(\Vec{v}_{jk} \times \Vec{v}_{kl} \right) }{\left|\Vec{v}_{ij} \times \Vec{v}_{jk} \right| \left| \Vec{v}_{jk} \times \Vec{v}_{kl}\right| }
		\end{equation}

In the TraPPE potentials of the ring-to-tail dihedrals, C$_1$-C$_0$-C$_6$-C$_{7}$, C$_1$-C$_0$-C$_6$-C$_{8}$ , C$_5$-C$_0$-C$_6$-C$_{7}$, C$_5$-C$_0$-C$_6$-C$_{8}$ all have the same values of $e_0$ and $e_1$ \cite{TRAppE2000}. A previous experimental study measured the infrared and raman spectra of deuterated and un-deuterated cumene \cite{Fishman2008} to determine the configurations of cumene. They found that only one possible configuration exists for the ring-tail dihedrals. They found the C-H bond for the C$_6$ is in the plane of the ring and the C$_7$ and C$_8$ are on either side of the ring plane. The TRAppE potential favors a different configuration in which both C$_7$, and C$_8$ prefer one side of the ring. The simulation values of the UA-model were changed to reflect the measurements of \Mycite{Fishman2008}. The simulation values are shown in table \ref{tab:Dihedral}

\begin{table}[h]
	\centering
	\begin{tabular}{l|llll}
	& $\frac{e_0}{k_b}$ [K] & Simulation $e_1$ [$^o$]  & TRAppE $e_1$ [$^o$]& type\\ \hline
	C$_0$-C$_1$-C$_2$-C$_{3}$  & 500 (rigid) & 0 &rigid  & 0         \\
	C$_1$-C$_2$-C$_3$-C$_{4}$  & 500 (rigid) & 0 &rigid  & 0          \\
	C$_2$-C$_3$-C$_4$-C$_{5}$  & 500 (rigid) & 0 &rigid  & 0          \\
	C$_3$-C$_4$-C$_5$-C$_{0}$  & 500 (rigid) & 0 &rigid  & 0          \\
	C$_4$-C$_5$-C$_0$-C$_{1}$  & 500 (rigid) & 0 &rigid  & 0          \\
	C$_5$-C$_0$-C$_1$-C$_{2}$  & 500 (rigid) & 0 &rigid  & 0          \\
	C$_4$-C$_5$-C$_0$-C$_{6}$  & 500 (rigid) & 0 &rigid  & 1          \\
	C$_2$-C$_1$-C$_0$-C$_{6}$  & 500 (rigid) & 0 & rigid  & 1          \\
	C$_1$-C$_0$-C$_6$-C$_{7}$  & 167  & 240  & 300  & 2          \\
	C$_5$-C$_0$-C$_6$-C$_{7}$  & 167  & 240  & 300  & 2          \\
	C$_1$-C$_0$-C$_6$-C$_{8}$  &  167 & -240 & 300   & 3          \\
	C$_5$-C$_0$-C$_6$-C$_{8}$  & 167  & -240 & 300   & 3          \\

	\end{tabular}
	\caption{The parameters for the torsion potential used in the UA model. TraPPE potentials also use rigid dihedral angles, which is implemented by making $e_0$ large. It is noted next to the simulation parameters}
	\label{tab:Dihedral}
\end{table}
		
\subsection{Searching for pseudo-isomorphs using MD-simulations}
		
\begin{figure}[h]
			\centering
			\includegraphics[width=0.65\textwidth]{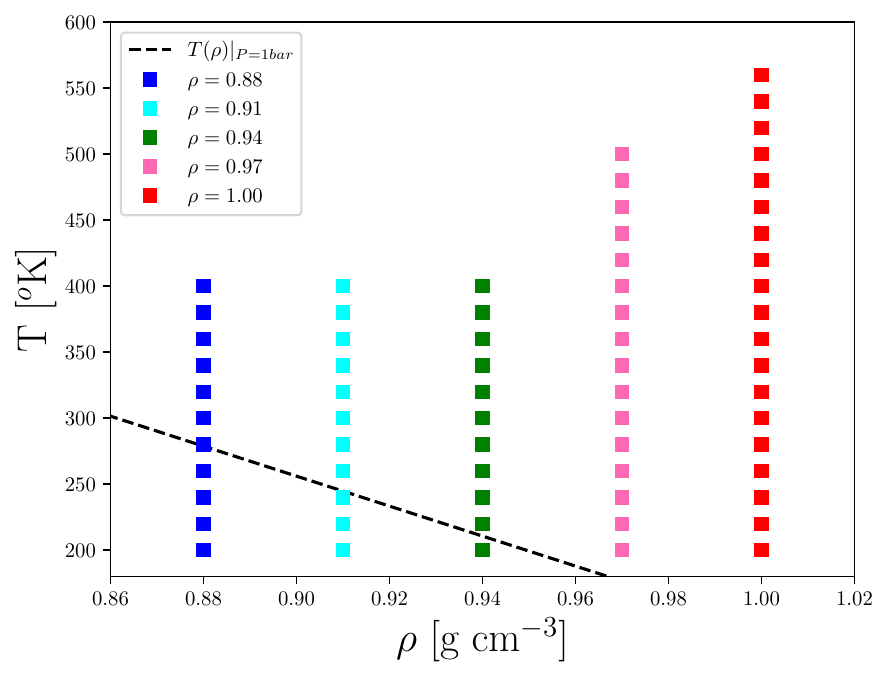}
			\caption{The $(\rho,T)$ phase diagram of the simulations presented in this study. The dashed line is the ambient pressure line, $T(\rho)|_{P=1\text{ bar}}$ calculated from equation \ref{eq:EoS_cumene}. Some simulated state points have negative pressure in the real world. }
			\label{fig:sim_phasediagram}
		\end{figure}

In this study, we simulated state points along 5 isochores with steps of 20 $^o$K in temperature. Every simulation starts from a configuration equilibrated at T=400 $^o$K, $\rho$ = 0.9 g cm$^{-3}$.  Each simulation comprises 1000 molecules (9000 particles). The time step used for each simulation is state point dependent and given by,

 \begin{equation}
 dt = dt_0\left(\frac{\rho}{\rho_0}\right)^{-\frac{1}{3}} \left( \frac{T}{T_0}\right)^{-\frac{1}{2}}  
\end{equation}

  where $dt_0 =0.0005$ ps, $\rho_0 = \rho(T=300K, P=atm) = 0.86136 $ g cm$^{-3}$, $T_0 = 300$ $^o$K. The reason for selecting a state point-dependent time step is to keep the dimensionless time step constant. The position of each simulated state point is shown in figure \ref{fig:sim_phasediagram}, along with the ambient pressure line calculated from equation \ref{eq:EoS_cumene}.

\begin{figure}[H]
			\begin{subfigure}[t]{0.48\textwidth}
				\centering
				\includegraphics[width=0.95\textwidth]{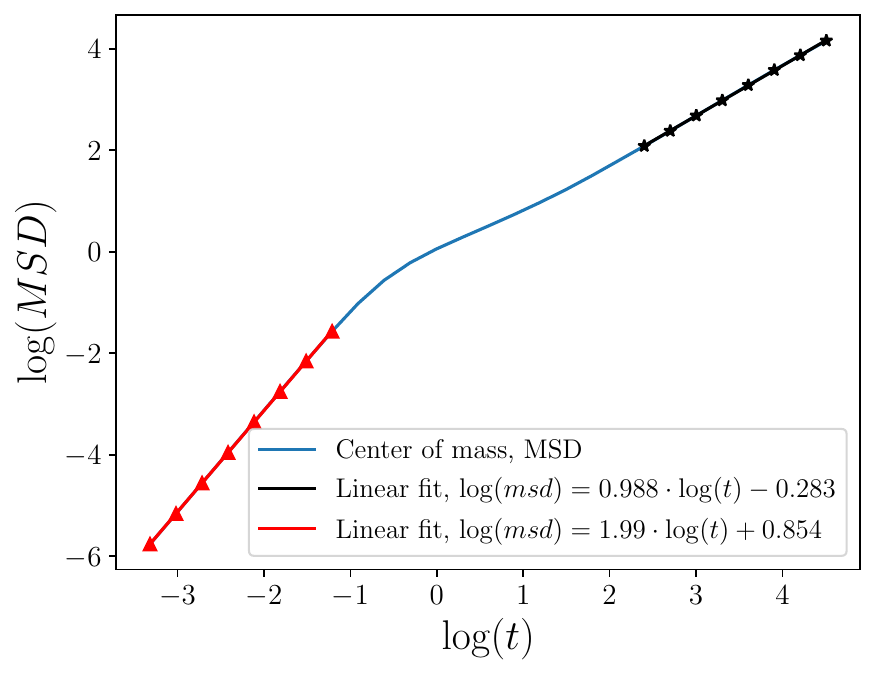}
				
				\caption{}
				\label{fig:MD_MSD_fits_a}
			\end{subfigure}\hfill
			\begin{subfigure}[t]{0.48\textwidth}
				\centering
				\includegraphics[width=\textwidth]{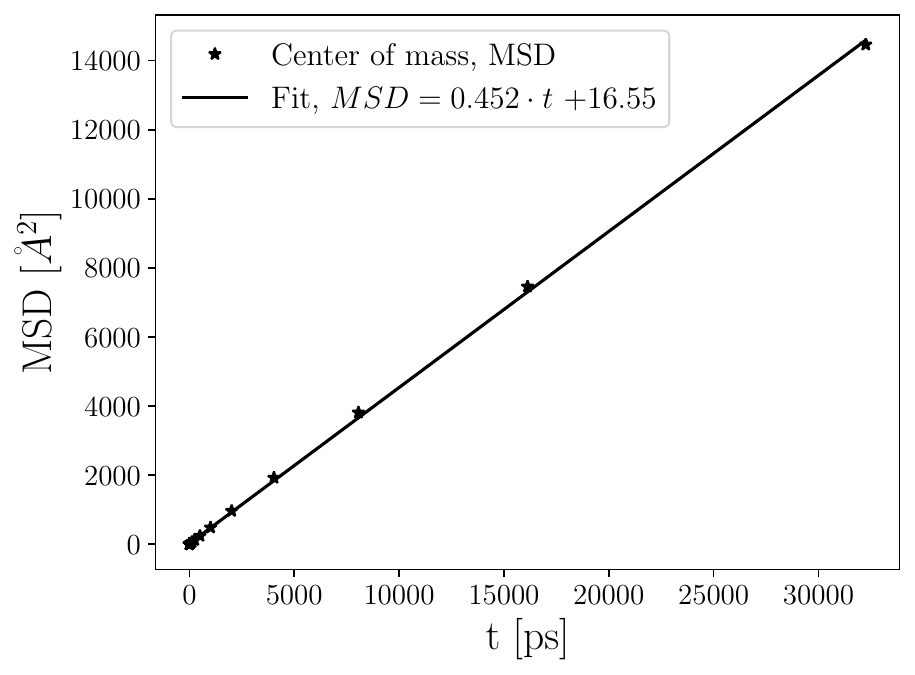}
				
				\caption{ }
				\label{fig:MD_MSD_fits_b}
			\end{subfigure}	
			
			\caption{The mean square displacement at T=300 $^o$K, $\rho = 0.97$ g cm$^{-3}$. In \ref{fig:MD_MSD_fits_a} the center-of-mass MSD is plotted against time in a log-log plot. The ballistic regime is plotted in red, while the diffusive regime in black. In \ref{fig:MD_MSD_fits_b} the center-of-mass MSD is fitted linearly to calculate the diffusion coefficient.     }
			\label{fig:MD_MSD_fits}
		\end{figure}

		In subfigure \ref{fig:MD_MSD_fits_a} the center-of-mass mean squared displacement(MSD)  is plotted as a function of time in a log-log plot for a single state point at $T = 300$ $^o$K, $\rho = 0.97$ g cm$^{-3}$. At the long time scales, the movement of the molecule is similar to Brownian motion. This is called the \textit{diffusive regime}. For particles with Brownian motion, the MSD can be calculated from the diffusion coefficient:
		
		\begin{equation}
			MSD = 6Dt
		\end{equation}
		where D is the diffusion coefficient and t is time. When plotted in a log-log plot, diffusive motion has a slope of 1. In subfigure  \ref{fig:MD_MSD_fits_a}, a fit to the diffusive regime is shown in black. At short time scales, the motion of the molecules is dominated by the initial velocity of the molecule.  This is called the \textit{ballistic regime} where $MSD \approx v^2t^2$. In the log-log plot, this is characterized by a slope of 2, and the fit to the ballistic regime is shown in red.
		
\begin{figure}[H]
			\centering
			\includegraphics[width=0.65\textwidth]{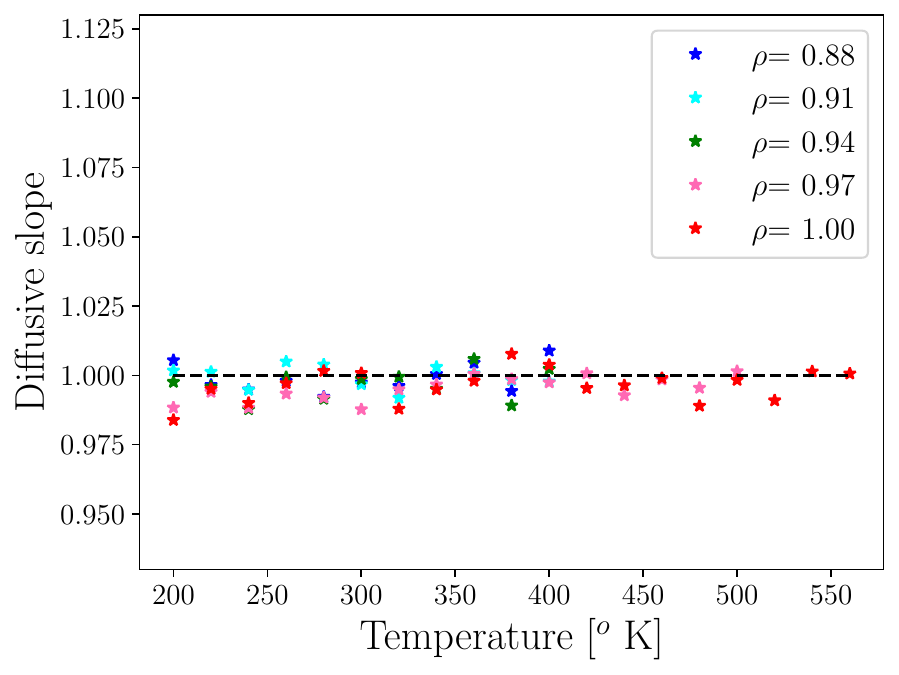}
			\caption{The slope of the diffusive part for every single state point simulated. The points that were fitted were every saved MSD data point, where $\log(t)>2.2$. If this deviates strongly from 1, we must simulate a longer time to enter the diffusive regime.   }
			\label{fig:MD_diffusive_regime}
		\end{figure}

		The diffusion coefficient is calculated by a linear fit to the MSD as a function of time, where the slope of the fit is $6D$. An example is shown in figure \ref{fig:MD_MSD_fits_b}. To test whether the simulations have  sufficient runtime, we determine whether we have entered the diffusive regime of the mean square displacement. The points that were fitted were every saved MSD data point, where $\log(t)>2.2$, and the results are shown in figure \ref{fig:MD_diffusive_regime}.

		The diffusion coefficient can be calculated by linearly fitting the center-of-mass mean square displacement as a function of time, as shown in figure \ref{fig:MD_MSD_fits_b}. For all state points  we are clearly in the diffusive regime. The center-of-mass diffusive coefficient is used to find our candidates for pseudo-isomorph: The isochrone. 
		
		\subsubsection{Finding isochrones:}
		
\begin{figure}[H]
			\centering
			\includegraphics[width=0.65\textwidth]{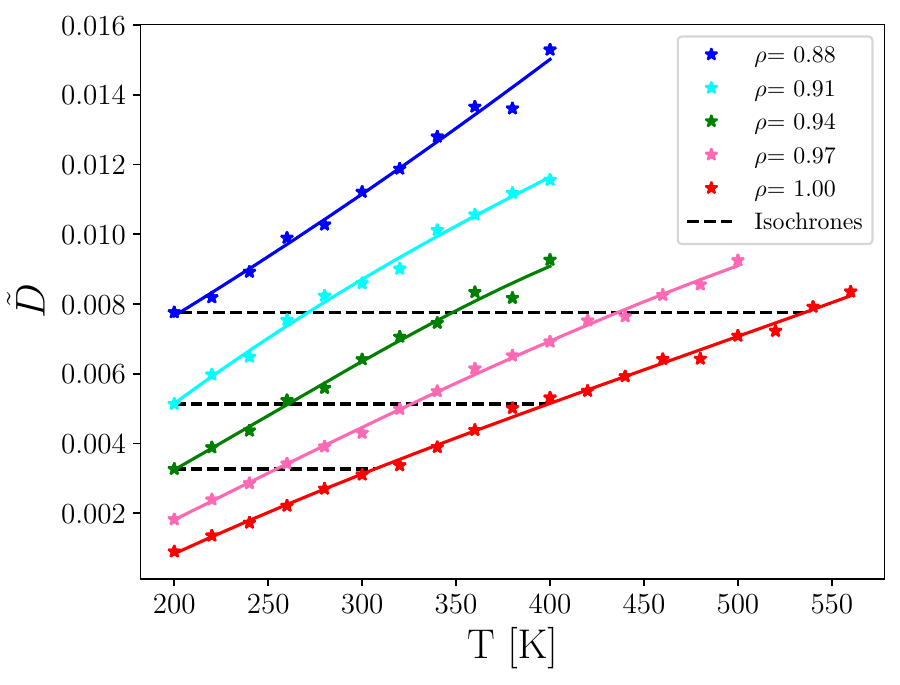}
			\caption{The reduced diffusive coefficient plotted against temperature, $\tilde{D} = D \rho^{\frac{1}{3}} T^{\frac{1}{2}}$. Here, each dot is a simulated state point, and the solid line is a fit with a third-order polynomial. The dashed black lines represent isochrones, and the intersections between isochrones and fits are shown in table \ref{tab:isochrones}.  }
			\label{fig:Cumene_find_isochrone}
		\end{figure}

		In the figure \ref{fig:Cumene_find_isochrone} the reduced diffusive coefficient is plotted for each state point against the temperature. The procedure for finding isochrones has been described in ref. \cite{Knudsen2024}. The reduced diffusive coefficient along each isochore was fitted with a third-order polynomial. The intersection between the isochoric fits and a specified value of $\tilde{D}$, gives $(\rho,T)$ values along that isochrone. The dashed lines in figure \ref{fig:Cumene_find_isochrone} are the three isochrones we can calculate directly. In table \ref{tab:isochrones} are the state points along each isochrone.

		\begin{table}[H]
			\centering
			\begin{tabular}{l|lll}
				& IC1    & IC2    & IC3    \\ \hline
				$\rho = 0.88 $ g cm$^{-3}$  & 200  $^o$K   & x      & x      \\
				$\rho = 0.91 $ g cm$^{-3}$ & 271.27 $^o$K & 200 $^o$K    & x      \\
				$\rho = 0.94 $ g cm$^{-3}$ & 348.35 $^o$K & 260.46 $^o$K & 200 $^o$K   \\
				$\rho = 0.97 $ g cm$^{-3}$ & 436.04 $^o$K & 326.13 $^o$K & 254.35 $^o$K \\
				$\rho = 1.00 $ g cm$^{-3}$ & 535.86 $^o$K & 398.92 $^o$K& 306.67 $^o$K
			\end{tabular}
			\caption{The state points of the three isochrones shown in figure \ref{fig:Cumene_find_isochrone}. For IC2 and IC3, the isochrone did not cross all isochores. }
			\label{tab:isochrones}
		\end{table}
		
\subsubsection{Using power-law density scaling:}

\begin{figure}[H]
	\centering
	\includegraphics[width=0.6\textwidth]{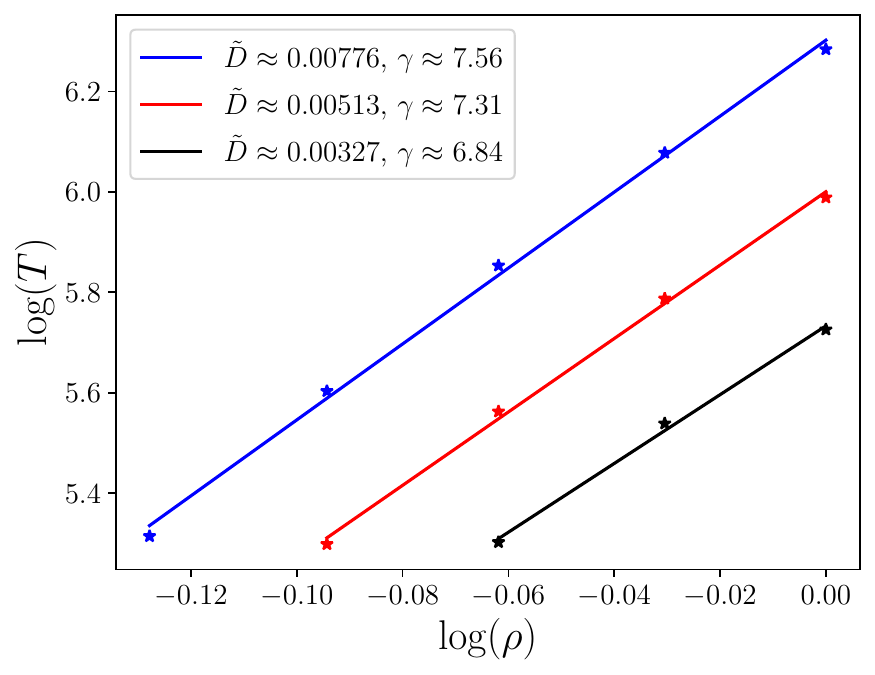}
	\caption{Power-law density scaling coefficents of the isochrones. Each color represents an isochrone, the dots represent the data points from table \ref{tab:isochrones}, and the solid line represent the linear fit. The $\gamma$ values calculated from each of the fits are higher than the real-world value of 4.8. There seems to be a tendency for $\gamma$ to be smaller along more viscous isochrones; however, the amount of data is limited. The behavior of gamma was also observed in a UA-model of an ionic liquid. }
	\label{fig:Cumene_density_scaling_MD}
\end{figure}

\begin{figure}[H]
	\begin{subfigure}[t]{0.48\textwidth}
		\includegraphics[width=\textwidth]{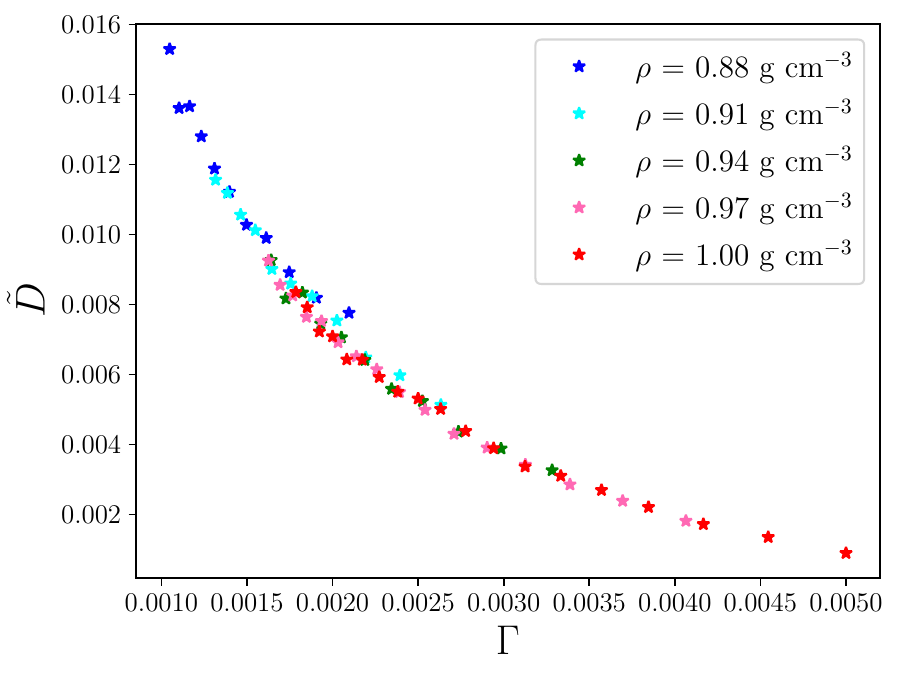}
		\caption{$\gamma = 6.8$}
		\label{fig:Cumene_density_scaling_a}
	\end{subfigure}\hfill
	\begin{subfigure}[t]{0.48\textwidth}
		\centering
		\includegraphics[width=\textwidth]{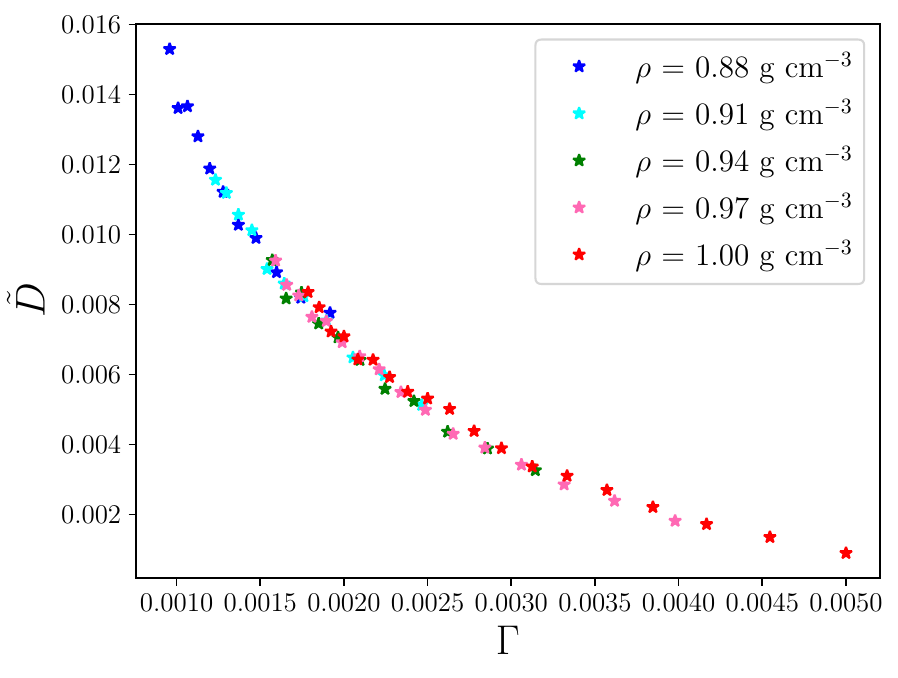}
		\caption{$\gamma =7.5$ }
		\label{fig:Cumene_density_scaling_b}
	\end{subfigure}	
	
	\caption{ Testing density scaling with the lowest and highest values of $\gamma$ from figure \ref{fig:Cumene_density_scaling_MD}. Both $\gamma$ values can cause the reduced diffusion coefficient to collapse reasonably well. The different $\gamma$ values make different parts collapse better. The $\gamma = 6.8$ makes the more viscous values collapse better, while $\gamma=7.5$ makes the more fluid data points collapse better. }
	\label{fig:Cumene_density_scaling}
\end{figure}

Experimentally, cumene shows power-law density scaling of dynamics with a $\gamma =4.8$, which has been verified over a large pressure and dynamical range \cite{Ransom2017,Wase2018,HansenHenrietteW.2017Cbfm}. The density scaling coefficient, $\gamma$, can be obtained by fitting $\log(T)$ as a function of $\log(\rho)$ along an isochrone. In figure \ref{fig:Cumene_density_scaling_MD} $\log(T)$ and $\log(\rho)$ is plotted for each point along the isochrones. The $\gamma$ value along each isochrone is the slope. The $\gamma$ values along each isochrone are higher than the 4.8 observed experimentally. Both of the $\gamma = 6.8$, and $\gamma = 7.5$ can be used to collapse the reduced center-of-mass diffusion coeffient data reasonably well, as shown in figure \ref{fig:Cumene_density_scaling}.  Although this is limited data, there seems to be a tendency for $\gamma$ to be lower along more viscous isochrones. The same picture is seen in \Mycite{Knudsen2024}, where the authors simulated a UA model of an ionic liquid. In their study, they observed that $\gamma$ was higher than those of the experimental value, and that their $\gamma$ values decreased at more viscous state points. The potential used in \Mycite{Knudsen2024} for nonbonded interactions is also the 12-6 Lennard-Jones potential, however the pseudo-atoms had charges, given that it was a model of an ionic liquid.

All three $\gamma$ values calculated from each isochrone can collapse the reduced diffusive coefficients from figure \ref{fig:Cumene_find_isochrone} into a single curve. This is shown in figure \ref{fig:Cumene_density_scaling}. In figure \ref{fig:Cumene_density_scaling_a}, the reduced diffusive coefficient are plotted against $\Gamma = \frac{\rho^\gamma}{T}$ for $\gamma =6.8$.  $\gamma=6.8$ can make the more viscous parts collapse quite well, but the data from the $\rho =0.88$ g cm$^{-3}$ and $\rho =0.91$ g cm$^{-3}$ isochores do not collapse as well. The explanation for this is that the $\gamma$-value is calculated using state points from the other three isochores. In figure \ref{fig:Cumene_density_scaling_b} the reduced diffusive coefficients are collapsed using $\gamma =7.5$. The $\gamma = 7.5$ value is found by fitting the isochrone that covers all five isochores. $\gamma = 7.5$ gives an even collapse of the data, but at more viscous diffusive coefficients the collapse is better with $\gamma =6.8$.   

Experimentally, the $\gamma$-value of cumene is around 4.8, and very independent of the state points \cite{Ransom2017}\cite{Wase2018}. For the UA-atom model, this is not the case, as the $\gamma$ changes with the fitted isochrone. The power-law approximation of $\Gamma = \frac{\rho^\gamma}{T}$ with $\gamma$ being constant, seems very good for the experimental data, but for the MD simulations, a different density-dependent function could capture the state point dependence of the collapse.

\subsection{The structure of cumene along an isochrone} 

\begin{figure}[h]
	\centering
	\includegraphics[width=0.75\textwidth]{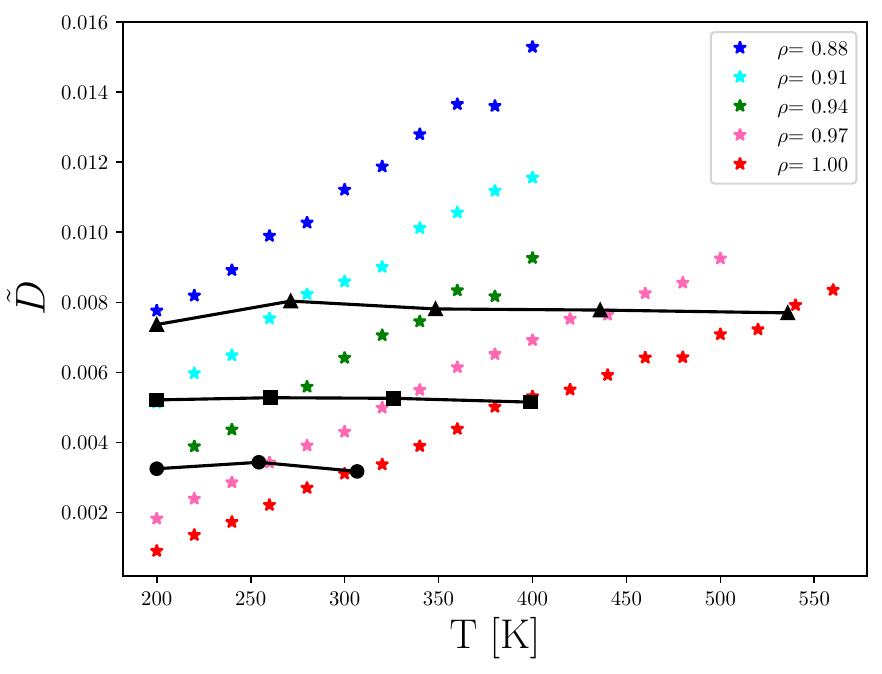}
	
	\caption{The simulation state points of the three isochrones we study. The state points are shown in table \ref{tab:isochrones}. Each symbol represents a state points along a different isochrone, and the black lines are to guide the eye. }
	\label{fig:Cumene_MD_isochrones}
\end{figure}

In the following subsection, the structure of cumene along an isochrone is studied first in real space and then in inverse space. In table \ref{tab:isochrones}, three isochrone have been found to find the density scaling exponent. We simulated at the same 12 state point as stated in table \ref{tab:isochrones}, and the diffusion coefficients in reduced units are plotted in figure \ref{fig:Cumene_MD_isochrones}.

\subsubsection{Real space}
\begin{figure}[h]
	\centering
	\includegraphics[width=0.75\textwidth]{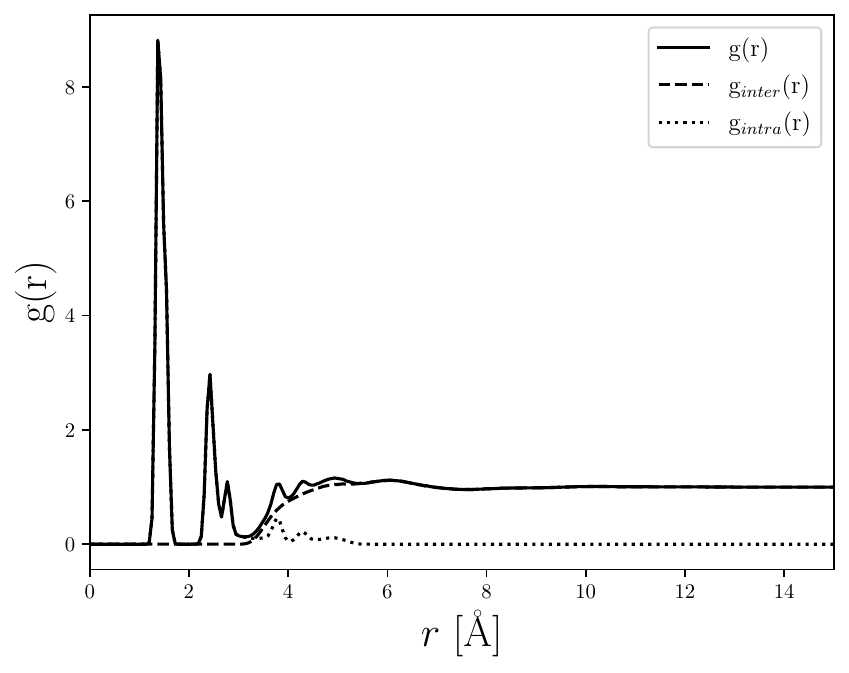}
	
	\caption{The pair distribution function of the UA-cumene model at $\rho$ = 1.00 g cm$^{-3}$, T = 535.86 K.  The solid black line is $g(r)$, the dashed line is $g_{inter}(r)$, while the dotted line is $g_{intra}(r)$. When $r < 3.5$ Å, the intramolecular structure dominates $g(r)$. The intramolecular structure comes from the bonds in the molecules, resulting in sharp peaks in $g(r)$, where the width of the peak comes from vibrations in the bonds.  When $ r \in [3.5; 5.5] $ the overlap between the intramolecular and intermolecular contributions to $g(r)$ can be observed. The small peaks in that range arise from intramolecular distances. At around 6 Å there is a peak in $g(r)$, which is the nearest-neighbor distance between molecules. }
	\label{fig:g(r)_cumene}
\end{figure}

In MD simulations, the structural measurement that is easiest to calculate is the pair distribution function $g(r)$, as introduced in chapter \ref{chapter:Structure}:

\begin{equation}
	g(r) = \frac{1}{4\pi N \rho_0 r^2} \sum_i \sum_j \left<\delta(r - r_{ij}) \right>
\end{equation}

where $\rho_0$ is the atom number density [Å$^{-3}$], N is the number of particles, and r$_{ij}$ [Å] is the distance between the i'th and j'th atoms. It is clear from the definition that it is possible to separate $g(r)$ into intra- and intermolecular pair distribution functions. In figure \ref{fig:g(r)_cumene} $g(r)$, $g_{inter}(r)$, and $g_{intra}(r)$ at a single state point for the UA-cumene model is shown. In the figure, the solid line represents the total $g(r)$, $g_{inter}(r)$ is the dashed line, and the dotted line is $g_{intra}(r)$. In the subsequent figures, this is not explicitly stated in the figure legend. At small r values, $g(r)$ is dominated by the intramolecular bonds in the molecule. The first sharp peak is a split-double peak corresponding to the bonds in the ring (r $\approx$ 1.4 Å) and the bonds in the propylgroup (r $\approx$ 1.54 Å). The further peaks arise from 2nd and 3rd nearest neighbor lengths. When $ r \in [3.5; 5.5] $ Å the intramolecular and intermolecular contributions to $g(r)$ start to overlap. At around 6 Å there is a weak peak corresponding is the nearest neighbor distance between molecules. 

\begin{figure}[H]
	\begin{subfigure}[t]{0.48\textwidth}
		\centering
		\includegraphics[width=\textwidth]{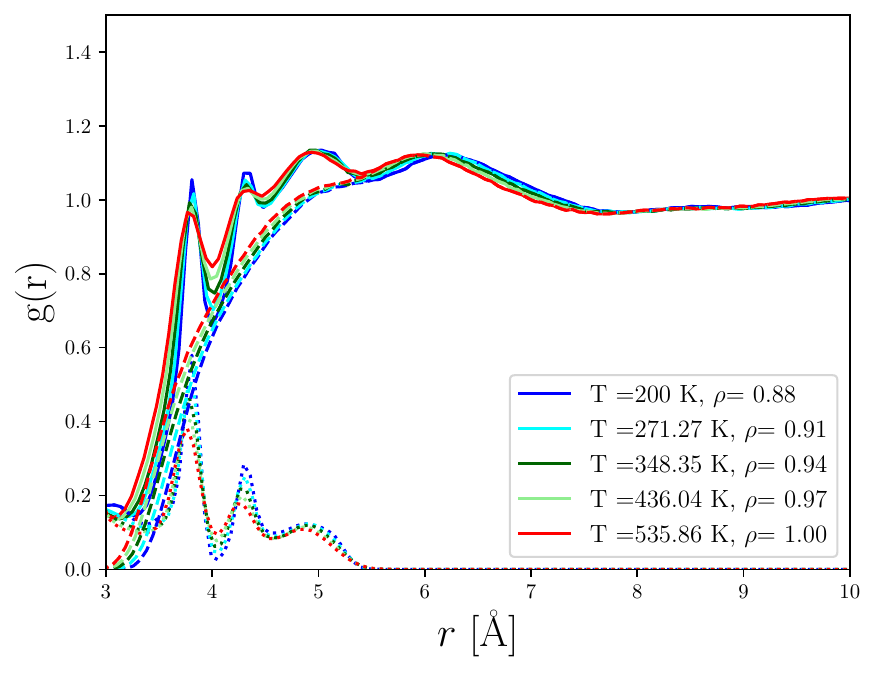}
		\caption{}
		\label{fig:cumene_RDF_IC1_a}
	\end{subfigure}\hfill
	\begin{subfigure}[t]{0.48\textwidth}
		\centering
		\includegraphics[width=\textwidth]{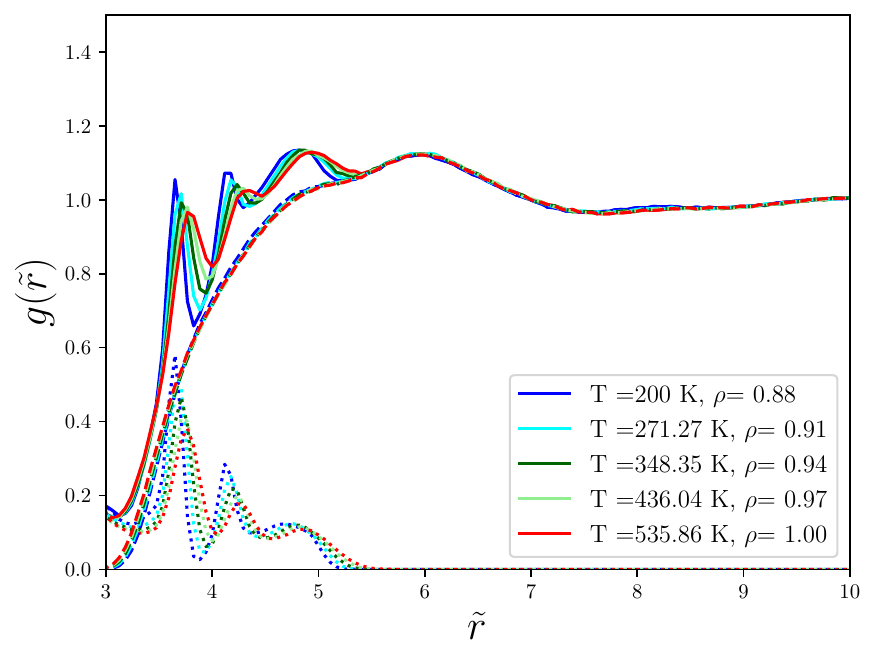}
		\caption{}
		\label{fig:cumene_RDF_IC1_b}
	\end{subfigure}\hfill
	\caption{The structure along an isochrone in (a) experimental and (b) reduced units. $g(r)$ is plotted as solid lines, $g_{intra}(r)$ as  dotted lines, and $g_{inter}(r)$ as dashed lines. Colors indicate different state points. This figure focuses on the $r$-values where the intramolecular and intermolecular contributions overlap. The five simulated state points are shown under IC1 in table \ref{tab:isochrones}.  The peaks observed in $g(r)$ before 5 Å are due to intramolecular contributions, most likely third- and fourth-nearest neighbors.  In figure \ref{fig:cumene_RDF_IC1_b}, the structure at the same five state points is shown in reduced units, $\tilde{r} = r \rho^{\frac{1}{3}}$.  The intermolecular structure collapses almost perfectly into a single curve.  The intramolecular structure does not scale using the reduced units, and this scaling deviation from the intramolecular structure is evident in the total $g(r)$.  }
	\label{fig:cumene_RDF_IC1}
\end{figure}

In figure \ref{fig:cumene_RDF_IC1} the structure along an isochrone is plotted in experimental units (figure \ref{fig:cumene_RDF_IC1_a}), and in reduced units (figure \ref{fig:cumene_RDF_IC1_b}). The state points plotted are shown in table \ref{tab:isochrones}. Most of the peaks in $g(r)$ are from the intramolecular structure, expect around 6 Å, where the peak is purely intermolecular. The height of intramolecular peaks are temperature dependent; The vibrations of the bond are temperature-dependent, and the less the bond vibrates, the narrower the peak in $g(r)$ will be. In figure \ref{fig:cumene_RDF_IC1_b} the structure along an isochrone is shown in reduced units. The intermolecular structure collapses into a single curve in reduced units. The reduced units system that can scale the intermolecular structure does not scale the intramolecular structure. As mentioned in the introduction, this is also shown by \Mycite{Veldhorst2014}. This \textit{scaling deviation} has a visible effect on the peaks in $g(r)$. The main intermolecular peak at $\tilde{r} \approx 6$ is unaffected by the scaling deviation and collapses perfectly in reduced units. The pair distribution function is a function of $r$, so the intramolecular and intermolecular contributions are roughly sorted in $r$. For all molecular glass-formers the intramolecular contributions are capped by the length of the molecules. There are r values where $g(r)$ is purely intramolecular and other $r$ values where $g(r)$ is purely intermolecular. Cumene is a relatively short molecule, making the overlap between intramolecular and intermolecular contributions a smaller interval.

\begin{figure}[H]
	\begin{subfigure}[t]{0.48\textwidth}
		\centering
		\includegraphics[width=\textwidth]{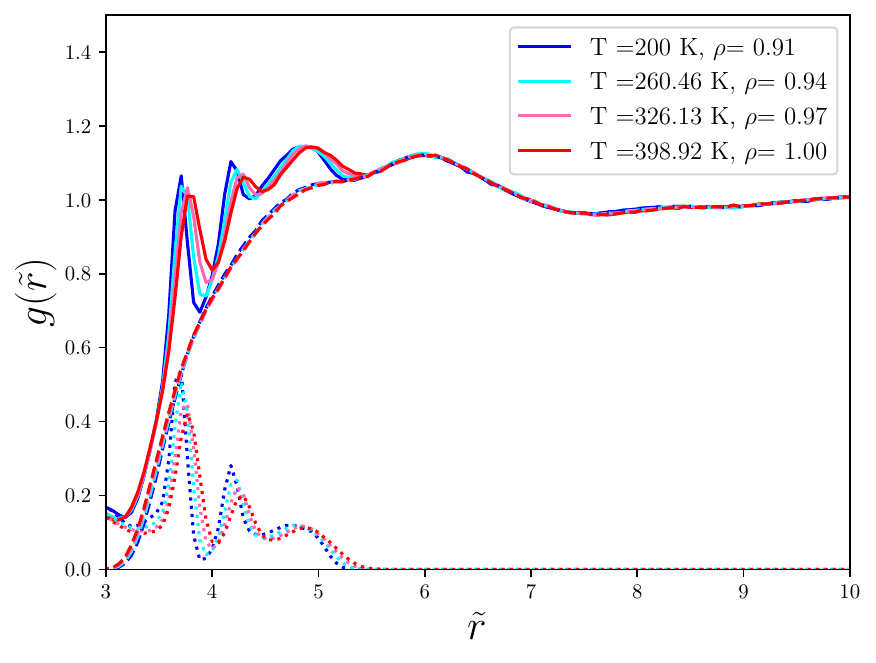}
		\caption{IC2}
		\label{fig:Cumene_gr_IC2_IC3_a}
	\end{subfigure}\hfill
	\begin{subfigure}[t]{0.48\textwidth}
		\centering
		\includegraphics[width=\textwidth]{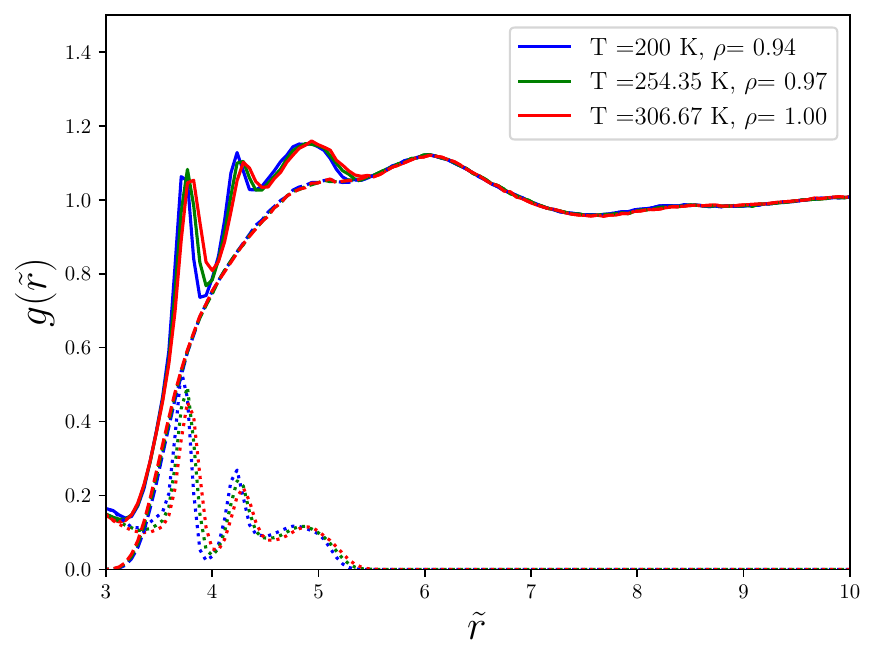}
		\caption{IC3}
		\label{fig:Cumene_gr_IC2_IC3_b}
	\end{subfigure}\hfill
	\caption{The structure along (a) IC2 and (b) IC3 from table \ref{tab:isochrones} presented in reduced units, $\tilde{r} = r\rho^{\frac{1}{3}} $. In each figure $g(r)$ is plotted as solid lines, $g_{intra}(r)$ as dotted lines, and $g_{inter}(r)$ as dashed lines. Just as the case with IC1, figure \ref{fig:cumene_RDF_IC1}, the intermolecular structure collapsed along each isochrone when presented in reduced units. The density changes along the isochrones are smaller than for IC1, so the effect of the scaling deviation is smaller for IC2 and IC3.}
	\label{fig:Cumene_gr_IC2_IC3}
\end{figure}

In figure \ref{fig:Cumene_gr_IC2_IC3} the structure along IC2 and IC3 from table \ref{tab:isochrones} are shown in reduced units. Just as we saw in figure \ref{fig:cumene_RDF_IC1}, for IC2 and IC3, the intermolecular structure along each isochrone collapses when presented in reduced units.

\begin{figure}[H]
	\centering
	\includegraphics[width=0.8\textwidth]{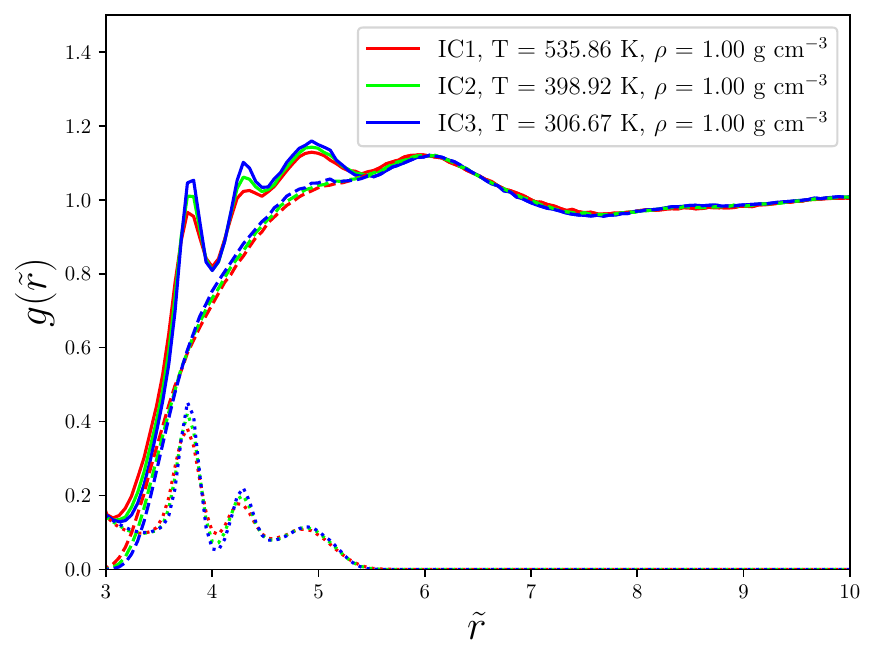}
	\caption{The structure along an isochore, $\rho = 1.00$ g cm$^{-3}$, crossing all three isochrones. $g(r)$ is plotted as a solid line, $g_{intra}(r)$ as a dotted line, and $g_{inter}(r)$ as a dashed line. The structure along each isochrone varies when presented in reduced units. In real-space, the measured structure is the distance between nearest-neighbor molecules, and the changes observed in the structure are quite small. The peak around $ \tilde{r} \approx 6$ is the distance between nearest-neighbor molecules, and even there, it seems that there is a small difference between the structures along each isochrone. }
	\label{fig:cumene_gr_isochrone_compare}
\end{figure}

In figure \ref{fig:cumene_gr_isochrone_compare}, $g(r)$ is separated into intramolecular and intermolecular contributions, plotted for all three isochrones. The three state points are isochoric, so the temperature dependence of the intramolecular structure can be clearly observed. At low temperature state points, the sharp intramolecular peaks are narrower and more well defined. The intermolecular structures along the three isochrones are different from each other. 

In real-space, the structure described by the pair-distribution function concerns distances between neighboring atoms and neighboring molecules. The changes we observe in the structure are between nearest-neighbor molecules, and these changes are very small. Still, in real space, the simulations show that the real space structure varies when moving from isochrone to isochrone, and that the intermolecular structure along an isochrone is invariant when presented in reduced units. To compare the results to the experimental data, the following section will examine the structure in reciprocal space.

\subsubsection{Structure in reciprocal space}

In experiments, we measure in reciprocal space instead of real space, and in this section, we present the structure factors as seen in an x-ray scattering experiment. By Fourier-transforming the partial-pair distribution functions, we can calculate the Faber-Ziman partial structure factors. In chapter \ref{chapter:Structure} the definition of the static structure factor and the relationship between the pair distribution function and static structure factor are presented. The main points are restated here. The static structure factor is calculated from the pair distribution function by:

\begin{equation}\label{eq:gr_to_sq_cumene}
S(q) -1 = \frac{4\pi  \rho_0}{q} \int_{0}^{\infty} r \left(g(r)-1\right)\sin\left(qr\right) \text{d}r
\end{equation}

In scattering theory the $-1$ term, comes from self scattering. It is often consider neither intramolecular nor intermolecular \cite{Fischer2006}. As shown in chapter \ref{chapter:Structure} $S(q)-1$, can be separated into an intramolecular and intermolecular contribution. 

\begin{equation} \label{eq:sq_seperation}
	S(q) -1 = S_{intra}(q)+ S_{inter}(q) 
\end{equation}

As mentioned, the Fourier-transform is additive, so one can calculate $S_{inter}(q)$ simply by Fourier-transforming $g_{inter}(r)$.  In this section $S(q)-1$ and $S_{inter}(q)$ are calculated by Fourier-transformation of $g(r)$ and $g_{inter}(r)$, as shown in equation \ref{eq:gr_to_sq_cumene}:

\begin{equation}
	S_{inter}(q) -1 = \frac{4\pi  \rho_0}{q} \int_{0}^{\infty} r \left(g_{inter}(r)-1\right)\sin\left(qr\right) \text{d}r
\end{equation}

$S_{intra}(q)$ is calculated as follows:

\begin{equation}
	S_{intra}(q) = S(q)-1 -S_{inter}(q)
\end{equation}

In this section when the static structure factor is decomposed into its intramolecular and intermolecular contributions, the results are presented as $S(q)-1$, $S_{intra}(q)$, and $S_{inter}(q)$. When the static structure factor is calculated, we do not have to account for the x-ray scattering lengths, if we assume that each atom has the same scattering length. The process of accounting for this would be calculated from the Faber-Ziman partial structure factors:

\begin{equation}
	S(q) =  \sum_\alpha \sum_\beta \frac{c_\alpha c_\beta f_\alpha f_\beta}{\left(\sum_\alpha c_\alpha f_\alpha \right)^2} S_{\alpha \beta} (q)
\end{equation}

where $\alpha$ and $\beta$  represent atom types, $c_\alpha$ is the atomic fraction for atom type $\alpha$,  $f_\alpha$ is the x-ray form factor for atom $\alpha$, and $S_{\alpha \beta} (q)$ is the Faber-Ziman partial structure factor for atom types $\alpha$ and $\beta$. For our model of cumene, the only atom type is carbon, and if we assume that each pseudo-atom has the same scattering length, they cancel out:

\begin{equation}
	S(q) =  \frac{c_C c_C f_C f_C}{\left( c_C f_C \right)^2} S_{C-C} (q) =  S_{C-C} (q)
\end{equation}

\begin{figure}[H]
\begin{subfigure}[t]{0.48\textwidth}
\centering
\includegraphics[width=\textwidth]{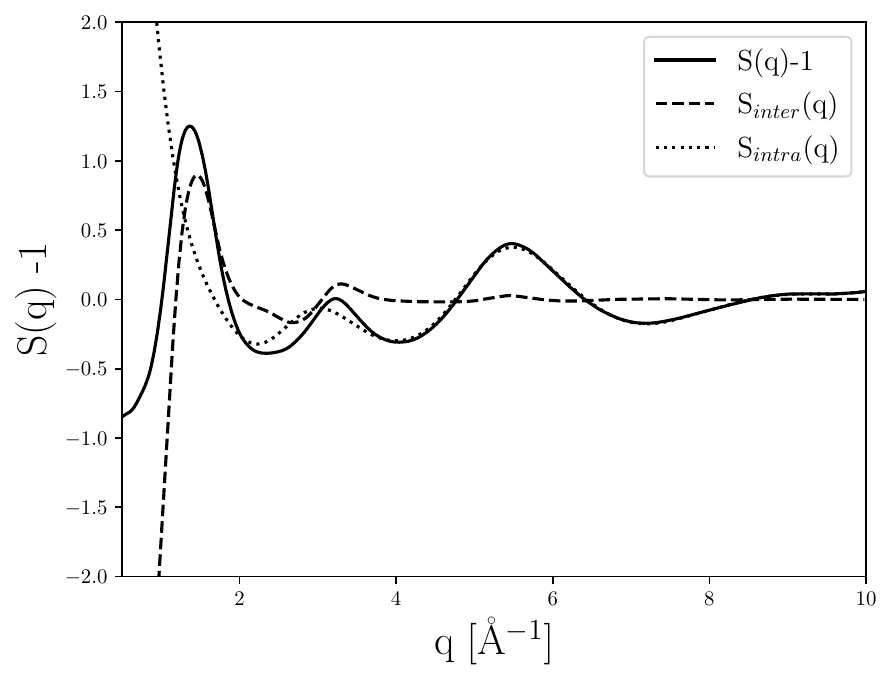}
\caption{}
\label{fig:Cumene_sq_example_cumene_a}
\end{subfigure}\hfill
\begin{subfigure}[t]{0.48\textwidth}
\centering
\includegraphics[width=\textwidth]{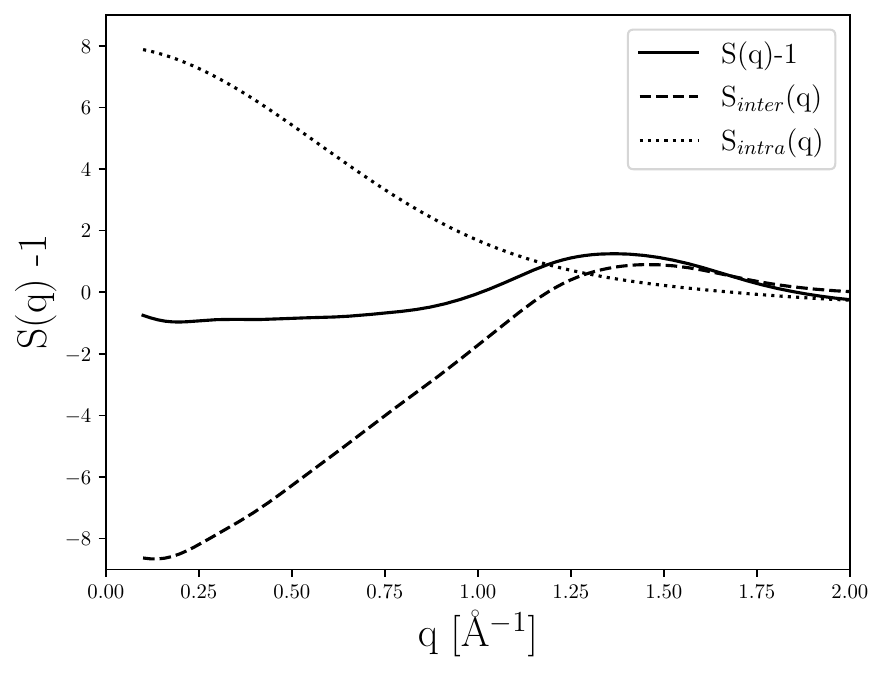}
\caption{}
\label{fig:Cumene_sq_example_cumene_b}
\end{subfigure}\hfill
\caption{The static structure factor separated into intramolecular and intermolecular contributions for a state point at $T = 535.86$ $^oK$, $\rho = 1.00$ g cm$^{-3}$. The solid black line is $S(q)-1$, the dashed line is $S_{inter}(q)$, and $S_{intra}(q)$ is the dotted line. Subfigure \ref{fig:Cumene_sq_example_cumene_a} shows $S(q)-1$ and its intramolecular and intermolecular contributions over a wide q range. At high q values, the oscillations in $S(q)-1$ arise from intramolecular contributions to the structure factor. The shape and position of the first peak are clearly affected by both intramolecular and intermolecular contributions. Figure  \ref{fig:Cumene_sq_example_cumene_b} shows $S(q)-1$ and its intramolecular and intermolecular contributions around the first peak in $S(q)-1$. The limits of $S(q)$, $S_{intra}(q)$ and $S_{inter}(q)$ when $q \rightarrow 0$ are discussed in the text and in chapter \ref{chapter:Structure}. The real-space representation of this state point is shown in figure \ref{fig:g(r)_cumene}. }
\label{fig:Cumene_sq_example_cumene}
\end{figure}

In figure \ref{fig:Cumene_sq_example_cumene} the static structure factor is separated into intramolecular and intermolecular contributions. In the following figures, The solid line is $S(q)-1$, the dashed line is $S_{inter}(q)$, and $S_{intra}(q)$ is the dotted line. Note that this will not always be explicitly stated in the figure legends, but then it will be stated in the figure captions. The separation of intra- and intermolecular contributions give rise to several interesting observations. At higher q-values, the oscillations in $S(q)-1$ originate from intramolecular contributions, whereas the oscillations from $S_{inter}(q)$ die out quickly. As shown in equation \ref{eq:sq_liquds_cumene} $S(q)$ is the sum of sinc functions, $\text{sinc}(qr_{ij})$, where $r_{ij}$ is the distance between atoms $i$ and $j$. The longer the distance between particles the quicker $\text{sinc}(qr_{ij})$ will go to 0 as a function of q. In general, intermolecular particle distances are longer than intramolecular particle distances; thus, intermolecular contributions to $S(q)$ will fade more quickly than intramolecular contributions.

\begin{equation}\label{eq:sq_liquds_cumene}
	S(q) -1 = \frac{1}{N \left<f^2\right> } \sum^N_{i\neq j} f_if_j \frac{\sin(qr_{ij})}{qr_{ij}}
\end{equation}

A second interesting observation is the effect that the intramolecular and intermolecular contributions have on the first peak of $S(q)$. In real-space the intramolecular and intermolecular contributions are relatively easy to separate. In inverse space, clearly both the intramolecular and intermolecular contributions affect $S(q)$ in almost all of q-space. In figure \ref{fig:Cumene_sq_example_cumene_b} the intramolecular and intermolecular contributions to the first peak are shown.

The behavior of $S_{intra}(q)$ and $S_{inter}(q)$ in the $q \rightarrow 0$ limit was discussed in chapter \ref{chapter:Structure}, but we restate the results. When $q \rightarrow 0$, $S(q)$ has the limit $S(0) = \rho_m k_b T \xi_T$. When $q \rightarrow 0$, we showed that the limit of $S_{intra}(q) \rightarrow 8$, where 8 is the number of atoms in a UA-cumene model molecule minus one. This is shown in equation \ref{eq:sq_intra_limit}. It follows that:

\begin{equation}
	 \lim_{q\rightarrow 0} S_{inter}(q) = \left(\rho_m k_b T \chi_T -1\right) -8
\end{equation}

 It is clear that this behavior affects both the peak position and width of the main structure peak. The peak position and width are characteristics that we used to describe our experimental measurement of the first peak in section \ref{sec:Cumene_exp_results}. The separation of intramolecular and intermolecular contributions to $S(q)-1$ and the behavior of $S_{intra}(q)$ and $S_{inter}(q)$ is consistent with previous works \cite{LehenyRL1996Ssoa}.

\begin{figure}[H]
	
	\centering
	\begin{subfigure}[t]{0.48\textwidth}
		\includegraphics[width=\textwidth]{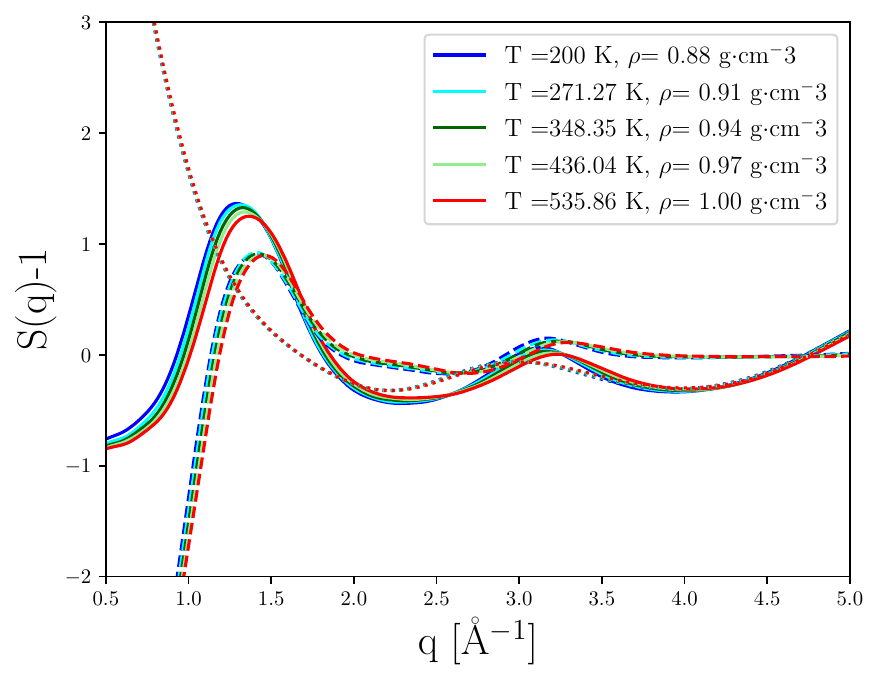}
		\caption{}
		\label{fig:Cumene_MD_sq_a}
	\end{subfigure}\hfill
	\begin{subfigure}[t]{0.48\textwidth}
		\centering
		\includegraphics[width=\textwidth]{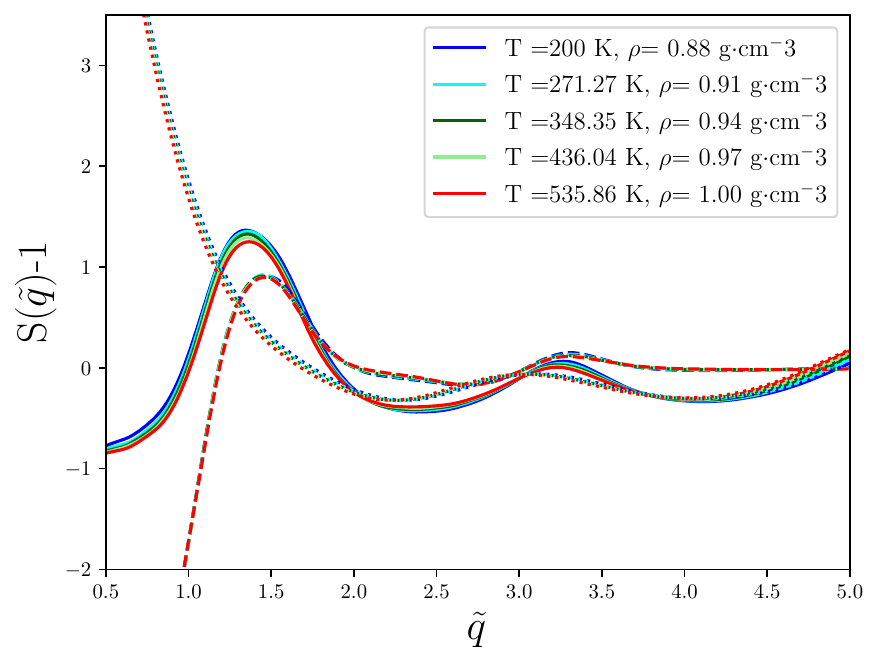}
		\caption{}
		\label{fig:Cumene_MD_sq_b}
		
	\end{subfigure}	
	\begin{subfigure}[t]{0.48\textwidth}
		\includegraphics[width=\textwidth]{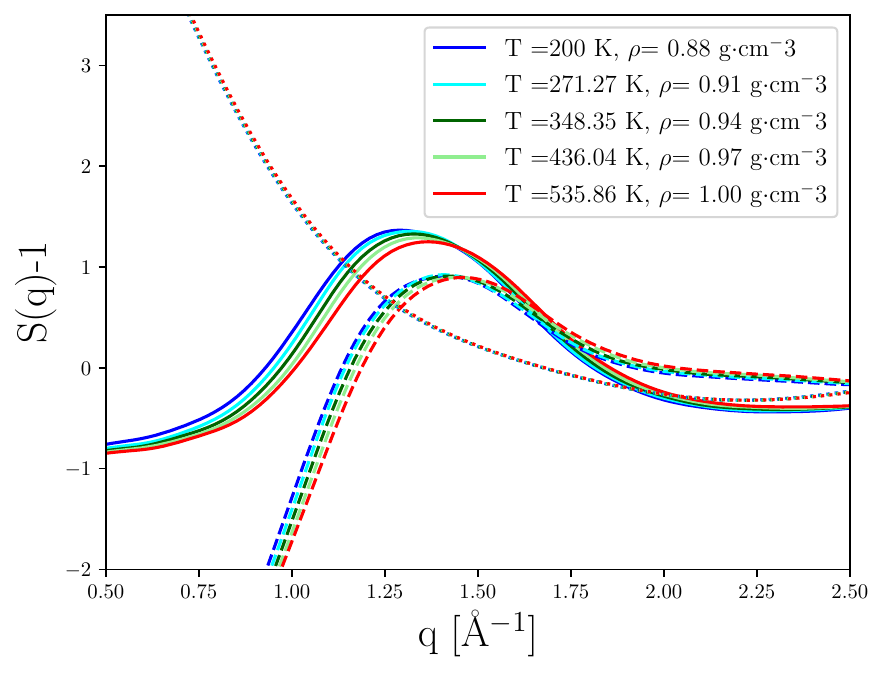}
		\caption{}
		\label{fig:Cumene_MD_sq_c}
	\end{subfigure}\hfill
	\begin{subfigure}[t]{0.48\textwidth}
		\centering
		\includegraphics[width=\textwidth]{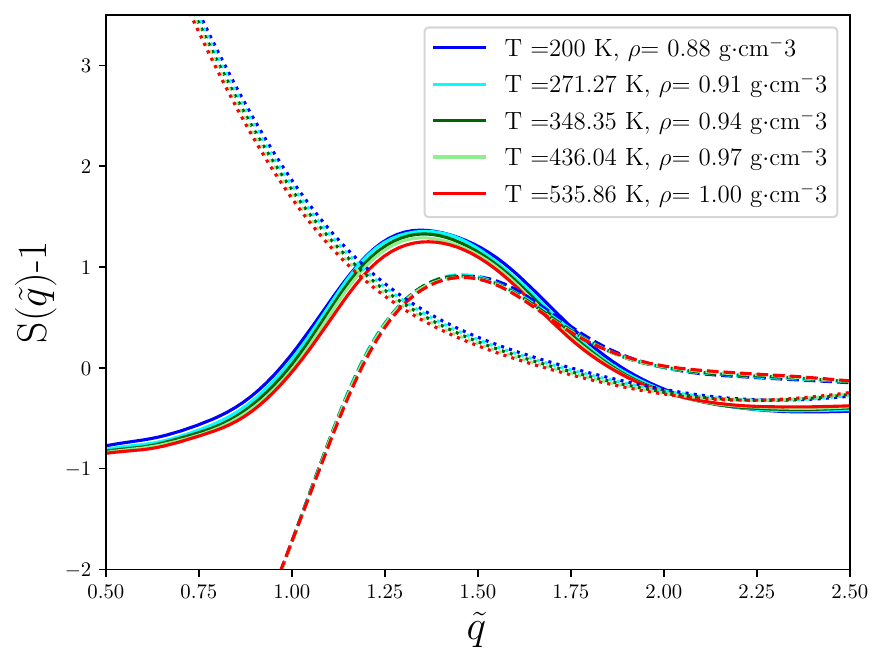}
		\caption{}
		\label{fig:Cumene_MD_sq_d}
	\end{subfigure}
	\caption{$S(q)$ along an isochrone along IC1 from table \ref{tab:isochrones}. The solid line is $S(q)-1$, the dashed line is $S_{inter}(q)$, and $S_{intra}(q)$ is the dotted line. In figure \ref{fig:Cumene_MD_sq_a} $S(q)-1$, $S_{inter}(q)$, and $S_{intra}(q)$ are shown along an isochrone in the experimental units. The intramolecular structures are basically invariant, while the intermolecular structure changes. In figure \ref{fig:Cumene_MD_sq_b}, the same five structure factors are plotted in reduced units, $\tilde{q} = q \rho^{-\frac{1}{3}} $. The reduced unit system that can scale the intermolecular structure does not scale the intramolecular structure, causing a scaling deviation in the total $S(q)$. figures \ref{fig:Cumene_MD_sq_c} and \ref{fig:Cumene_MD_sq_d} are close-ups on the first peak of $S(q)$ in the experimental and reduced units respectively. The same state points are shown in real-space in figure \ref{fig:cumene_RDF_IC1}.}¨
	\label{fig:Cumene_MD_sq}
\end{figure}

In figure \ref{fig:Cumene_MD_sq}, $S(q)-1$ along IC1 from table \ref{tab:isochrones} is plotted with the intramolecular and intermolecular contributions to $S(q)$. In subfigure \ref{fig:Cumene_MD_sq_a} $S(q)-1$, $S_{inter}(q)$, and $S_{intra}(q)$ are plotted in experimental units. The intramolecular structure is invariant, and changes to $S(q)-1$ along an isochrone arises from intermolecular changes. The measured intermolecular changes includes the effect of density. In subfigure \ref{fig:Cumene_MD_sq_b} $S(\tilde{q})-1$, $S_{inter}(\tilde{q})$, and $S_{intra}(\tilde{q})$ are all plotted in reduced units,  $\tilde{q} = q \rho^{-\frac{1}{3}} $. When plotted in the reduced unit system, the intermolecular structure collapsed almost perfectly into a single curve, whereas the intramolecular structure no longer does because it is scaled by $\rho$, but does not change with $\rho$. This \textit{scaling deviation} from the intramolecular contribution is affecting the total $S(q)-1$ over the entire q-space.

In figures \ref{fig:Cumene_MD_sq_c} and \ref{fig:Cumene_MD_sq_d}, the first peak of $S(q)$ is shown in the experimental and reduced units, respectively. The change from experimental to reduced units leading to the main peak of $S(q)$ gives a reasonable, but not perfect collapse of $S(q)$. The effect of the scaling deviation from $S_{intra}(q)$ is the origin of this, and clearly affects both the peak position and width of the peak. In subfigure \ref{fig:Cumene_MD_sq_d} the width of the red high density, high temperature $S(q)$ is narrower than the blue low density, low temperature $S(q)$.

\begin{figure}[H]
	
	\centering
	\begin{subfigure}[t]{0.48\textwidth}
		\includegraphics[width=\textwidth]{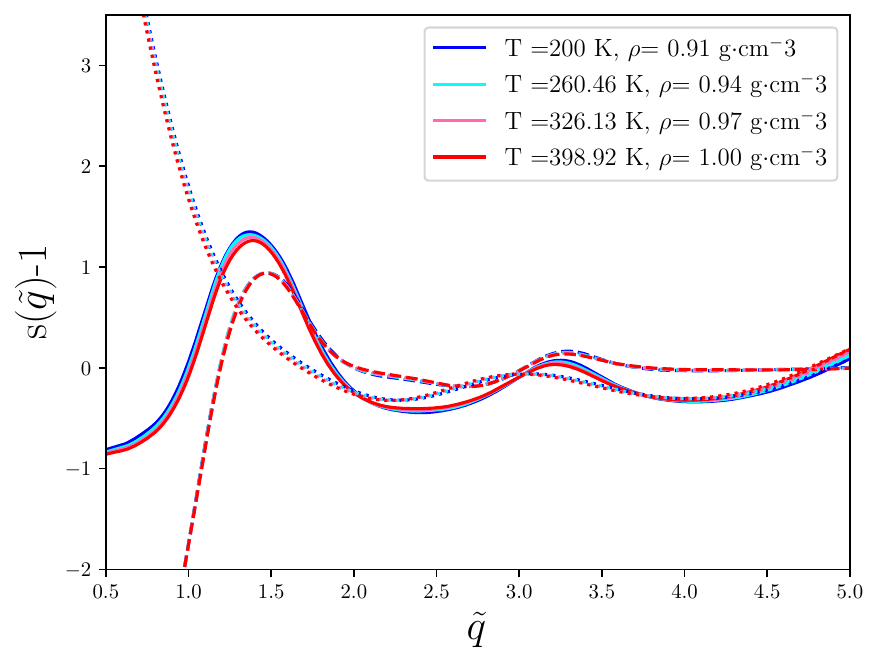}
		\caption{IC2}
		\label{fig:Cumene_MD_sq_IC2}
	\end{subfigure}\hfill
	\begin{subfigure}[t]{0.48\textwidth}
		\centering
		\includegraphics[width=\textwidth]{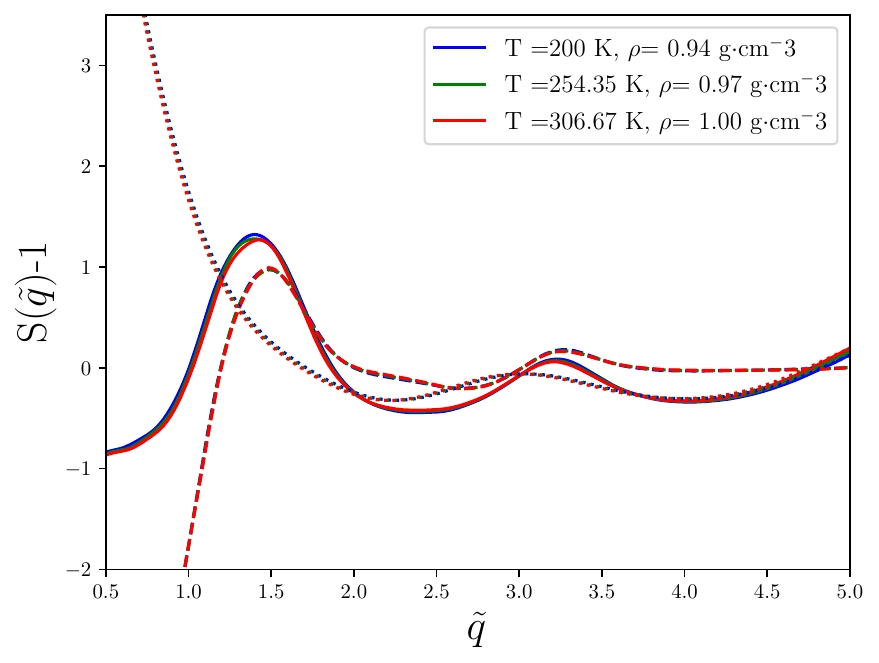}
		\caption{IC3}
		\label{fig:Cumene_MD_sq_IC3}
	\end{subfigure}
	\caption{The structure along IC2, subfigure \ref{fig:Cumene_MD_sq_IC2},  and IC3, subfigure \ref{fig:Cumene_MD_sq_IC3}, from table \ref{tab:isochrones} presented in reduced units. Just like in figure \ref{fig:Cumene_MD_sq}, the intermolecular structure collapsed along each isochrone when presented in reduced units. The effect of the scaling deviation from the intramolecular contributions, which arises when they are presented in reduced units, is smaller for IC2 and IC3, as the density changes are smaller than for IC1. The structure along each isochrone is shown in real-space in figure \ref{fig:Cumene_gr_IC2_IC3}.}
	\label{fig:Cumene_MD_sq_IC2_and_IC3}
\end{figure}

In figure \ref{fig:Cumene_MD_sq_IC2_and_IC3} the structure along IC2 and IC3 from table \ref{tab:isochrones} are shown in reduced units. The intermolecular structure along each isochrone also collapsed for the other two isochrones. The density differences along IC2 and IC3 are smaller than those along IC1, so the effect of the scaling deviation is smaller for IC2 and IC3. However, even for IC3, for which the density change is approximately $6 \%$, the effect of the scaling deviation is still visible on $S(\tilde{q})-1$. A 6 \% change in density was comparable to the largest density change along an isochrone in the experimental data.

\begin{figure}[H]
	\centering
	\includegraphics[width=0.65\textwidth]{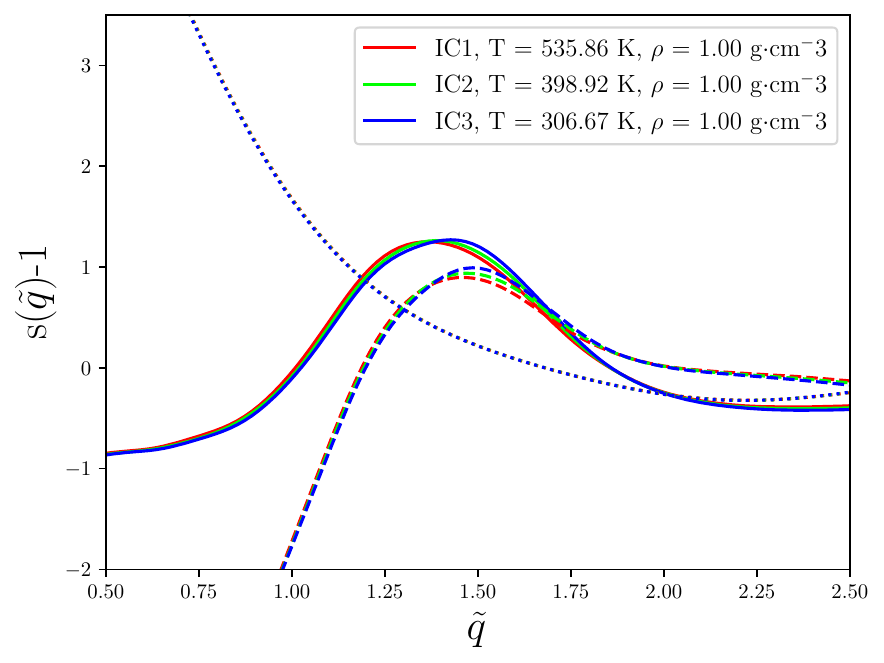}
	\caption{Isochoric comparison between the structure along the three isochrones. The intramolecular contributions were the same for all three simulations in reduced units because the density was the same for all simulations, $\tilde{q} = q \rho^{-\frac{1}{3}}$. The intermolecular structures differ for different isochrones and these different intermolecular structures affect the total $S(q)$. A similar figure in real space is shown in figure \ref{fig:cumene_gr_isochrone_compare}. }
	\label{fig:Cumene_MD_sq_compare}
\end{figure}

In figure \ref{fig:Cumene_MD_sq_compare} the structure along each isochrone is compared with that along an isochore in reduced units. The intramolecular contributions in reduced units are the same for all three state points  because the structures are compared along an isochore. This can be observed from the definition of $\tilde{q}$, since the density is the same for all three state points. The intermolecular structure clearly differs along each of the three isochrones, and the changes that happen to $S(\tilde{q})-1$ are purely intermolecular.

Using the UA-cumene model, we gained important insight into the microscopic structure of cumene. As shown in figures \ref{fig:Cumene_MD_sq} and \ref{fig:Cumene_MD_sq_IC2_and_IC3}, the intermolecular structure along an isochrone is invariant when presented in reduced units, as predicted by the isomorph theory. In figure \ref{fig:Cumene_MD_sq_compare} it is shown that the intermolecular structure changes when moving from isochrone to isochrone. Experimentally, we saw a good, but not perfect collapse of the structural descriptors. This can most likely be explained by the scaling deviation from the intramolecular structures. The fact that both the reduced diffusion coefficient and the intermolecular structure are invariant along the same lines in the phase diagram provides strong evidence for the existence of pseudo-isomorphs.

\subsection{Comparison between the MD-model and experiment}

In this subsection, the goal is to compare the first peak obtained from the MD simulations with the measured peak from the x-ray experiment. To compare the peak from every single state point, we fitted the peak from the MD-simulations with a spilt-pseudo Voigt function in the range $q \in [1,1.7]$. This is the same procedure followed when the experimental data was fitted.

\begin{figure}[H]
	\centering
	\begin{subfigure}[b]{0.495\textwidth}
		\includegraphics[trim = 30mm 80mm 40mm 80mm, clip=true,width=0.99\textwidth]{Chapters/Figures/rho_qmax_scaled.pdf}
		\caption{Experimental data, reprint of figure \ref{fig:fit_density_Gamma_compare_a}}
		\label{fig:simulation_experiment_compare_rho_a}	
	\end{subfigure}
	\begin{subfigure}[b]{0.495\textwidth}
		\includegraphics[width=0.99\textwidth]{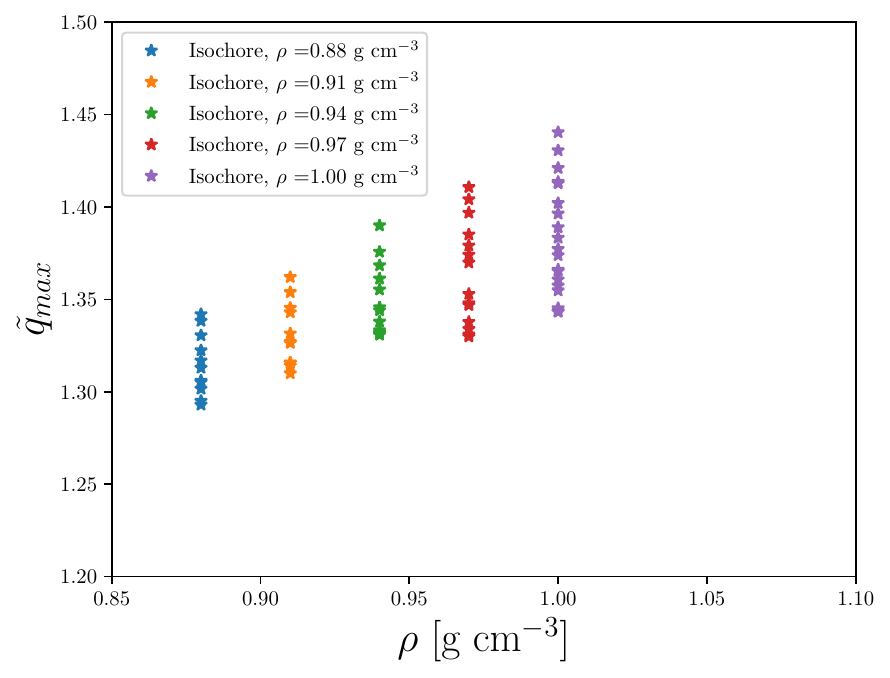}
		\caption{Simulation data}
		\label{fig:simulation_experiment_compare_rho_b}		
	\end{subfigure}
	
	\begin{subfigure}[b]{0.495\textwidth}
		\includegraphics[trim = 30mm 80mm 40mm 80mm, clip=true,width=0.99\textwidth]{Chapters/Figures/rho_FWHM_scaled.pdf}	
		\caption{Experimental data, reprint of figure \ref{fig:fit_density_Gamma_compare_c}}	
		\label{fig:simulation_experiment_compare_rho_c}
	\end{subfigure}
	\begin{subfigure}[b]{0.495\textwidth}
		\includegraphics[width=0.99\textwidth]{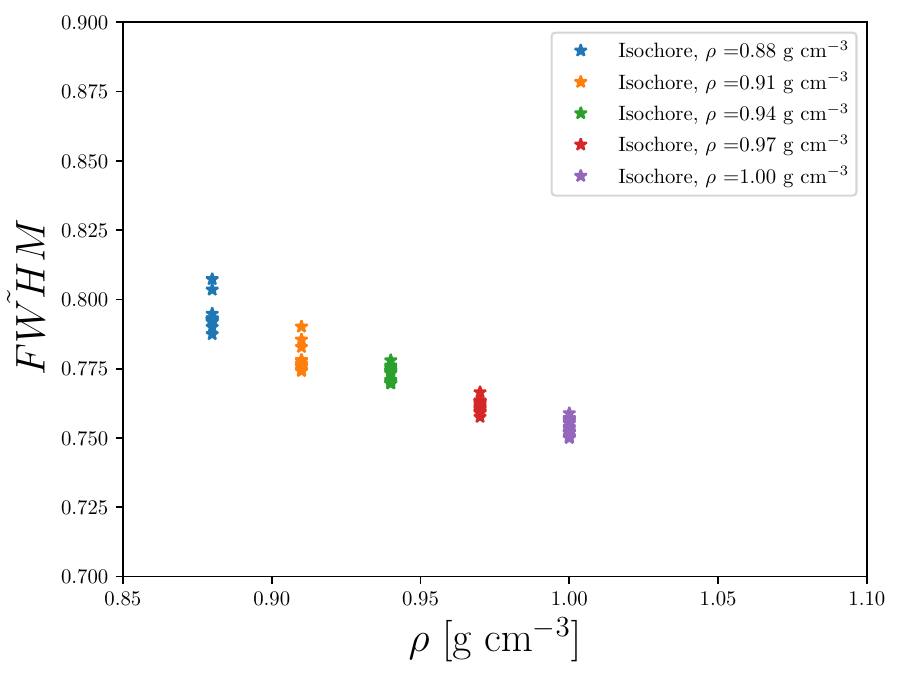}	
		\caption{Simulation data}
		\label{fig:simulation_experiment_compare_rho_d}
	\end{subfigure}
	\begin{subfigure}[b]{0.495\textwidth}
		\includegraphics[trim = 30mm 80mm 40mm 80mm, clip=true,width=0.99\textwidth]{Chapters/Figures/rho_skewness.pdf}		
		\caption{Experimental data, reprint of figure \ref{fig:fit_density_Gamma_compare_e} }
		\label{fig:simulation_experiment_compare_rho_e}
	\end{subfigure}
	\begin{subfigure}[b]{0.495\textwidth}
		\includegraphics[width=0.99\textwidth]{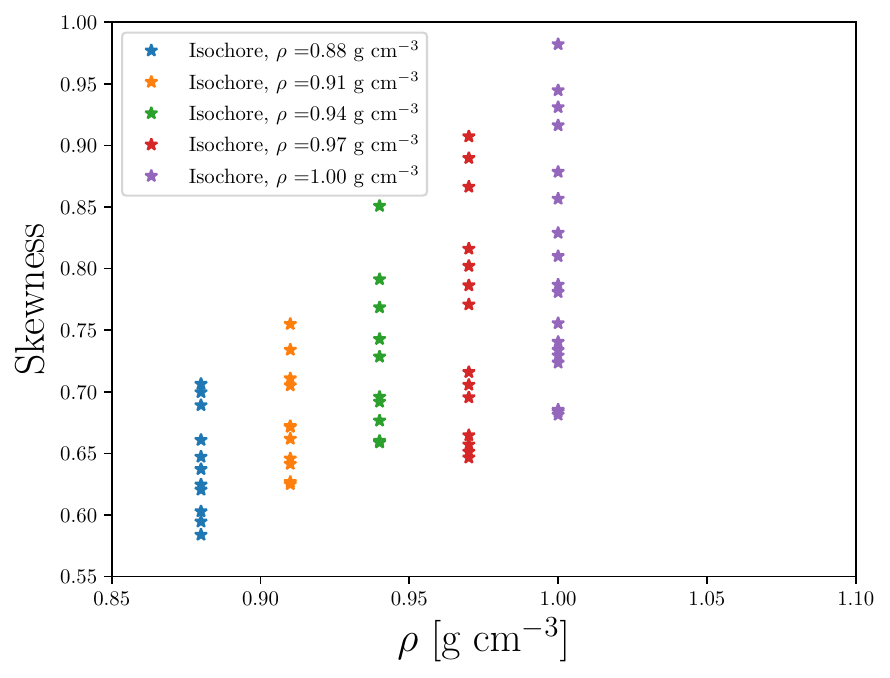}		
		\caption{Simulation data}
		\label{fig:simulation_experiment_compare_rho_f}
	\end{subfigure}
	\caption{Comparison between the density dependences of the three descriptors, $\tilde{q}$, $\tilde{FWHM}$, and skewness in the experimental result and MD simulations. The MD simulations were performed along five isochors and at a larger temperature range than the experiments. 	}
	\label{fig:simulation_experiment_compare_rho}
\end{figure}

In figure \ref{fig:simulation_experiment_compare_rho} the density dependence of the three descriptors of the peak, $q_{max}$, FWHM, and skewness of the peak from experiments and MD simulations are compared. In the MD simulations, we have simulated along five isochores, while experimentally, we measured along isotherms and isobars. For the experimental data there are therefore at most 4-5 state points with the same density, so when comparing, we will have to look at the tendency of the behavior. In general, the fitting results for the MD simulations were similar to the experimental results. The FWMH tended to be relatively invariant for both experiment and simulations, but the FWHM seem to be smaller for the MD simulations.  For the simulations, the FWHM is clearly affected from the scaling deviation, causing the FWHM to increase with smaller densities. 

\begin{figure}[H]
	\centering
	\begin{subfigure}[t]{0.495\textwidth}
		\includegraphics[trim = 30mm 80mm 40mm 80mm, clip=true,width=0.99\textwidth]{Chapters/Figures/Gamma_qmax_scaled.pdf}
		\caption{Experimental data,$\gamma_{exp}$= 4.8, reprint of figure \ref{fig:fit_density_Gamma_compare_b}}	
		\label{fig:simulation_experiment_compare_Gamma_a}
	\end{subfigure}\hfill
	\begin{subfigure}[t]{0.495\textwidth}
		\includegraphics[width=0.99\textwidth]{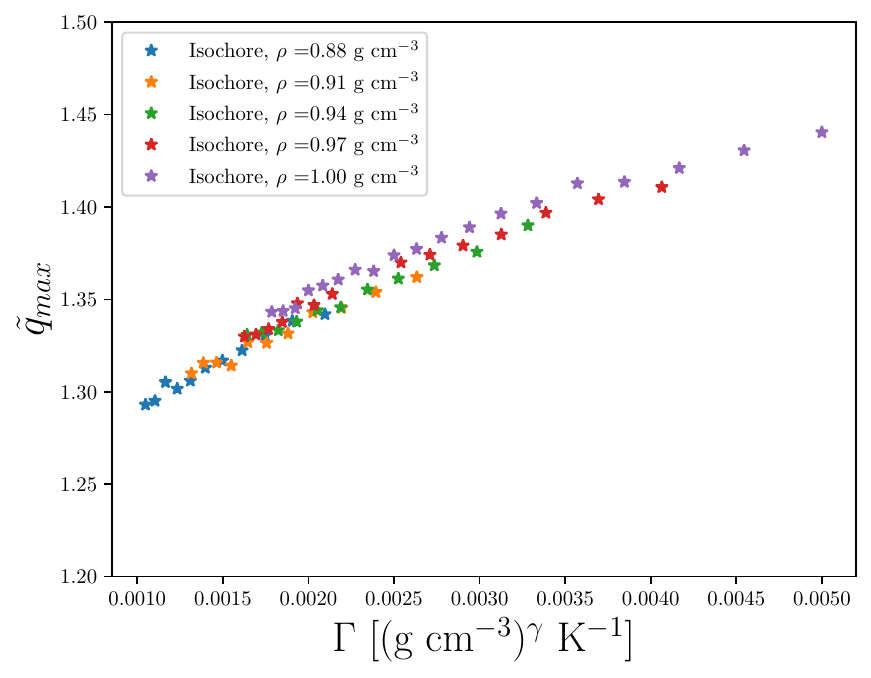}
		\caption{Simulation data, $\gamma_{MD}=6.8$ }	
		\label{fig:simulation_experiment_compare_Gamma_b}	
	\end{subfigure}
	
	\begin{subfigure}[t]{0.495\textwidth}
		\includegraphics[trim = 30mm 80mm 40mm 80mm, clip=true,width=0.99\textwidth]{Chapters/Figures/Gamma_FWHM_scaled.pdf}	
		\caption{Experimental data, $\gamma_{exp}$= 4.8, reprint of figure \ref{fig:fit_density_Gamma_compare_d}}
		\label{fig:simulation_experiment_compare_Gamma_c}	
	\end{subfigure}\hfill
	\begin{subfigure}[t]{0.495\textwidth}
		\includegraphics[width=0.99\textwidth]{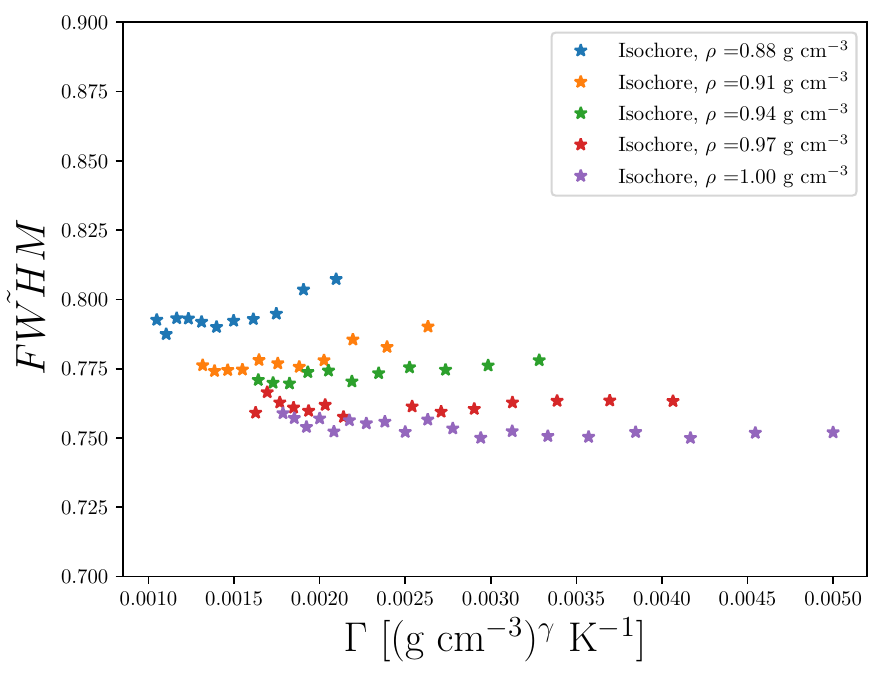}	
		\caption{Simulation data,$\gamma_{MD}=6.8$}
		\label{fig:simulation_experiment_compare_Gamma_d}
	\end{subfigure}
	\begin{subfigure}[t]{0.495\textwidth}
		\includegraphics[trim = 30mm 80mm 40mm 80mm, clip=true,width=0.99\textwidth]{Chapters/Figures/Gamma_skewness.pdf}		
		\caption{Experimental data,$\gamma_{exp}$= 4.8, reprint of figure \ref{fig:fit_density_Gamma_compare_f}}
		\label{fig:simulation_experiment_compare_Gamma_e}
	\end{subfigure}\hfill
	\begin{subfigure}[t]{0.495\textwidth}
		\includegraphics[width=0.99\textwidth]{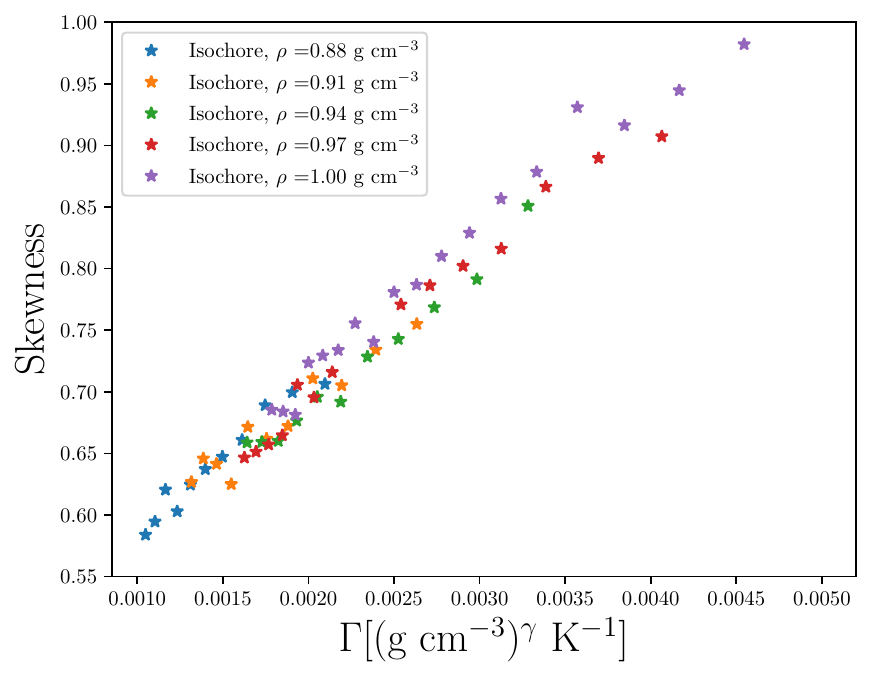}		
		\caption{Simulation data,$\gamma_{MD}=6.8$}
		\label{fig:simulation_experiment_compare_Gamma_f}
	\end{subfigure}
	\caption{Comparison between the $\Gamma$ dependence of the three descriptors, $\tilde{q}$, $\tilde{FWHM}$, and skewness in the experimental result and MD simulations. The $\Gamma$ values cannot be directly compared, since $\gamma_{MD} = 6.8$ and $\gamma_{exp} = 4.8$. In general, the behavior of the peak seemed to behave similarly between the experiment and simulations.	}
	\label{fig:simulation_experiment_compare_Gamma}
\end{figure}

In figure \ref{fig:simulation_experiment_compare_Gamma} the three descriptors of the peak are plotted against $\Gamma$. As shown in figure \ref{fig:Cumene_density_scaling_MD}, the $\gamma$ value used for the MD simulations is, $\gamma_{MD} =6.8$, while for the experimental model $\gamma_{exp}=4.8$. As a consequence of the different $\gamma$ values for the model and real molecule, $\Gamma_{exp}(\rho,T) \neq \Gamma_{MD}(\rho,T)$.  When we compare experiment and model, it is best to consider the state points in the experimental and simulation phase diagrams. In the experiments, we are generally able to be at much more viscous state points, while in simulations, we can reach higher temperatures. In general, the behavior of the peak as function of $\Gamma$ is similar in experiment and model. We observed a pretty good, but not perfect, collapse of the peak position and skewness as a function of $\Gamma$, but the scaling deviation from the intramolecular contributions affected both the experiment and the simulations. Interestingly, when comparing figure \ref{fig:simulation_experiment_compare_rho_d} and \ref{fig:simulation_experiment_compare_Gamma_d} the $\tilde{FWHM}$ of the peak did not change a lot, but when it did, it seemed to be as a function of density rather than $\Gamma$. The scaling deviation effect is the most likely cause of this. As shown in figure \ref{fig:Cumene_MD_sq_compare}, the scaling deviation along an isochore is constant since $\tilde{q} = q\rho^{- \frac{1}{3}}$.

\section{Structural changes when moving in the phase diagram} \label{sec:strucral changes}

Thus far, we have shown the collapse of the intermolecular structure along an isochrone when presented in reduced units. We have also discussed the role of the scaling deviation arising from the intramolecular contributions to the structure factor. The goal of this section is to express what is changing in the main peak of $S(\tilde{q})$ when moving between state points.

A very simple approach is simply to write up the change seen in $S(\tilde{q})$ between two state points. Here are the changes one would see when subtracting S(q) at two state points, $S^{(1)}\left(\tilde{q}\right)$ at $\left(\rho_1,T_1\right)$  and $S^{(2)}\left(\tilde{q}\right)$ at $\left(\rho_2,T_2\right)$ : 

\begin{equation}\label{eq:sq_isolate_inter}
	S^{(1)}\left(\tilde{q}\right) - S^{(2)}\left(\tilde{q}\right) = 	S^{(1)}_{intra}\left(\rho_1^{-\frac{1}{3}}q\right) - S^{(2)}_{intra}\left(\rho_2^{-\frac{1}{3}}q\right)  + S^{(1)}_{inter}\left(\rho_1^{-\frac{1}{3}}q\right) - S^{(2)}_{inter}\left(\rho_2^{-\frac{1}{3}}q\right)
\end{equation}

\begin{figure}[h]
	\centering
	\includegraphics[width=0.65\textwidth]{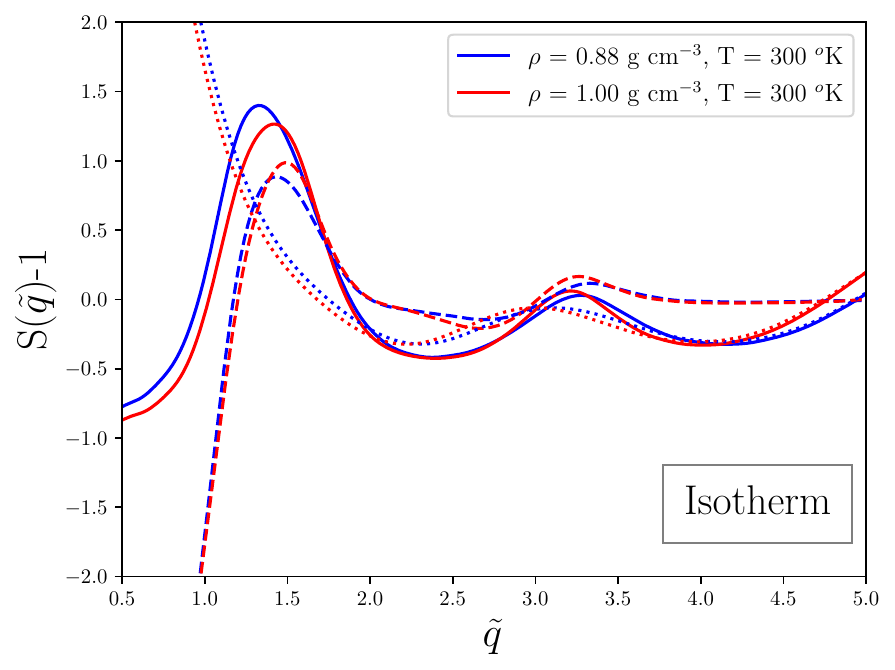}
	
	\caption{The structure for two state points along an isotherm from the UA-model. The solid line is $S(\tilde{q})-1$, the dashed line is $S_{inter}(\tilde{q})$, and $S_{intra}(\tilde{q})$ is the dotted line. The measured changes in $S(\tilde{q})$ are from both real intermolecular structural changes and the scaling deviation arising from the intramolecular structure that does not scale with $\rho$. }
	\label{fig:delta_sq_isotherm}
\end{figure}

Along an isotherm, the reduced intermolecular structure changes since we are moving between pseudo-isomorphs. The density also changes, so there is a scaling deviation from using the reduced units. The structural change seen in $S(q)$ is a mixture of real intermolecular structural changes and the scaling deviation from the intramolecular structure. In figure \ref{fig:delta_sq_isotherm} this is shown. 

Under isochoric conditions, $\rho_1 = \rho_2$, the term $S^{(1)}_{intra}\left(\rho_1^{-\frac{1}{3}}q\right) - S^{(2)}_{intra}\left(\rho_2^{-\frac{1}{3}}q\right)  \approx  0$ in equation \ref{eq:sq_isolate_inter}. If the two state points in are subtracted in the experimental units:  $	S^{(1)}\left(q\right) - S^{(2)}\left(q\right) $, one can also isolate the intermolecular change, but this intermolecular change also includes the effect of density changes. Only along along an isochore is it possible to isolate an signal independent of density.

\begin{figure}[H]
	\begin{subfigure}[t]{0.48\textwidth}
		
		\includegraphics[width=\textwidth]{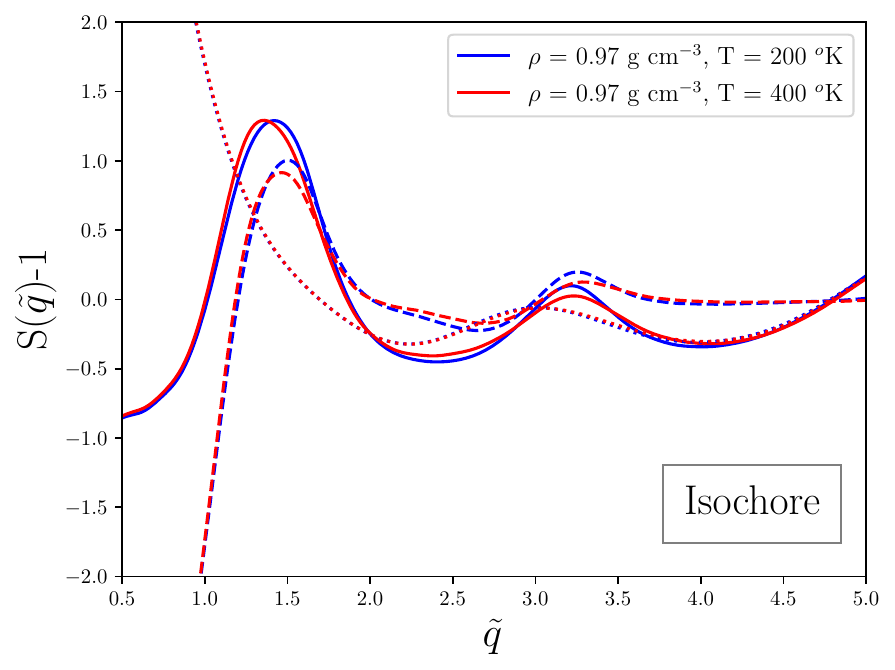}
		\caption{}
		\label{fig:delta_sq_isochore_a}
	\end{subfigure}\hfill
	\begin{subfigure}[t]{0.48\textwidth}
		
		\includegraphics[width=\textwidth]{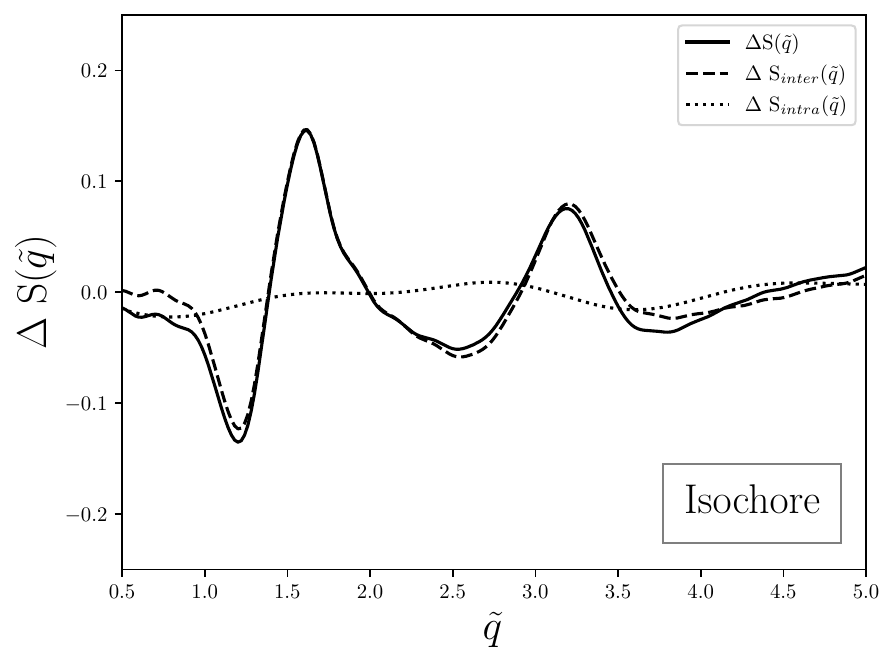}
		\caption{ }
		\label{fig:delta_sq_isochore_b}
	\end{subfigure}\hfill
	
	\caption{Isolating a intermolecular signal along an isochor. The results of the united-atom MD simulations are used as an example. The solid lines shows $S(\tilde{q})-1$, the dashed lines shows $S_{inter}(\tilde{q})$, and $S_{intra}(\tilde{q})$ indicated with the dotted lines. In figure \ref{fig:delta_sq_isochore_a} the structure factors for two state points along an isochore are plotted, along with the intermolecular and intramolecular contributions to $S(\tilde{q})$. The intramolecular structure is invariant in reduced units, whereas there is a clear intermolecular change to the structure arising from something other than density. In figure \ref{fig:delta_sq_isochore_b}, $\Delta S(\tilde{q}) = 	S^{(1)}\left(\tilde{q}\right) - S^{(2)}\left(\tilde{q}\right)$,  along with  $\Delta S_{intra}(\tilde{q})$ and $\Delta S_{inter}(\tilde{q})$ are plotted for the two state points. It can be seen that $\Delta S(\tilde{q}) \approx \Delta S_{inter}(\tilde{q})$ for a temperature jump of a 200 K. This shows that experimentally, it is possible to isolate intermolecular changes in the structures that are independent of density. For this to work, there are high demands to the quality of the experimental data.
	}
	\label{fig:delta_sq_isochore}
\end{figure}

In figure \ref{fig:delta_sq_isochore} structural change along an isochore in reduced unit is illustrated for cumene. Figure \ref{fig:delta_sq_isochore_a} shows  $S(q)-1$,  $S_{inter}(q)$, $S_{intra}(q)$ for two state points on an isochore. Figure \ref{fig:delta_sq_isochore_b} shows the change for $S(q)-1$,  $S_{inter}(q)$, $S_{intra}(q)$ between the two state points. The structural change in $S(q)$ along an isochor is almost exclusively intermolecular and independent of density. This "trick" for isolating density independent intermolecular structural changes assumes that $S^{(1)}_{intra}\left(\rho_1^{-\frac{1}{3}}q\right) - S^{(2)}_{intra}\left(\rho_2^{-\frac{1}{3}}q\right)  =  0$. Cumene is a small and relatively rigid molecule with few internal degrees of freedom. For a different sample this assumption might work less well.

Being able to separate the intramolecular and intermolecular contributions to $S(\tilde{q})$ experimentally would be very interesting. There are several approaches to obtain microscopic information directly from scattering data. Reverse Monte Carlo (RMC) techniques like Emperical Potential Structure Refinement (EPSR) can be used for this purpose \cite{SoperAlan2017}. For a good result, they require both, x-ray and several types of isotopic substituted neutron measurements. For two  isochoric measurements it is possible to isolate an purely intermolecular changes independent of density changes; However, the practical use of this signal is difficult to see.

\begin{figure}[H]
	\begin{subfigure}[t]{0.48\textwidth}
		
		\includegraphics[width=\textwidth]{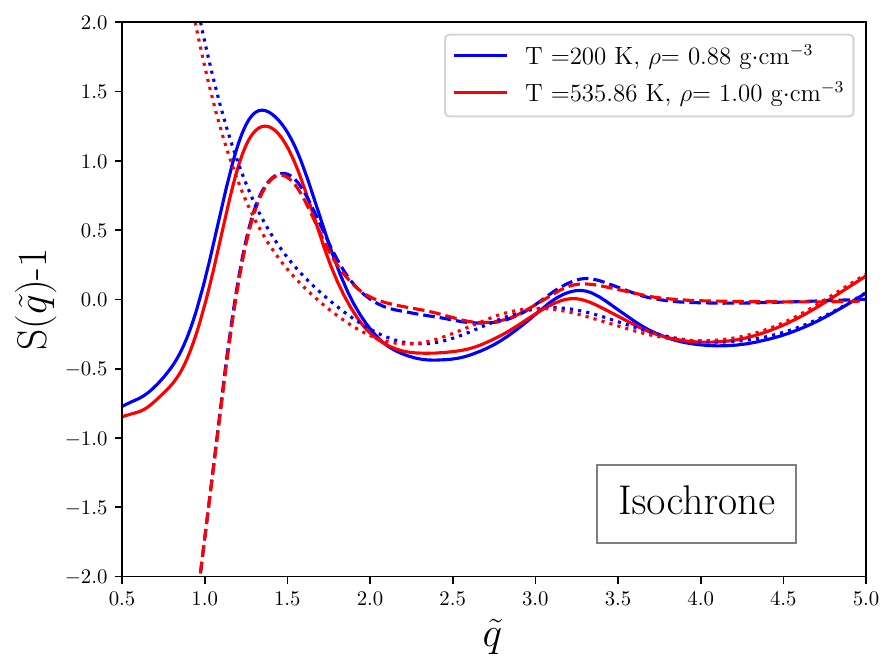}
		\caption{}
		\label{fig:delta_sq_isochrone_a}
	\end{subfigure}\hfill
	\begin{subfigure}[t]{0.48\textwidth}
		
		\includegraphics[width=\textwidth]{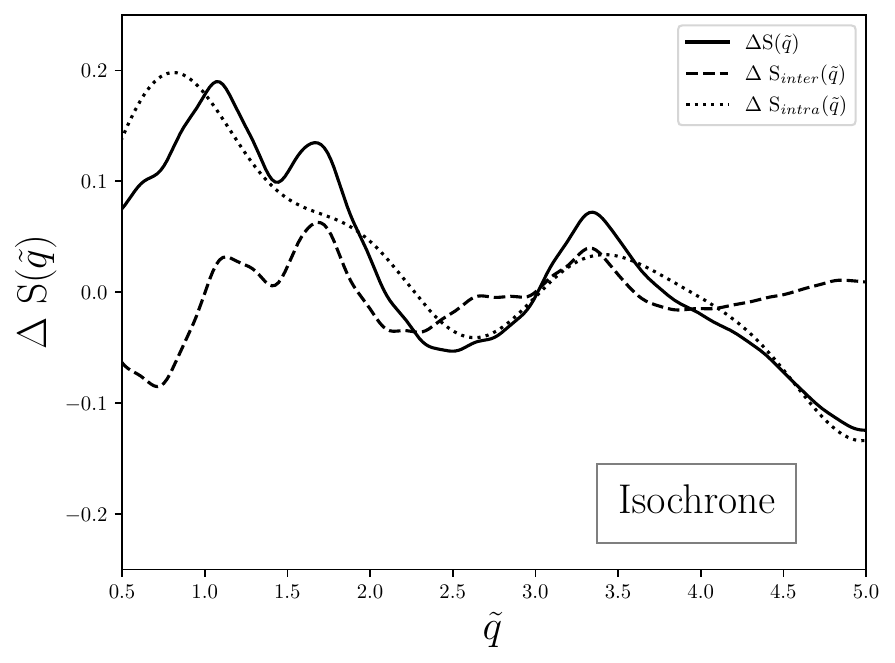}
		\caption{ }
		\label{fig:delta_sq_isochrone_b}
	\end{subfigure}\hfill
	
	\caption{The changes seen in $S(q)$ along an isochrone.  In figure \ref{fig:delta_sq_isochore_a} the structure factors for two state points along an isochrone are plotted along with the intermolecular and intramolecular contributions to $S(\tilde{q})$. The solid lines is $S(\tilde{q})-1$, the dashed lines is $S_{inter}(\tilde{q})$, and $S_{intra}(\tilde{q})$ is the dotted lines.  In figure \ref{fig:delta_sq_isochore_b} $\Delta S(\tilde{q})$ is plotted along with $\Delta S_{intra}(\tilde{q})$ and $\Delta S_{inter}(\tilde{q})$ for the two state points. The changes in $S(\tilde{q})$ follow the scaling deviation originating from the intramolecular structure not scaling with the density.}
	\label{fig:delta_sq_isochrone}
\end{figure}

Now we consider the structural changes we would measure along an isochrone in reduced units. Along an isochrone, we would expect that the intermolecular contribution to $S(\tilde{q})$ is invariant. So the term $S^{(1)}_{inter}\left(\rho_1^{-\frac{1}{3}}q\right) - S^{(2)}_{inter}\left(\rho_2^{-\frac{1}{3}}q\right)  \approx  0$ in equation \ref{eq:sq_isolate_inter}. The change in $S(q)$ we see is almost purely the scaling deviation from the intermolecular structure not scaling with density. This is shown in figure \ref{fig:delta_sq_isochrone_a}, where the structure at two state points along the same isochrone is plotted. In figure \ref{fig:delta_sq_isochrone_b} the changes in structure between the two state points are plotted. Along an isochrone, the measured change in $S(\tilde{q})$ arises from the scaling deviation. The intermolecular difference does not seem systematic and seems to fluctuate around 0. The intermolecular difference and are smaller than the intramolecular difference. It is also worth noting that it is very difficult to construct an perfect isochrone, defined here as state point for which the reduced diffusion coefficient is the same. In figure \ref{fig:Cumene_MD_isochrones}, the reduced diffusion coefficients are plotted for the two state points. While the two state points have approximately the same reduced diffusion coefficient there are small differences.

The goal of this section was to illustrate the structural change seen in $S(\tilde{q})$ for pseudo-isomorphs when comparing the structure between two state points in reduced units. Changing the density results in a scaling deviation from the intramolecular structure when presented in reduced units.  The effect of the scaling deviation is dependent on the density change between the state points and the form of $S_{intra}(\tilde{q})$. The scaling deviation for cumene has the biggest effect in same q-range as the main peak of $S(\tilde{q})$, this is not necessarily the case for every molecule. Moving between pseudo-isomorphs causes the intermolecular structure to change in reduced units. If these two state points are on an isochore, there is no scaling deviation from the reduced units. How much the intermolecular structure changes between the two state points is most likely also different for different liquids.

	\clearpage
	\section{Discussion and conclusion}
	
	In this chapter, the results of one of the first systematic tests of the existence of pseudo-isomorphs have been discussed. The studied liquid, cumene, is a small organic glass-former for which the dynamical prediction of isomorph theory has been shown to be true \cite{Ransom2017, Wase2018, Wase2018_nature, HansenHenrietteW.2017Cbfm,}. In section \ref{sec:Cumene_exp_results} we presented the results of a high-pressure experiment measuring the first peak of $S(q)$ as a function of pressure and temperature. The structure of cumene was compared along isotherms, isobars, isochores, and isochrones, and the best collapse was clearly observed along isochrones. When $S(q)$ curves along isochrones were plotted on top of each other, the structure collapsed nicely into a single curve, see figure \ref{fig:rainbow_zoom}. When the peak was fitted and various parameters were plotted, we also observed the best collapse along isochrones, see figure \ref{fig:fit_density_Gamma_compare}. The collapse along isochrones was good, but not perfect. 
	
	Using a united atom MD-model, we obtained a microscopic understanding of the liquid structure of cumene. We observed that the intermolecular structure was invariant in reduced units along the isochrones. The intramolecular structure in the model was not invariant in reduced units. This gave a density-dependent scaling deviation when plotting the intramolecular structure in reduced units, $\tilde{q} = \rho^{-\frac{1}{3}} q$. In real space, there are some overlap between inter- and intramolecular structures in the range $r \approx$ 3-5 Å. The scaling deviation was also visible in the total $g(r)$. The sharp peaks from the third and fourth nearest neighbors are clearly shifted from the scaling deviation, see figure \ref{fig:cumene_RDF_IC1}. However, in real space, the effect of the scaling error is limited to a relatively small $r$ range, up to $\approx$ 5 Å. In reciprocal space, the scaling deviation is not confined to a single part of $S(q)$. The effect of the scaling deviation is visible on the first peak of $S(q)$, where, in the MD simulations, it can affect both the width and position of the first peak. The scaling deviation is the most likely cause of the good, but not perfect collapse we observed experimentally.
	
	In section \ref{sec:Cumene_rho_vs_Gamma} we compared the structure along isochores and isochrones. We have shown that cumene has pseudo-isomorphs, so there exist lines in the phase diagram where the intermolecular structure is invariant in reduced units. In section \ref{sec:Cumene_rho_vs_Gamma} and \ref{sec:strucral changes} we discuss the role of density when measuring the structure. When the measurements are presented in experimental units, the density changes cause the structure to change. If the effect of density is scaled out, by using reduced units, then there are two origins to the changes seen in $S(q)$. Firstly, there are intermolecular structural changes that occur when moving from pseudo-isomorph to pseudo-isomorph. Secondly, there is a scaling deviation arising from scaling the intramolecular structure with density. The size of the scaling deviation depends on the density change between the two compared state points. If the path in the phase diagram is not an isochrone or an isochore, both effects are present. The results of the MD simulations for cumene showed that the intramolecular structure can be assumed to be invariant, regardless of the state point, in experimental units. For other molecules, the intramolecular structure might be less rigid. The effect of the scaling deviation on larger more flexible molecules would be interesting to study in future work. An important result presented in this chapter has been to illuminate what is measured in $S(q)$ when moving around the phase diagram.
	
	\Mycite{Chen2021} studied the temperature dependence of the structure of the three hydrogen-bonding glass formers, glycerol, xylitol, and D-sorbitol. They found that the temperature dependence of the position of the main peak was close to invariant when presented in reduced units. For these glass-formers, it seems that the effect of the scaling deviation was much less visible than for Cumene. Also, it seems that the  structural changes originating from something else than density changes, are larger for cumene then glycerol, xylitol, and D-sorbitol. The origin of the scaling deviation is very clear, and what is the origin of the other non-density structure changes is an interesting research question to ask.  
	
	\Mycite{Headen2019} studied the ambient pressure, temperature dependent changes of benzene using EPSR and isotopic substitution. The chemical structures of benzene and cumene are not the same, but they are relatively similar. For neighboring benzene molecules there are several structural motives that tend to repeat themselves in the liquid. For benzene these structural motives are amongst others, off-set parallel stacking, stacking where the plane of one benzene-ring is perpendicular to the plane of the other benzene-ring, and a trimer structure where the angle between planes are 60$^o$ \cite{Headen2019}.  When the temperature is increased, the disorder in the system increases, causing these structural motifs to be washed out. With decreasing temperature, these key, energy-favorable, structural motifs appears more often in liquid and less energy favorable structural motifs decrease. It is interesting that some of these structural motifs seen in liquid benzene also occur in crystalline benzene. The crystalline structure of Cumene is not yet solved, but an interesting study would be comparing the crystalline and liquid local structures in Cumene. \Mycite{Headen2010} compare the local structures of benzene and toluene. They found that benzene appeared more structured, compared to toluene. The presence of the methyl group on toluene affects the ability of the molecules to pack. \Mycite{Falkowska2016} studied five structurally similar molecules, benzene, toluene, cyclohexane, cyclohexene, and methylcyclohexane to study the effect of the methyl groups and pi-pi interactions on the local structure. They come to the same conclusion for the role of the methyl group,  namely that it affects the molecule's ability to pack. For cumene, there is a propyl group attached to the ring, and it would most likely have a similar effect.
	
	

 Using MD simulations, we could isolate the intramolecular and intermolecular contributions to $S(q)$ and observed that the intermolecular contributions collapsed along isochrones in reduced units. The intramolecular contribution caused a scaling deviation because the intramolecular structure is invariant in experimental units. This is shown in figure \ref{fig:Cumene_MD_sq_d}. Another interesting observation seen in figure \ref{fig:Cumene_MD_sq_d} is that the scaling deviation seems to have a bigger effect at lower q-values. This is also shown very clearly in figure \ref{fig:delta_sq_isochrone}. As mentioned in the introduction \Mycite{Wase2020}, showed that for the ionic liquid Pyr14TFSI the "structure peak" collapses along isochrones while the "charge peak" did not seem to collapse. The structure peak was located at approximately 1.3-1.4 Å$^{-1}$, while the "charge peak" was approximately 0.8-0.9 Å$^{-1}$. I can only speculate on how the intramolecular contributions to $S(q)$ would look for Pyr14TFSI, but it is possible that it would affect the "charge peak" more than the "structure peak". It would be interesting to test this with the simulations of \Mycite{Knudsen2024}. In \Mycite{Knudsen2024} they also found that the structure peak did not collapse with a larger density change along the isochrone. The structural origin of the "charge peak", is from charge ordering of ion causing longer range structures in the liquid. However, the effect of the scaling deviation would most likely be larger in that q-range. The ionic liquid system is much more complex than cumene. This is also seen in the intramolecular structure of the two molecules of the ionic liquid, however, it would be interesting to test the role of the intramolecular contribution to $S(q)$ for the ionic liquid. 
	
	
	In this chapter, we have shown that for cumene, there exist lines in the phase diagram where both structure and dynamics are invariant in reduced units. This is direct experimental evidence for the existence of pseudo-isomorphs. The following chapter expands the search for pseudo-isomorphic liquids by examining additional glass-forming liquids.
	


	\chapter[Pseudo-Isomorphs, Part II]{Experimental evidence of Pseudo-Isomorphs, part II} \label{chapter:Diamond}
	
The goal of this chapter is to extent one of the first systematic experimental tests
of pseudo-isomorphs. In Chapter \ref{chapter:Cumene} the structure of the van der Waals liquid, cumene, was measured as function of pressure and temperature. The measured structure was compared along both isotherms, isobars, isochores, and the experimental candidates for pseudo-isomorphs, isochrones. The results show that the first peak of $S(q)$ measured along an isochrone collapsed on top of each other when presented in reduced units, and that it is not invariant along isotherms, isobars and isochrones. Further more, microscopic information of the structure was obtained using MD-simulations. It was possible to separate the structure into intra- and intermolecular contributions to $S(q)$. Along an isochrone the intermolecular structure collapsed into a single curve when presented in reduced units. The intramolecular structure did not change in experimental units. There was a scaling deviation arising when $S(q)$ presented in reduced units from contributions from the intramolecular structure. The results of chapter \ref{chapter:Cumene} have shown the existence of pseudo-isomorph in a real-world liquid. The next natural step is extent the search of pseudo-isomorph to other possible candidates. In this chapter we will test for pseudo-isomorphs in three van der Waals liquids, DC704, 5PPE, and Squalane and a hydrogen-binding liquid DPG. The hydrogen-binding liquid is chosen as possible counter-example, since we do not expect DPG to have pseudo-isomorphs.
		
	

	
	\section{Experimental setup}
	
	The experiment was performed at the Diamond Light Source at the SAXS/WAXS beamline I22 \cite{I22Paper}, experimental number SM-34190. At the I22 beamline, the SAXS and WAXS signals can be measured simultaneously. The SAXS signal is measured using a Pilatus P3-2M detector and the WAXS signal is measured using a Pilatus P3-2M-DLS-L detector. Only the WAXS signal was used in the experiment. The WAXS detector was placed 0.17 m from the sample. The count time for each state point was 2s, and the energy of the beam was 18.0 keV. The measured intensities were treated using the DAWN software \cite{DAWNsoftware}.
	
	The pressure cell used in the experiment was the P-jump cell  shared between beamline I15 and I22. The cell was designed for x-ray diffraction measurements of soft condensed-matter samples \cite{PjumpCell}. The working pressure range for the cell is 0-500 MPa, and the working temperature range is -20 $^o$C - 120 $^o$C. Pressure was applied using water as a pressure liquid. The sample cell was a polycarbonate capillary, sealed with epoxy glue, with a small plastic bead in between glue and sample to avoid contamination. The sample was placed in a capillary sample holder and carrier designed to ensure that the capillary was placed in the beam. 
	
	\begin{figure}[H]
		\centering
		\includegraphics[width=0.5\textwidth]{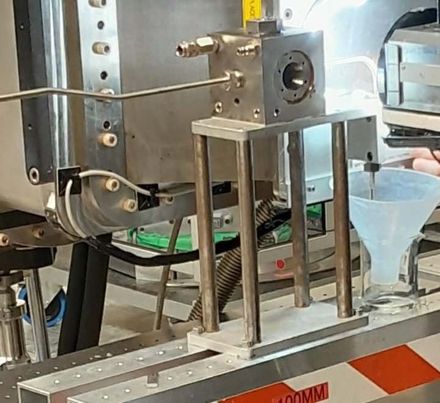}
		\caption{The P-jump cell.}
	\end{figure}
	
	The experiment was conducted on four different samples, each presented in detail in Section \ref{sec:Diamond_samples}. The chronological order of sample measurements was DC704, 5PPE, squalane, and DPG. The measuring protocol was to sample the phase diagram by measuring along four isotherms at approximately 19 $^o$C, 39 $^o$C, 57 $^o$C, and 76 $^o$C, over a pressure range of approximately 50 MPa - 400 MPa. The cell was prone to leakage when the pressure was completely released, therefore to minimize leakage, we only measured under pressure. The chronology of the isotherms was 19 $^o$C $\rightarrow$ 57 $^o$C $\rightarrow$ 76 $^o$C $\rightarrow$ 39 $^o$C.
	
	\section{The samples}\label{sec:Diamond_samples}
	
	The four liquids measured in this experiment were DC704, 5PPE, squalane, and DPG. The samples are described in detail in this section. The equation of state (EOS) for the four samples is introduced along with the phase diagram for all four liquids. A general introduction to the Tait equation of state was given in appendix \ref{app:EOS}. DC704, 5PPE, and squalane are all van der waals bonded liquids, where the dynamical predictions of isomorph theory have been  successfully verified \cite{GundermannDitte2011Ptde} \cite{XiaoWence2015Itpf} \cite{RolandCM2006Tsot}. The final sample, DPG, is a hydrogen-bonded liquid that was chosen as a counter example because we did not expect the predictions of the isomorph theory to hold true for hydrogen-bonded liquids \cite{Ingebrigtsen2012_simple_liquid}. The procedure for finding isochrones was the same as that used in the Cumene experiment, see chapter \ref{chapter:Cumene}. We use $\Gamma$ from power-law density scaling $\Gamma = \frac{\rho^\gamma}{T}$, as a measure of which isochrones each state point is on. In chapter \ref{chapter:IDS}, we will revisited the structural measurement of DC704 and DPG, with isochronal density scaling and isochronal temperature scaling.
	
	\begin{figure}[H]
		\begin{subfigure}[t]{0.5\textwidth}
			\centering
			\includegraphics[width=0.9\textwidth]{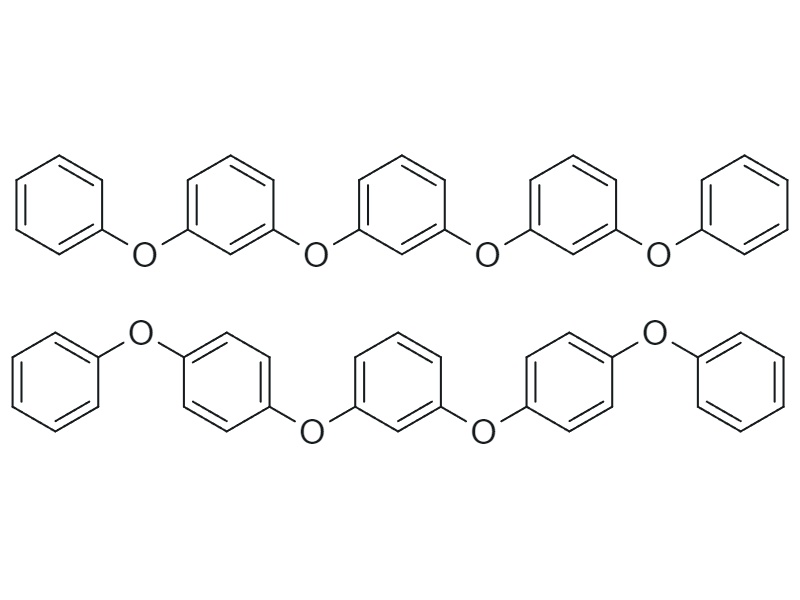}
			\caption{The isomers of 5PPE}		
		\end{subfigure}
		\begin{subfigure}[t]{0.5\textwidth}
			\centering
			\includegraphics[width=0.9\textwidth]{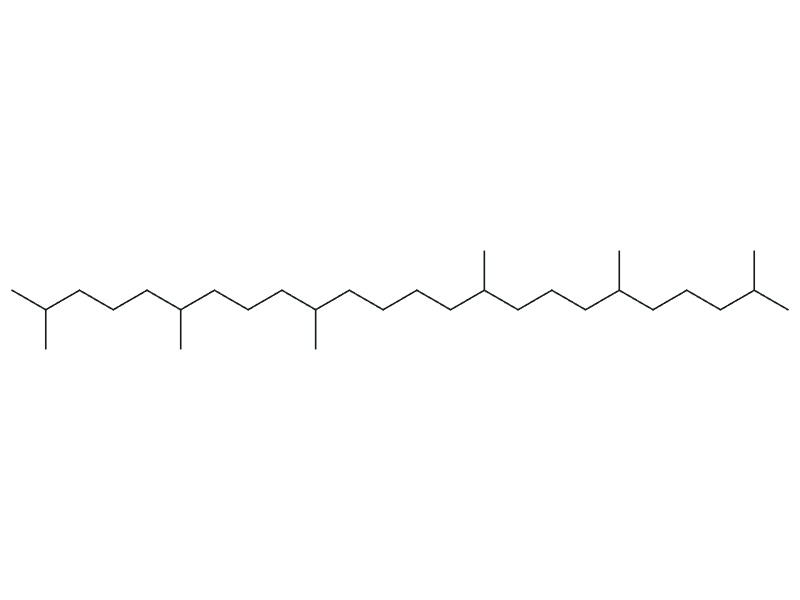}
			\caption{Squalane}		
		\end{subfigure}
		\begin{subfigure}[t]{0.5\textwidth}
			\centering
			\includegraphics[width=0.8\textwidth]{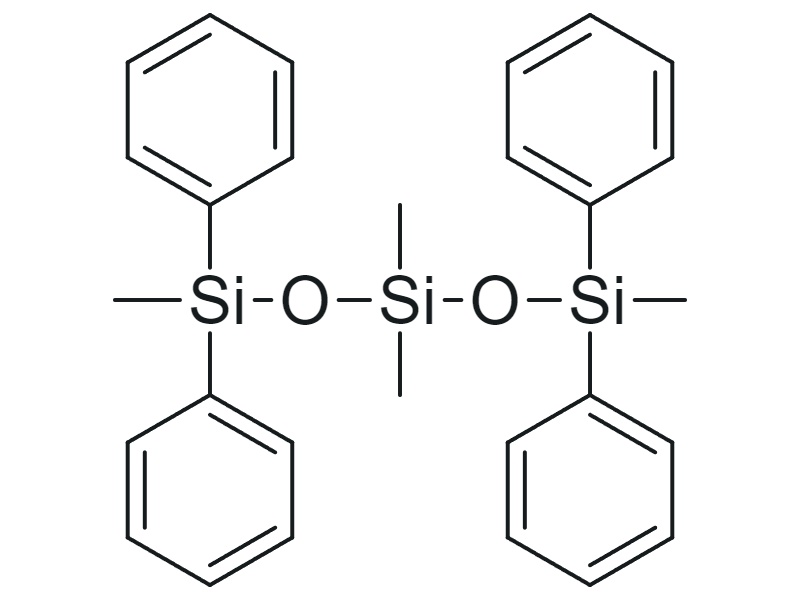}
			\caption{DC704}
		\end{subfigure}
		\begin{subfigure}[t]{0.5\textwidth}
			\centering
			\includegraphics[width=0.8\textwidth]{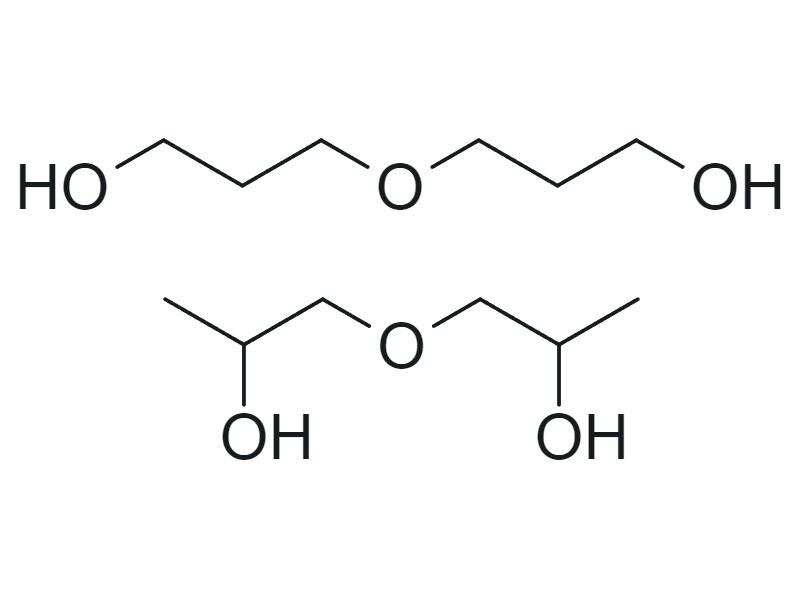}
			\caption{The isomers of DPG}
		\end{subfigure}
		\caption{The four samples measured. The four samples are the three van der waals liquids 5-phenyl-4-ether (5PPE), squalane, tetramethyl-tetraphenyl-trisiloxane (DC704) and the hydrogen bonded liquid dipropylene glycol (DPG). }
		\label{fig:samples_diamond}
	\end{figure}

	\subsection{DC704}  \label{sec:DC704_background}

	Tetramethyl-tetraphenyl-trisiloxane (DC704), C$_{28}$H$_{32}$O$_2$Si$_3$, is a silicone oil often used as a standard sample for the Glass and Time group \cite{GundermannDitte2011Ptde} \cite{HecksherTina2013Msog} \cite{HecksherTina2022Rmft}. DC704 has a glass transition temperature of 210 K and an ambient pressure fragility index of  $m_p \approx 80$ \cite{HecksherTina2013Msog}. DC704 is a van der Waals-bonded liquid, and \Mycite{GundermannDitte2011Ptde} have shown experimentally that DC704 is a R-simple system, by calculating $R$ experimentally directly from the Prigogine-Defay ratio. They found that DC704 has an R = 0.9 $\pm 0.1$ . Therefore, we expect the structure along isochrones to be invariant when presented in reduced units. A equation of state for DC704 is given by \Mycite{Ditte_PHD}:
	
	\begin{equation} \label{eq:EoS_DC704}
		\rho(T,P) = \left(V_0\exp\left(\alpha_0 T\right) \left[1 - C \ln 	\left(1 + \frac{P}{b_0 \exp\left(-b_1T\right)}\right)\right]\right)^{-1}
	\end{equation}
	
	In table \ref{tab:DC704_eos} the fit parameters is shown.
	The equation of state was fitted on PVT data in the temperature range 20$^oC$-75$^oC$ and pressure range 10 MPa -200 MPa. 
	
	\begin{figure}[H]
		\centering
		\includegraphics[width=0.7\textwidth]{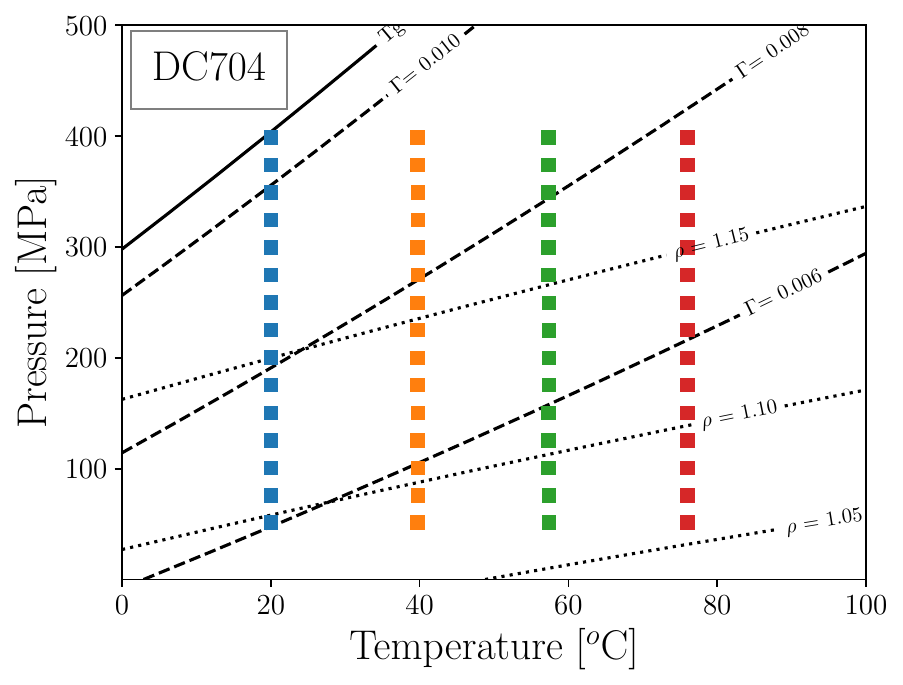}	
		\caption{The experimental phase diagram of DC704. The solid line represents the glass transition, calculated as a line of constant $\Gamma = \frac{\rho^\gamma}{T}$, with $\gamma=6.2$, extrapolated from the $\Gamma$ value at $T_g$. The dotted lines are isochores. The dashed lines are lines of constant $\Gamma$. The squares represent the state points at which measurements were made and the colors represent different isothermal measurement series. The unit of density used in this figure is g cm$^{-3}$.}
		\label{fig:DC704_phasediagram}
	\end{figure}
	
	The density scaling coefficient $\gamma$ have been reported in the literature to be 6.1-6.2   \cite{GundermannDitte2011Ptde} \cite{AdrjanowiczKarolina2017PNDo}. \Mycite{Ransom_DC704} measured the relaxation time of DC704 over a very large temperature and pressure range and showed that power-law density scaling with a constant $\gamma$ is not true in their experimental temperature and pressure range. In the previous chapter (chapter \ref{chapter:Cumene}) power-law density scaling was shown to work in large pressure and temperauture range for Cumene. This seems not to be the case for DC704. For the data analysis in section \ref{sec:Diamond_fit} we will use $\gamma =6.2$.
	
	\Mycite{Ransom_DC704} fit a equation of state in a larger pressure and temperature range than used in the equation of state from \Mycite{Ditte_PHD}. \Mycite{Ransom_DC704} fitted the Tait EoS in a temperature range of -45 $^o$C to 100 $^o$C  and from 0 - 900 MPa. The fit parameters to the Tait equation from \Mycite{Ransom_DC704} and \Mycite{Ditte_PHD} are compared in table \ref{tab:DC704_eos}. 
	
	\begin{table}[h]
		\centering
		\begin{tabular}{l|l|l}
			& \Mycite{Ditte_PHD}             & \Mycite{Ransom_DC704}                 \\ \hline
			$V_0$      & $0.92 $   g$^{-1}$ cm$^{3}$    & $0.9199  $    g$^{-1}$ cm$^{3}$                       \\
			$\alpha_0$ & $7.1 * 10^{-4}$ $^o$C$^{-1}$& $7.334 * 10^{-4}$ $^o$C$^{-1}$\\
			$C$        & $0.088   $                         & $0.0939  $                           \\
			$b_0$      & $188    $  MPa                        & $204  $    MPa                          \\
			$b_1$      & $4.8*10^{-3}$ $^o$C$^{-1}$   & $5.13 * 10^{-3} $ $^o$C$^{-1}$
		\end{tabular}
		\caption{The fitting parameters to the Tait Equation of state from \Mycite{Ditte_PHD} and \Mycite{Ransom_DC704} }
		\label{tab:DC704_eos}
	\end{table}

\begin{figure}[H]
		\begin{subfigure}[b]{0.5\textwidth}
			\centering
			\includegraphics[width=0.9\textwidth]{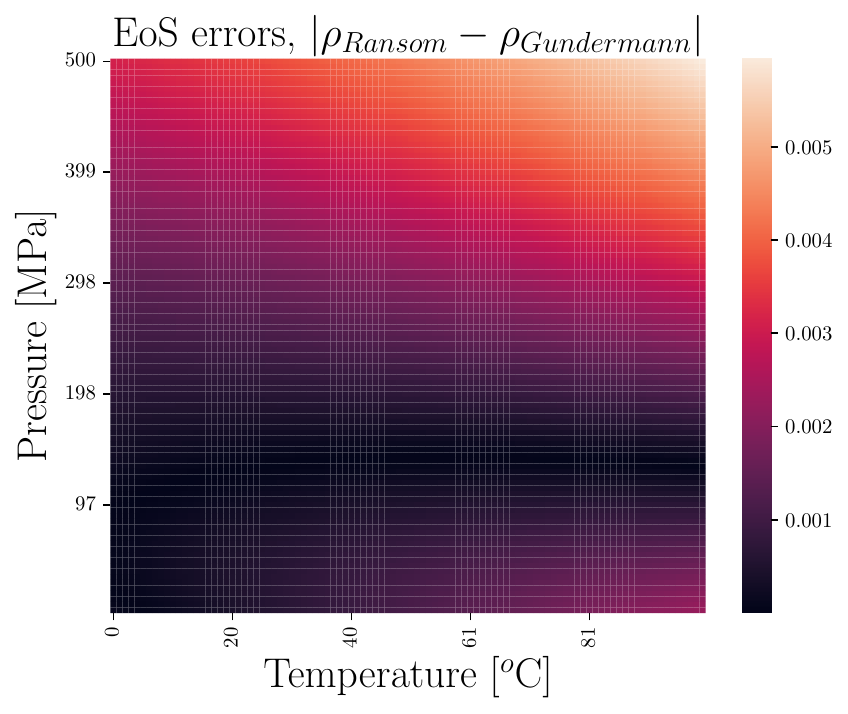}
			\caption{}		
		\end{subfigure}
		\begin{subfigure}[b]{0.5\textwidth}
			\centering
			\includegraphics[width=0.9\textwidth]{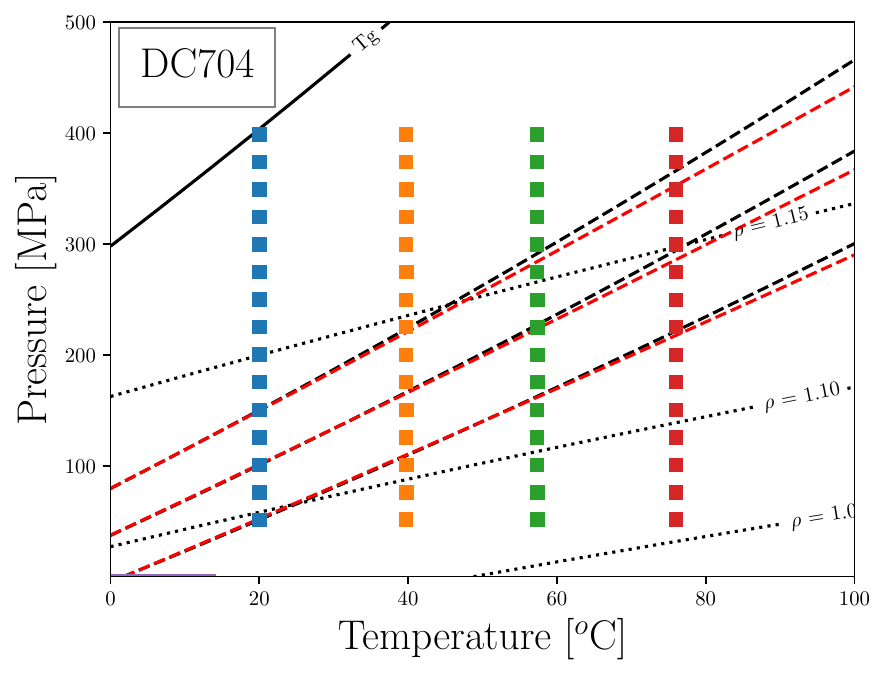}
			\caption{}		
			\label{fig:DC704_heatmap_b}
		\end{subfigure}
		
		\caption{Heat map for the difference between the two fits to the Tait Equation of state from \Mycite{Ditte_PHD} and \Mycite{Ransom_DC704}.  In figure (a) the difference between the calculated densities, and in figure (b) the resulting $\Gamma = \frac{\rho^\gamma}{T}$, $\gamma = 6.2$ from the two different EOS, with $\Gamma_{Gundemann}$ in black and $\Gamma_{Ransom}$ in red. In the experimental range, the difference between the two equations of state are less than 0.005 g cm$^{-3}$. In practice this give an difference in density less that 0.5 \% as also reported by \Mycite{Ransom_DC704}. Still this small difference in density gives different predictions of the isochrones. }
		\label{fig:DC704_heatmap}
	\end{figure}

	In figure \ref{fig:DC704_heatmap} the absolute value of the difference between $\rho_{Ransom}$ and $\rho_{Gundermann}$. The difference between the two EoS was less than 0.005 g cm$^{-3}$ in our experimental range. In our experimental range, this resulted in a difference in density of less that 0.5 \%  as also reported by \Mycite{Ransom_DC704}.  In figure \ref{fig:DC704_heatmap_b} effect of the small density difference between the two EoS is show. Calculating the line of constant $\Gamma = \frac{\rho^{6.2}}{T}$, from the two equations of states, we see that the two equation of state predicts different isochrones. The chosen examples are the most extreme cases in our experimental range, but the different EoS give different predictions. For the data treatment in section \ref{sec:Diamond_fit} the EoS from \Mycite{Ditte_PHD} was chosen.
	In chapter \ref{chapter:IDS} we revisit the results using two new types of scaling, isochronal density scaling and isochronal temperature scaling. In chapter \ref{chapter:IDS} we will use the EoS of \Mycite{Ransom_DC704}. 
	
	For DC704 there have in the literature been reported different values of $\gamma$ \cite{GundermannDitte2011Ptde,AdrjanowiczKarolina2017PNDo}, and also a breakdown of power-law density scaling \cite{Ransom_DC704}. There are also two different equation of states reported in the literature and they predict different "isochrones". So which EoS and $\gamma$ should we use and when is this important? The small difference in density between the two EoS, will only have an small effect on the reduced units, since the maximum difference between, $\rho_{Ransom}$ and $\rho_{Gundermann}$ is 0.005 g cm$^{-3}$. This small uncertainty is acceptable with respect to the reduced units. The most important aspect is, can we predict isochrones using power-law density scaling, and are we comparing structure along isochrones? This question is first answered in the next chapter of the thesis, chapter \ref{chapter:IDS}. However, we will show that in the experimental range, using power-law density scaling is an good approximation in this case. 
	
	
	
	\subsection{Dipropylene Glycol (DPG)} \label{sec:DPG_background}
	
	Dipropylene Glycol (DPG), C$_{6}$H$_{14}$O$_3$, is a hydrogen-bonded mixture of isomers that easily forms a glass. The glass transition temperature of DPG is $T_g =195 ^oK$. Previous results have shown mixed results on whether power-law density scaling of dynamics is possible for DPG. \Mycite{Wase2018} have shown that in the pressure range of 0-400 MPa, DPG shows power-law density scaling of dynamics with $\gamma=1.5$ \cite{Wase2018}. However, \Mycite{ChatKatarzyna2019Tdsi} find that for both $\gamma=1.5$ and the best fit value of $\gamma=1.9$ it is not possible to use power-law density scaling on the relaxation time of DPG. The reason for choosing DPG a sample was that it most likely would be a counterexample. Given that it is a hydrogen-bonded liquid, we do not expect the predictions of the isomorph theory to hold.
	The EoS for DPG was taken from  \Mycite{CasaliniRiccardo2003Dαai} which is fitted to PVT-data in the temperature range of 310K - 430K, and a pressure range of 0-200 MPa:  
	
	\begin{equation}
		V(T,P) = V(T,0)\left(1-0.0894\ln\left[1+ \frac{P}{B(T)}\right]\right)
	\end{equation}
	
	where $\rho(T,P) = V(T,P)^{-1}$. $V(T,0)$ is the specific volume at zero pressure and is described by a 2nd order polynomial: 
	
	\begin{equation}
		V(T,0) = v_0+ v_1T +v_2T^2
	\end{equation} 
	
	where, $v_0 = 0.901$ cm$^3$g$^{-1}$, $v_1 = 6.79 \cdot 10^{-4}$ cm$^3$g$^{-1}$C$^{-1}$, $v_2 = 1.339 \cdot 10^{-6}$ cm$^3$g$^{-1}$C$^{-2}$. $B(T)$ is given by
	
	\begin{equation}
		B(T) = b_0 \exp(-b_1 T)
	\end{equation}
	
	where $b_0 = 184$ MPa and $b_1 = 6.08*10^{-3}$ C$^{-1}$. 
	
	\begin{figure}[h]
		\centering
		\includegraphics[width=0.7\textwidth]{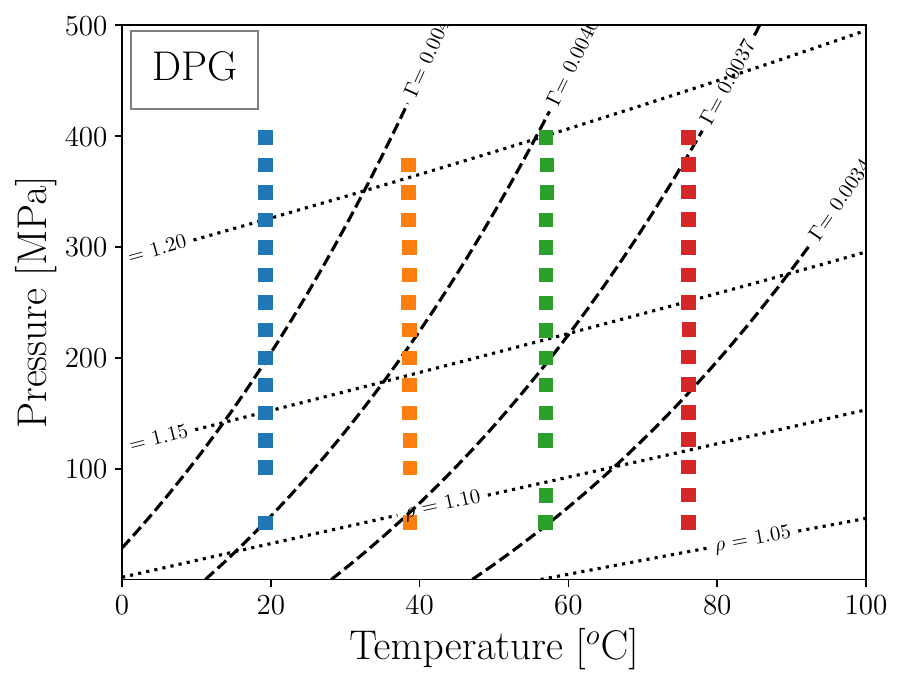}	
		\caption{The experimental phase diagram of DPG. The dotted line are isochores. The dashed lines are lines where $\Gamma = \frac{\rho^{1.5}}{T}$ is constant. The dashed lines are our best guesses for an isochrone. The squares represent the measured state points and the colors represent different measurement series. The unit of density in the figure is g cm$^{-3}$. }
		\label{fig:DPG_phasediagram}
	\end{figure}
	
	As mentioned, the equation of state was fitted to data in the temperature range of 310 K - 430 $^o$K (37 $^o$C -157 $^o$C), and a pressure range of 0-200 MPa. Unfortunately the fitted EoS was fitted in a smaller range than that of our pressure measurements, and therefore we have to extrapolate the EoS to double the fitted pressure range. However, the same equation of state was used by \Mycite{ChatKatarzyna2019Tdsi}, in a pressure range of 0.1 - 450 MPa and a temperature range 200-270 K. 
	In chapter \ref{chapter:IDS} the results of \Mycite{ChatKatarzyna2019Tdsi} are revisited in the context of isochronal density scaling and isochronal temperature scaling, so for consistency we choose the same equation of state. \Mycite{Burk2021} argue that the pressure dependence of the Tait model is stable and can safely be extrapolated to up to about twice the upper regression value. This is also discussed in the squalane section of this chapter. The regression values are not reported by \Mycite{ChatKatarzyna2019Tdsi}, but our experimental range is within 2 times the fitted range.
	
	
	\subsection{5PPE}
	
	5-phenyl-4-ether (5PPE), C$_{30}$H$_{22}$O$_4$, is a mixture of isomers often used as a standard sample for the glass and time group \cite{HecksherTina2013Msog} \cite{XiaoWence2015Itpf} \cite{HansenHenrietteW.2017Cbfm}. 5PPE is a van der Waals-bonded liquid with a glass transition of 245 K \cite{HecksherTina2013Msog}. Because of the relatively high glass transition temperature, it was possible for us to experimentally enter the glass via pressure for both measurements along the 19  $^o$C, and 39 $^o$C isotherm. 5PPE has an ambient pressure fragility index of $m_p \approx 80$ \cite{HecksherTina2013Msog}, and a density scaling coefficient of $\gamma = 5.5$ \cite{GundermannDitte2011Ptde}\cite{XiaoWence2015Itpf}. \Mycite{HansenHenrietteW.2017Cbfm} showed that 5PPE exhibits isochronal superposition of the alpha relaxation peak over many decades of dynamics. 
	Given that 5PPE is a van der waals-bonded liquid that obeys power-law density scaling and isochronal superposition of thealpha relaxation peak, we expect that 5PPE obeys the predictions of isomorph theory.	The equation of state for 5PPE is given by \Mycite{Ditte_PHD}:
	
	\begin{equation}
		\rho(T,P) = \left(V_0\exp\left(\alpha_0 T\right) \left[1 - C \ln 	\left(1 + \frac{P}{b_0 \exp\left(-b_1T\right)}\right)\right]\right)^{-1}
	\end{equation}
	
	where $V_0 = 0.82$ cm$^3$g$^{-1}$, $\alpha_0 = 6.5 \cdot 10^{-4}$ $^o$C$^{-1}$, $C=9.4 \cdot 10^{-2}$, $b_0 = 286$ MPa, and $b_1 =4.4 \cdot 10^{-3}$ $^o$C$^{-1}$. 	The equation of state has previously been used for measurement up to 250 MPa \cite{Wase2018}. In this study we will be extrapolating the equation of state beyond where it has been fitted. Using an equation of state beyond its fitted range is a weakness, but we have no alternative. For DPG, and squalane we also extrapolate the EoS, beyond the fitted range. To restate \Mycite{Burk2021} argue that the pressure dependence of the Tait model is stable and can safely be extrapolated to up to about twice the upper regression value. The upper regression value is not stated by \Mycite{Ditte_PHD}, but we are within twice the fitted range.
	

	\begin{figure}[H]
		\centering
		\includegraphics[width=0.7\textwidth]{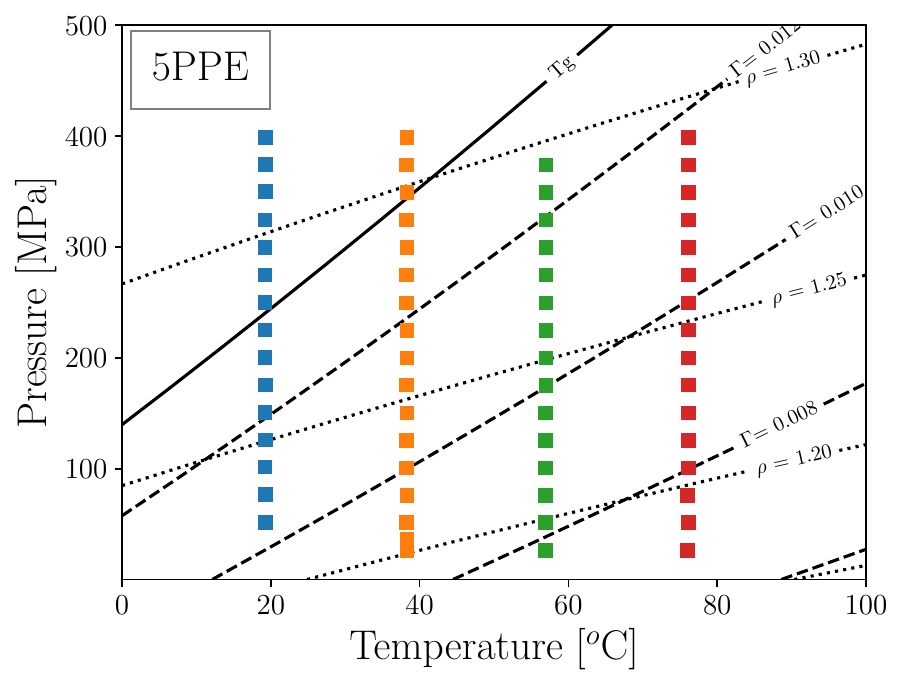}	
		\caption{The experimental phase diagram of 5PPE. The solid line represents the glass transition. The dotted line are isochores. The dashed lines are lines of constant $\Gamma$, for $\gamma=5.5$. The squares represents the measured state points, and the colors represent different measurement series. The unit of density in this figure is g cm$^{-3}$.}
		\label{fig:5PPE_phasediagram}
	\end{figure}

	\subsection{Squalane}
	
	Squalane is a van der Waals bonded liquid that easily forms a glass. It has a glass transition temperature of approximately 168 K and an ambient pressure fragility index of $m_p = 75$ \cite{RichertR2003Dogl}. It has been reported at that squalane has a density scaling coefficient $\gamma = 4.2$  \cite{PensadoAlfonsoS2008RbVC}\cite{RolandCM2006Tsot}. Squalane is a sample that we expect to have pseudo-isomorphs. The equation of state  used for squalane used is from  \Mycite{Burk2021}. The equation of state from \Mycite{Burk2021} is an extension of a previously published EoS by \Mycite{SchmidtKurtA.G2015NEDa}. Both the equations of state of \Mycite{Burk2021} and \Mycite{SchmidtKurtA.G2015NEDa} uses the Tait EoS to describe the density:
	
	\begin{equation}
		\rho(P,T) = \frac{\rho_0(T)}{1 - C\log\left(\frac{B(T) + P }{B(T)  + P_0}\right)}
	\end{equation}

	where $\rho_0$ is the temperature dependent density at the reference pressure, $P_0 = 0.1 $ MPa and $B(T)$ is a temperature dependent variable used to describe compressibility the correlation, $C = 8.76434 \cdot 10^{-2}$ is a constant. For the EoS, the temperature dependency of $\rho_0(T) $ and $B(T)$ is described by a second order polynomial:
	\begin{align}
		\rho_0(T) &= a_0 +a_1T +a_2T^2\\
		B(T) &= b_0 + b_1T + b_2T^2  
	\end{align}
	
	where, $a_0 =982.517$ kg m$^{-3}$, $a_1 = -0.559566$ kg m$^{-3}$ T$^{-1}$, $a_2 =- 1.18953 \cdot 10^{-4}$ kg m$^{-3}$ T$^{-2}$, and $b_0 = 363.316$ MPa, $b_1 = -1.05977$ MPa T$^{-1}$, $b_2 = 8.04430 \cdot 10^{-4}$ MPa T$^{-2}$.
	
	The equation of state was fitted to data in the temperature range of 273 - $\approx$ 561 K and pressure range up to $\approx$ 250 MPa. The regression analysis of the model states an upper pressure limit of the model 244.8 MPa; however, the authors argue that the pressure dependence of the Tait model is stable and can safely be extrapolated to up to about twice the upper regression value\cite{Burk2021}. The experimental upper pressure limit was 400 MPa, so was within 2 times the upper regression value of 244.8 MPa. 
	
	\begin{figure}[H]
		\centering
		\includegraphics[width=0.7\textwidth]{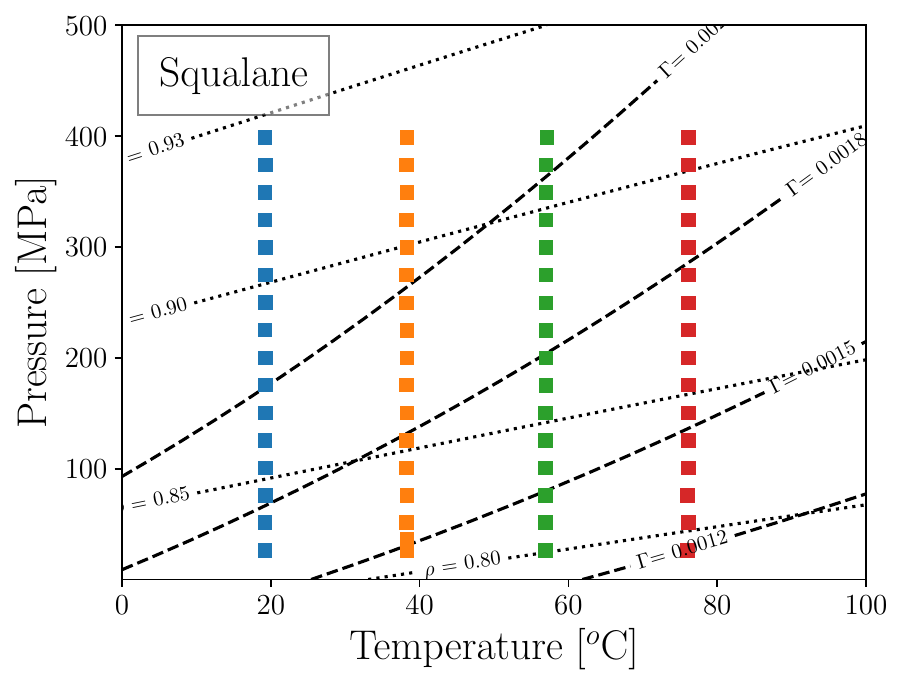}	
		\caption{The experimental phase diagram of squalane. The dotted line are isochores. The dashed lines are lines of constant $\Gamma$. The squares represent the measured state points, and the colors represent different measurement series. The unit of density in this figure is g cm$^{-3}$.}
		\label{fig:Squalane_phasediagram}
		
	\end{figure}

	\section{Background subtraction }\label{sec:background}
	
\begin{figure}[H]
		\begin{subfigure}[b]{0.5\textwidth}
			\centering
			\includegraphics[width=0.99\textwidth]{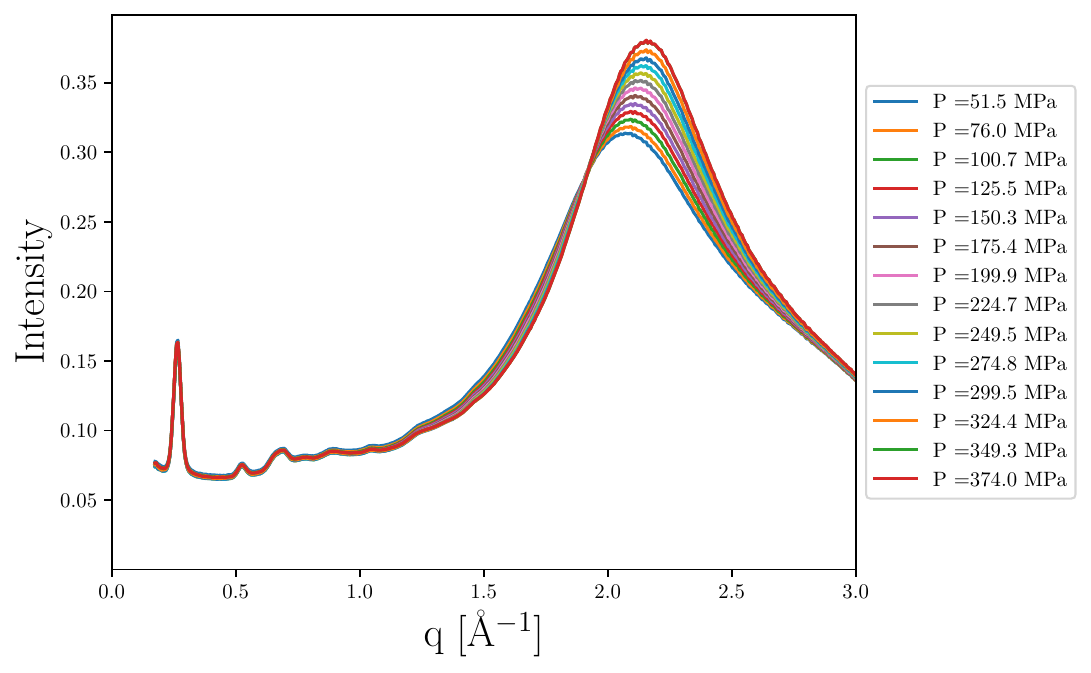}
			\caption{}		
		\end{subfigure}
		\begin{subfigure}[b]{0.5\textwidth}
			\centering
			\includegraphics[width=0.99\textwidth]{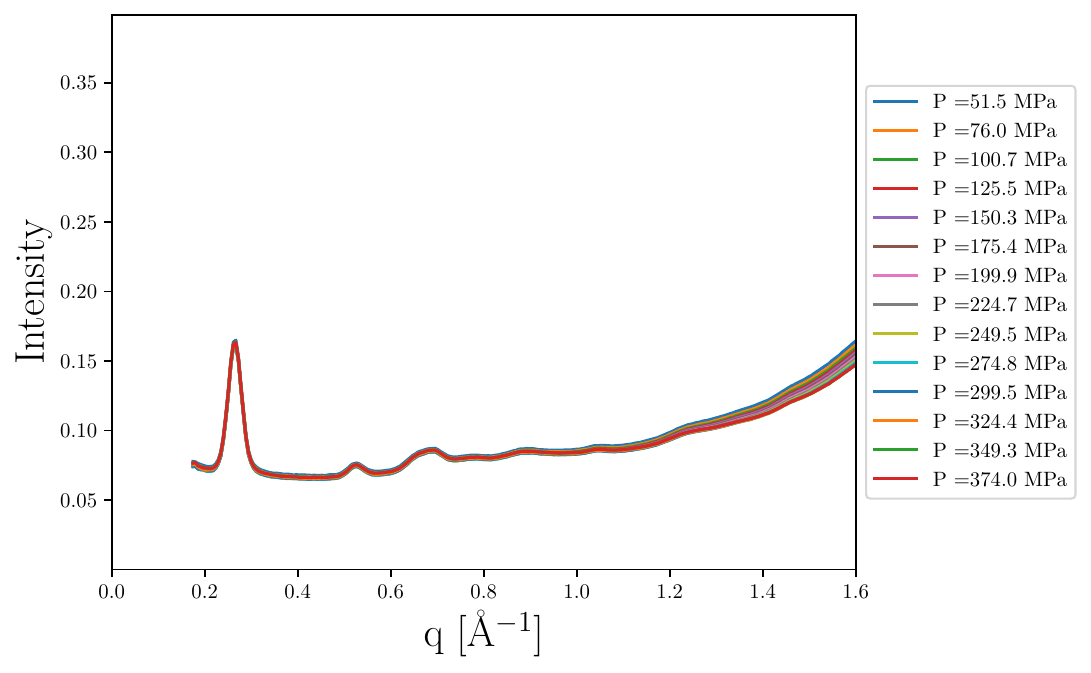}
			\caption{}
		\end{subfigure}
		\caption{Measurements of only water in the cell as a function of pressure at T = 57 $^o$C. In figure (a) the measurements are shown in the entire measured q-range. The water have a peak around $q = 2.0 -2.5$ Å$^{-1}$. Figure (b) shows a close up of the measurement in a range of 0 Å$^{-1}$- 1.6 Å$^{-1}$. In the q-range of 0 Å$^{-1}$- 1.6 Å$^{-1}$ the background contribution from the water is almost independent of pressure.  }
		\label{fig:Water No_capillary}
	\end{figure}

	Background subtraction for the measurements is not trivial in this setup. The beam passes through both the diamond windows on front and behind the sample, the pressure liquid (water), all around the capillary, both sides of the plastic capillary and the sample itself. These contributions to the scattered intensity do not necessarily add up linearly and, of course, can be dependent on pressure and temperature. The water used as pressure liquid is the contribution to the background that is most pressure and temperature dependent. To gain a better understanding of the behavior of the water background, we measured both bulk water in the cell and water in a plastic capillary along a few isotherms. The results are shown in figure \ref{fig:Water No_capillary}. The water has a peak  at approximately  $q = 2.0 -2.5$ Å$^{-1}$ depending on the state point. When increasing the temperature and/or pressure the peak moves to higher q. This counterintuitive behavior of the first peak of water when increasing temperature has been reported in literature \cite{SoperAlan2017,Sellberg2014}. A short explanation is that water has a low-density phase and a high-density phase and liquid water can be seen as mixture of the two phases \cite{Gallo2016}.  
	
\begin{figure}[H]
		\begin{subfigure}[b]{0.5\textwidth}
			\centering
			\includegraphics[width=0.99\textwidth]{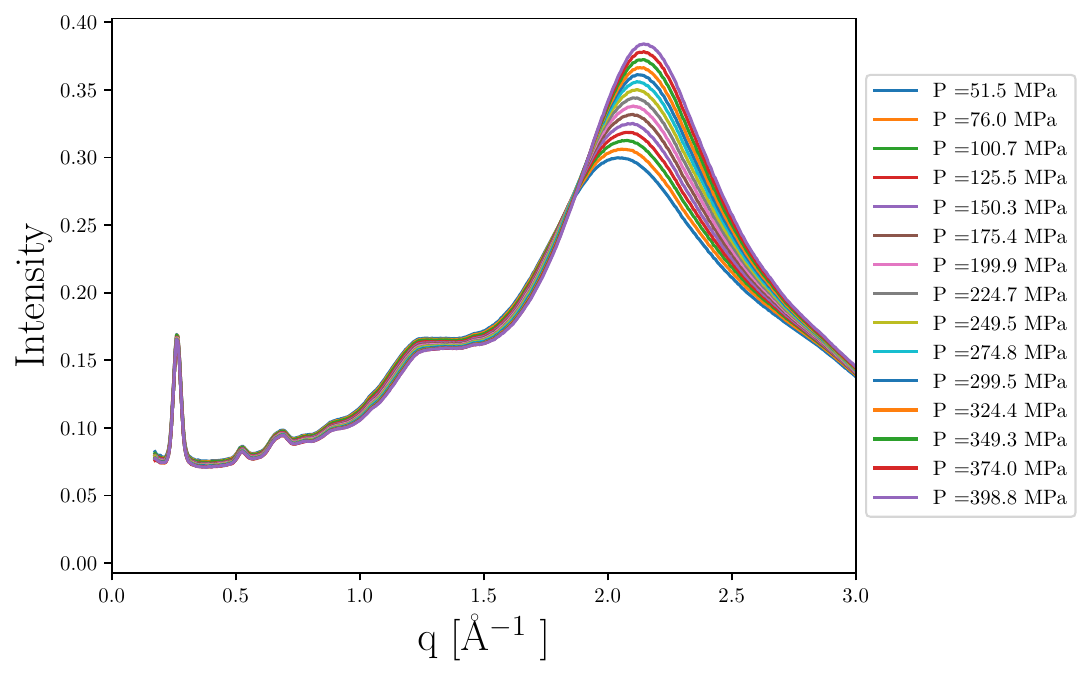}
			\caption{}		
		\end{subfigure}
		\begin{subfigure}[b]{0.5\textwidth}
			\centering
			\includegraphics[width=0.99\textwidth]{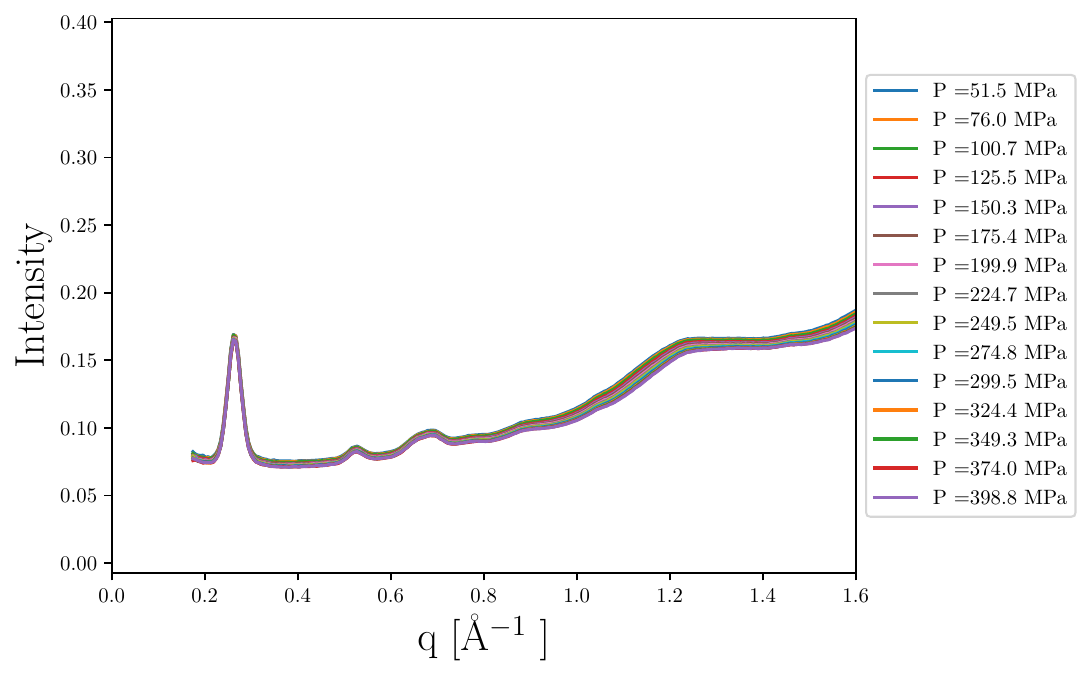}
			\caption{}
			\label{fig:Water capillary_b}
		\end{subfigure}
		\caption{Water in a capillary measured at T = 57 $^o$C and in a pressure range of 50-400 MPa. In figure (a) the measurements are shown in the entire measured q-range. The water have a peak around $q = 2.0 -2.5$ Å$^{-1}$, and the capillary cause a shoulder around $q = 1.2 -1.4$ Å$^{-1}$. Figure (b) shows a close up of the measurement in the range of 0 Å$^{-1}$- 1.6 Å$^{-1}$.  However, the background contributions from water and capillary are almost independent of pressure until $ Q \approx 1.6$ Å$^{-1}$  }
		\label{fig:Water capillary}
	\end{figure}

	Most of the samples we studied have the main diffraction peak in the range of $1.2-1.5$ Å$^{-1}$, so there is not a direct overlap between the q-ranges of interest. In figure \ref{fig:Water capillary},  measurement of water in a capillary as a function of pressure is shown. The pressure effect of the water signal is relatively limited until $ q \approx 1.6$ Å$^{-1}$, and with increasing pressure and temperature, the water signal  moves further out of the window of interest. The shoulder in the data around $q = 1.0 -1.5$ Å$^{-1}$ is from the capillary.

	\subsection[5PPE and Squalane]{Background subtraction for 5PPE and Squalane measurements} \label{sec:5ppe_background}
	
\begin{figure}[H]
		\begin{subfigure}[b]{0.49\textwidth}
			\centering
			\includegraphics[width=0.9\textwidth]{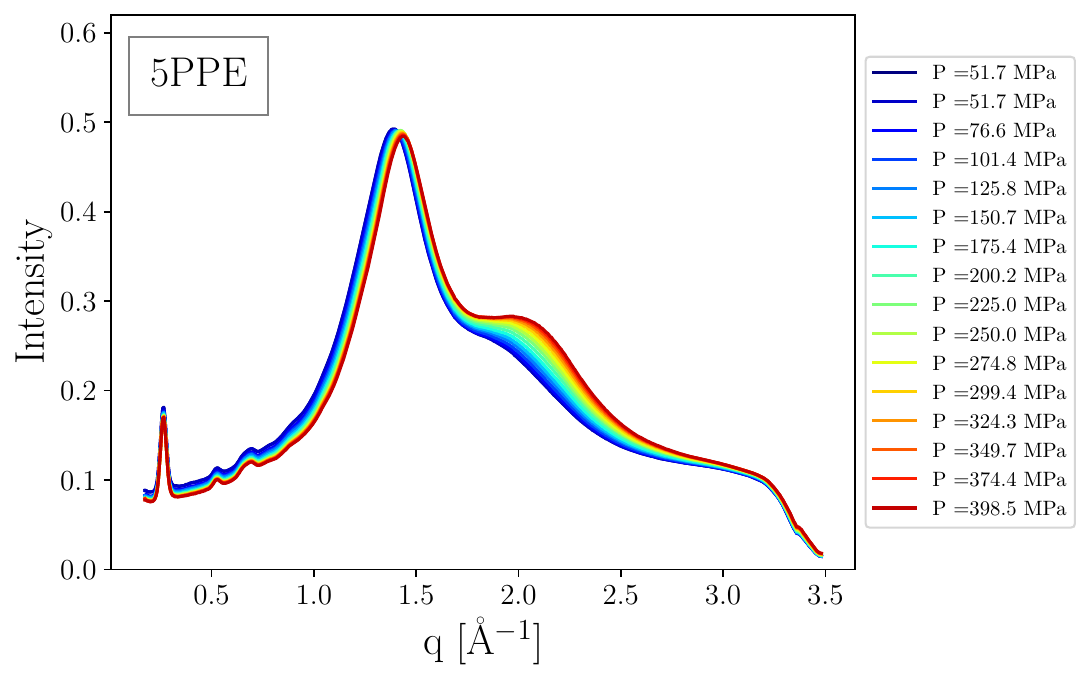}
			\caption{T = 19.3 $^o$C}	
					\label{fig:hopeless_intensity_a}	
		\end{subfigure}
		\begin{subfigure}[b]{0.49\textwidth}
			\centering
			\includegraphics[width=0.9\textwidth]{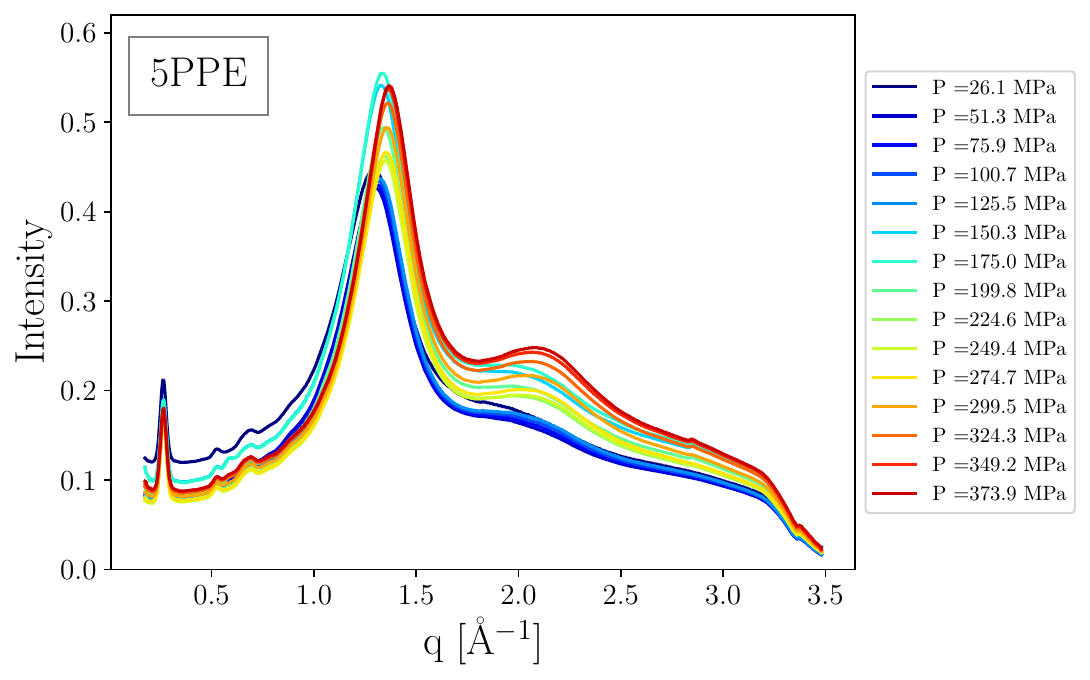}
			\caption{T = 56.9 $^o$C}		
					\label{fig:hopeless_intensity_b}
		\end{subfigure}	
		\begin{subfigure}[b]{0.49\textwidth}
			\centering
			\includegraphics[width=0.9\textwidth]{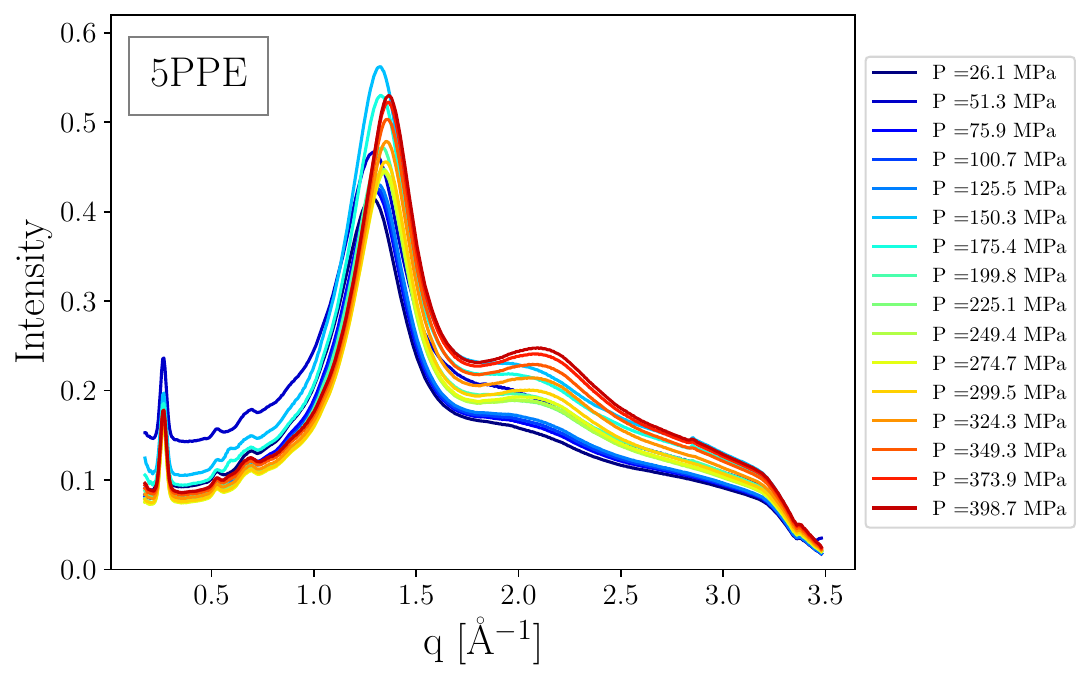}
			\caption{T = 76.0 $^o$C}		
					\label{fig:hopeless_intensity_c}
		\end{subfigure}
		\begin{subfigure}[b]{0.49\textwidth}
			\centering
			\includegraphics[width=0.9\textwidth]{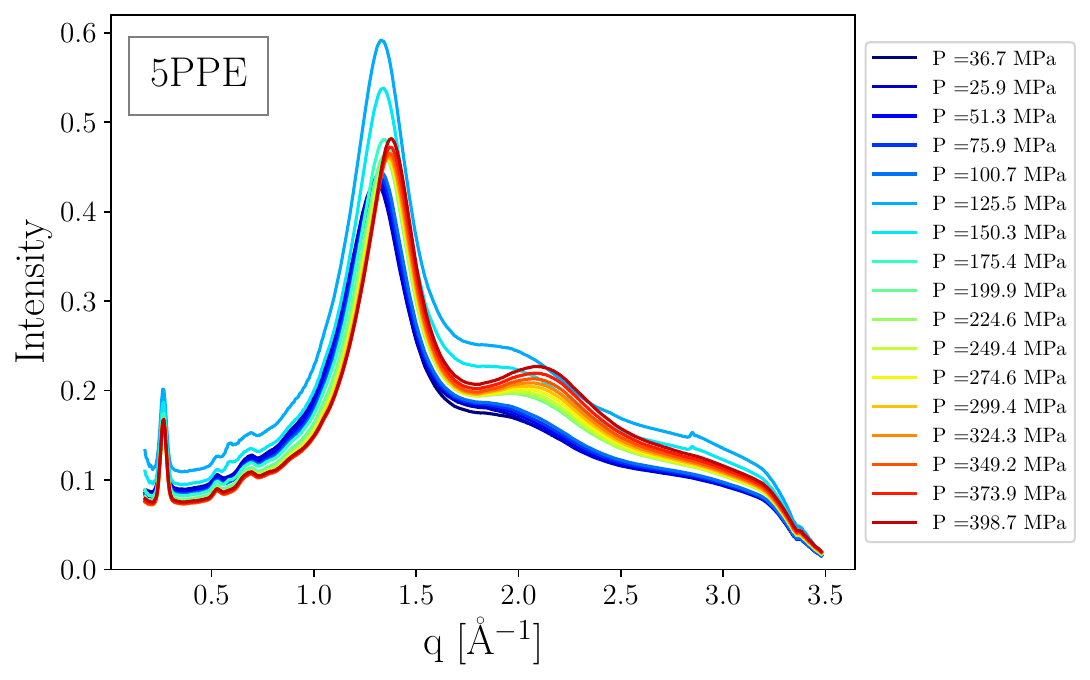}
			\caption{T = 38.3 $^o$C}
					\label{fig:hopeless_intensity_d}
		\end{subfigure}
		\caption{The measurement along the four isotherms in chronological order: 19.3 $^o$C $\rightarrow$ 56.9 $^o$C $\rightarrow$ 76.0 $^o$C $\rightarrow$ 38.3 $^o$C. The color scale shows the pressure at each measurement, with blue being low pressures and red being high. Between the measurements at 19.3 $^o$C and 56.9 $^o$C leaks at the detector side window caused the measured intensity to fluctuated wildly and unsystematically. The leak caused a droplet to be formed, and most likely the edge of the droplet caused the fluctuations in maximum intensity.}
		\label{fig:hopeless_intensity}
	\end{figure}

	Another practical issue with background subtraction was that during the 5PPE measurements we experienced that the measured intensity fluctuated wildly and unsystematically, as shown in Figure \ref{fig:hopeless_intensity}. The chronology of the measurements was 19.3 $^o$C $\rightarrow$ 56.9 $^o$C $\rightarrow$ 76.0 $^o$C $\rightarrow$ 38.3 $^o$C, and the intensity fluctuations began between the 19.3 $^o$C and 56.9 $^o$C measurement. This can be seen in figures \ref{fig:hopeless_intensity_b}, \ref{fig:hopeless_intensity_c}, and \ref{fig:hopeless_intensity_d}. If we compare the three figures with the measurement at 19.3 $^o$C, figure \ref{fig:hopeless_intensity_a}, we see a change in behavior. In figure \ref{fig:hopeless_intensity_a}, the color scale shows the pressure at each measurement, with blue being low pressures and red being high. With increasing pressure the maximum intensity at the main structure peak is almost constant. In the other three subfigures there are clear fluctuations in the maximum intensity at the peak around 1.3-1.4 Å$^{-1}$.
	
\begin{figure}[H]
		\begin{subfigure}[b]{0.49\textwidth}
			\centering
			\includegraphics[width=0.9\textwidth]{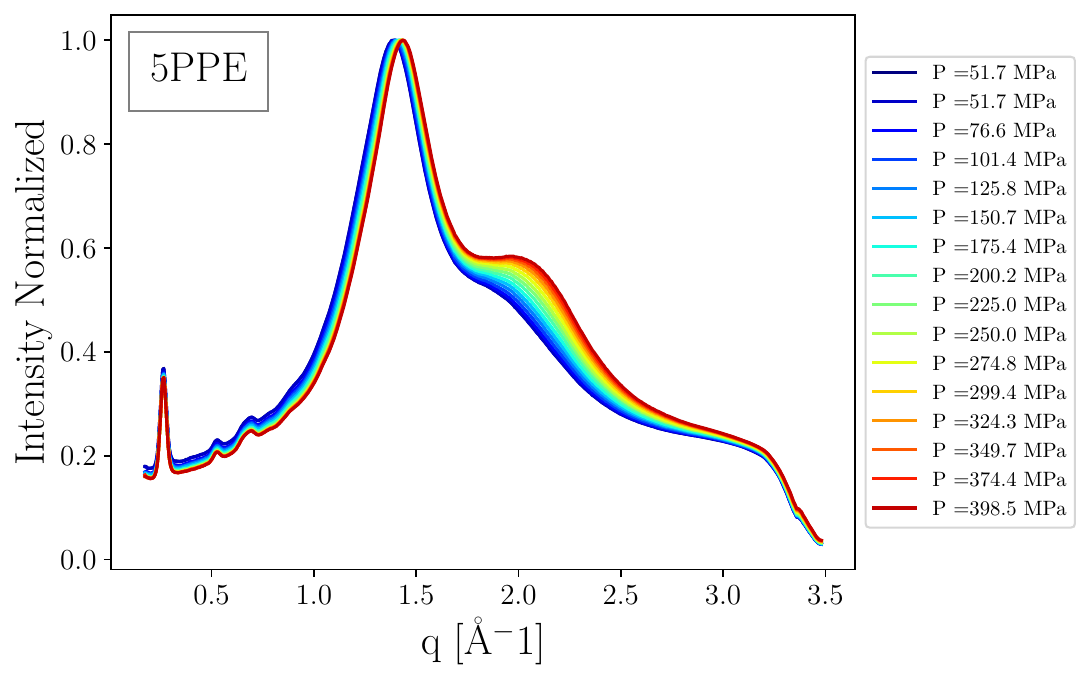}
			\caption{T = 19.3 $^o$C}		
		\end{subfigure}
		\begin{subfigure}[b]{0.49\textwidth}
			\centering
			\includegraphics[width=0.9\textwidth]{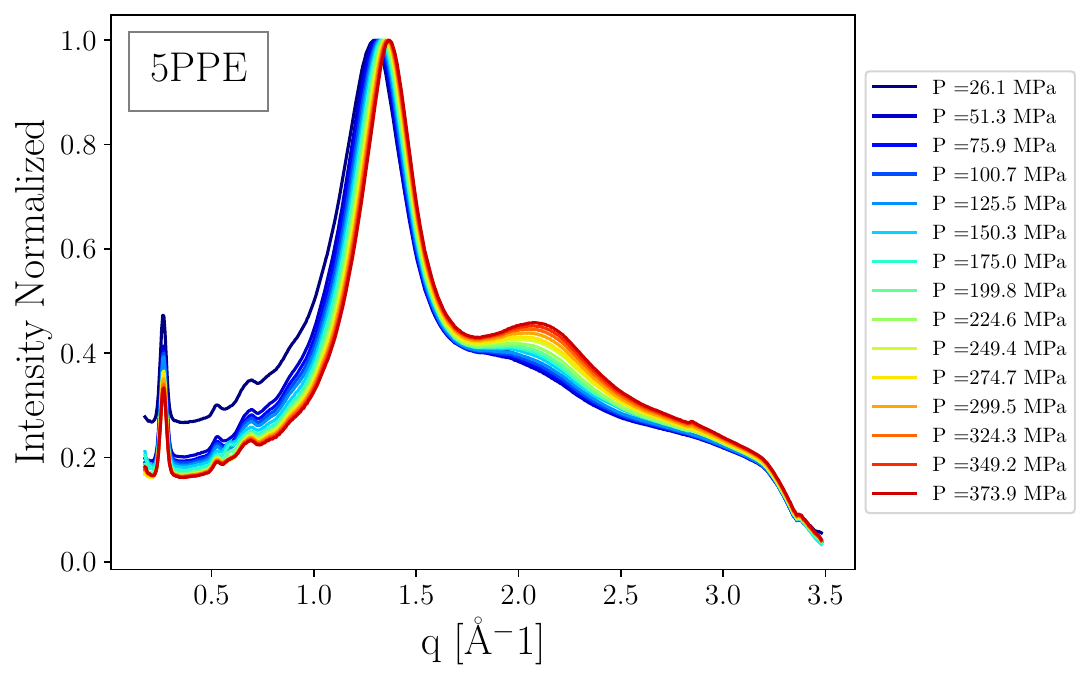}
			\caption{T = 56.9 $^o$C}		
		\end{subfigure}	
		\begin{subfigure}[b]{0.49\textwidth}
			\centering
			\includegraphics[width=0.9\textwidth]{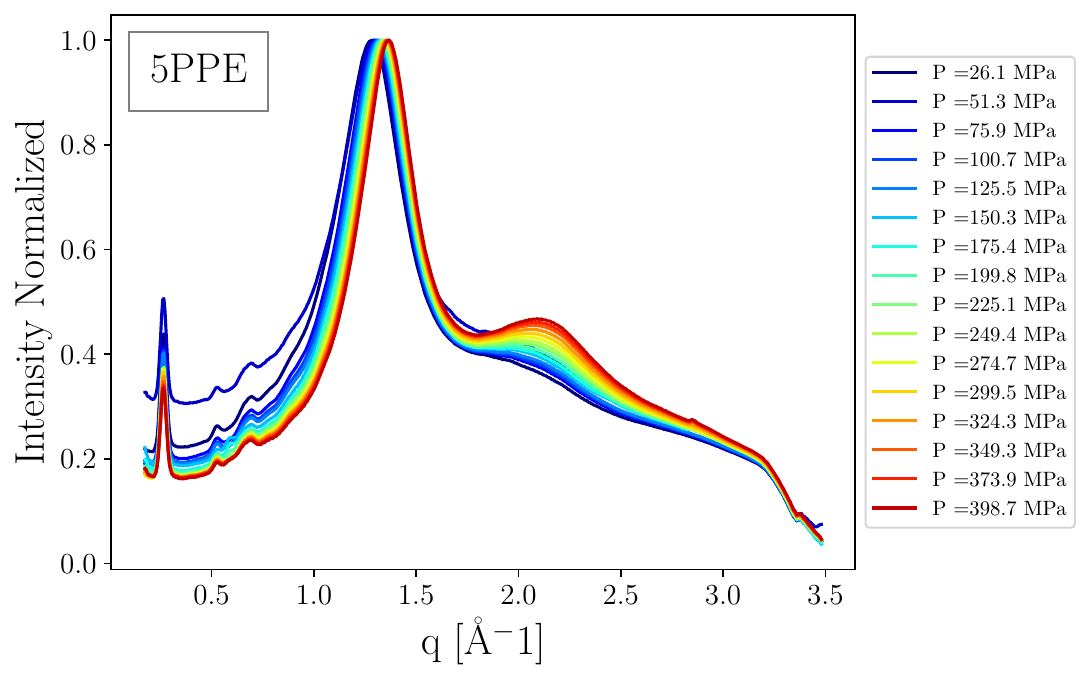}
			\caption{T = 76.0 $^o$C}		
		\end{subfigure}
		\begin{subfigure}[b]{0.49\textwidth}
			\centering
			\includegraphics[width=0.9\textwidth]{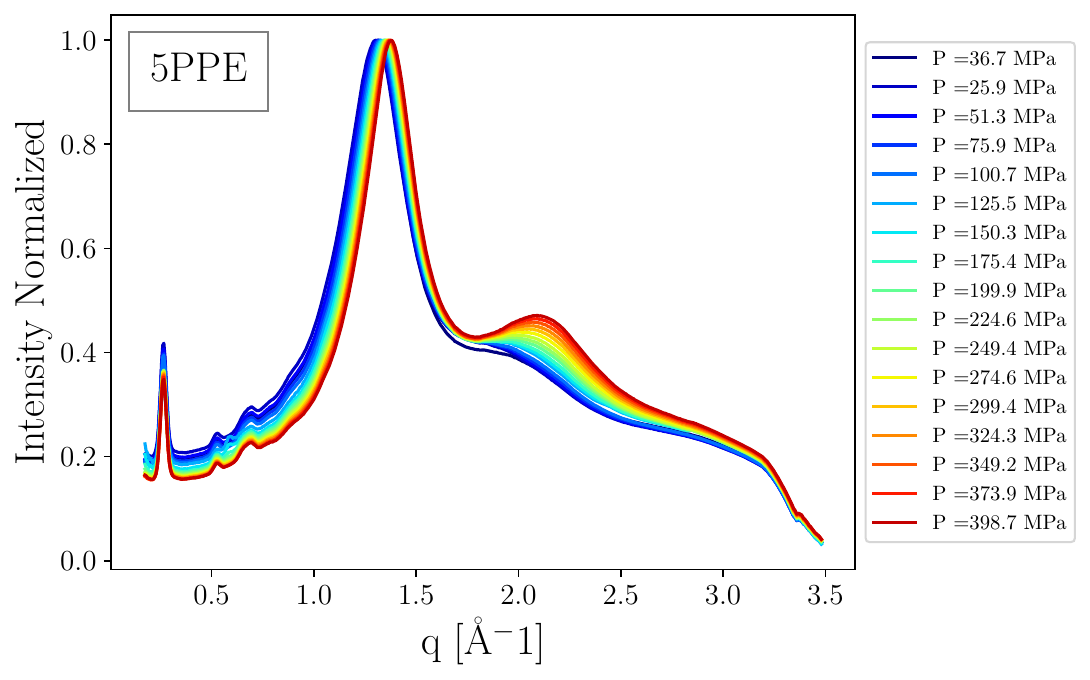}
			\caption{T = 38.3 $^o$C}
		\end{subfigure}
		\caption{The measurement along the four isotherms normalized with peak intensity. The chronological order of the measurements was 19.3 $^o$C $\rightarrow$ 56.9 $^o$C $\rightarrow$ 76.0 $^o$C $\rightarrow$ 38.3 $^o$C. }
		\label{fig:hopeless_intensity_scaled}
	\end{figure}

\begin{figure}[H]
		\centering
		\includegraphics[width=0.5\textwidth]{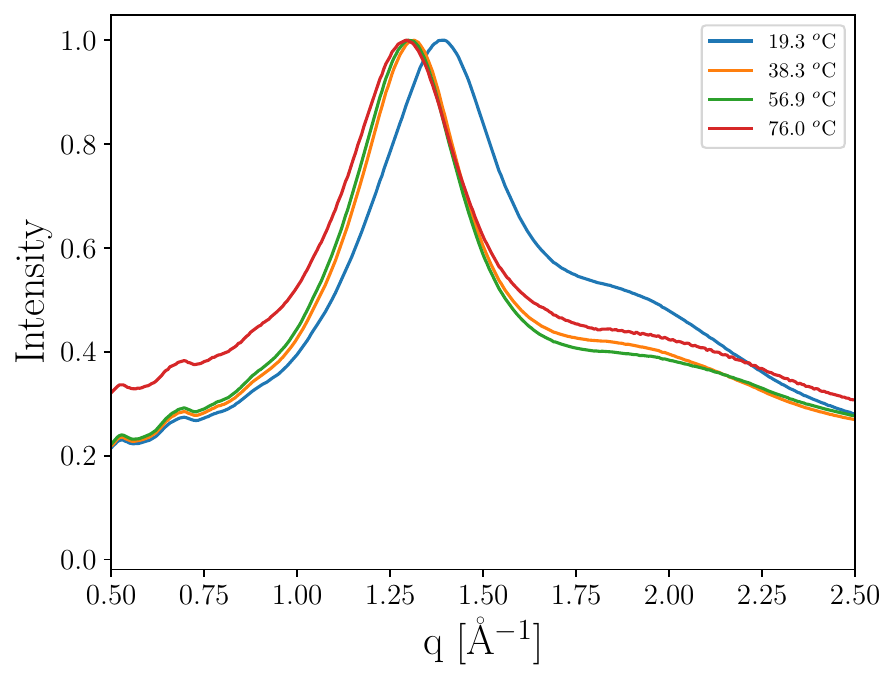}
		\caption{The normalized intensity along an isobar, $P =51$ MPa. The intensity jumps began  during the measurements on 5PPE between the 19.3 $^o$C and 56.9 $^o$C measurements . The jumps in intensity caused the peak to be shifted most likely due to a signal from water from a leak on the detector side window of the cell. Given that the peak shape and position are so different before and after the jumps in intensity, it is difficult to trust the data. The analysis for 5PPE and squalane are therefore only preliminary. }
		\label{fig:5PPE_isobar_normalized}
	\end{figure}

	The most likely explanation is a water droplet forming on the diamond window nearest the detector, and the jumps are due to scattering from the edge of the droplet. In figure \ref{fig:hopeless_intensity_scaled} we scale out the intensity fluctuations by normalizing with the maximum intensity. The peak intensity of the first peak  equals 1 in figure \ref{fig:hopeless_intensity_scaled}. The shape of the scattering intensity is consistent within the isotherms and behaves as expected. The peak moves to higher q with increasing pressure and the signal from the water peak at q $\approx$ 2 Å$^{-1}$ increases with pressure. If we compare before and after the jumps intensity figure \ref{fig:hopeless_intensity_scaled} (a) - (b-d), the measured spectrum changes because of the leaks. This is also shown in figure \ref{fig:5PPE_isobar_normalized} were isobaric measurements are compared. This shift in peak position made it difficult for us to trust the data and test our hypothesis. The hypothesis is that the structure is invariant when presented in dimensionless units, not in the experimental units. The extra signal that shifts the peak will give an error when scaling the density out, similar to what we observe from the intramolecular contribution to the structure factor. 
	
	The consequence of the jumps in intensity is that background subtraction is very difficult. Even if we try to normalize the intensity there will be an unaccounted background from the leaking windows that clearly affects the peak. This error affected both the 5PPE and Squalane measurements. We will perform a preliminary analysis of the raw data to provide inspiration and an indication of the results of further work. It is worth stressing that any results for squalane and 5PPE should be retested. However, we were able to subtract the background for DC704 and DPG.

	\subsection[DC704 and DPG]{Choosing an appropriate background for DC704 and DPG:}
	
\begin{figure}[H]
		\begin{subfigure}[b]{0.49\textwidth}
			\centering
			\includegraphics[width=0.9\textwidth]{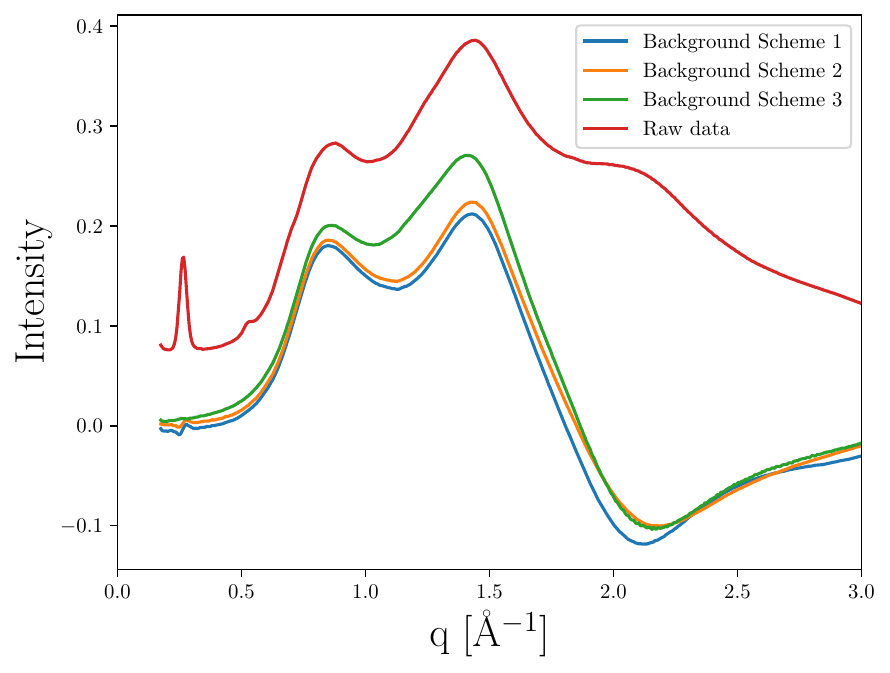}
			\caption{}		
		\end{subfigure}
		\begin{subfigure}[b]{0.49\textwidth}
			\centering
			\includegraphics[width=0.9\textwidth]{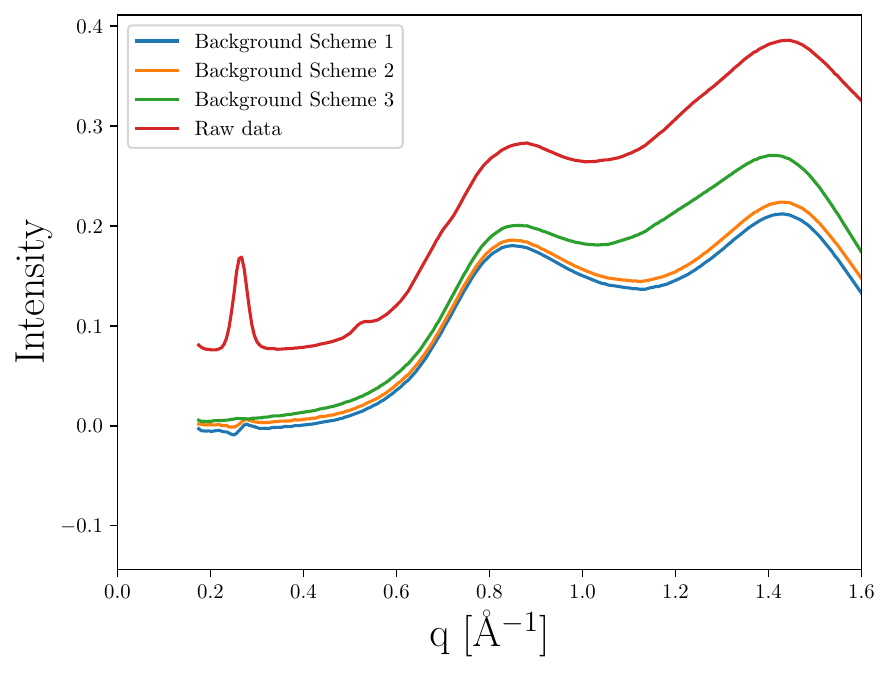}
			\caption{}
		\end{subfigure}
		\caption{The same measurement of DC704 for the three different background schemes and the raw data for a state point at T = 39 $^o$C, P = 250 MPa . Background scheme 1 is subtracting water in a capillary from two different isotherms. Background scheme 2 is using water in capillary measured along the 57 $^o$C isotherm. Background scheme 3 is only subtracting the water signal. }
		\label{fig:Background_compare}
	\end{figure}

	To choose an appropriate background, we tested three different background schemes: water in the cell + water in capillary at two different temperatures, water in the cell + water in a capillary, and only water in the cell. For each background scheme, we subtract too much water(see Figure \ref{fig:Background_compare}), however in the q-range of interest, this has a limited effect. Each scheme has different advantages and disadvantages. 
	
	\textbf{Water in a Capillary, Scheme 1:} For the water in a capillary, we performed measurements along the T = 19 $^o$C and T = 57 $^o$C isotherms in the pressure range from $50 - 400$ MPa. For this background: we subtract T = 19 $^o$C background, from T = 19 $^o$C and T = 39 $^o$C, and the T = 57 $^o$C background from T = 57 $^o$C and T = 76 $^o$C. The advantage of this scheme is that it allows for an estimation of the temperature dependence of the background. 
	
	\textbf{Water in a Capillary, Scheme 2:} We only use the T = 57 $^o$C water in capillary measurement as background. The background will only account for the pressure-dependence of the water and the capillary. Any temperature dependence of the background will be systematic.
	
	\textbf{Pure Water Background, Scheme 3:} For pure water, we performed measurements along all four isobars in the pressure range from 100 - 400 MPa. The reduced pressure range compared to the sample measurements was due to leakage when pressure was decreased, narrowing the window we could measure. Due to the smaller pressure range it is not possible to subtract a background for all measurements. The advantage of this scheme is that we can account for both the temperature and pressure dependence of the water signal. The disadvantage is that we are not able to subtract any contribution from the capillary. The capillary  gives a contribution around $q = 1.0 -1.5$ Å$^{-1}$, as shown in figure \ref{fig:Water capillary_b}. This overlaps with the structure peaks for the sample.
	
	\textbf{Raw data:} The final scheme is simply analyzing the raw data. The disadvantages of the scheme is clear: the contributions of both the water and capillary are unaccounted for. However, as shown in figure \ref{fig:Water capillary} the contributions of both are relatively constant as a function of pressure in the region of interest. 
	
	In summary, four possible schemes are itemized. The background subtraction schemes written in bold are the schemes used for analysis the data:
	
	\begin{itemize}
		\item \textit{Background scheme 1:} Water in a Capillary Scheme 1, Water in Capillary background using data from two different isotherms
		\item \textbf{Background scheme 2:} Water in a Capillary Scheme 2, Water in Capillary background using data from a single isotherm. This scheme is used for DC704 and DPG data.
		\item \textit{Background scheme 3:} Pure water background 
		\item \textbf{Raw Data:} No background subtraction. This scheme is used for the 5PPE and the Squalane measurements.
	\end{itemize}
	
	For 5PPE and Squalane, we only fitted the raw data, whereas for DC704 and DPG data we used Background scheme 2. For 5PPE and Squalane,  it is only possible to analyze the data without background subtraction because of the jumping intensities. For DC704 and DPG, the reasons for choosing background scheme 2 are several. For the pure water background, while we have a background for all of the isotherms, we only have measurements in a smaller pressure range and we cannot account for contributions from the capillary. Although ideally we would like to have measurement on water in a capillary along each isotherm, In practice, we only have measured along two isotherms. Background scheme 1 creates two different levels of intensities as shown in Figure \ref{fig:background_scheme_compare_a}. For the measurements that can be compared, the results seem unaffected by whatever scheme is used for subtracting a background. The robustness of each background scheme is tested in the following section for DC704.

	\subsubsection{Testing different schemes}

\begin{figure}[H]
		
		\begin{subfigure}[t]{0.49\textwidth}
			\centering
			\includegraphics[width=0.9\textwidth]{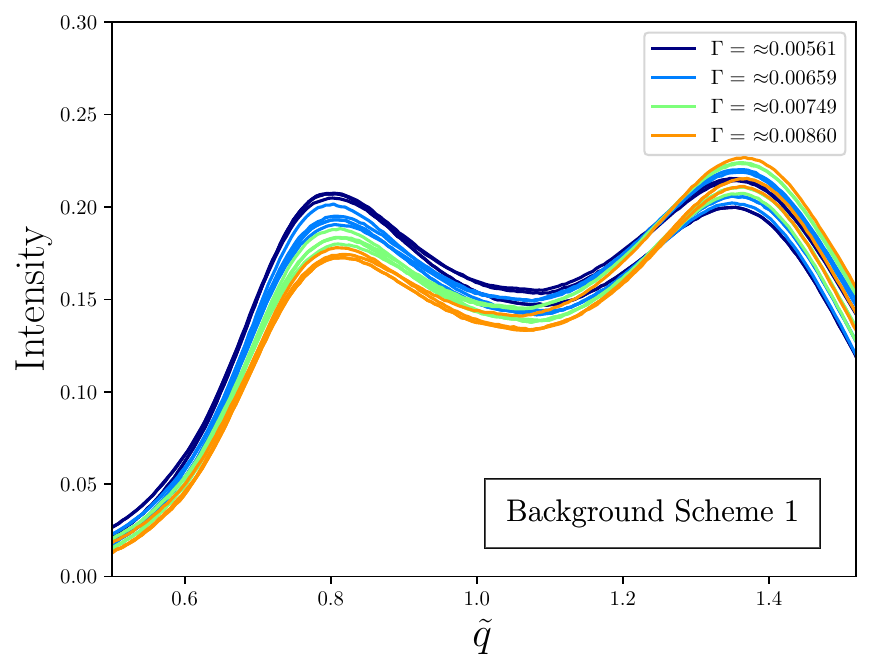}
			\caption{Background scheme 1. The water in the cell + water in capillary at two different temperatures - scheme.}		
			\label{fig:background_scheme_compare_a}
		\end{subfigure}		
		\begin{subfigure}[t]{0.49\textwidth}
			\centering
			\includegraphics[width=0.9\textwidth]{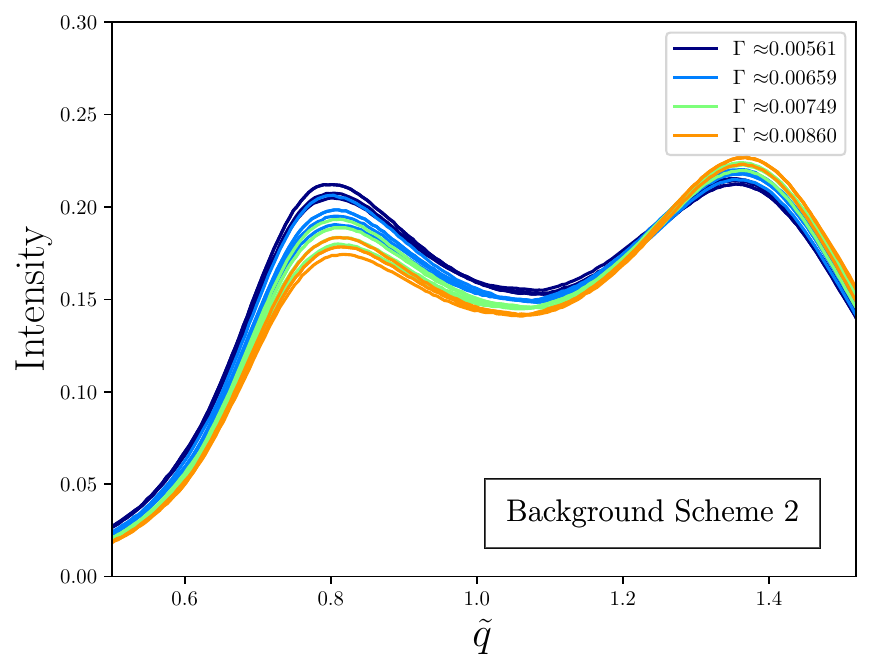}
			\caption{Background scheme 2. The water in the cell + water in capillary at one different temperatures - scheme.}
			\label{fig:background_scheme_compare_b}		
		\end{subfigure}
		\begin{subfigure}[t]{0.49\textwidth}
			\centering
			\includegraphics[width=0.9\textwidth]{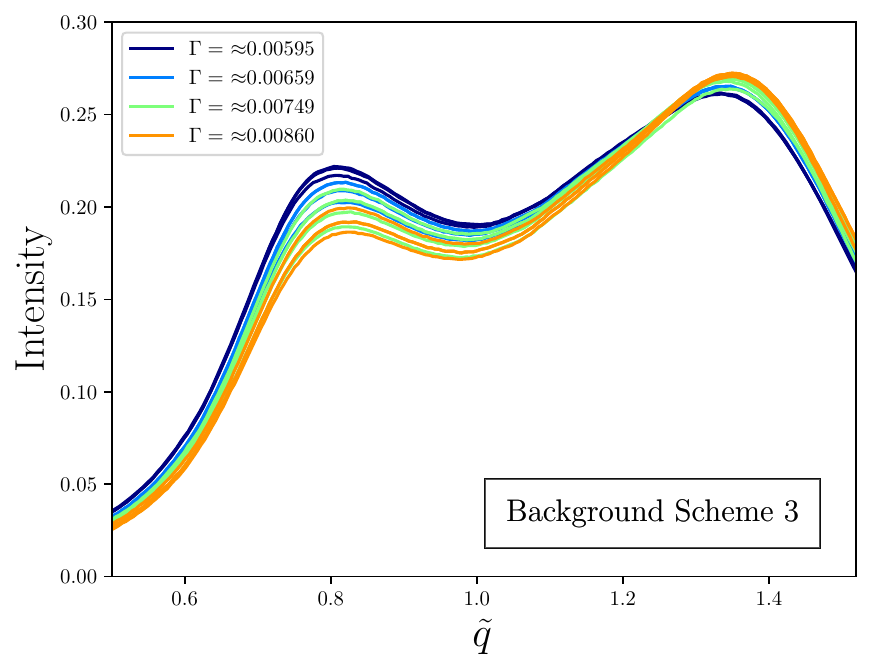}
			\caption{Background scheme 3. The only water - scheme. Note! The measurements under 100 MPa are missing from this figure}
			\label{fig:background_scheme_compare_c}		
		\end{subfigure}		
		\begin{subfigure}[t]{0.49\textwidth}
			\centering
			\includegraphics[width=0.9\textwidth]{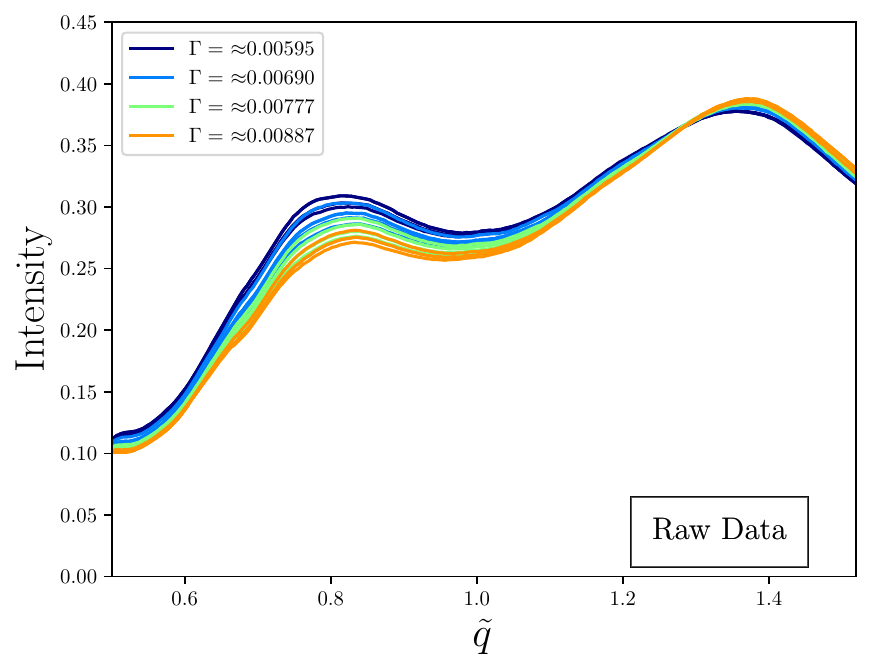}
			\caption{Raw data scheme. The measured intensity, with no background subtraction.}		
						\label{fig:background_scheme_compare_d}
		\end{subfigure}
		\begin{subfigure}[t]{0.99\textwidth}
			\centering
			\includegraphics[width=0.65\textwidth]{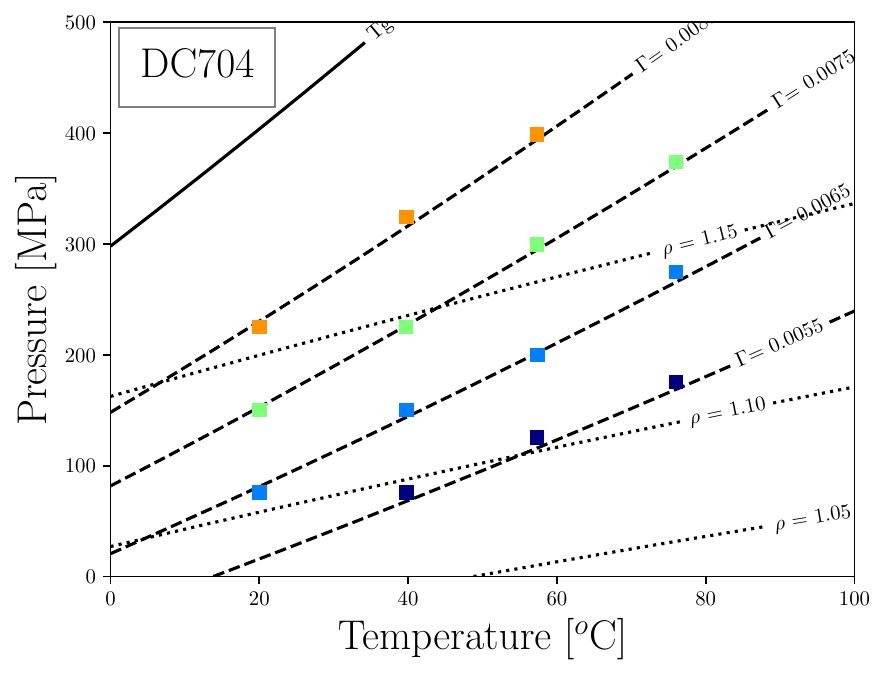}
			\caption{}
			\label{fig:background_scheme_compare_e}
		\end{subfigure}
		\caption{ (a)-(d)  the measurements for DC704 along four different isochrones. Each figure shows the same measurements with a different scheme for background subtraction. Note that the measurements under 100 MPa are missing from figure \ref{fig:background_scheme_compare_c}. The data are presented in reduced units $\tilde{q} = q \rho^{- \frac{1}{3}}$.  Each color represent a different line of constant $\Gamma$. In figure (e), the color and position of each data point is shown in the phase diagram.}
		\label{fig:background_scheme_compare}
	\end{figure}

\begin{figure}[H]
		\begin{subfigure}[b]{0.49\textwidth}
			\centering
			\includegraphics[width=0.9\textwidth]{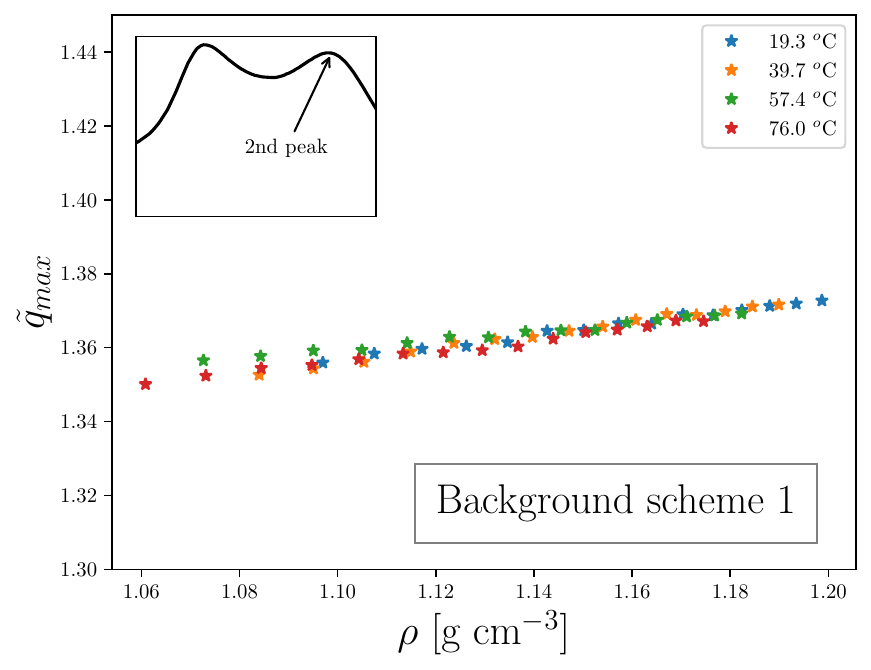}
			\caption{Position of the second peak, as function of $\rho$, background scheme 1}		
		\end{subfigure}
		\begin{subfigure}[b]{0.49\textwidth}
			\centering
			\includegraphics[width=0.9\textwidth]{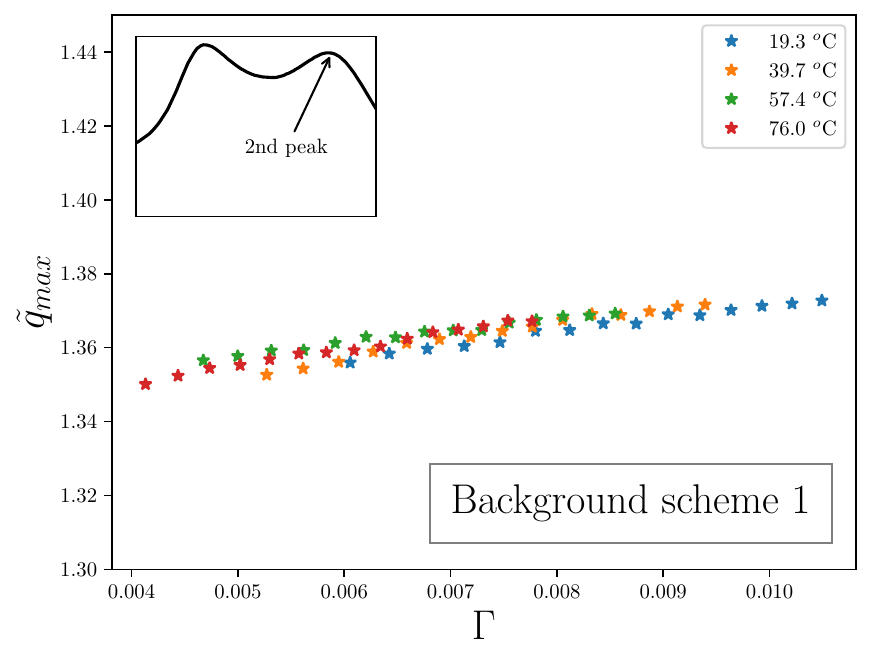}
			\caption{Position of the second peak, as function of $\Gamma$, background scheme 1}		
		\end{subfigure}
		\begin{subfigure}[b]{0.49\textwidth}
			\centering
			\includegraphics[width=0.9\textwidth]{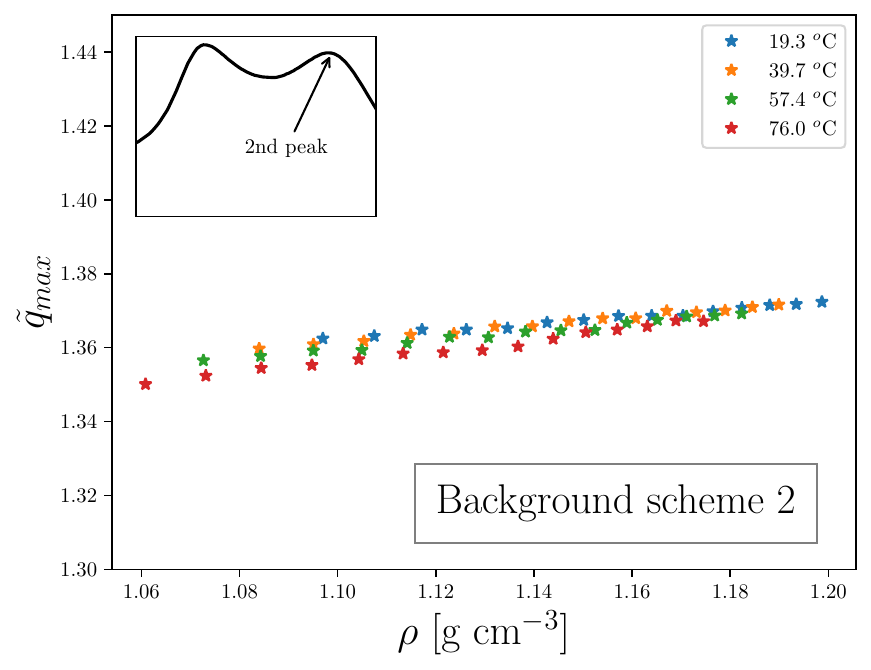}
			\caption{Position of the second peak, as function of $\rho$, background scheme 2.}		
		\end{subfigure}
		\begin{subfigure}[b]{0.49\textwidth}
			\centering
			\includegraphics[width=0.9\textwidth]{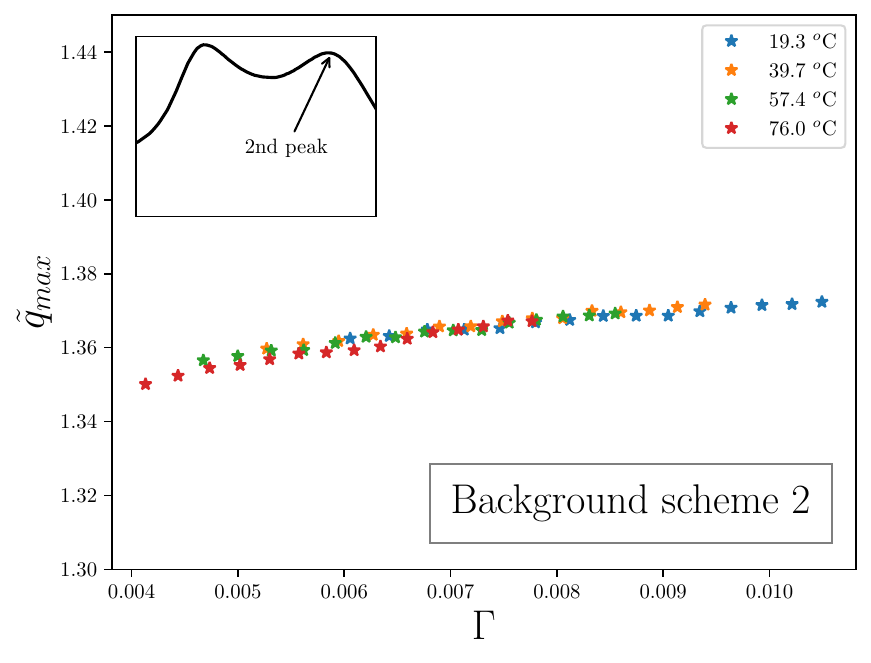}
			\caption{Position of the second peak, as function of $\Gamma$, background scheme 2.}		
		\end{subfigure}
		\begin{subfigure}[b]{0.49\textwidth}
			\centering
			\includegraphics[width=0.9\textwidth]{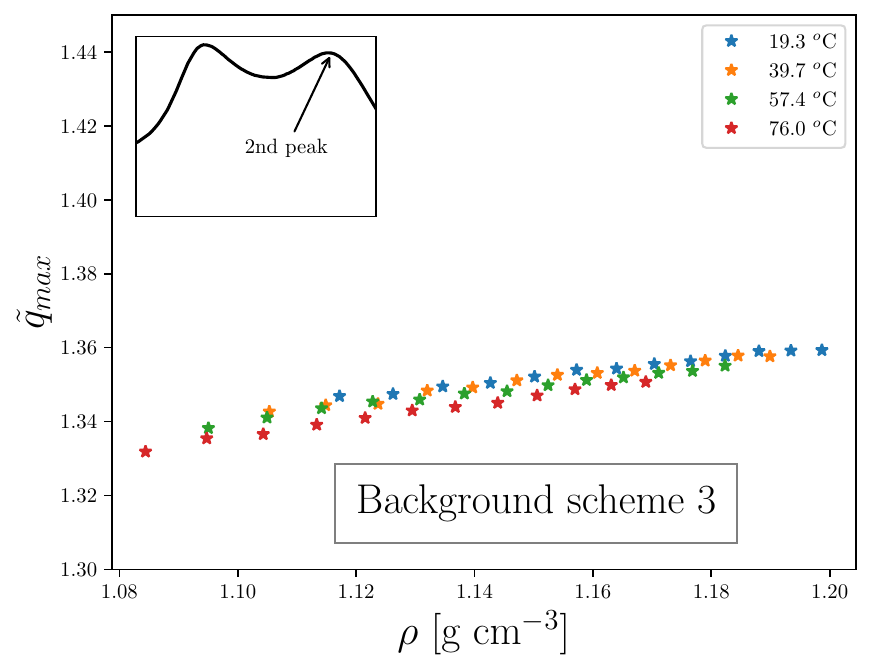}
			\caption{Position of the second peak, as function of $\rho$, background scheme 3.}		
		\end{subfigure}
		\begin{subfigure}[b]{0.49\textwidth}
			\centering
			\includegraphics[width=0.9\textwidth]{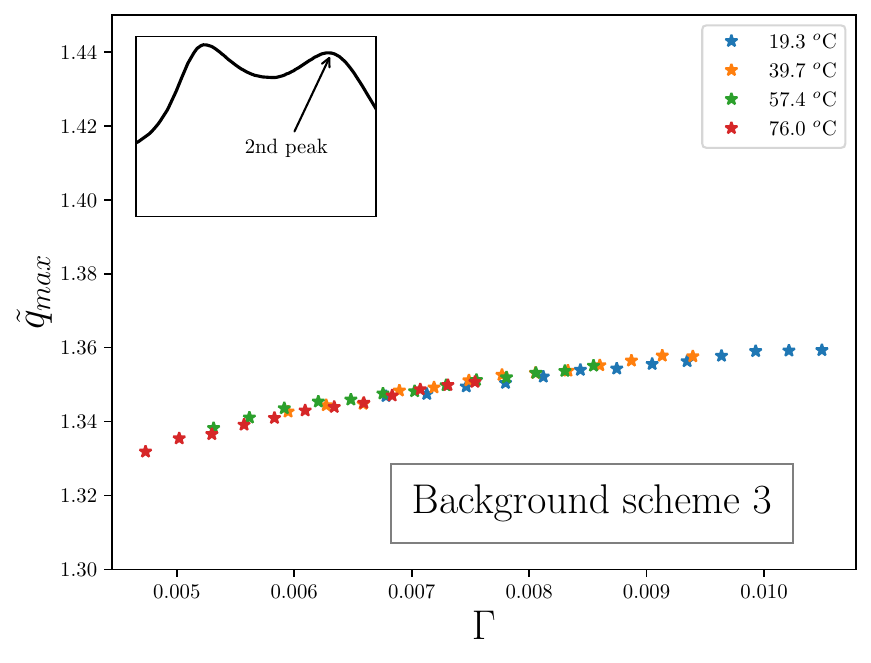}
			\caption{Position of the second peak, as function of $\Gamma$, background scheme 3.}		
		\end{subfigure}
		\caption{The peak position of the second peak for different background subtraction schemes. The second peak is fitted with a split-pseudo Voigt distribution as described in section \ref{sec:Split_Voigt}. The results are discussed in detail in section \ref{sec:DC704_fit_results}. We see that the peak position of the 2nd peak, regardless of the background schemes, seem to collapse along lines of constant $\Gamma$, subfigures (b), (d), (f), better than isochores, subfigures (a), (c), (e). }
		\label{fig:fitting_2nd_peak_all_schemes}
	\end{figure}

\begin{figure}[H]
		
		\begin{subfigure}[b]{0.49\textwidth}
			\centering
			\includegraphics[width=0.9\textwidth]{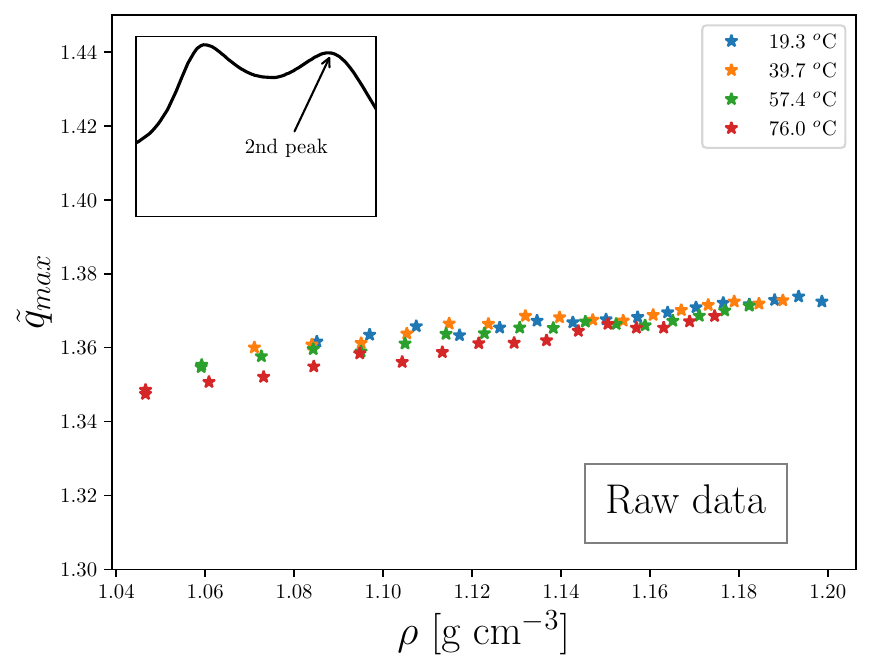}
			\caption{Position of the second peak, as function of $\rho$. No background subtraction. }		
		\end{subfigure}		
		\begin{subfigure}[b]{0.49\textwidth}
			\centering
			\includegraphics[width=0.9\textwidth]{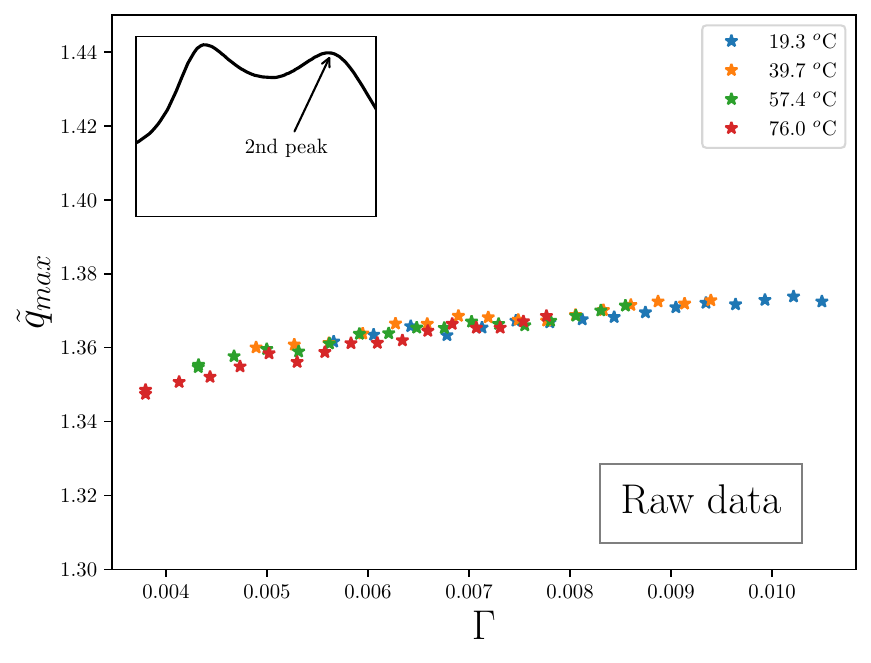}
			\caption{Position of the second peak, as function of $\Gamma$. No background subtraction.}		
		\end{subfigure}
		\caption{The position of the second peak without background subtraction. Even when no background was subtracted the results are comparable with background subtraction. }
		\label{fig:fitting_2nd_peak_raw}
	\end{figure}

	The results for different background schemes are compared in Figure \ref{fig:background_scheme_compare}, where the structure along four different isochrones is plotted as a function of  $\tilde{q} =q \rho^{-\frac{1}{3}}$.  It is worth noting that while the results are fairly consistent, for the second peak using background scheme 1, there are clearly two different intensities, depending on whether the 19 $^o$C or 57 $^o$C background was subtracted. In figure \ref{fig:fitting_2nd_peak_all_schemes} the results of the fitting are shown for the three different schemes and in Figure \ref{fig:fitting_2nd_peak_raw} fits on the raw data are shown. The details of the fit will be discussed in a subsequent section, but it is worth noting that the results are very consistent regardless of background scheme. In figure \ref{fig:fitting_2nd_peak_all_schemes} and \ref{fig:fitting_2nd_peak_raw} we see that the peak position regardless of background schemes collapses better as a function of $\Gamma$ than as a function of density. The robustness of the results, regardless of the background scheme, is very promising for the conclusions of this study. For the DC704 and DPG background scheme 2 is used for the data analysis, while for the preliminary results of 5PPE and squalane the Raw data is fitted.

	\section[DC704 \& DPG]{Testing for pseudo-isomorphs}\label{sec:Diamond_fit}
	
	This section will present the result of fitting the two peaks of DC704  and the structural measurements and fitting of DPG . Part of the results for DC704 have already been introduced as part of the robustness test of the background scheme, in section \ref{sec:background}. In figure \ref{fig:background_scheme_compare}, we have shown that the first peak of DC704 clearly varies along isochrones. The second peak seems to collapse along isochrones. It is similar to what we saw for cumene, a pretty good, but not perfect collapse. Firstly, the temperature and pressure dependence of the two peaks of DC704 are presented, then each peak is fitted and the position of the two peaks are compared as a function of density and $\Gamma$. Then DPG will be examined using the same procedure, first the peak behavior as a function of pressure and temperature, then as function of density and $\Gamma$.
	
	\subsection{DC704} \label{sec:DC704_fit_results}
	
\begin{figure}[h]
		\begin{subfigure}[b]{0.49\textwidth}
			\centering
			\includegraphics[width=0.99\textwidth]{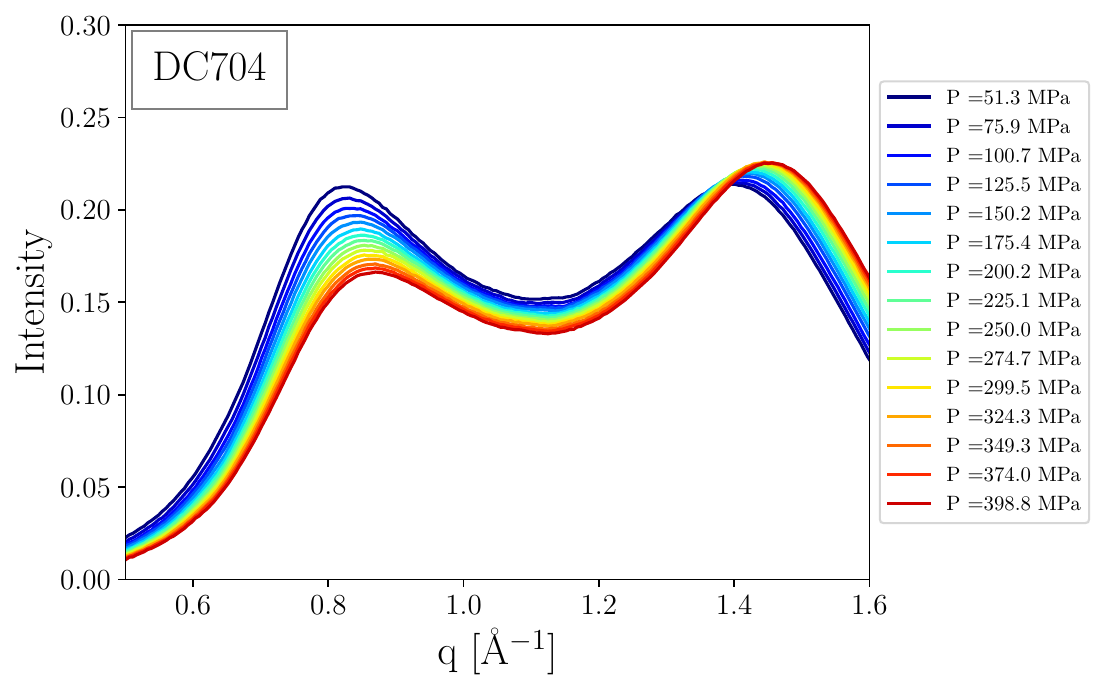}
			\caption{T = 19.3 $^o$C}		
		\end{subfigure}
		\begin{subfigure}[b]{0.49\textwidth}
			\centering
			\includegraphics[width=0.99\textwidth]{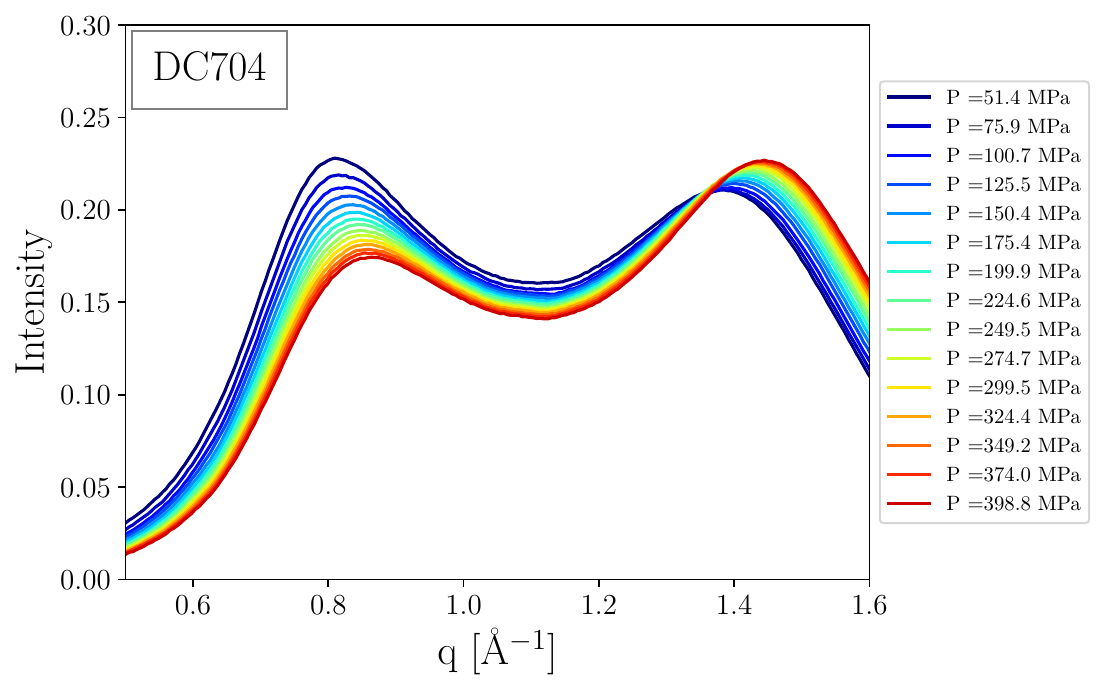}
			\caption{T = 57.4 $^o$C}		
		\end{subfigure}	
		\begin{subfigure}[b]{0.49\textwidth}
			\centering
			\includegraphics[width=0.99\textwidth]{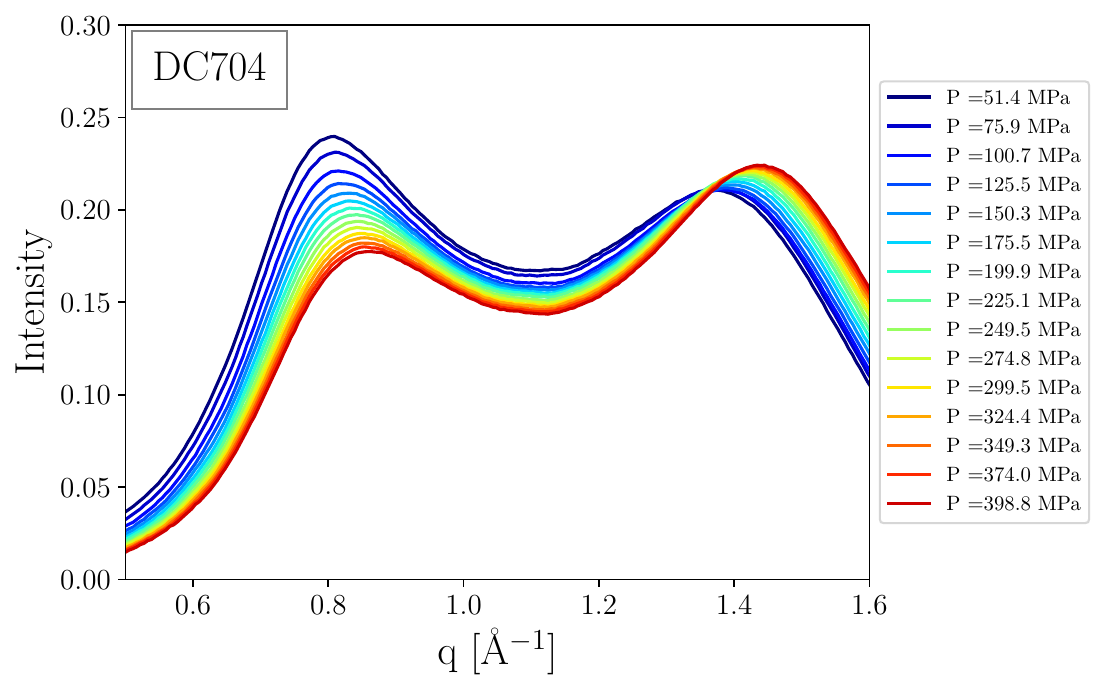}
			\caption{T = 76.0 $^o$C}		
		\end{subfigure}
		\begin{subfigure}[b]{0.49\textwidth}
			\centering
			\includegraphics[width=0.99\textwidth]{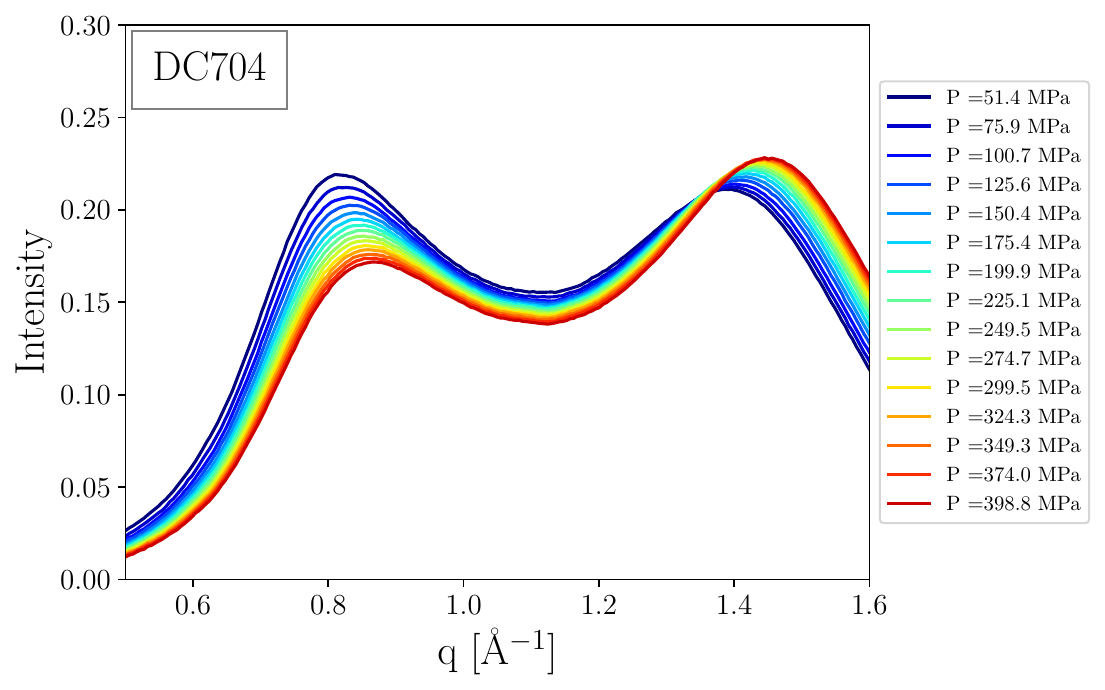}
			\caption{T = 39.7 $^o$C}
		\end{subfigure}
		\caption{DC704 measurement along the four isotherms in chronological order: 19.3 $^o$C $\rightarrow$ 57.4 $^o$C $\rightarrow$ 76.0 $^o$C $\rightarrow$ 39.7 $^o$C. The color scale from blue to red denotes increasing pressure. }
		\label{fig:DC704_data}
	\end{figure}

	The measured intensities along the four isotherms are shown for DC704 in Figure \ref{fig:DC704_data}. DC704 has a double peak: a first peak with a maximum around $q \approx 0.8$ Å$^{-1}$ and a second peak with a maximum around $q \approx 1.4$ Å$^{-1}$. The two peaks exhibit different behavior with increasing pressure. The absolute intensity of the first peak decrease with increasing pressure, whereas that of the second peak increases. However, with increasing pressure, the maxima of both peaks move to higher q. If we compare Figure \ref{fig:DC704_data} (a) and \ref{fig:DC704_data} (c), the absolute intensity of first peak increases, whereas that of the second peak decreases with increasing temperature. The structural origin of DC704's double peak is so far unknown, but the behavior of both peak with changing temperature and pressure is similar to those of other studied glass formers in this thesis. Increasing the density by changes to the pressure or temperature tends to moves the peak higher q.  It is easy to speculate about the structural origin of DC704's double peak and their movement, but only with microscopic knowledge from MD and/or Reverse Monte Carlo simulation studies can any hypotheses be confirmed or denied. Therefore, in figure \ref{fig:DC704_MD} we show work in progress on MD simulations of DC704 from \Mycite{DC704MD}. Our approach is to fit the peaks, with the only goal of describing each peak. The two peaks are too entangled to find any width; thus, we will fit to obtain the peak position. We fit using a split pseudo-Voigt distribution, as described in section \ref{sec:Split_Voigt}.

	\subsubsection{Peak analysis}
	
\begin{figure}[H]
		\begin{subfigure}[b]{0.49\textwidth}
			\centering
			\includegraphics[width=0.9\textwidth]{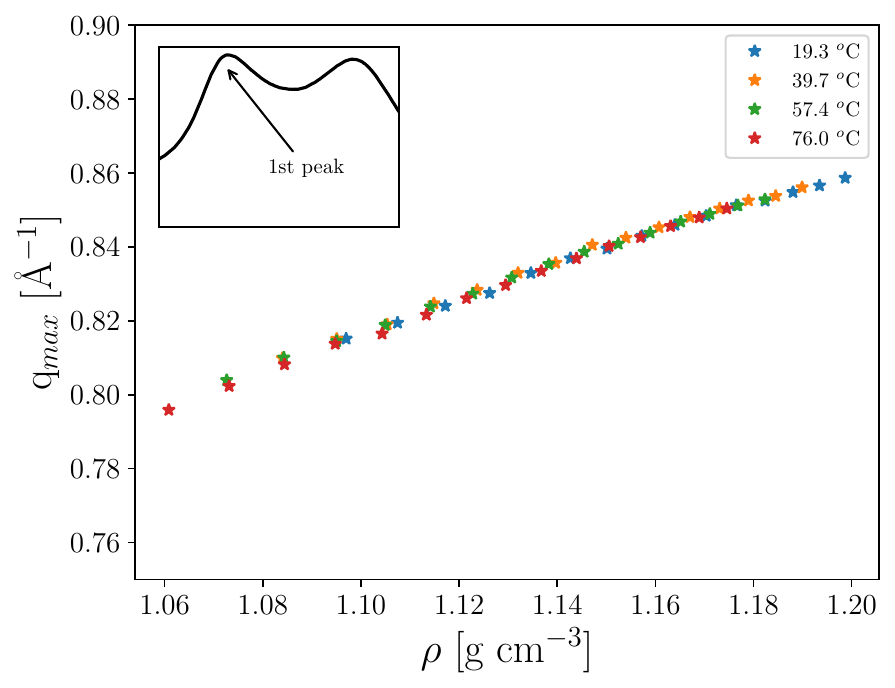}
			\caption{q$_{max}$ for the first peak as a function of $\rho$}		
						\label{fig:DC704_1st_fit_a}
		\end{subfigure}
		\begin{subfigure}[b]{0.49\textwidth}
			\centering
			\includegraphics[width=0.9\textwidth]{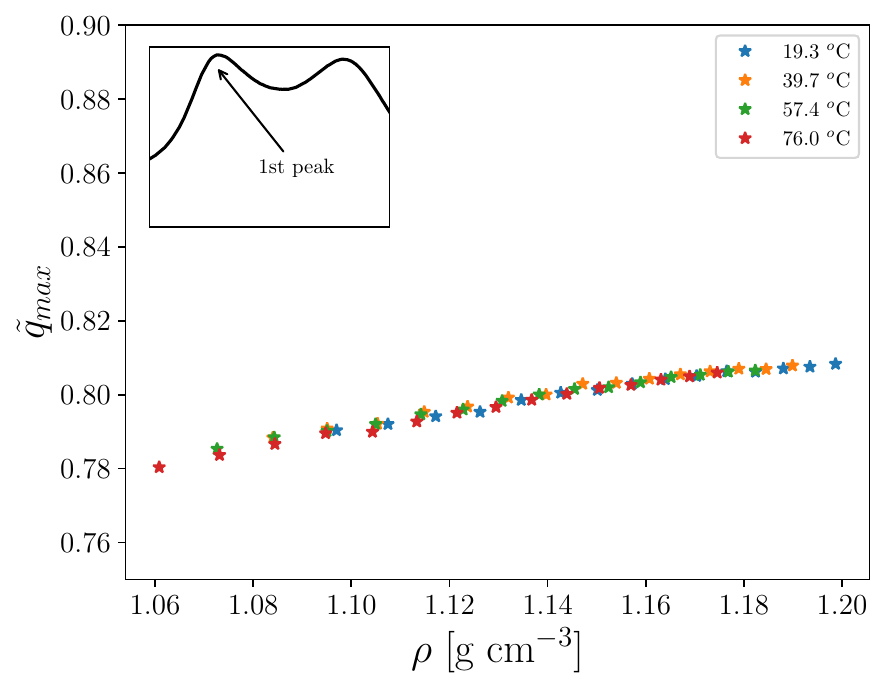}
			\caption{$\tilde{q}_{max}$ for the first peak as a function of $\rho$ }
						\label{fig:DC704_1st_fit_b}
		\end{subfigure}
		
		\begin{subfigure}[b]{0.49\textwidth}
			\centering
			\includegraphics[width=0.9\textwidth]{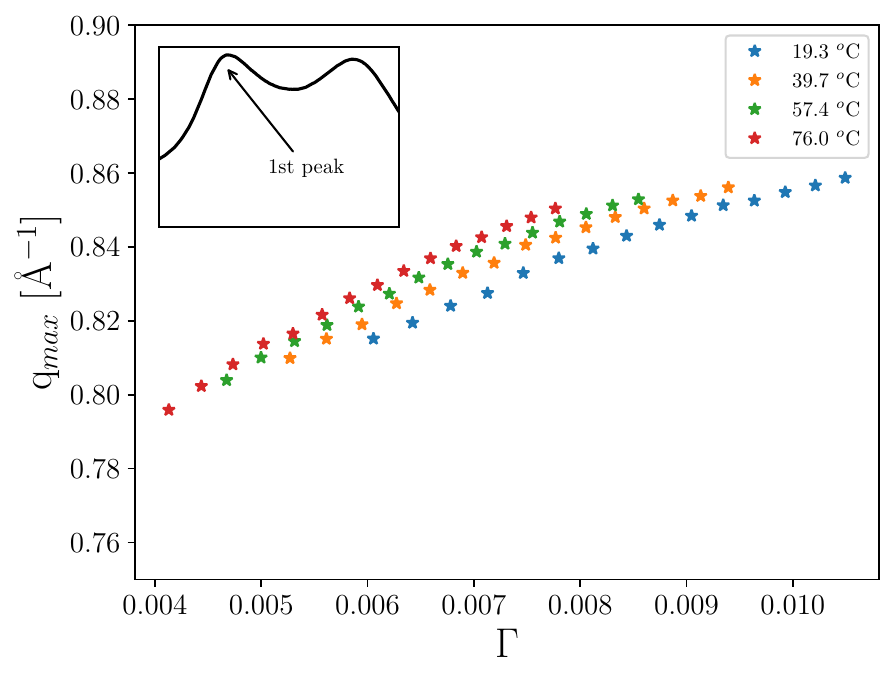}
			\caption{q$_{max}$ for the first peak as a function of $\Gamma$}
						\label{fig:DC704_1st_fit_c}		
		\end{subfigure}
		\begin{subfigure}[b]{0.49\textwidth}
			\centering
			\includegraphics[width=0.9\textwidth]{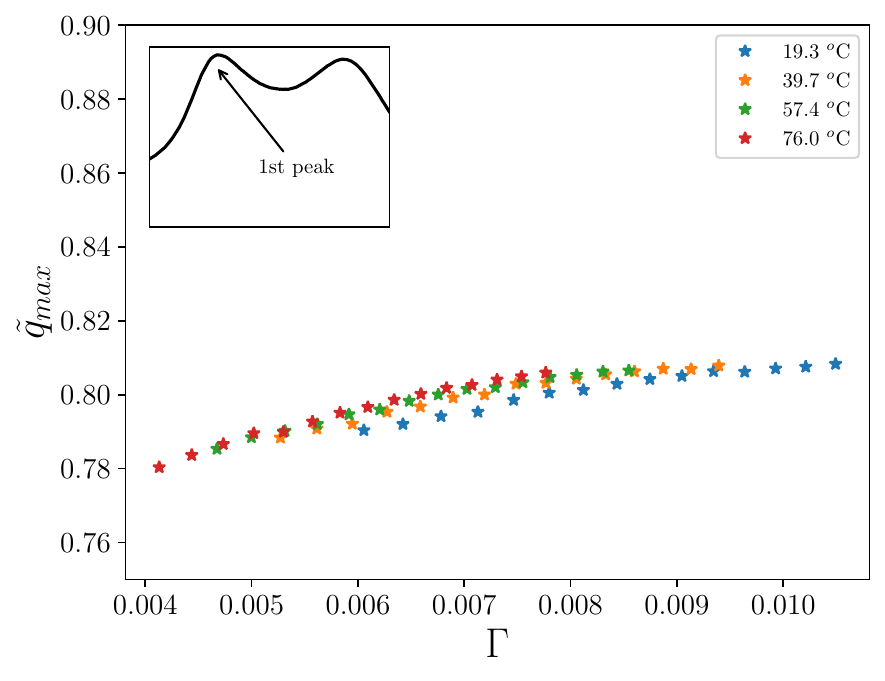}
			\caption{ $\tilde{q}_{max}$ for the first peak as a function of $\rho$ }
			\label{fig:DC704_1st_fit_d}
		\end{subfigure}	
		\caption{The position of the first peak of DC704 as a function of $\rho$, (a) and (b), and as a function of $\Gamma$, (c) and (d). In (b) and (d) the peak position is plotted in dimensionless units $\tilde{q}_{max} = $ q$_{max} \rho^{-\frac{1}{3}}$.}
		\label{fig:DC704_1st_fit}
		
	\end{figure}

	In Figure \ref{fig:DC704_1st_fit} the results of the fitting for the first peak is shown as a function of both $\rho$, figure \ref{fig:DC704_1st_fit_a} and \ref{fig:DC704_1st_fit_b}, and as a function of $\Gamma$, figure \ref{fig:DC704_1st_fit_c} and \ref{fig:DC704_1st_fit_d}. In Figure \ref{fig:DC704_1st_fit_a} and \ref{fig:DC704_1st_fit_c}, the peak position, q$_{max}$, is shown in Å$^{-1}$, while in \ref{fig:DC704_1st_fit_b} and \ref{fig:DC704_1st_fit_d}  q$_{max}$ is plotted in reduced units, $\tilde{q}_{max} = $ q$_{max} \rho^{-\frac{1}{3}}$. The first peak position seems to collapse very well as a function of density, and even when scaling out the effect of density it does not collapse as a function of $\Gamma$. 
	
\begin{figure}[H]
		\begin{subfigure}[b]{0.49\textwidth}
			\centering
			\includegraphics[width=0.9\textwidth]{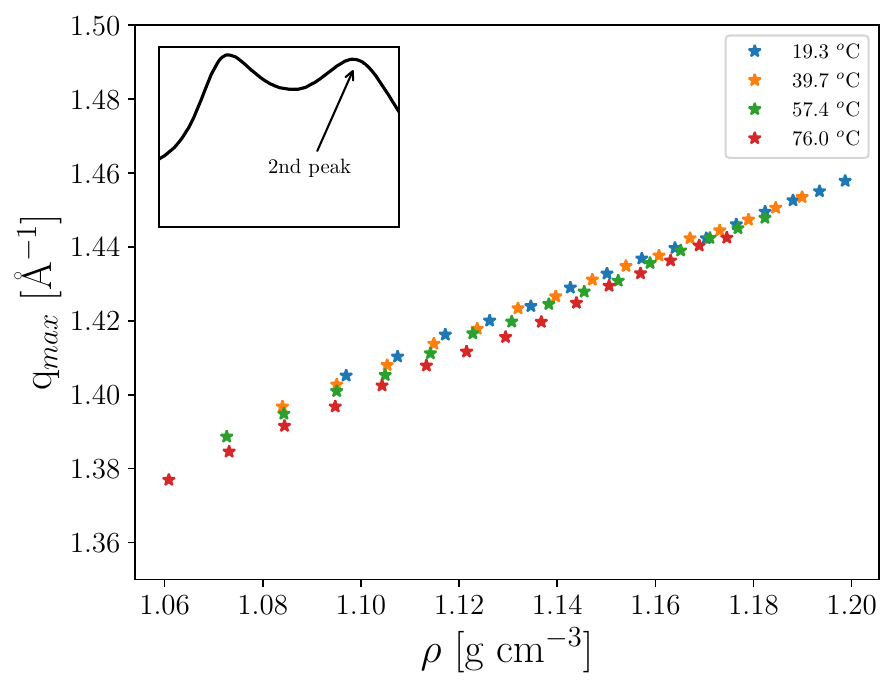}
			\caption{q$_{max}$ for the 2nd peak as a function of $\rho$}		
						\label{fig:DC704_2nd_fit_a}
		\end{subfigure}
		\begin{subfigure}[b]{0.49\textwidth}
			\centering
			\includegraphics[width=0.9\textwidth]{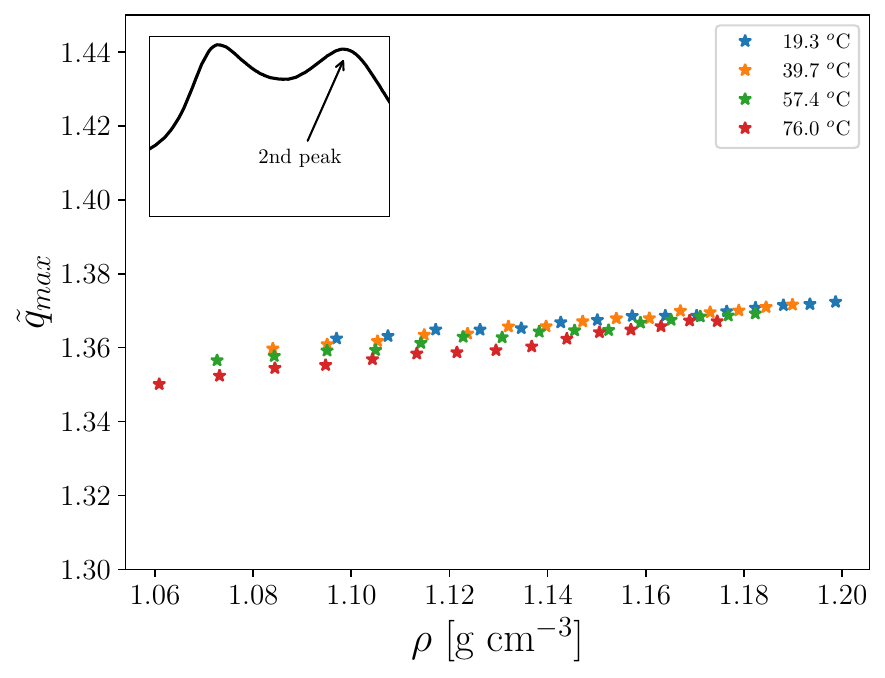}
			\caption{$\tilde{q}_{max}$ for the 2nd peak as a function of $\rho$  }
						\label{fig:DC704_2nd_fit_b}
		\end{subfigure}
		
		\begin{subfigure}[b]{0.49\textwidth}
			\centering
			\includegraphics[width=0.9\textwidth]{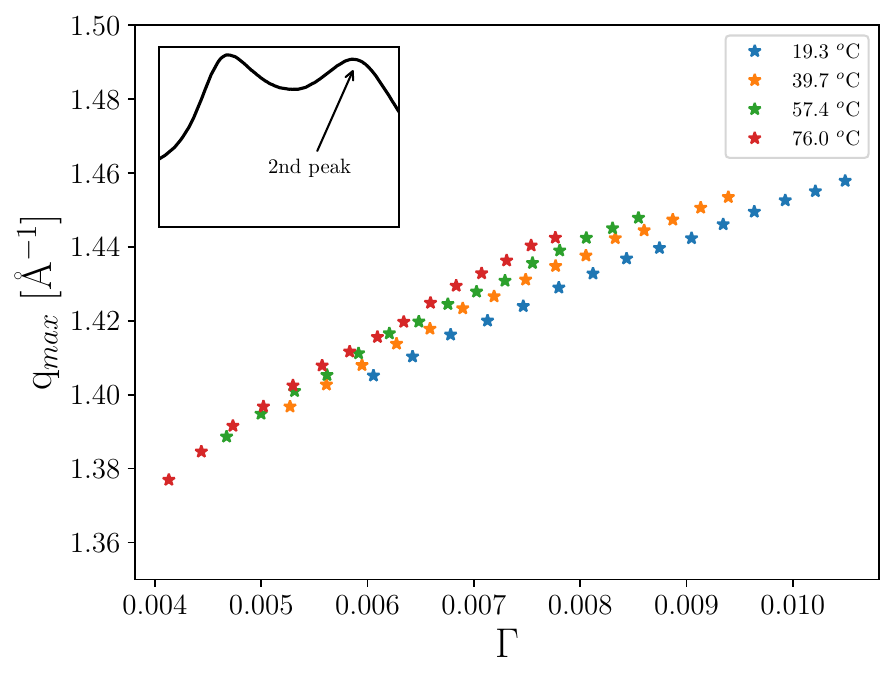}
			\caption{q$_{max}$ for the 2nd peak as a function of $\Gamma$}	
						\label{fig:DC704_2nd_fit_c}	
		\end{subfigure}
		\begin{subfigure}[b]{0.49\textwidth}
			\centering
			\includegraphics[width=0.9\textwidth]{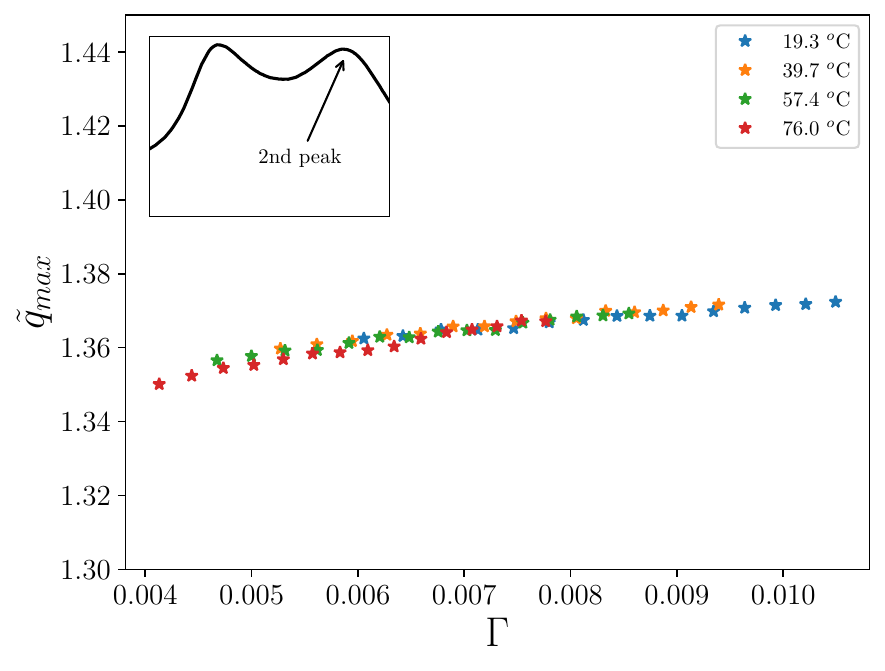}
			\caption{$\tilde{q}_{max}$ for the 2nd peak as a function of $\Gamma$  }
			\label{fig:DC704_2nd_fit_d}
		\end{subfigure}	
		\caption{The position of the second peak of DC704, plotted as a function of $\rho$, (a) and (b), and as a function of $\Gamma$, (c) and (d). In (b) and (d) the peak position is plotted in dimensionless units $\tilde{q}_{max} = $ q$_{max} \rho^{-\frac{1}{3}}$.}
		\label{fig:DC704_2nd_fit}
		
	\end{figure}

	The fitting for the second peak is shown Figure \ref{fig:DC704_2nd_fit}. In Figure \ref{fig:DC704_2nd_fit} the second peak is shown as a function of both $\rho$, \ref{fig:DC704_2nd_fit_a} and \ref{fig:DC704_2nd_fit_b}, and as a function of $\Gamma$, \ref{fig:DC704_2nd_fit_c} and \ref{fig:DC704_2nd_fit_d}. Surprisingly, the two peaks do not appear to behave similarly when compared in reduced units. The second peak collapses better as a function of $\Gamma$,  Figure \ref{fig:DC704_2nd_fit_d}, than as a function of $\rho$, Figure  \ref{fig:DC704_2nd_fit_b}. 
	

	If DC704 has pseudo-isomorph, we expect that intermolecular structure in reduced units to be invariant along lines of constant $\Gamma$ in the phase diagram. We do not expect that the intramolecular structure will change with density. The intramolecular contributions span the entire q-space. There will be a scaling deviation from using the reduced units and that deviation would be present in the whole of q-space. This is illustrated in further detail in section \ref{sec:strucral changes}. 
	
	Depending on the place in q-space and the phase diagram, the size of the deviation will be different, but it will have an influence on $S(q)$ in the whole q-range. For the UA-Cumene model we observed that while  $S_{\text{inter}}(q)$ collapses well along an isochrone, the influence of intramolecular contributions cause the total $S(q)$ to not collapse perfectly, see figure \ref{fig:Cumene_MD_sq}. For DC704, we do not have the full set of MD simulations, but I have been permitted to share some work in progress data for DC704 \cite{DC704MD}. \Mycite{DC704MD} has a single state point simulated using a UA-model for DC704, and separated the inter-molecular/intra-molecular contributions to the total $S(q)$. 
	
\begin{figure}[H]
		\begin{subfigure}[b]{0.49\textwidth}
			\centering
			\includegraphics[trim = 30mm 80mm 40mm 80mm, clip=true,width=0.99\textwidth]{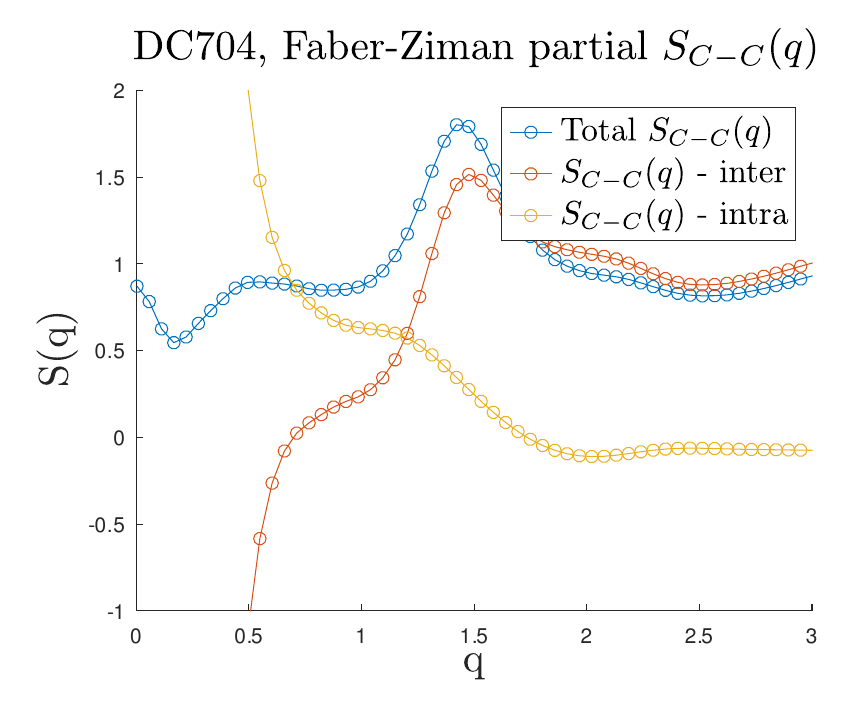}
			
			\caption{}
			\label{fig:DC704_MD_a}
		\end{subfigure}	
		\begin{subfigure}[b]{0.49\textwidth}
			\centering
			\includegraphics[trim = 30mm 80mm 40mm 80mm, clip=true,width=0.99\textwidth]{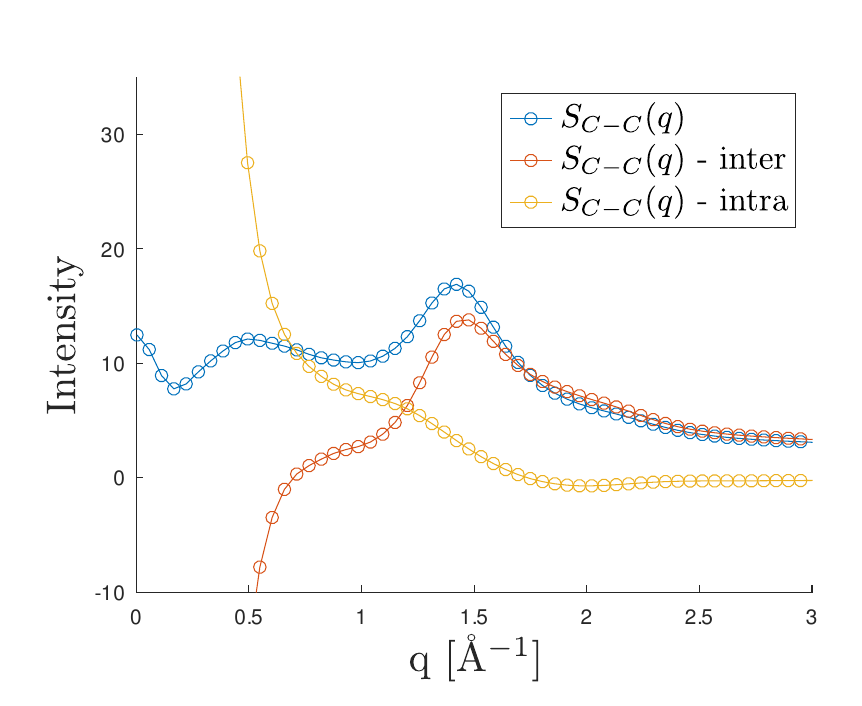}
			\caption{ }
			\label{fig:DC704_MD_b}
			
		\end{subfigure}	
		\caption{Work in progress MD simulation of DC704 \cite{DC704MD}, showing the Faber-Ziman $C-C$ partial structure factor. Figure \ref{fig:DC704_MD_b} show the Faber-Ziman $C-C$ partial structure factor in an approximate intensity calculated from equation \ref{eq:DC704 intensity}.  }
		\label{fig:DC704_MD}
		
	\end{figure}

	In figure \ref{fig:DC704_MD_a} the Faber-Ziman partial structure factor for C-C contributions is shown. It is calculated by Fourier transformation of C-C contributions to $g(r)$. We observe that the intramolecular structure has a contribution, especially on the first peak In figure \ref{fig:DC704_MD_b} $S(q)$ is converted into approximate intensity by the definition of $S(q)$, equation \ref{eq:sq_def}:
	
	\begin{equation}\label{eq:DC704 intensity}
		I_{C-C}(q) \propto S_{C-C}(q)*\left< f_c(q)\right>^2 \quad 
	\end{equation}
	
	where $f_c(q)$ is the x-ray form factor of atom species carbon, taken from \Mycite{Waasmaier1995}. From this, it seems that the first peak in $S(q)$ is a shoulder, that the q-dependent form factors expand into the double peak we measure.

	So why does the second peak collapse as a function of $\Gamma$, and the first peak as a function of $\rho$?
	In the last chapter, section \ref{sec:strucral changes}, we discussed the structural changes seen in $S(\tilde{q})$ when moving around the phase diagram. When we measure structure along an isochore in dimensionless units, the changes we measure in $S(\tilde{q})$ are the intermolecular structural changes. When we measure along an isochrone, the changes we measure in $S(\tilde{q})$ originate from the density dependent scaling deviation to the intramolecular structure. It is clear from figure \ref{fig:DC704_MD_b}  that intramolecular contributions to the structure factor have a larger affect on the first peak than the second. When we compare the structure along isochores, we are measuring a purely intermolecular change, and when we compare the structure along isochrones the structural change is the scaling deviation. The first peak of DC704 seems much more sensitive to the effect of the scaling deviation, than the any intermolecular structural change. This would explain why the first peak being invariant along an isochore, instead of isochrones. For the second peak the picture seem to the opposite. The second peak seems barely affected by the scaling deviation, while the intermolecular changes are more pronounced in that $q$-range. This would explain why we see a collapse along isochrones of the second peak, but not the first.


	\subsection{DPG}
\begin{figure}[H]
		\begin{subfigure}[b]{0.49\textwidth}
			\centering
			\includegraphics[width=0.9\textwidth]{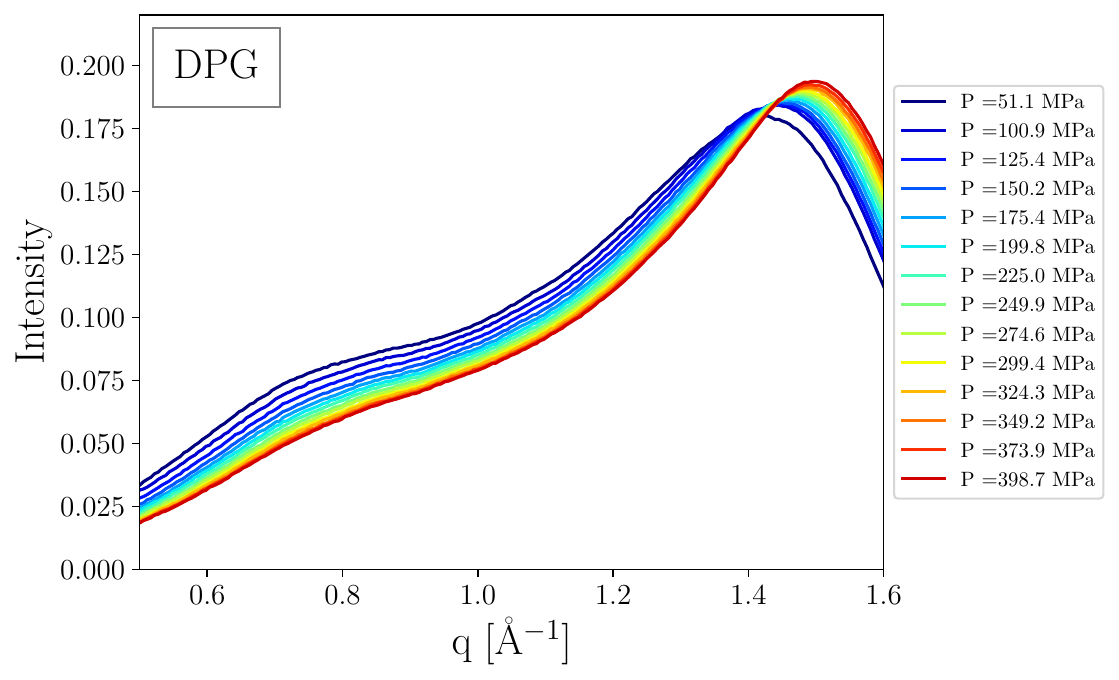}
			\caption{T = 19.3 $^o$C}		
					\label{fig:DPG_data_a}
		\end{subfigure}
		\begin{subfigure}[b]{0.49\textwidth}
			\centering			\includegraphics[width=0.9\textwidth]{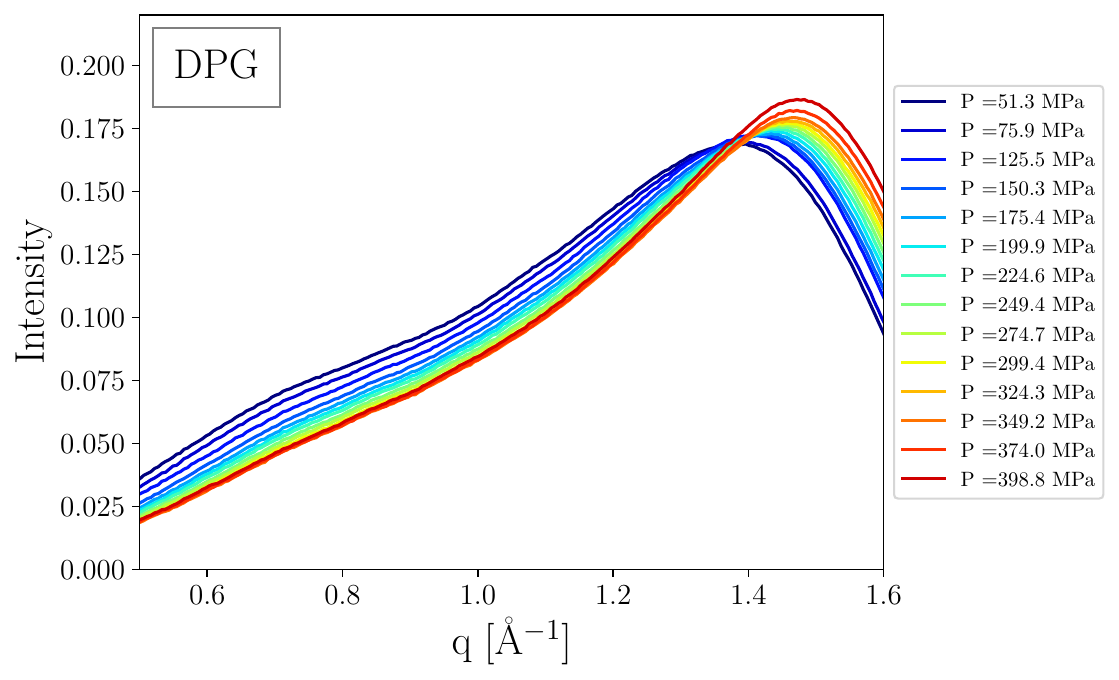}
			\caption{T = 56.9 $^o$C}		
					\label{fig:DPG_data_b}
		\end{subfigure}	
		\begin{subfigure}[b]{0.49\textwidth}
			\centering
			\includegraphics[width=0.9\textwidth]{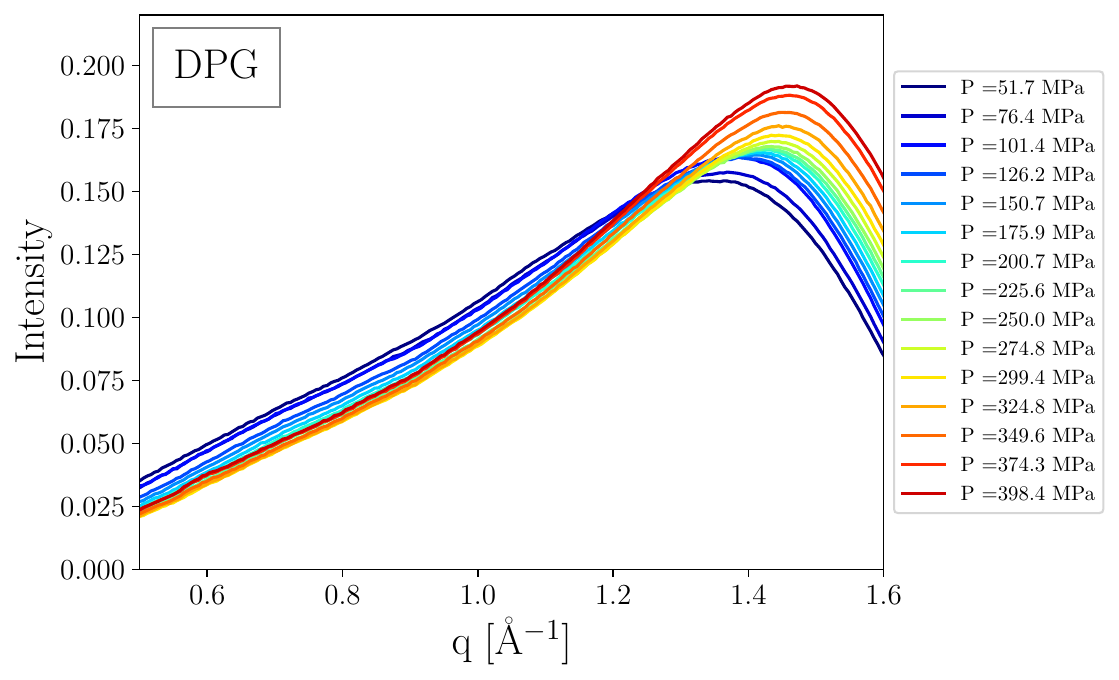}
			\caption{T = 76.1 $^o$C}		
					\label{fig:DPG_data_c}
		\end{subfigure}
		\begin{subfigure}[b]{0.49\textwidth}
			\centering
			\includegraphics[width=0.9\textwidth]{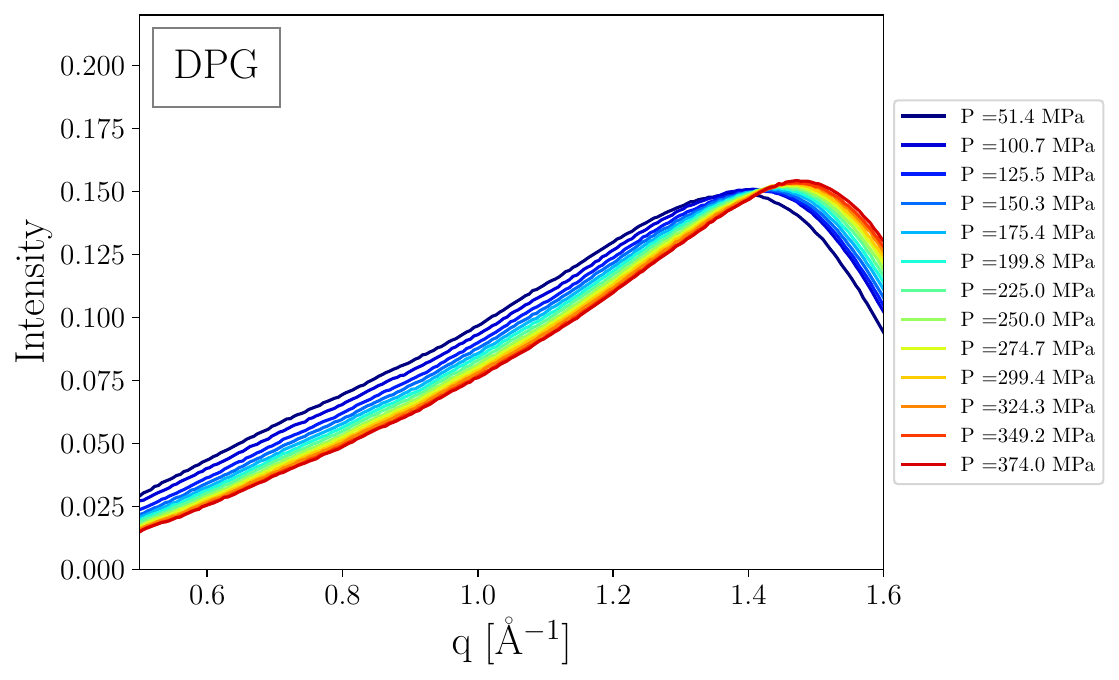}
			\caption{T = 38.7 $^o$C}
					\label{fig:DPG_data_d}
		\end{subfigure}
		\caption{DPG measurements along the four isotherms in chronological order: 19.3 $^o$C $\rightarrow$56.9 $^o$C $\rightarrow$ 76.1 $^o$C $\rightarrow$ 38.7 $^o$C. For the measurement at 38.7 $^o$C there is a loss in intensity due to a leak in the capillary. }
		\label{fig:DPG_data}
	\end{figure}

	The measured intensities along four isotherms for DPG are shown in Figure \ref{fig:DPG_data}, in a q range from 0.5-1.6  Å$^{-1}$. For the 38.7 $^o$C measurement there is clearly a loss in intensity for the peak. The shoulder  around 0.8 Å$^{-1}$, is also more pronounced in the 56.9 $^o$C measurement.  This can be seen by comparing the peak intensity, in figure \ref{fig:DPG_data_d} with figure \ref{fig:DPG_data_b} and \ref{fig:DPG_data_a}.  After the measurement we observed that the capillary was broken during the measurements. Due to the sudden drop in intensity that occurred during the 38.7 $^o$C measurement series, we believe that it broke during those measurements. The 38.7 $^o$C measurement series was excluded from the analysis.

	The DPG peak shape has a main peak around 1.5 Å$^{-1}$, with a shoulder around 0.8 Å$^{-1}$. The shoulder is characteristic of some hydrogen-bonded liquids, where hydrogen bonds can cause clustering on longer length scales than nearest-neighbor distances \cite{Bolle2020}. The shoulder does not disappear, but it can be suppressed with increasing pressure and temperature. 
	
	\subsubsection{Peak analysis}
	
\begin{figure}[H]
		\begin{subfigure}[b]{0.49\textwidth}
			\centering
			\includegraphics[width=0.9\textwidth]{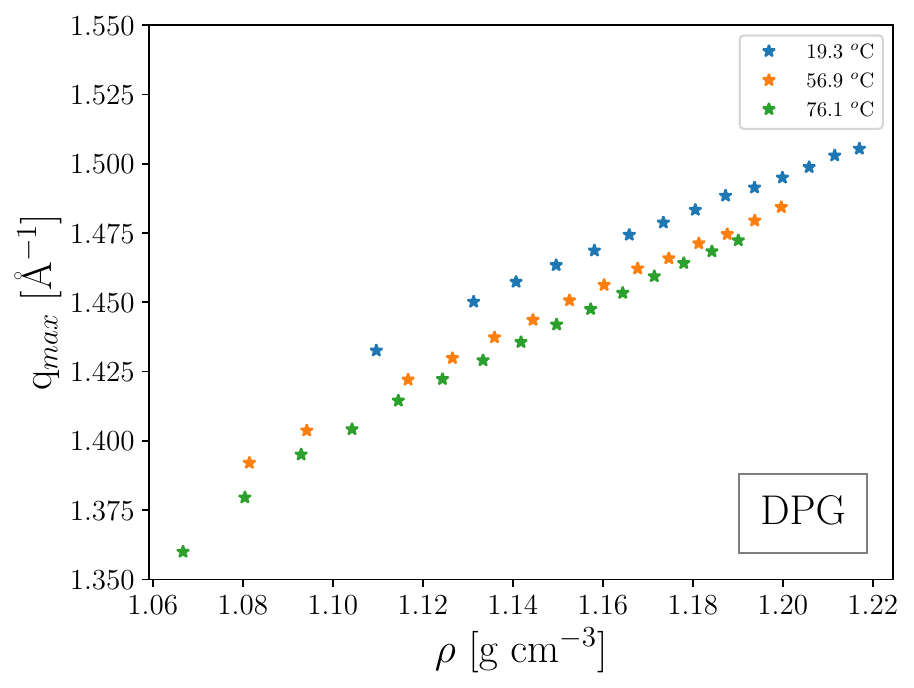}
			\caption{}		
		\end{subfigure}
		\begin{subfigure}[b]{0.49\textwidth}
			\centering
			\includegraphics[width=0.9\textwidth]{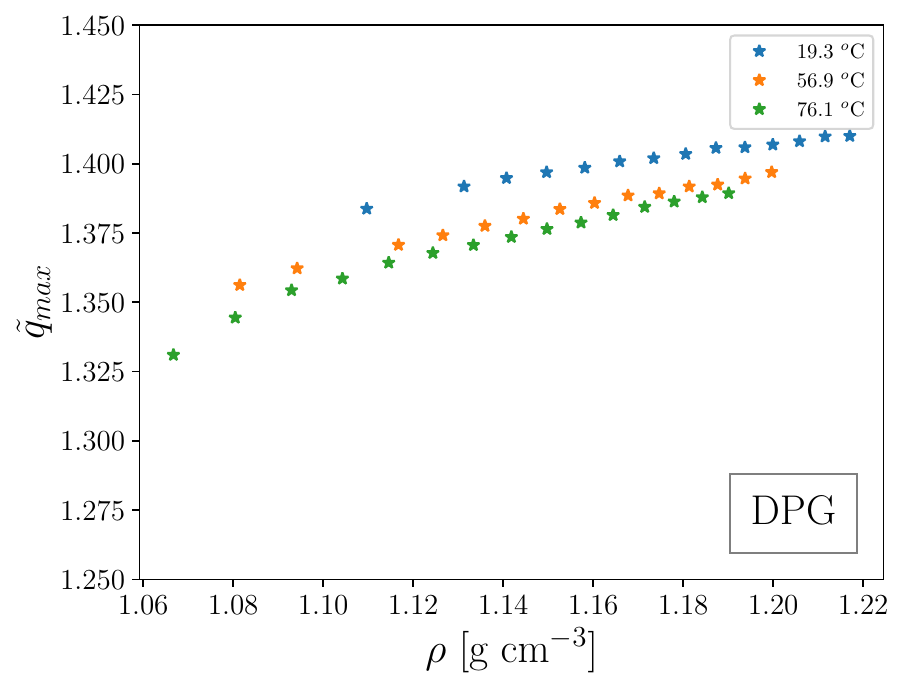}
			\caption{ }
		\end{subfigure}
		
		\begin{subfigure}[b]{0.49\textwidth}
			\centering
			\includegraphics[width=0.9\textwidth]{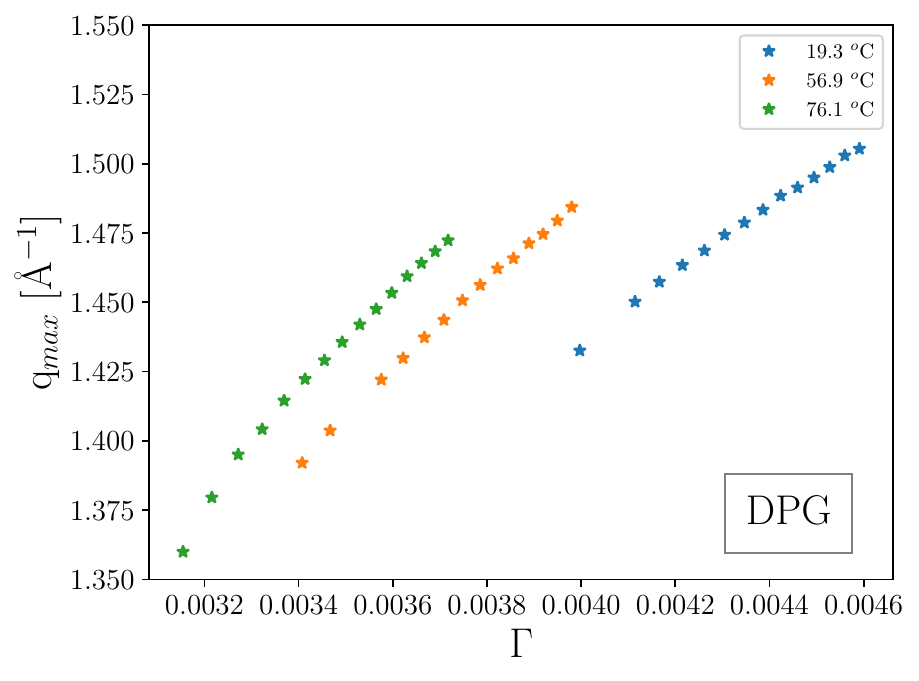}
			\caption{}		
		\end{subfigure}
		\begin{subfigure}[b]{0.49\textwidth}
			\centering
			\includegraphics[width=0.9\textwidth]{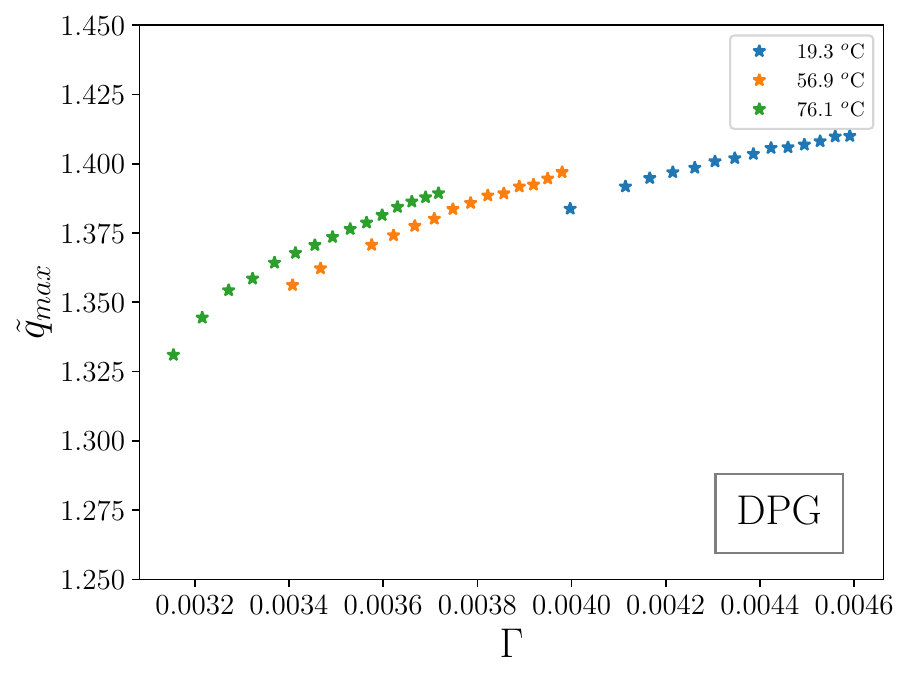}
			\caption{ }
			\label{fig:DPG_fit_d}
		\end{subfigure}	
		\caption{The peak position of the main structure peak of DPG as a function of both $\rho$, (a) and (b), and as a function of $\Gamma$, (c) and (d). For the (c) and (d) plots $\gamma =1.5$ is used as density scaling coefficient. In (b) and (d) the peak is plotted in dimensionless units $\tilde{q}_{max} = $ q$_{max} \rho^{-\frac{1}{3}}$. }
		\label{fig:DPG_fit}
	\end{figure}

	In figure \ref{fig:DPG_fit} the results of the fitting are shown. The sample DPG was chosen as a counter-example, because we did not expect DPG to have pseudo-isomorphs.  In figure \ref{fig:DPG_fit} (b), when scaling out the effect of density, the structure does not collapse. For DPG the structure does not collapse along lines of constant density or relaxation time, while for van der Waals bonded liquids both lines give a much better collapse, see figure \ref{fig:fitting_2nd_peak_all_schemes}. 
	
\begin{figure}[H]
		\centering
		\begin{subfigure}[b]{0.49\textwidth}
			\includegraphics[width=0.9\textwidth]{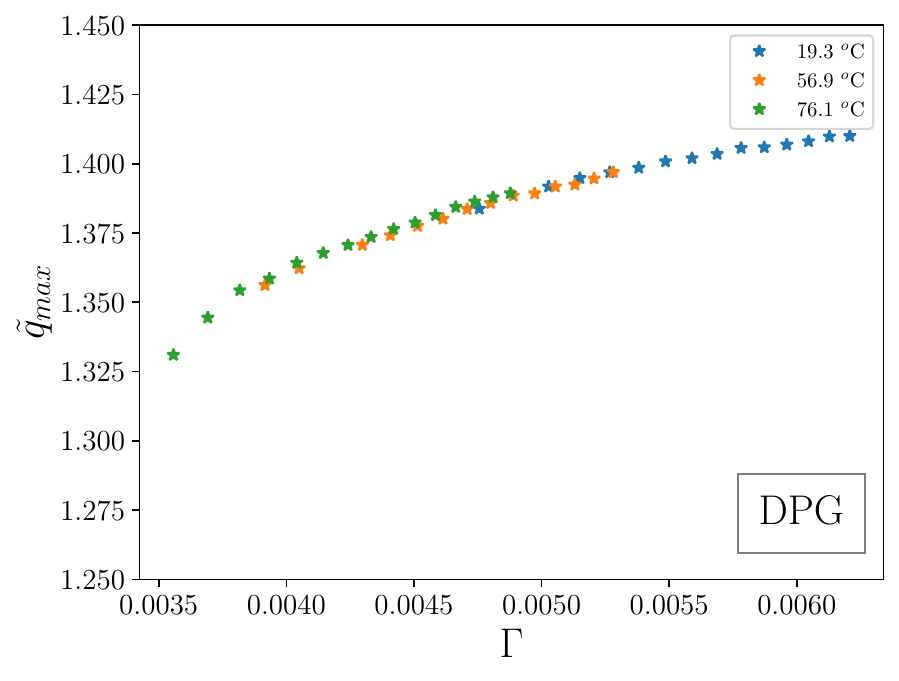}
			\caption{}
		\end{subfigure}	
		\begin{subfigure}[b]{0.49\textwidth}
			\includegraphics[width=0.9\textwidth]{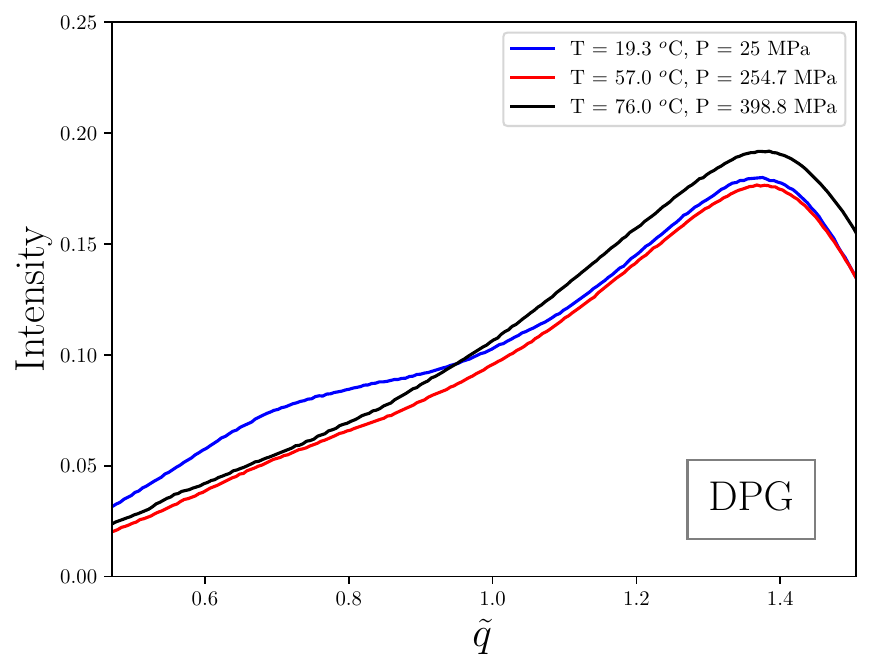}
			\caption{}
		\end{subfigure}	
		\caption{It is possible to make the peak position of the main structure peak of DPG collapse as a function of $\Gamma$ using $\gamma = 2.9$. Previous studies reported values of the density scaling coefficient $\gamma=1.5$, $1.9$ \cite{ChatKatarzyna2019Tdsi,Wase2018}. It is however only the peak position that collapses, the intensities and peak shapes are very different when $\Gamma = \frac{\rho^{2.9}}{T}$ is constant.}
		
		\label{fig:DPG_fit_29}
	\end{figure}

	DPG is a hydrogen-bonded liquid, and there has been some debate on whether density scaling works for DPG,  and if it does, in what pressure range. The assumption that constant $\Gamma$ implies constant relaxation time is therefore not necessarily true. In Figure \ref{fig:DPG_fit_29} it is shown that choosing a $\gamma= 2.9 $ can cause the peak position to collapse. Previous suggested values for the density scaling coefficient,  $\gamma= 1.5$, $1.9$,\cite{Wase2018} \cite{ChatKatarzyna2019Tdsi}, are relatively far away from  $\gamma = 2.9$. This should only be seen as a measure of how far the peak position are from a collapse. It is also worth noting that is only the peak position that collapses, the intensities and peak shapes are different.
	
	\section[5PPE \& Squalane]{Preliminary results of 5PPE and Squalane}
	In section \ref{sec:5ppe_background} the effect of fluctuations in intensity was discussed. The fluctuations in intensity were most likely due to leaks in the detector side window, and as shown in figure \ref{fig:5PPE_isobar_normalized} there is a clear change in the shape of the peak before and after the leaks. Therefore, the results of the data treatment in this section are only preliminary. 
	\subsection{5PPE}
	
\begin{figure}[H]
		\begin{subfigure}[b]{0.49\textwidth}
			\centering
			\includegraphics[width=0.9\textwidth]{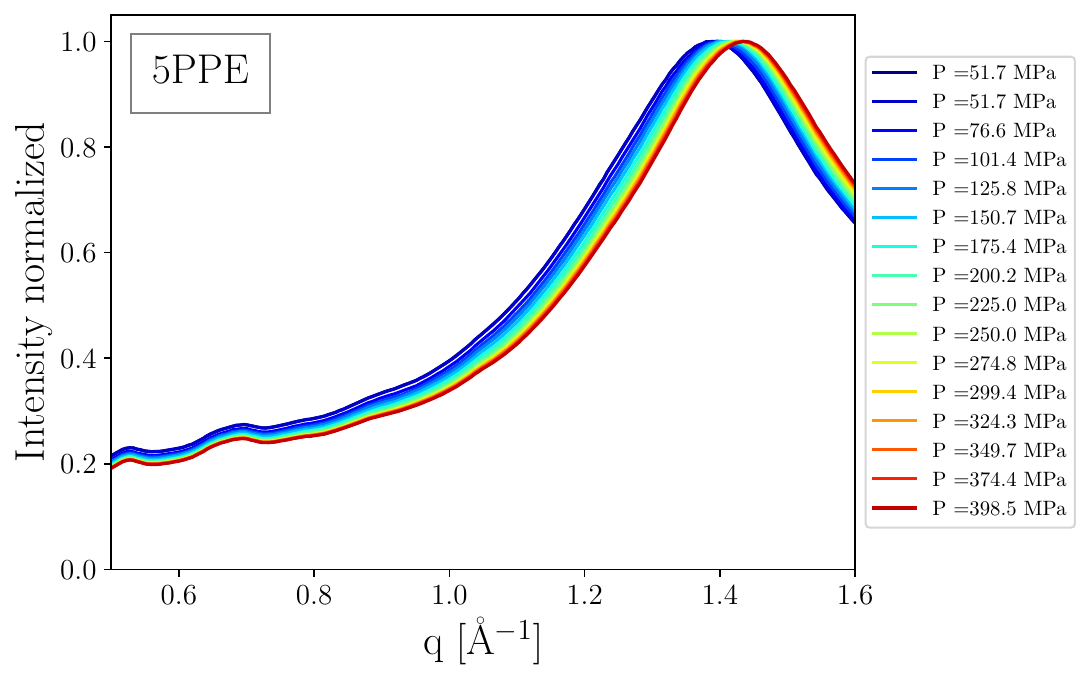}
			\caption{T = 19.3 $^o$C}		
		\end{subfigure}
		\begin{subfigure}[b]{0.49\textwidth}
			\centering
			\includegraphics[width=0.9\textwidth]{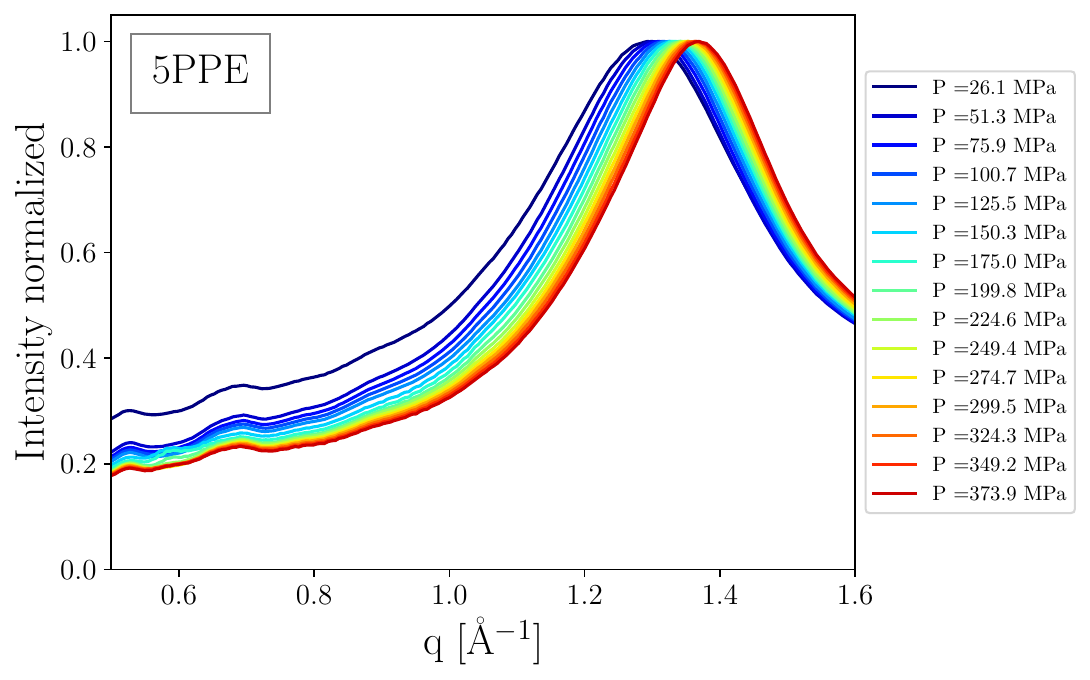}
			\caption{T = 56.9 $^o$C}		
		\end{subfigure}	
		\begin{subfigure}[b]{0.49\textwidth}
			\centering
			\includegraphics[width=0.9\textwidth]{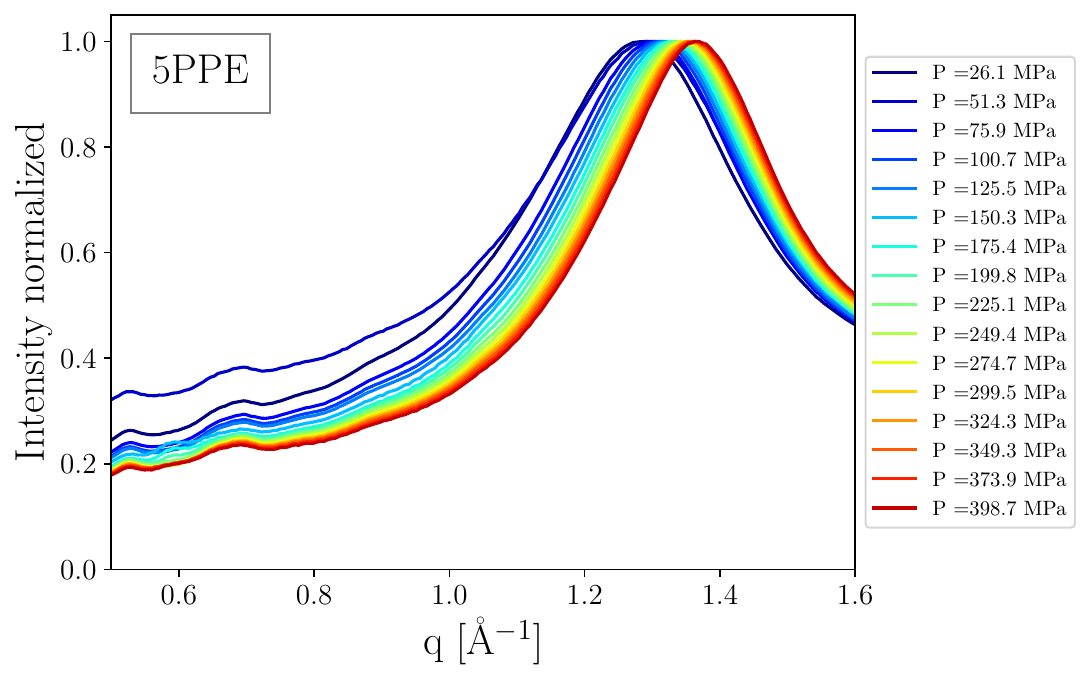}
			\caption{T = 76.0 $^o$C}		
		\end{subfigure}
		\begin{subfigure}[b]{0.49\textwidth}
			\centering
			\includegraphics[width=0.9\textwidth]{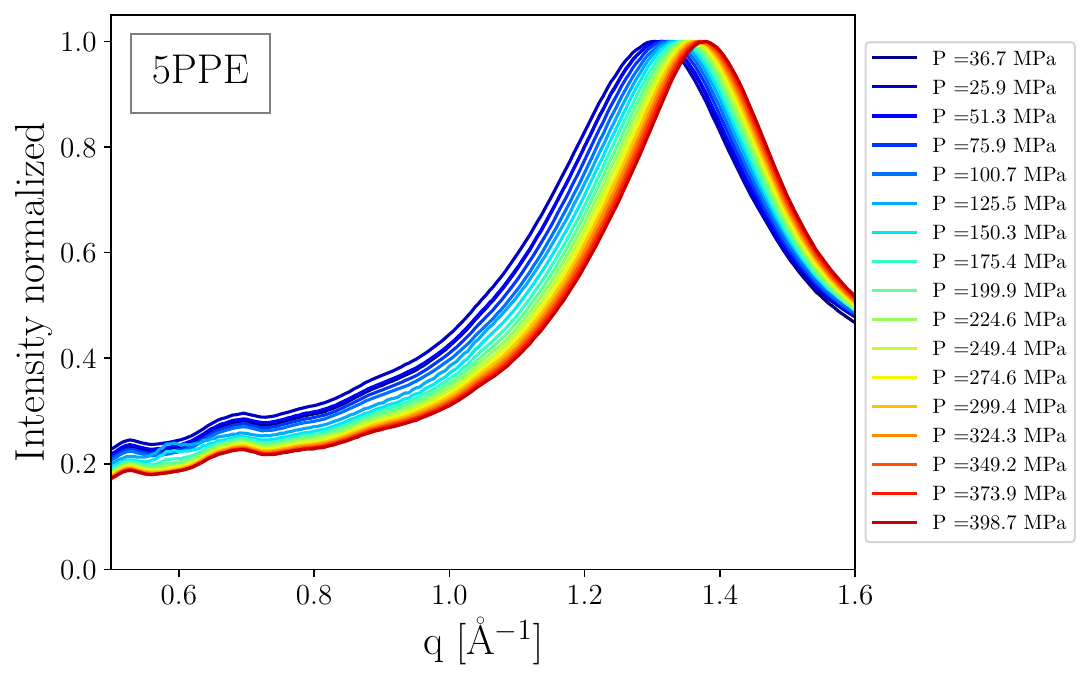}
			\caption{T = 38.3 $^o$C}
		\end{subfigure}
		\caption{The measurements along the four isotherms normalized to peak intensity. The chronological order of the measurements: 19.3 $^o$C $\rightarrow$ 56.9 $^o$C $\rightarrow$ 76.0 $^o$C $\rightarrow$ 38.3 $^o$C. }
		\label{fig:5PPE_data}
	\end{figure}

	The normalized intensities along four isotherms are shown for 5PPE in figure \ref{fig:5PPE_data}, in a q-range from 0.5-1.6  Å$^{-1}$. As discussed in the section \ref{sec:background}, between the 19.3 $^o$C measurement and the 56.9 $^o$C measurement, water from leaks in the detector side window caused jumps in the intensity. It is clear when comparing Figure \ref{fig:5PPE_data}(a) with \ref{fig:5PPE_data}(b)-(d)f that we are measuring something more than just 5PPE in the latter. Although the results for the 39, 57, and 76 $^o$C isotherms might be internally consistent, they cannot be compared with 19 $^o$C isotherm. It is also clear that the leaks affect the widths of the peaks.  
	
	The scattered intensity of 5PPE has a single peak around 1.4  Å$^{-1}$, that moves to higher q-values with increasing pressure and decreasing temperature. In the following, we only show the results of the fit for q$_{max}$. 
	
	\subsubsection{Peak analysis}
	
\begin{figure}[H]
		\begin{subfigure}[b]{0.49\textwidth}
			\centering
			\includegraphics[width=0.9\textwidth]{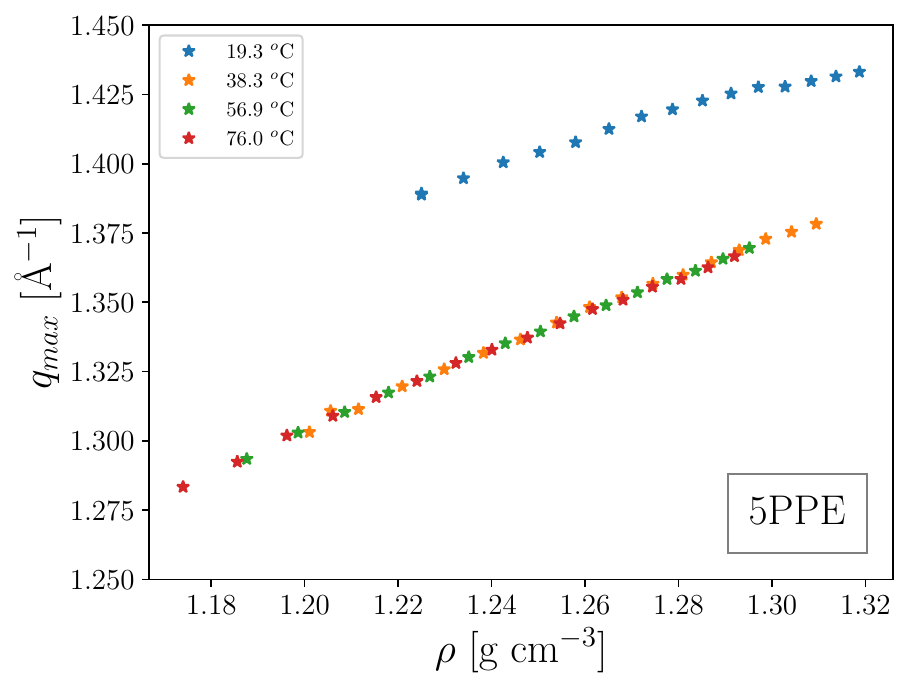}
			\caption{}		
		\end{subfigure}
		\begin{subfigure}[b]{0.49\textwidth}
			\centering
			\includegraphics[width=0.9\textwidth]{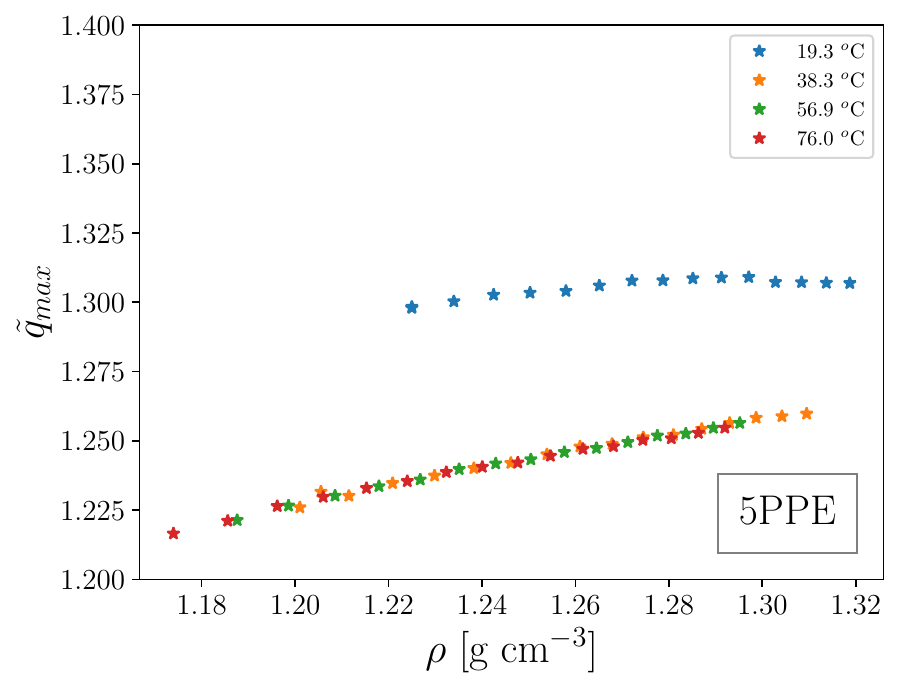}
			\caption{ }
		\end{subfigure}
		
		\begin{subfigure}[b]{0.49\textwidth}
			\centering
			\includegraphics[width=0.9\textwidth]{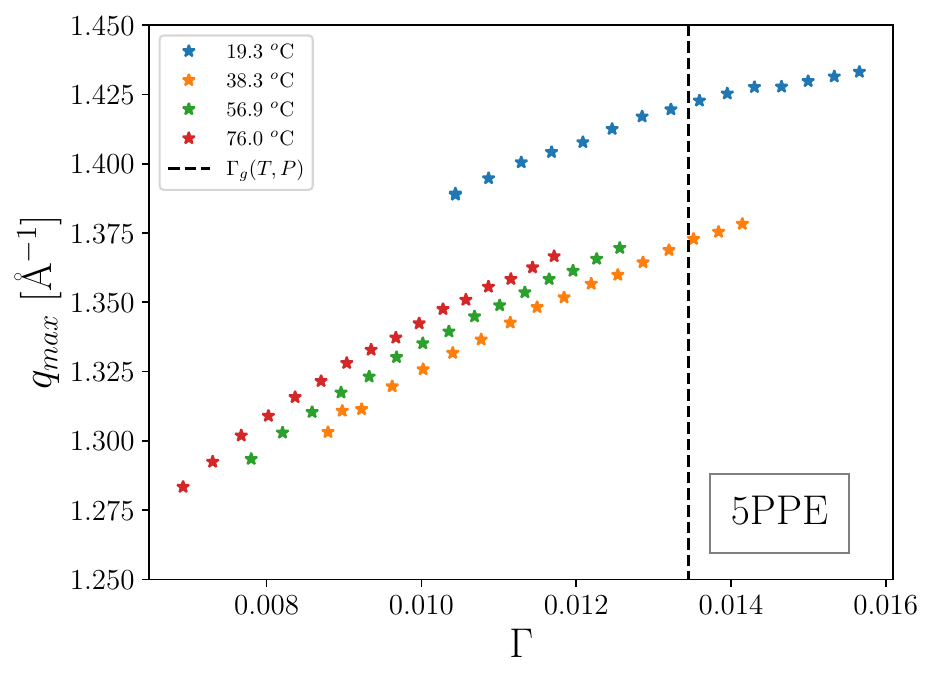}
			\caption{}		
		\end{subfigure}
		\begin{subfigure}[b]{0.49\textwidth}
			\centering
			\includegraphics[width=0.9\textwidth]{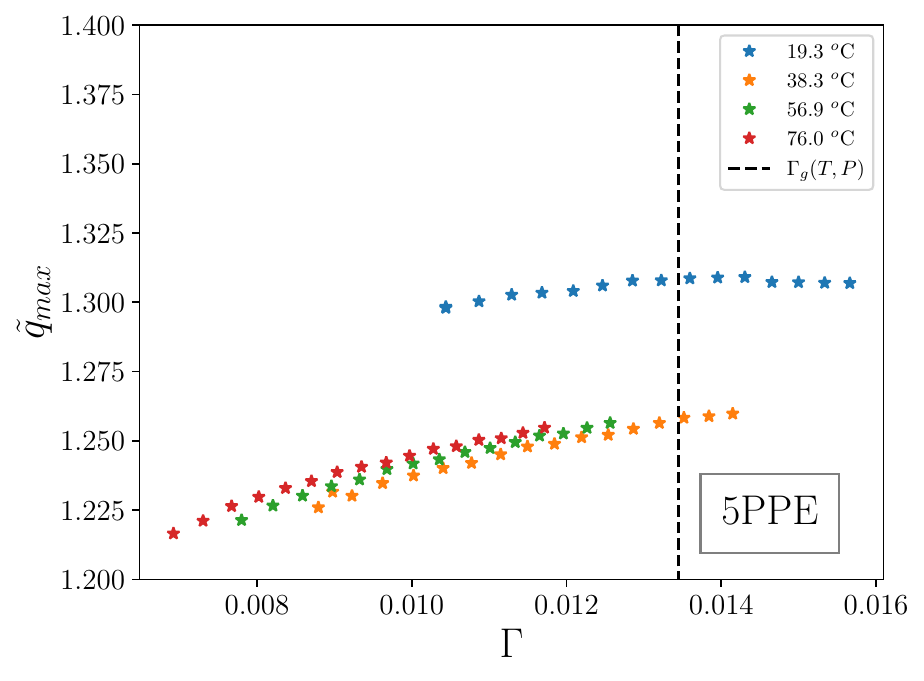}
			\caption{ }
		\end{subfigure}
		\caption{The position of the main structure peak of 5PPE, shown as a function of both $\rho$, (a) and (b), and as a function of $\Gamma$, (c) and (d). The dashed line in (c) and (d) indicates the glass transition. In (b) and (d), the peak is plotted in dimensionless units $\tilde{q}_{max} = $ q$_{max} \rho^{-\frac{1}{3}}$. The huge difference between the results of 19 $^o$C isotherm, and the other isotherms is likely due to water leakage on the detector side of the cell.}
		\label{fig:5PPE_fit}
	\end{figure}

	In Figure \ref{fig:5PPE_fit} we see the peak position of the main structure peak of 5PPE as a function of both $\rho$,(a) and (b) and, as a function of $\Gamma$, (c) and (d). As shown in Figure \ref{fig:5PPE_phasediagram}, we can enter the glass phase in some parts of the phase diagram we measure. This is relevant for about three state points in the 39 $^o$C isotherm, and seven state points in the 19 $^o$C isotherm. From a visual expectation, it seems that there is a change in slope near the literature value for the glass transition. Experimentally, it can be difficult to apply pressure evenly throughout the sample very close to the glass transition, because of the high viscosity at those state points. This is a nice confirmation that the actual pressure is close to the applied pressure.
	
	Comparing the fits of the peak positions before and after the intensity spikes, the slope in Figure \ref{fig:5PPE_fit} (a) and (b) are very different. This is probably an effect of the extra background resulting from the droplets forming on the detector side window of the pressure cell.
	
	The peak position collapses surprisingly along lines of constant density, and not it does not collapse as well as a function of $\Gamma$.
	If the structural change in the first peak was purely an effect of density, then in Figure \ref{fig:5PPE_fit} (b) and (d), the evolution of the first peak in reduced units should not change at all when moving around the phase diagram. A big part of the structural change in experimental units is from the density, but when this is scaled out there are still structural changes. This picture is close to the behavior of the 19 $^o$C isotherm when it enters the glass, however we should not trust the EoS in that area. Given the quality of the data, one should be careful not to jump to conclusions, but the nearly perfect collapse along the isochores is striking.
	
	\clearpage
	
	\subsection{Squalane}
\begin{figure}[H]
		\begin{subfigure}[b]{0.49\textwidth}
			\centering
			\includegraphics[width=0.9\textwidth]{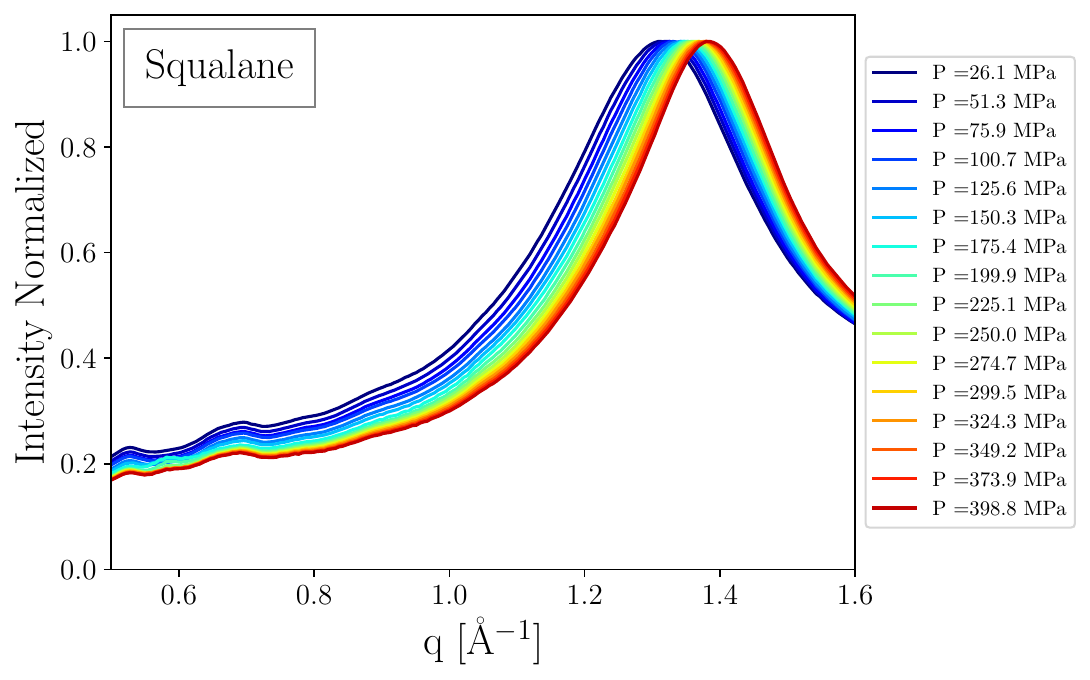}
			\caption{T = 19.2 $^o$C}		
		\end{subfigure}
		\begin{subfigure}[b]{0.49\textwidth}
			\centering
			\includegraphics[width=0.9\textwidth]{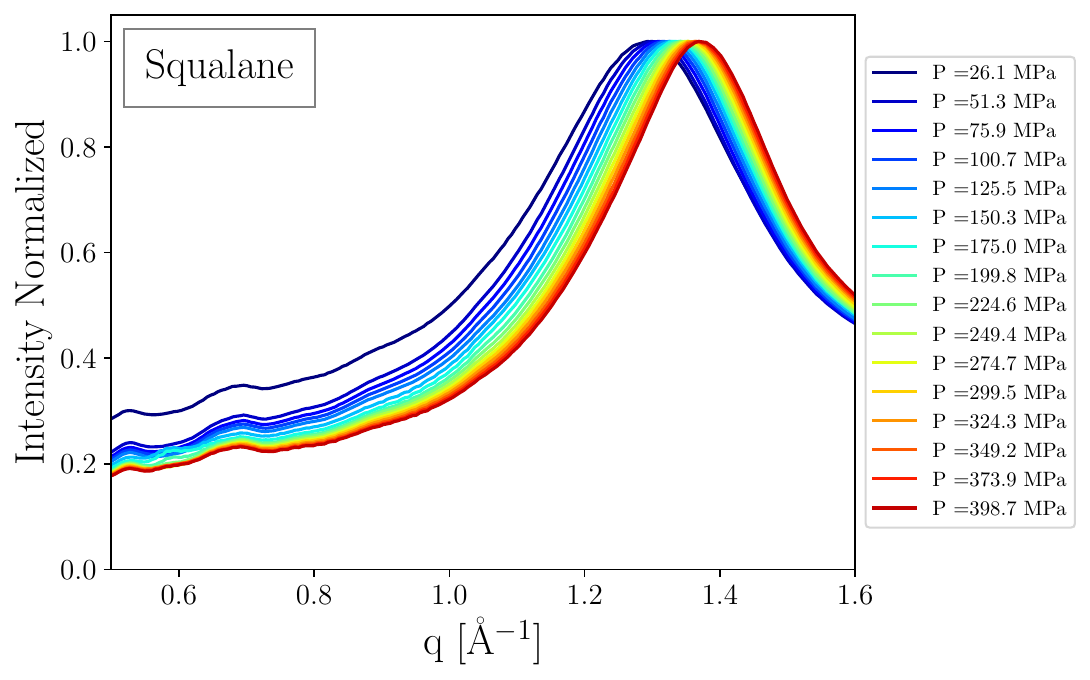}
			\caption{T = 56.9 $^o$C}		
		\end{subfigure}	
		\begin{subfigure}[b]{0.49\textwidth}
			\centering
			\includegraphics[width=0.9\textwidth]{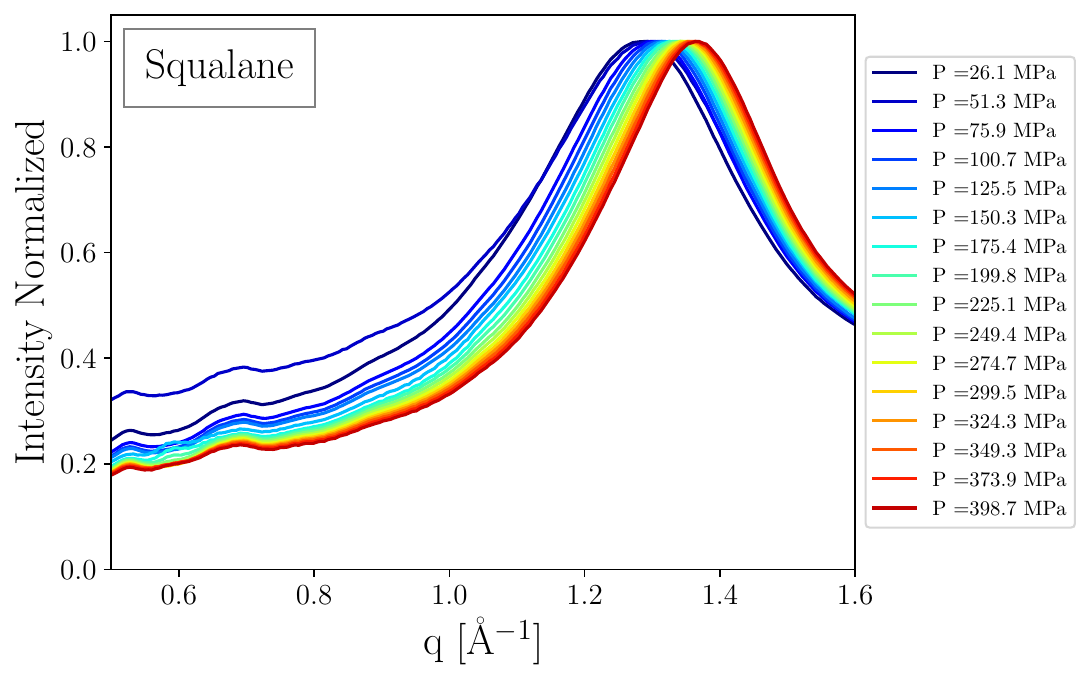}
			\caption{T = 76.0 $^o$C}		
		\end{subfigure}
		\begin{subfigure}[b]{0.49\textwidth}
			\centering
			\includegraphics[width=0.9\textwidth]{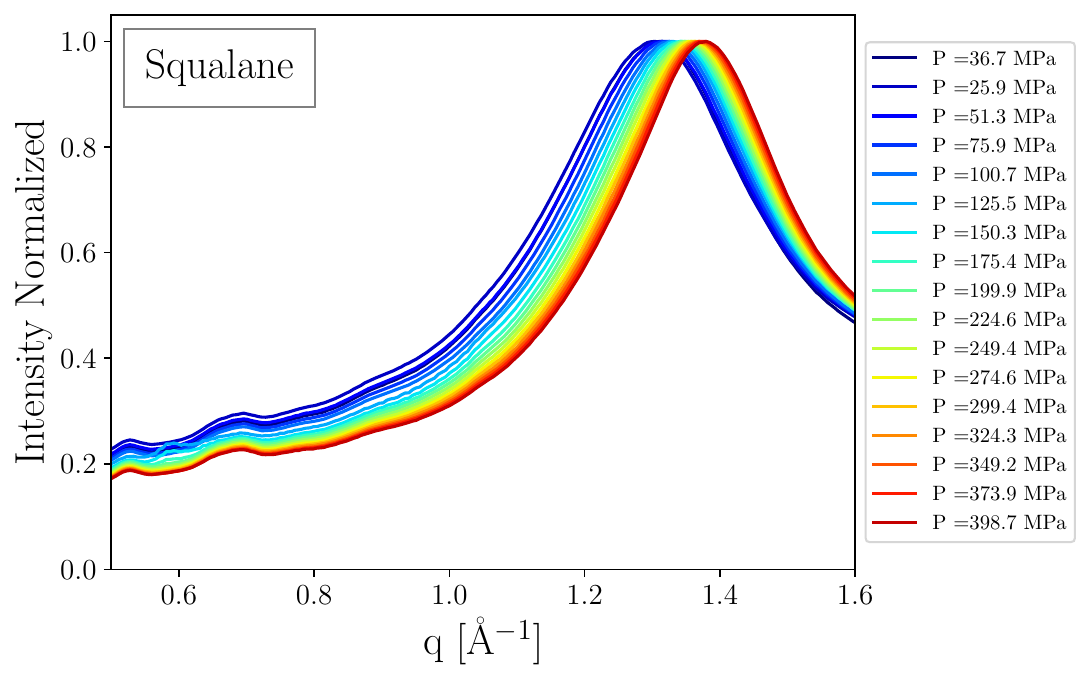}
			\caption{T = 38.3 $^o$C}
		\end{subfigure}
		\caption{The measured spectrum for squalane along four isotherms normalized to peak intensity. The chronological order of the measurements was:  19.2 $^o$C $\rightarrow$ 56.9 $^o$C $\rightarrow$ 76.0 $^o$C $\rightarrow$ 38.3 $^o$C. }
		\label{fig:Squalane_data}
	\end{figure}

	The normalized intensities along the four isotherms are shown for squalane in figure \ref{fig:Squalane_data}, in a q range from 0.5-1.6  Å$^{-1}$. The squalane data were also affected by leaks from the detector-side window of the cell; thus, the data is analyzed without any background subtraction, and with intensities normalized to the peak intensity. The measured intensities for Squalane and 5PPE look very similar. Both spectra have a single peak at approximately 1.3-1.4 Å$^{-1}$, and do not appear to evolve any extra features like shoulders or pre-peaks when moving around the phase diagram. The molecular structures of squalane and 5PPE both include long carbon chains; however, squalane has much more flexibility, given that there are no phenyl-rings. The intramolecular structure could be very different between the two liquids. Squalane has the flexibility to curl into a ball-like shape \cite{Squalane_Sim_2020}, while for 5PPE one could speculate that it would be less flexible. The fitting results are shown in Figure \ref{fig:Squalane_fit}.
	
	\subsubsection{Peak analysis}
	
\begin{figure}[H]
		\begin{subfigure}[b]{0.49\textwidth}
			\centering
			\includegraphics[width=0.9\textwidth]{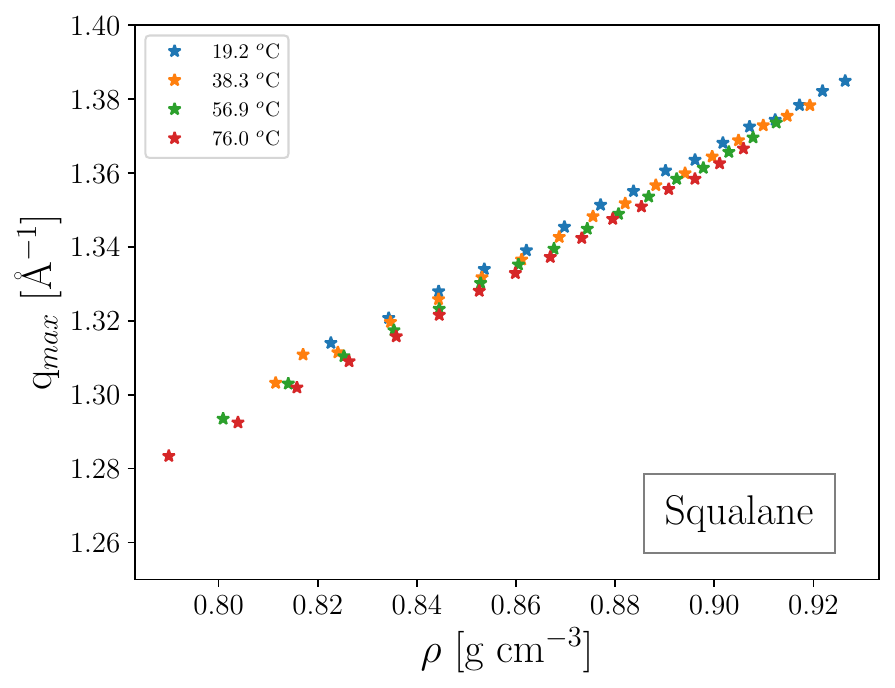}
			\caption{}		
		\end{subfigure}
		\begin{subfigure}[b]{0.49\textwidth}
			\centering
			\includegraphics[width=0.9\textwidth]{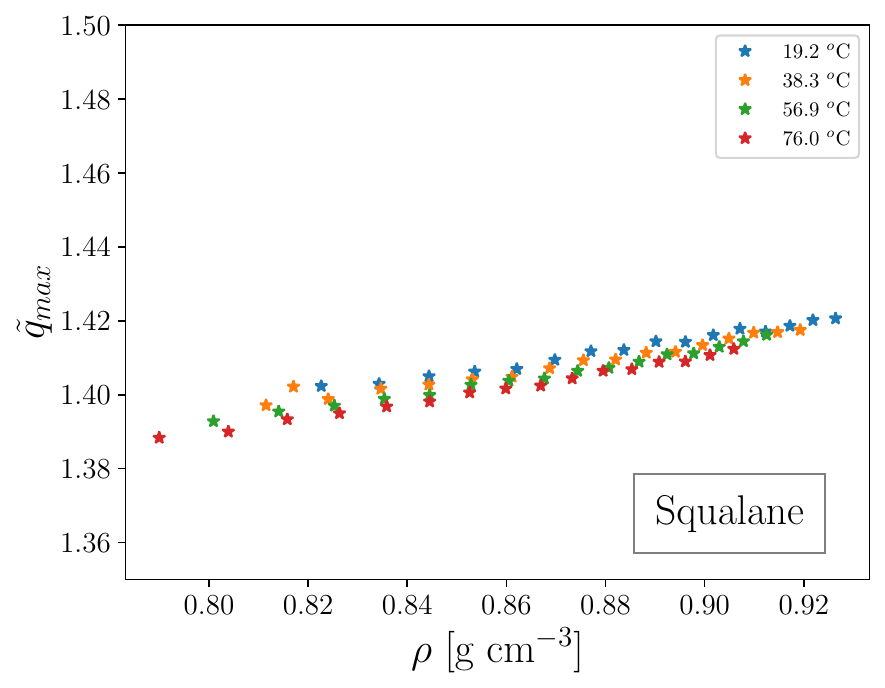}
			\caption{ }
		\end{subfigure}
		
		\begin{subfigure}[b]{0.49\textwidth}
			\centering
			\includegraphics[width=0.9\textwidth]{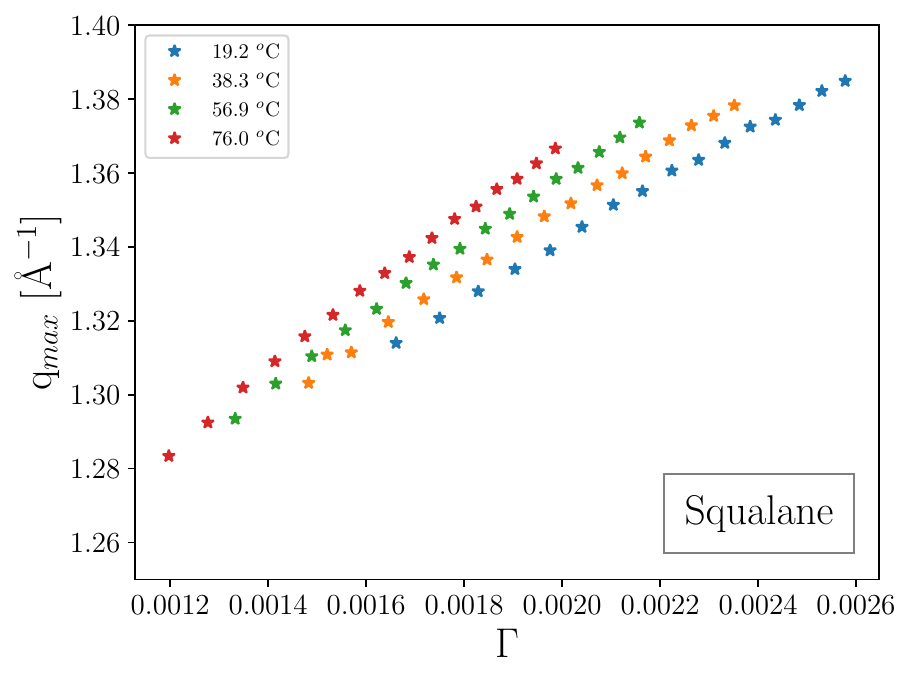}
			\caption{}		
		\end{subfigure}
		\begin{subfigure}[b]{0.49\textwidth}
			\centering
			\includegraphics[width=0.9\textwidth]{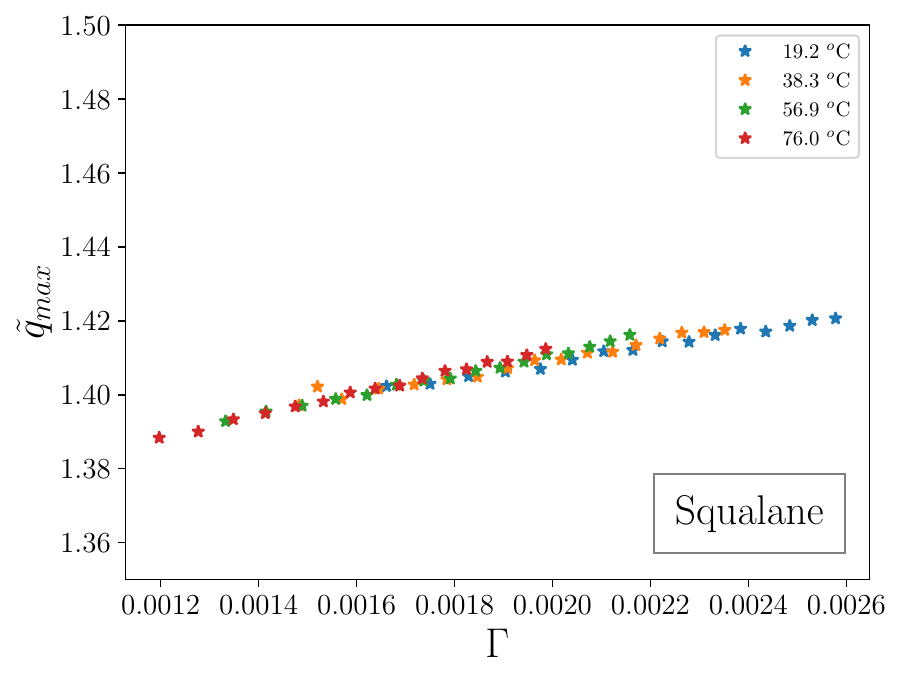}
			\caption{ }
		\end{subfigure}	
		
		\caption{The position the main structure peak of squalane plotted as a function of $\rho$, (a) and (b), and as a function of $\Gamma$, (c) and (d). In (b) and (d) the peak is plotted in dimensionless units $\tilde{q}_{max} = $ q$_{max} \rho^{-\frac{1}{3}}$.}
		\label{fig:Squalane_fit}
	\end{figure}

	In Figure \ref{fig:Squalane_fit} the peak position of the main structure peak of Squalane is plotted as a function of both $\rho$,(a) and (b) and, as a function of $\Gamma$, (c) and (d). When scaling out the effect of density, for squalane, the peak position collapses well as a function of $\Gamma$ and not as a function of $\rho$. Given the similarities between 5PPE and squalane, it is surprising that the results are different for the two liquids. They are both long carbon-dominated molecules with very similar peak shapes in the same q-range. Even though no background subtracted and the intensities have been normalized to account for the jumps in intensity, if there is something extra that would affect the measurement, it should affect 5PPE and Squalane equally.
	
	\subsubsection{5PPE and Squalane comparison along isochrones}
	
\begin{figure}[H]
		\begin{subfigure}[b]{0.49\textwidth}
			\centering
			\includegraphics[width=0.9\textwidth]{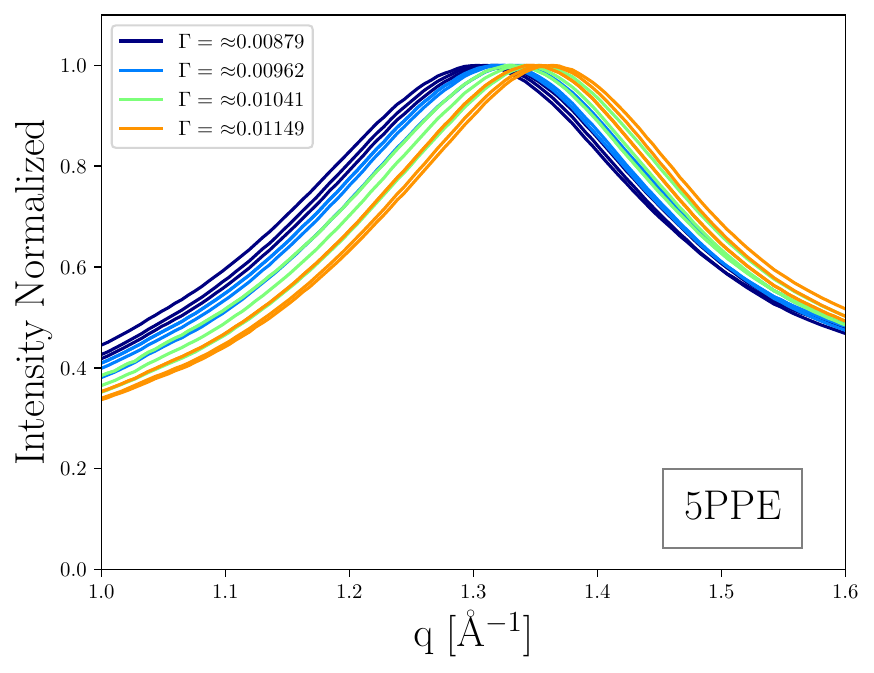}
			\caption{}		
		\end{subfigure}
		\begin{subfigure}[b]{0.49\textwidth}
			\centering
			\includegraphics[width=0.9\textwidth]{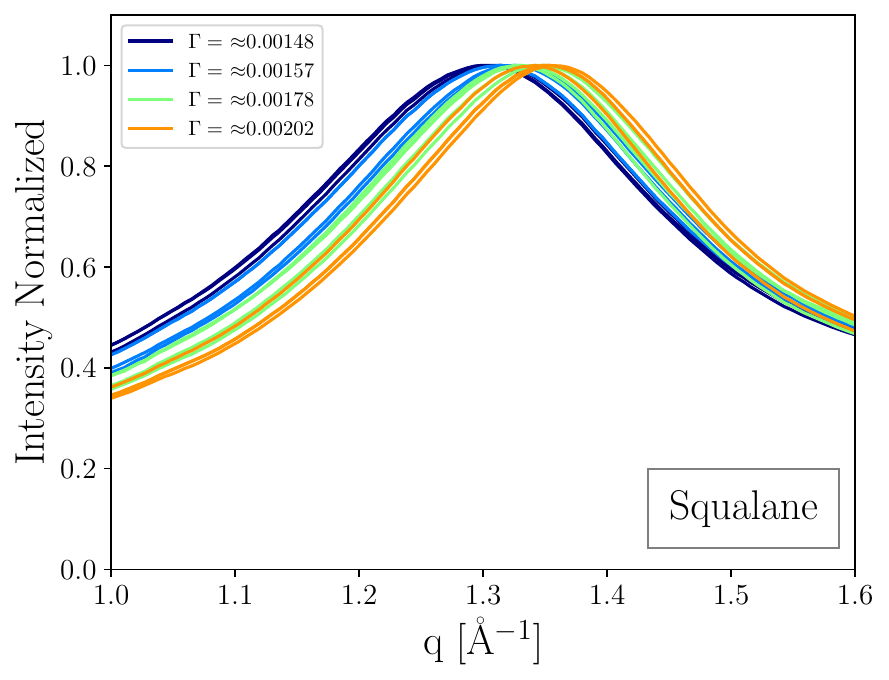}
			\caption{}		
		\end{subfigure}
		\begin{subfigure}[b]{0.49\textwidth}
			\centering
			\includegraphics[width=0.9\textwidth]{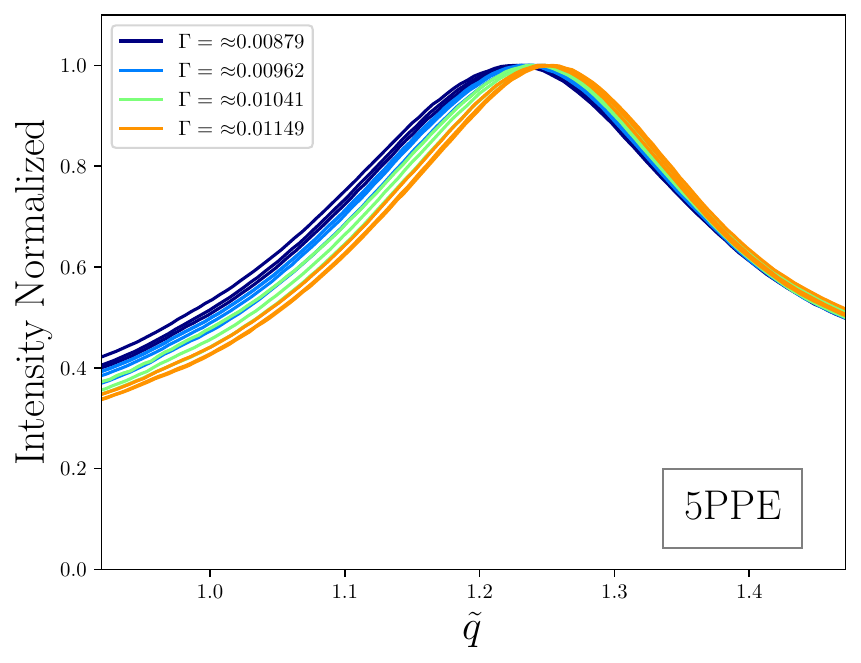}
			\caption{}		
		\end{subfigure}
		\begin{subfigure}[b]{0.49\textwidth}
			\centering
			\includegraphics[width=0.9\textwidth]{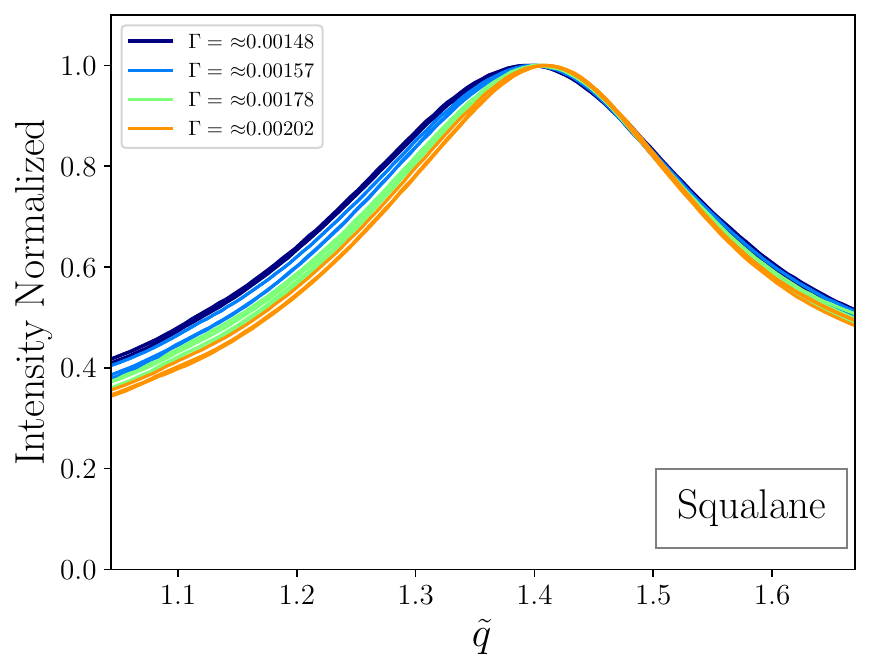}
			\caption{}		
		\end{subfigure}
		\begin{subfigure}[b]{0.49\textwidth}
			\centering
			\includegraphics[width=0.9\textwidth]{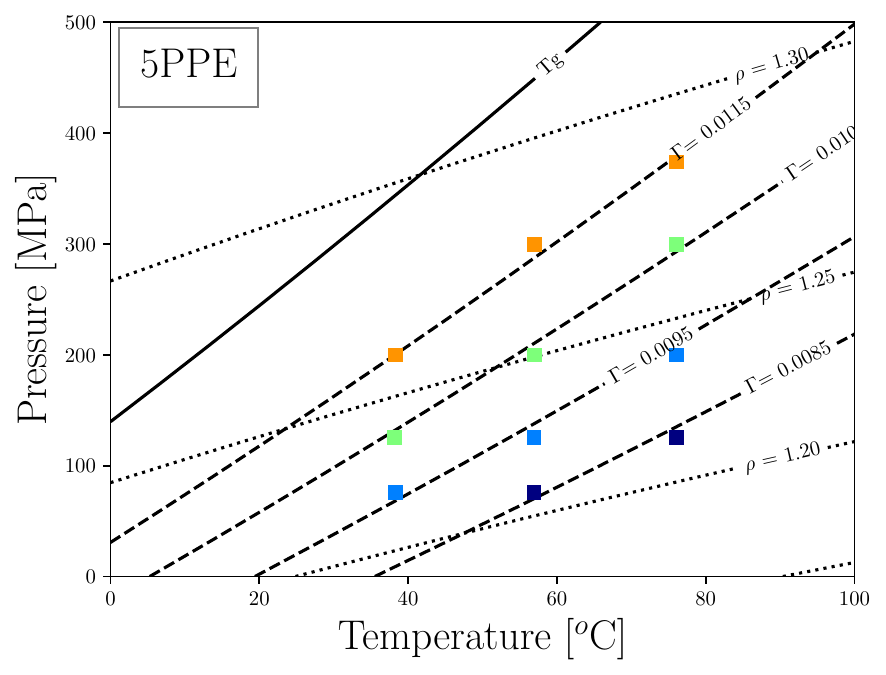}
			\caption{}		
		\end{subfigure}
		\begin{subfigure}[b]{0.49\textwidth}
			\centering
			\includegraphics[width=0.9\textwidth]{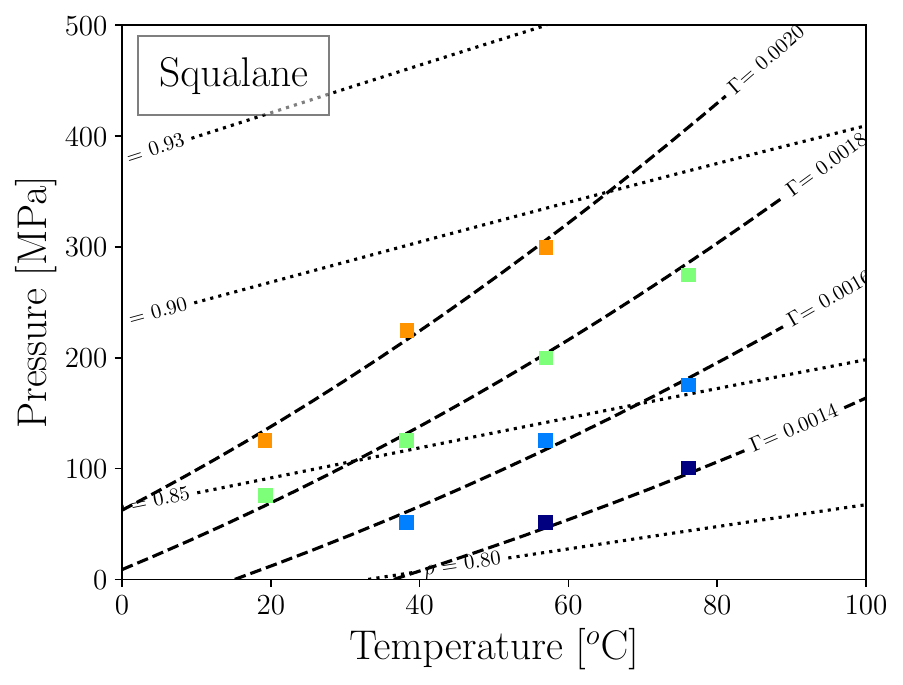}
			\caption{}		
		\end{subfigure}		
		\caption{The measured intensities along four isochrones for 5PPE, left side, and squalane, right side. Figures (a-b) show the normalized measured intensities as a function of q, (c-d) show the normalized measured intensities as a function of $\tilde{q}$, while (e-f) show the phase diagram where each measurement was taken. For 5PPE in total 11 state points along four different isochrones are compared, while for Squalane, a total of 12 state points along four different isochrones are compared. }
		\label{fig:5PPE_Squalane_compare}
	\end{figure}

	In Figure \ref{fig:5PPE_Squalane_compare} the measured structure along four isochrones of both 5PPE (left side) and squalane (right side) are plotted. Figure \ref{fig:5PPE_Squalane_compare} (a)-(b) shows the normalized measured intensities as a function of q, (c)-(d) shows the normalized measured intensities as a function of $\tilde{q}$, while (e)-(f) show the phase diagram where each measurement was taken.
	
	If we look at the scaled structure along isochrones, Figure \ref{fig:5PPE_Squalane_compare} (c)-(d), we see that for both 5PPE and squalane, the structure appears to collapse much better on the right side of the peak, than on the left side of the peak position. For squalane, the left side of the peak appeared to collapse better than for 5PPE. On the right side of the peak, the structure appeared to collapse for Squalane. For 5PPE the structure collapsed alot better on the right side then the left. For both Squalane and 5PPE, it also seems that the structure along each line of constant $\Gamma$ is different from each other. As mentioned earlier the intramolecular structure of the two liquids could be very different. Squalane has the possibility to roll up into a ball, while the 5PPE is a more rigid molecule. The scaling deviation from the intramolecular contributions to the static structure factor could be very different for the two liquids. The intramolecular contributions to the static structure factor could affect the peak position of 5PPE more than Squalane. This could explain why the structure collapses better for squalane than 5PPE, but to confirm this hypothesis microscopic information of the structure of both liquids is needed. 	
	
	\section{Discussion and Conclusion}
	
	In this chapter, the structure of the four different glass formers was presented as we move around the phase diagram. We measured successfully on the van der Waals bonded liquid, DC704, where we expect the structure to collapse along isochrones. We also as a counter example measured on a hydrogen bonded liquid, DPG, where we do not expect the predictions of isomorph theory to hold true. The measurements of the two other van der Waals-bonded liquids, 5PPE, and squalane, were tainted by leaks from the cell resulting in jumps in the intensity. We performed a preliminary analysis of the normalized intensities, but the results should only be considered preliminary.
	

	For 5PPE and squalane, the data quality was worse that DC704 and DPG. It was not possible to subtract the background due to fluctuations in the measured intensities. The origin of these jumps in intensity was most likely leaks on the detector-side window of the cell, which caused the peak to change significantly. This makes any results preliminary, and they should be tested in future studies. For both Squalane and 5PPE, many similarities can be observed in the shape of the main diffraction peak. They have approximately the same peak position, and both molecules are based on long carbon chains. Surprisingly, for 5PPE, we see that the main peak appears to collapse along isochores, while for squalane, the position of the main structure peak collapses almost perfectly along isochrones. The intramolecular structure of 5PPE and squalane could be very different. Squalane is a flexible chain that can properly roll up into a ball, whereas 5PPE is a more rigid molecule due to its chains of phenyl rings. This is illustrated in figure \ref{fig:samples_diamond}. My guess is that the intramolecular scaling deviation is the origin of this difference. If the intramolecular structure is significantly different between squalane and 5PPE, then the scaling deviation would be different for the two liquids. The dynamics of 5PPE been shown to obey the predictions of isomorph theory \cite{XiaoWence2015Itpf}, and we would expect that the structure does as well. Hopefully, it will be possible for someone to further examine this in a future study.
	
	For both DPG and DC704, the peak positions as a function of density were compared with the peak positions as a function of $\Gamma$. When scaling out the effect of density, if the structural changes was only an effect of density, $\tilde{q}_{max}(\rho)$ would be a horizontal line. For all four samples, we see that in experimental units density changes is major part of the measured change. When the effect of density have been scaled out, there are still structural changes. Even for simple van der Waals-bonded liquids, there are some small changes to the structure that are not due to density changes.  An interesting observation is that for DPG and DC704 we see that $\tilde{q}_{max}$ does not change very much as a function of $\Gamma$, compared what we see for Cumene. This can be seen by compared by figure \ref{fig:fit_density_Gamma_compare} with figures \ref{fig:DC704_1st_fit}, \ref{fig:DC704_2nd_fit} and \ref{fig:DPG_fit}. \Mycite{Chen2021} found that the peak position of the hydrogen-bonded liquids glycerol, xylitol, and D-sorbitol was close to invariant as a function of temperature in reduced units. It is interesting how differently the structural change is for the different glass-formers. This could be interesting future work to find and compare the structural change for many glass-formers, to find why there are such differences in how much the structure changes.
	
	Before the experiment, the expectation was that the van der Waals bonded liquid would follow the expectations of the isomorph theory. For DC704, we observed that the first peak position collapsed better as a function of density, figure \ref{fig:DC704_1st_fit}, while the second peak's position collapsed very well as a function $\Gamma$, figure \ref{fig:DC704_2nd_fit}. In figure \ref{fig:background_scheme_compare}, we see the measured structure along an isochrone for the second peak collapses when presented in reduced units. For the first peak, clear differences in the measured intensities were observed for isochronal measurements. In figure \ref{fig:DC704_MD} we saw from simulations at a single state point that the intramolecular contributions to the structure factor influenced especially the first peak of DC704. The contributions to the intramolecular structure were significantly lower for the second peak. The size of the scaling deviation from the intramolecular contributions are likely much bigger for the first peak than the second peak. 
	
	The scaling deviation can explain why the position of the first peak collapses so well as a function of density. In section \ref{sec:strucral changes} we show that along the isochore the change in $S(\tilde{q})$ is purely intermolecular. If DC704 has pseudo-isomorphs then along the isochrone the change in $S(\tilde{q})$ is the scaling deviation arising from the intramolecular contribution. For the first peak, intramolecular contributions dominate, so the structural change we see from the scaling deviation is larger than the intermolecular structural changes going from isochrone to isochrone. Therefore, the peak position for the first peak collapses better as a function of $\rho$, than $\Gamma$. For the second peak, intermolecular contributions dominate, so the structural change we would see going from isochrone to isochrone is larger than the scaling error due to the intramolecular contributions to the structure factor. For the second peak, the peak position collapsed better as a function of $\Gamma$, than $\rho$.

	DPG was originally chosen as a counter-example, and the peak position behaves accordingly. The peak position does not collapse as a function of $\Gamma$ or density. It is possible to obtain a collapse of the peak position, when plotted against $\frac{\rho^{\gamma}}{T}$, with $\gamma =2.9$. This is only the peak position; the shape of the scattered intensities are not similar along these lines in the phase diagram.

	In summary, we observed that both the structure and dynamics for DC704 seem to collapse along the same lines in the phase diagram, making DC704 an pseudo-isomorph. Using results from MD-simulations at a single state point, we can see that the effect of the intramolecular contribution to $S(q)$ is larger for the first peak than the second peak. We have also presented preliminary results for the structures of 5PPE and Squalane. For DPG, we cannot use the dynamics to collapse the structure. Therefore, DPG is an excellent choice as a counter-example to the predictions of the isomorph theory.

	
	In the introduction of DC704 and DPG, we mentioned that have been reported a breakdown of power-law density scaling for both liquids. In the next chapter, we will examine the origin of the reported breakdowns. We will revisit the structural measurements presented in this chapter and in chapter \ref{chapter:Cumene}, with two new types of scaling, isochronal density scaling and isochronal temperature scaling.
	
	
	\clearpage
	

	\chapter{Isochronal Density Scaling and Isochronal Temperature Scaling}
	\label{chapter:IDS}
	
This chapter will revisit the results chapter of \ref{chapter:Cumene} and \ref{chapter:Diamond}, using isochronal density scaling and isochronal temperature scaling. To find candidates for pseudo-isomorphs, we have used lines of constant relaxation time, isochrones. The goal of this chapter is first to test if these new types of scaling, can be used to scale literature data from real molecular glass-formers. Second, to find the isochrones, we have previously used power-law density scaling to find lines of constant $\Gamma$ as we assumed that those are isochrones. Power-law density scaling is often a good approximation, but it has also been reported to breakdown for several glass-formers, like dibutyl phthalate and decahydroisoquinoline \cite{Bøhling2012}. Power-law density scaling has also been reported to break down for two of the liquids studied in chapter \ref{chapter:Diamond}, DC704 \cite{Ransom_DC704} and DPG \cite{AdrjanowiczKarolina2017PNDo}. We want to ensure that the candidates for isochrones used in chapter \ref{chapter:Diamond} are actually isochrones. The third goal of this chapter is to probe the origin of the breakdown of power-law density scaling for DC704 and DPG.  The structure of the chapter is a short recap of isochronal temperature scaling and isochronal density scaling from section \ref{sec:IDS_intro}, followed by revisiting the results for cumene, DC704 and DPG liquid separately.
	
Isochronal density scaling and isochronal temperature scaling are two new scaling concepts that result from the generalization of freezing temperature scaling and freezing density scaling. Both freezing temperature scaling and freezing density scaling have been used to scale several transport coefficients for model liquid systems \cite{Rosenfeld2000,Rosenfeld2001,KhrapakS.A.2024Fdso}. In section \ref{sec:IDS_intro} we showed how isochronal density scaling and isochronal temperature scaling can be derived by making assumptions about the temperature and density dependence of $\gamma$. To recap, isochronal temperature scaling assumes that $\gamma$ is independent of temperature. From this assumption one can also derive the original formulation of density scaling, $\frac{e(\rho)}{T}$. This is also shown in section \ref{sec:IDS_intro}. Isochronal temperature scaling as defined in equation \ref{eq:Isochronal_T_scaling} is given by
	
\begin{equation}\label{eq:Isochronal_T_scaling_later}
		\tilde{\tau}(\rho,T) =	 f\left(\frac{T}{T_{\text{target}}(\rho)}\right)
\end{equation}

Isochronal density scaling was derived by assuming that $\gamma$ was only dependent on the temperature. Isochronal density scaling was originally defined in equation \ref{eq:Isochronal_density_scaling}, and is given by
	
\begin{equation}\label{eq:Isochronal_density_scaling_later}
\tilde{\tau}(\rho,T) = f\left(\frac{\rho}{\rho_{\text{target}}(T)}\right)
\end{equation}	
	
\begin{figure}[H]
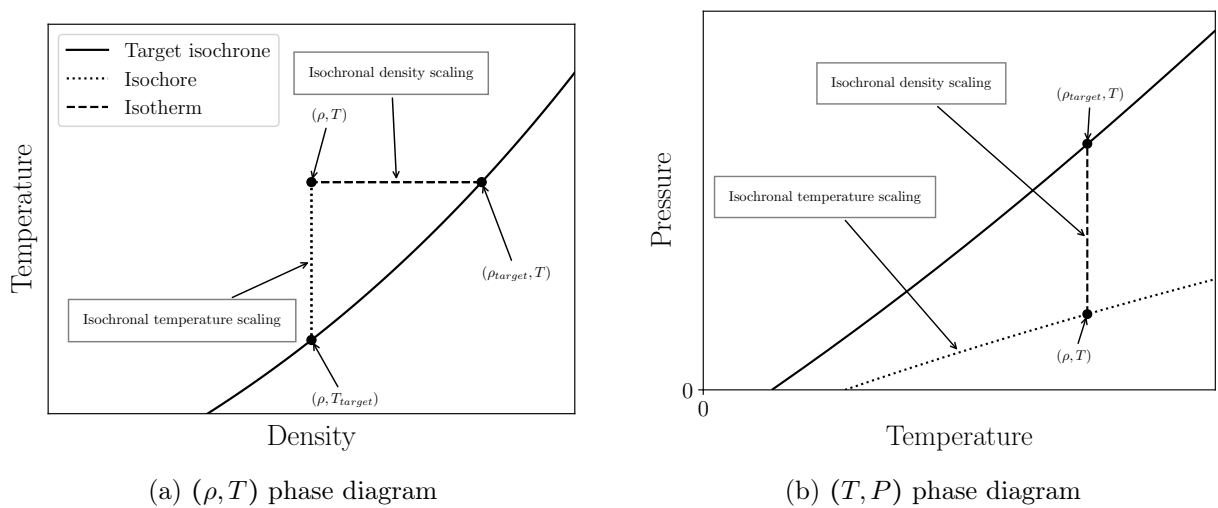

	\begin{subfigure}[t]{0.48\textwidth}
		\centering
		\includegraphics[width=0.99\textwidth]{Chapters/Figures/IDS_example_v4.pdf}
		\caption{  $(\rho,T)$ phase diagram}
		\label{fig:IDS_ITS_phase_a_again}
	\end{subfigure}\hfill
	\begin{subfigure}[t]{0.48\textwidth}
		\centering
		\includegraphics[width=0.99\textwidth]{Chapters/Figures/IDS_example_v5.pdf}
		\caption{$(T,P)$ phase diagram}
		\label{fig:IDS_ITS_phase_b_again}
	\end{subfigure}
	\caption{ Reprint of figure \ref{fig:IDS_ITS_phase}. Illustration of isochronal temperature scaling and isochronal density scaling in the $(\rho,T)$ phase diagram and in the $(T,P)$ phase diagram. In Practice, isochronal temperature scaling can run into the problem where $T_{target}(\rho)$ is at nonphysical negative pressures. This is illustrated in figure \ref{fig:IDS_ITS_phase_b_again}. }
	\label{fig:IDS_ITS_phase_again}
\end{figure}

For both isochronal density scaling and isochronal temperature scaling, we need a target isochrone to perform the scaling. This is illustrated in figure \ref{fig:IDS_ITS_phase_again}. A practical issue for isochronal temperature scaling is that the value of $T_{\text{target}}(\rho_0)$ can be at negative pressures. This becomes an issue for the Cumene data, as we will see in section \ref{sec:IDS_cumene}.

To conclude this short recap of isochronal temperature scaling and isochronal density scaling, here is a small practical "how to" guide for both isochronal temperature scaling and isochronal density scaling.

	
	\subsubsection{How to isochronal density scaling (a) and isochronal temperature scaling (b)} \label{sec:IDS_ITS_ process}
		\begin{enumerate}
		\item Identify a target (reduced units) isochrone to scale onto, $ \tilde{\tau}_{\text{target}}(T_{\text{target}},\rho_{\text{target}}) = \text{constant}$. 
		\item[2.a] For each state point, $(T_0,\rho_0)$, find the density, $\rho_{\text{target}}$ of the target isochrone at the same temperature. 
		\item[2.b] For each state point, $(T_0,\rho_0)$, find the temperature, $T_{\text{target}}$ of the target isochrone at the same density. 
		\item[3] Plot the dynamics as a function of $\rho/\rho_{\text{target}}(T)$  or $T/T_{\text{target}}(\rho)$: 
		\begin{equation*}\tag{\text{Isochronal Density Scaling}}
			\tilde{\tau}(T,\rho) =\tilde{\tau}\left(\frac{\rho}{\rho_{\text{target}}(T)}\right)
		\end{equation*}
		\begin{equation*}\tag{\text{Isochronal Temperature Scaling}}
	\tilde{\tau}(T,\rho) = \tilde{\tau}\left(\frac{T}{T_{\text{target}}(\rho)}\right)
\end{equation*}

	\end{enumerate}
	
As discussed many times in this thesis, the isochrone is our best experimental candidate for the pseudo-isomorph. Therefore, if a liquid has pseudo-isomorphs, like cumene and DC704, then many types of dynamics and the structure should also be invariant along the isochrone when presented in reduced units. For liquids with pseudo-isomorphs we can test isochronal density scaling and isochronal temperature scaling with other transport coefficients than relaxation time. The prediction of isomorph theory, is that there exist lines in the phase diagram where both structure and dynamics is invariant in \textit{reduced units} \cite{Dyre2014,IsomorphPaper4}. In section \ref{sec:thermology_of_isomorph_theory}, we argue that along an pseudo-isomorph the relaxation time in experimental unit is also approximately invariant. Therefore could the glass transition could also be used as a target isochrone.

	
\section{Cumene} \label{sec:IDS_cumene}
	
	In chapter \ref{chapter:Cumene} we showed that cumene has pseudo-isomorphs, e.g. lines in the phase diagram where both structure and dynamics are invariant when presented in reduced units. Power-law density scaling has been shown to work in a large temperature and pressure range for Cumene \cite{Ransom2017}. If a liquid have power-law density scaling then we expect both isochronal density scaling and isochronal temperature scaling to work, as shown in section \ref{sec:IDS_intro}. For cumene, we will test isochronal density scaling with literature data of the viscosity, using the glass transition as the target isochrone.

	 \Mycite{Ransom2017} measured the glass transition temperature of cumene at pressures ranging from ambient pressure to 4.55 GPa. They fit the pressure dependence of $T_g$ using the Andersson-Andersson equation \cite{AnderssonS.Peter1998RSoP}:
	
	\begin{equation}\label{eq:AnderssonAndersson}
		T_g(P) = T_g(P= 1 \text{ atm}) \left(1 + \frac{b}{c}P\right)^{\frac{1}{b}}
	\end{equation}
	
	where $T_g(P= 1 \text{ atm})$ is the glass transition temperature at ambient pressure, and $b$ and $c$ are fitting parameters. They found that the fitting values $T_g(P= 1 \text{ atm})$ = $129 \pm 2 $K, $b = 1.67 \pm 0.04$, and $c =1.16 \pm 0.05 $ GPa could describe the glass transition over the whole pressure range \cite{Ransom2017}. The glass transition is approximately an isochrone and they estimate the density scaling coefficient value $\gamma \approx 4.8$, by linearly fitting $\log(T_{g})$ versus $\log(\rho)$.
	
\begin{figure}[H]
		\centering
		\includegraphics[width=0.75\textwidth]{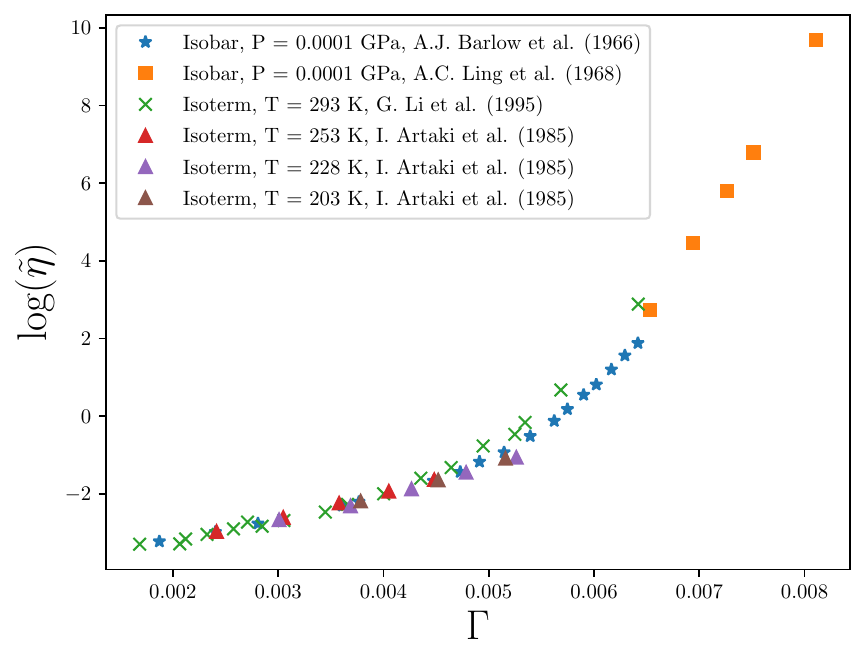}
		\caption{Recreation of figure 5 from \Mycite{Ransom2017}, showing power-law density scaling for cumene. A reference for the viscosity measurements are described in the text. All viscosities measured for cumene collapse when plotted against $\Gamma= \frac{\rho^\gamma}{T}$, where $\gamma=4.8$. }
		\label{fig:Cumene_dym_DS}
	\end{figure}

	Using the $\gamma$ calculated from $T_g(P)$, \Mycite{Ransom2017}  make several measurements of the viscosity of cumene from the literature collapse into a single curve. \Mycite{ARTAKI1985PEOT} measured the viscosity along three isotherms, at 253 K, 228 K, and 203 K, and at pressures up to 0.4 GPa. \Mycite{LiG1995Patd} measured along a single isotherm at 293 $^o$K and at pressures up to 1.37 GPa. \Mycite{BarlowA.J.1966Vbos} and  \Mycite{LINGAC1968VOSO} both measured the viscosity of Cumene along the ambient pressure isobar. Figure \ref{fig:Cumene_dym_DS} is a recreation of figure 5 from \Mycite{Ransom2017}, showing power-law density scaling for the above mentioned measurements of the viscosity for cumene.
	
\begin{figure}[H]
		\begin{subfigure}[t]{0.48\textwidth}
			
			\includegraphics[width=\textwidth]{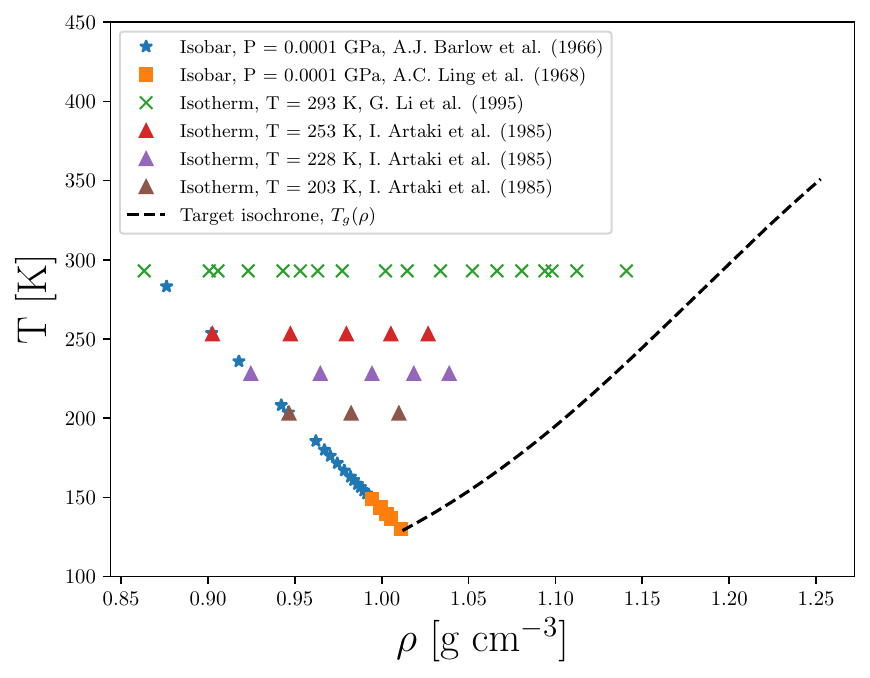}
			\caption{}
			\label{fig:Cumene_dym_IDS_a}
		\end{subfigure}\hfill
		\begin{subfigure}[t]{0.48\textwidth}
			\centering
			\includegraphics[width=\textwidth]{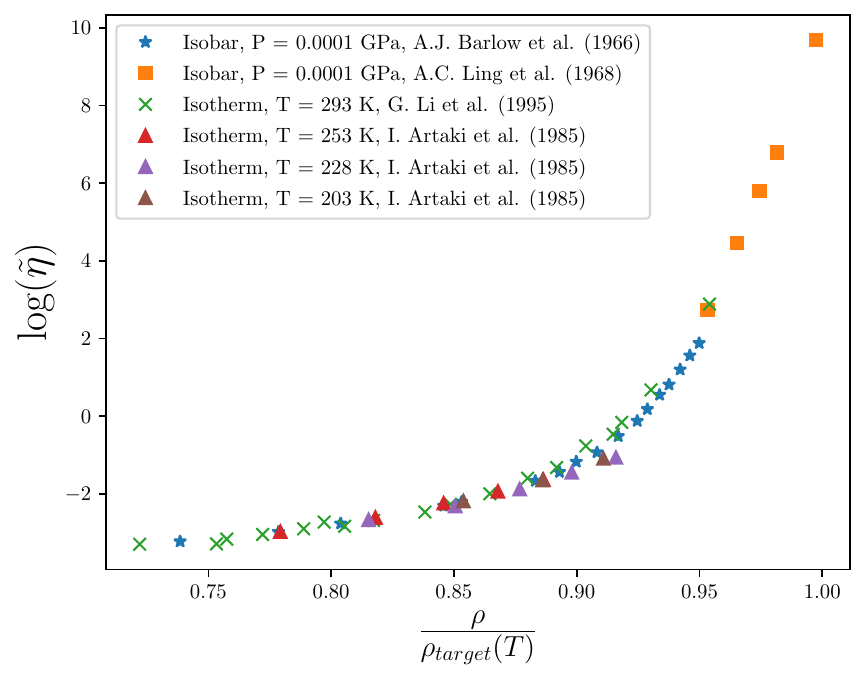}
			\caption{ }
			\label{fig:Cumene_dym_IDS_b}
		\end{subfigure}	
		\caption{Isochronal density scaling for the viscosity of cumene. In figure \ref{fig:Cumene_dym_IDS_a} the location of each viscosity measurement is plotted in the $(\rho,T)$ phase diagram. Each symbol is a different series from the literature. The dashed line is the target isochrone used for the scaling. The target isochrone is the glass transition line fitted to the Andersson-Andersson equation (equation \ref{eq:AnderssonAndersson}) from \Mycite{Ransom2017}. For many of the state points it is not possible to employ isochronal temperature scaling due to $T_{target}(\rho)$ being at nonphysical negative pressures.
		Figure \ref{fig:Cumene_dym_IDS_b} shows isochronal density scaling for the viscosity of cumene, using the same data. Cumene is a liquid for which power-law density scaling works well therefore we would expect both isochronal temperature scaling and isochronal density scaling to collapse the viscosity measurements.}
		\label{fig:Cumene_dym_IDS}
	\end{figure}

	As shown in equation \ref{eq:AnderssonAndersson}, \Mycite{Ransom2017} fitted the glass-transition temperature of Cumene to the Andersson-Andersson equation. The glass transition temperature is an isochrone in experimental units, so we can use it as a target isochrone. A practical issue for using the glass transition as the target isochrone is that isochronal temperature scaling is not possible because $T_{target}(\rho)$ will be at nonphysical negative pressures. This can also be seen in figure \ref{fig:Cumene_dym_IDS_a}.

	Figure \ref{fig:Cumene_dym_DS} shows that it is possible to apply power-law density scaling to collapse the viscosity measurements into a single curve. Isochronal density scaling also collapses the data very well, as shown in figure \ref{fig:Cumene_dym_IDS_b}. The collapse using isochronal density scaling seems to work slightly better then density scaling, but this is mainly due to the measurements from \Mycite{LiG1995Patd} (green crosses). Isochronal density scaling of the literature data gives a very promising collapse, making it obvious to revisit the results of the MD simulations. 
	
\begin{figure}[H]
		\begin{subfigure}[t]{0.48\textwidth}
			
			\includegraphics[width=\textwidth]{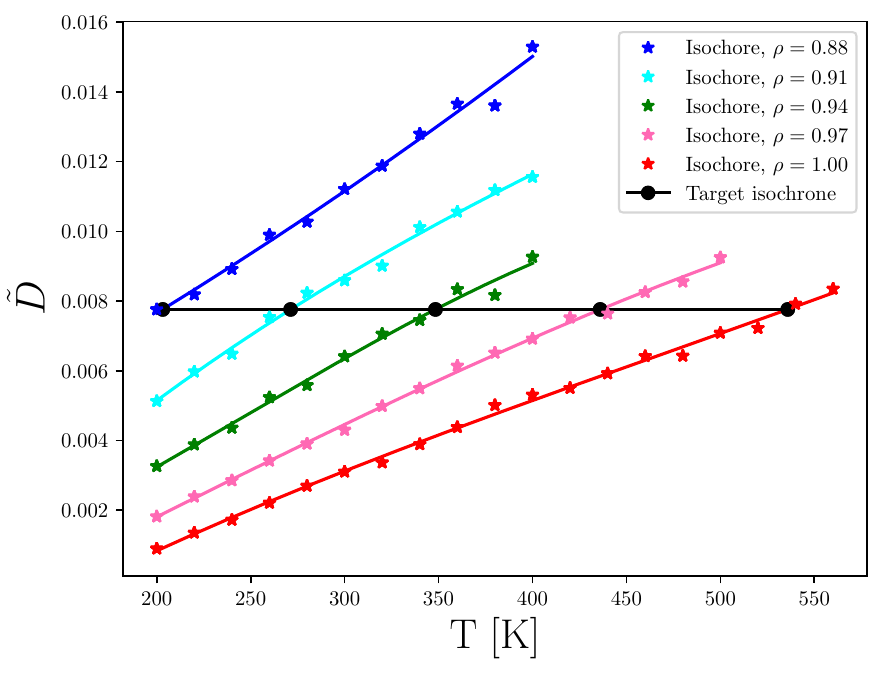}
			\caption{}
			\label{fig:Cumene_dym_MD_IDS_a}
		\end{subfigure}\hfill
		\begin{subfigure}[t]{0.48\textwidth}
			
			\centering
			\includegraphics[width=\textwidth]{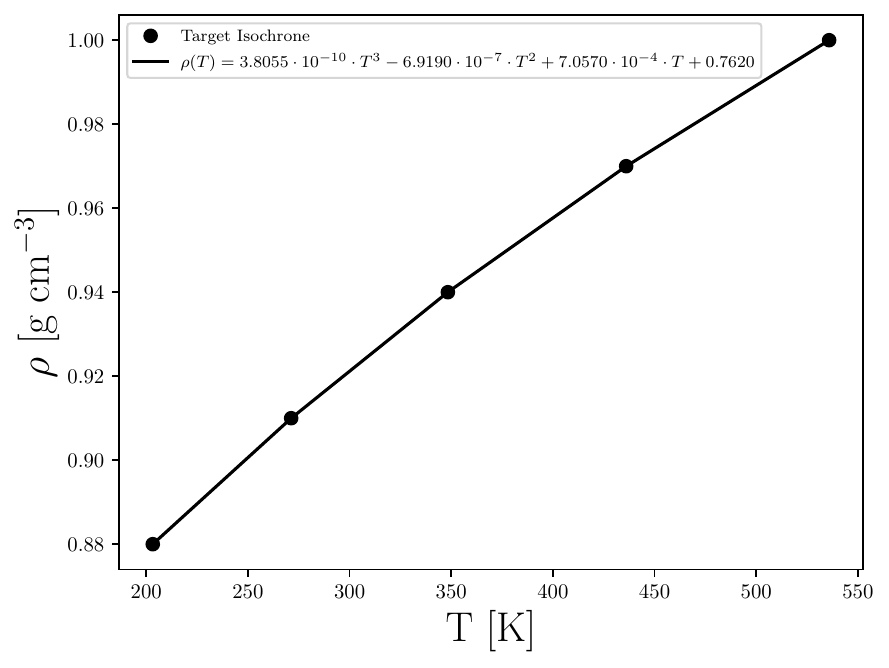}
			\caption{ }
			\label{fig:Cumene_dym_MD_IDS_b}
		\end{subfigure}
		\begin{subfigure}[t]{0.48\textwidth}
			
			\centering
			\includegraphics[width=\textwidth]{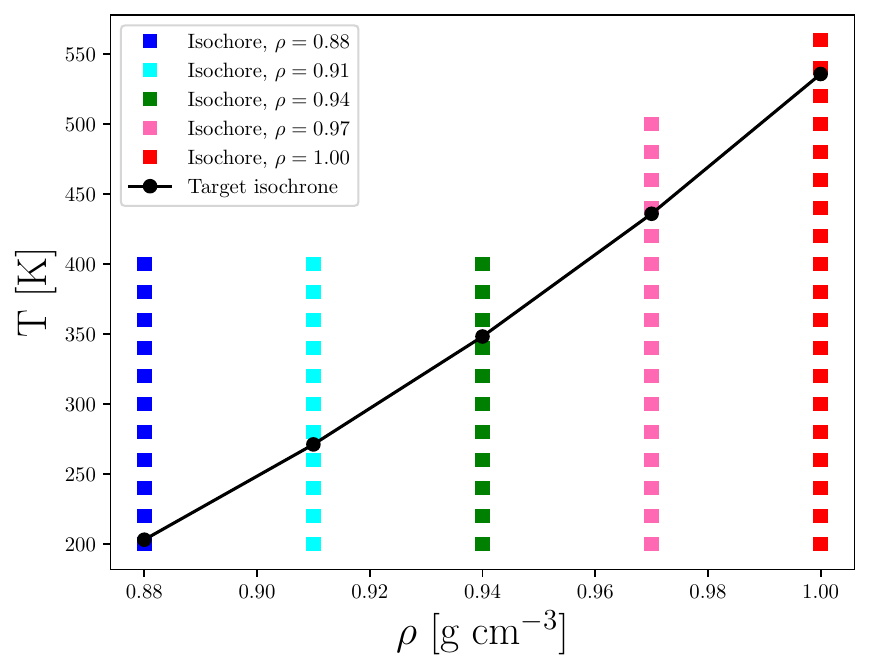}
			\caption{}
			\label{fig:Cumene_dym_MD_IDS_c}
		\end{subfigure}	\hfill
		\begin{subfigure}[t]{0.48\textwidth}
			
			\centering
			\includegraphics[width=\textwidth]{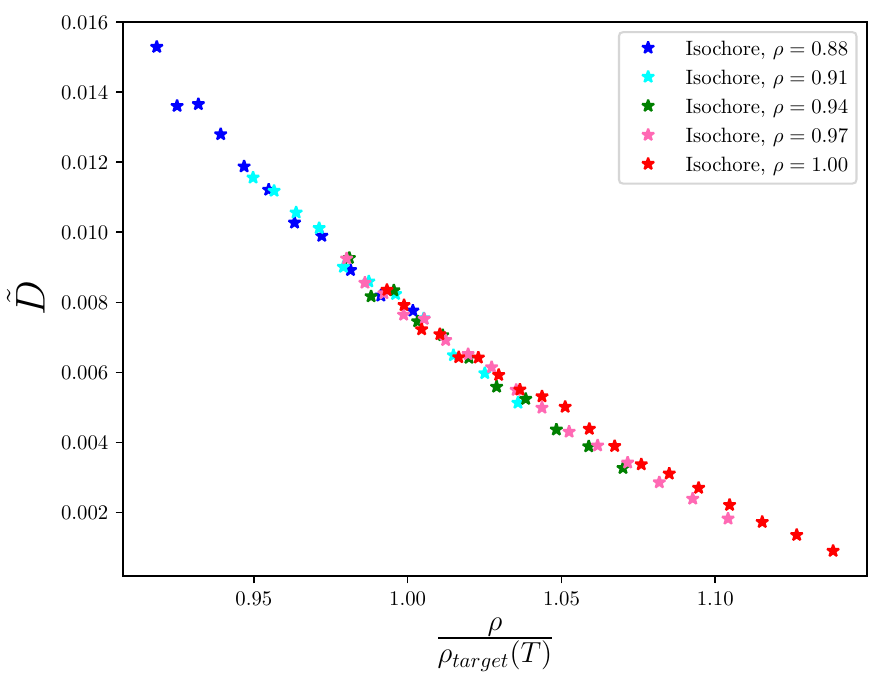}
			\caption{}
			\label{fig:Cumene_dym_MD_IDS_d}
		\end{subfigure}

		\caption{Isochronal density scaling of the UA-cumene model. In subfigure \ref{fig:Cumene_dym_MD_IDS_a} the reduced diffusion coefficient is plotted for all simulated state points. The solid lines represent the reduced diffusion coefficients for each isochor fitted with third-order polynomial. The black line is the target isochrone used for the fitting, the black dots are the intersection between the fits and target isochrone.  In figure \ref{fig:Cumene_dym_MD_IDS_b} the fit to the target isochrone is plotted. The target isochrone fitted with a 3rd order polynomial: $\rho_{\text{target}}(T) = 3.8055 \cdot 10^{-10}\cdot T^3 - 6.9190 \cdot 10^{-7} \cdot T^2 + 7.0570 \cdot 10^{-4}\cdot T + 0.7620 $.
		In figure \ref{fig:Cumene_dym_MD_IDS_c} the target isochrone in the $(\rho,T)$ phase diagram are shown. Each color symbolized an isochore while each square is a state point simulated. In figure \ref{fig:Cumene_dym_MD_IDS_d} $\tilde{D}$ collapse well on a single curve when plotted against $\rho / \rho_{\text{target}}(T)$ .    }
		\label{fig:Cumene_dym_MD_IDS}
		
	\end{figure}

	In figure \ref{fig:Cumene_dym_MD_IDS_a} the reduced diffusion coefficient, $\tilde{D}$, is plotted for all simulated state points. The simulation data is introduced in chapter \ref{chapter:Cumene}. The reduced diffusion coefficient for each isochore is fitted with a third-order polynomial, and the black line is the target isochrone used to scale the reduced diffusion coefficients. In figure \ref{fig:Cumene_dym_MD_IDS_b} $\rho_{\text{target}}(T)$ along the target isochrone is fitted with a third-order polynomial to interpolate between the isochores. In figure \ref{fig:Cumene_dym_MD_IDS_c} the target isochrone is shown in the $(\rho,T)$ phase diagram. For almost all state points, we can interpolate $\rho_{\text{target}(T)}$ from the fit. For high temperature state points on the 1.00 g cm$^{-3}$ isochore, we have to extrapolate using the fit. In figure \ref{fig:Cumene_dym_MD_IDS_d} $\tilde{D}$ is plotted against $\frac{\rho}{\rho_{\text{target}}(T)}$ for all simulated state points. Using isochronal density scaling it is possible to collapse all the reduced diffusion coefficients into a single curve. In chapter \ref{chapter:Cumene} we also able to collapse the data with power-law density scaling.

	\begin{figure}[H]
		\begin{subfigure}[t]{0.48\textwidth}
	\centering
	\includegraphics[width=\textwidth]{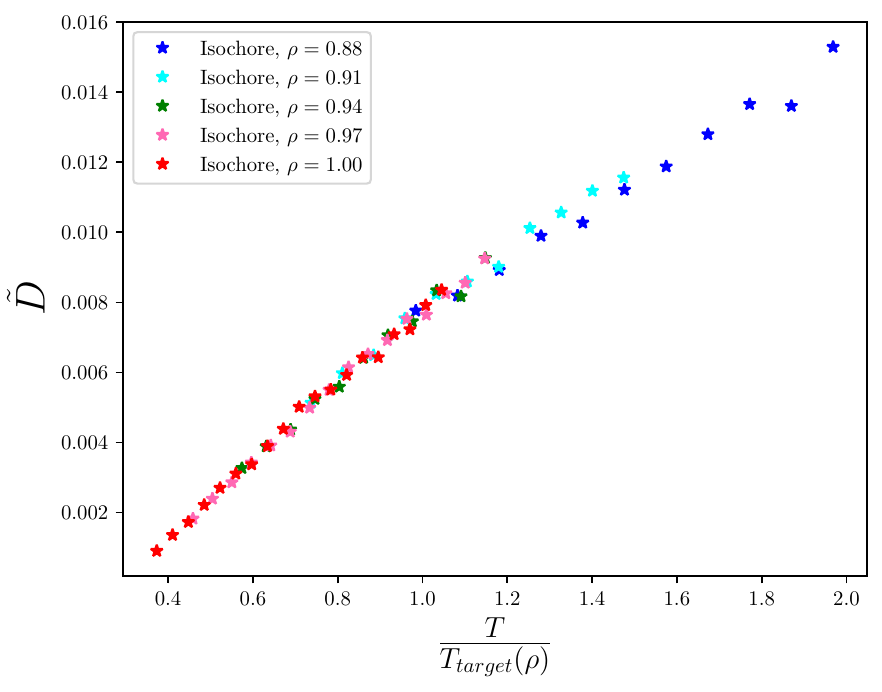}
	\caption{}
	\label{fig:Cumene_dym_ITS_a}
\end{subfigure}
\begin{subfigure}[t]{0.48\textwidth}
	\centering
	\includegraphics[width=\textwidth]{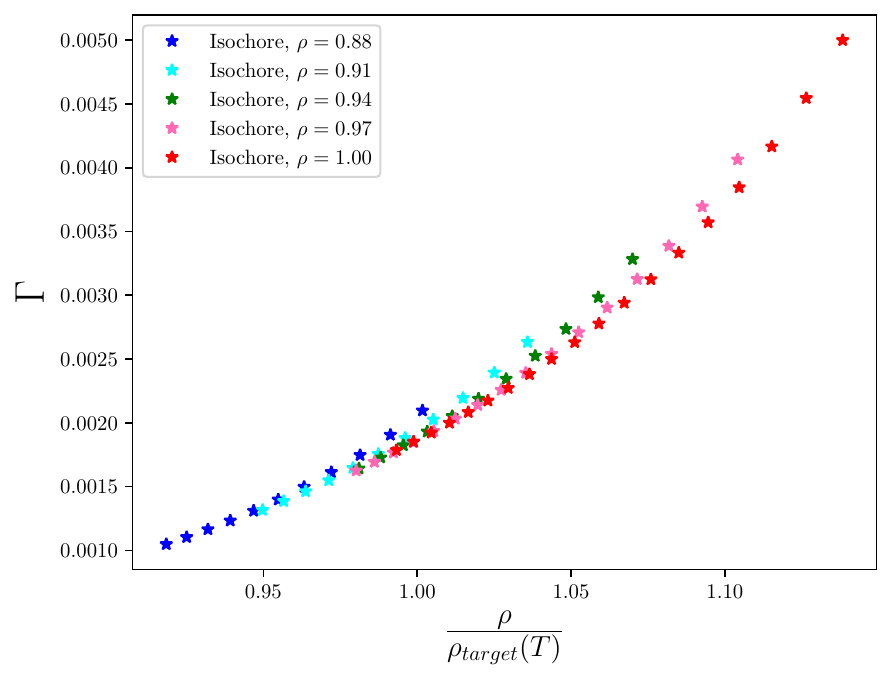}
	\caption{$\Gamma = \rho^\gamma / T$ where $\gamma = 6.8$}
	\label{fig:Cumene_dym_ITS_b}
\end{subfigure}		
\begin{subfigure}[t]{0.48\textwidth}
	\centering
	\includegraphics[width=\textwidth]{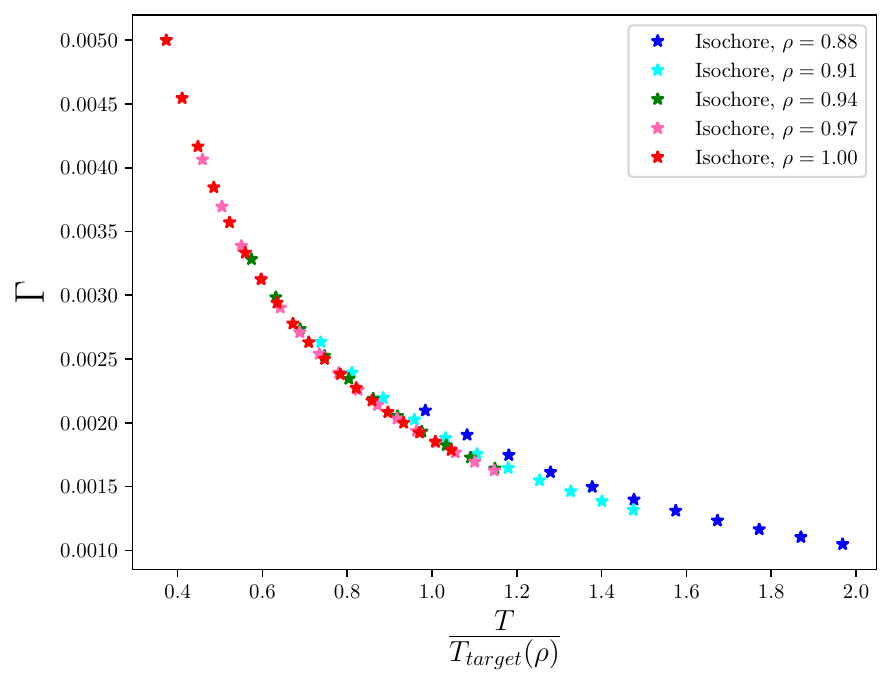}
	\caption{$\Gamma = \rho^\gamma / T$ where $\gamma = 6.8$}
	\label{fig:Cumene_dym_ITS_c}
\end{subfigure}		
\begin{subfigure}[t]{0.48\textwidth}
	\centering
	\includegraphics[width=\textwidth]{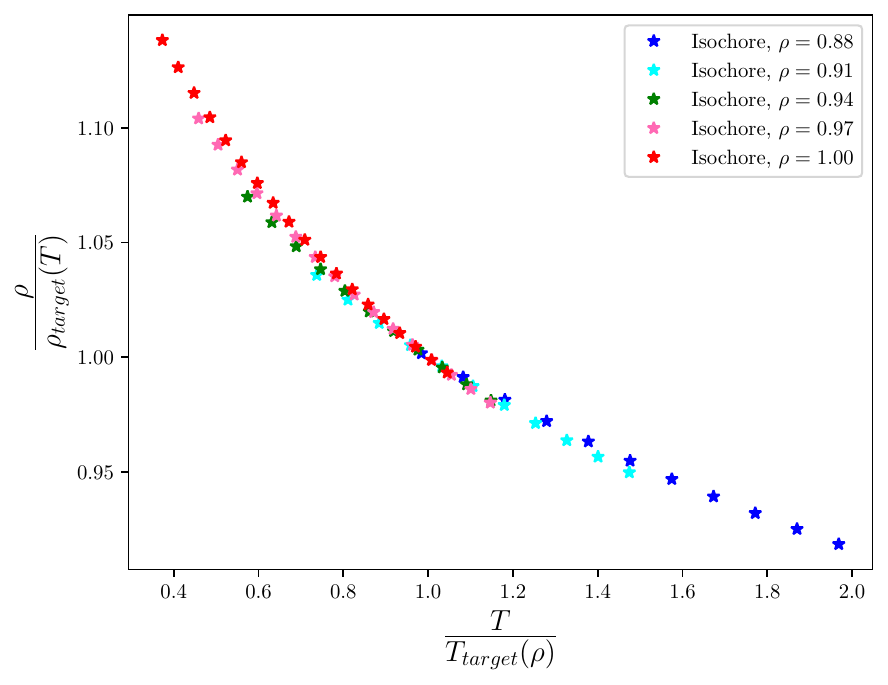}
	\caption{}
	\label{fig:Cumene_dym_ITS_d}
\end{subfigure}		
\caption{Isochronal temperature scaling for the UA-cumene model and comparison with $\Gamma$ and isochronal density scaling. In figure \ref{fig:Cumene_dym_ITS_a} isochronal temperature scaling is shown for UA-cumene model. In figure \ref{fig:Cumene_dym_ITS_b}  $\rho / \rho_{\text{target}}(T)$  is plotted against $\Gamma$, for $\gamma=6.8$. In figure \ref{fig:Cumene_dym_ITS_c}  $T/T_{\text{target}}(\rho)$ is plotted against $\Gamma$, for $\gamma=6.8$. In figure \ref{fig:Cumene_dym_ITS_d} $T/T_{\text{target}}(\rho)$ is plotted against $\rho / \rho_{\text{target}}(T)$. In general both isochronal temperature scaling and isochronal density scaling give a good collapse of the reduced diffusion coefficient. The behavior of the UA-Cumene model is consistent with $\gamma$ being weakly dependent on density an temperature.     }
\label{fig:Cumene_dym_ITS}
	\end{figure}

In figure \ref{fig:Cumene_dym_ITS} isochronal temperature scaling is shown, along with comparisons between $T/T_{\text{target}}(\rho)$, $\rho / \rho_{\text{target}}(T)$ and $\Gamma$. In chapter \ref{chapter:Cumene} we were able to collapse the reduced diffusion coefficient reasonable well using power-law density scaling (see figure \ref{fig:Cumene_density_scaling}). However the value of the power-law density scaling coefficient $\gamma$ changed between different isochrones. This could be an indication that $\gamma$ is not independent of the density and/or the temperature. Isochronal temperature scaling assumes that $\gamma$ is only a function of density, while isochronal density scaling assumes that $\gamma$ is only a function of temperature. In a way, comparing the two can be seen as a "stress test" of power-law density scaling.

For the UA-cumene model we find reasonable collapses with both isochronal density scaling and isochronal temperature scaling. It makes sense that we are also able to get a reasonable collapse using power-law density scaling. In the simulations it is clear that we change the temperature much more than compared to the density. 
In figure \ref{fig:Cumene_dym_ITS_b} we compare  $\rho / \rho_{\text{target}}(T)$ with $\Gamma$, and in figure \ref{fig:Cumene_dym_ITS_c} we compare $T/T_{\text{target}}(\rho)$ with $\Gamma$, for $\gamma = 6.8$. 

If we compare the range of temperatures and densities in the simulations, then the density range is 0.88-1 g cm$^{-3}$ and the temperature range is 200-560 K. In comparison then we change the temperature much more than the density. It is hard to determinate if isochronal temperature scaling or isochronal density scaling collapse the data better. If we take the range of temperatures and densities into account it seems that for the model, $\gamma$ is more dependent on density than temperature. If an additional set of simulations was made in larger density range, it may be possible to breakdown isochronal density scaling for the MD-model. However in the temperature and density range of these simulations, $\gamma$ only have a weak dependency of density and temperature.

For cumene, we have shown that isochronal density scaling can collapse both literature data on reduced viscosity (figure \ref{fig:Cumene_dym_IDS_b}) and the reduced diffusion coefficient calculated from simulations of united atom MD simulations (figure \ref{fig:Cumene_dym_MD_IDS_d}). When isochronal density scaling is compared with power-law density scaling, both can make the dynamical data collapse for both the MD model and the experimental data. 
For isochronal temperature scaling we were also able to scale the reduced diffusion coefficient calculated from simulation. We were not able to test isochronal temperature scaling on the experimental measurements of the viscosity. Another thing to note when comparing power-law density scaling with isochronal density scaling and isochronal temperature scaling is the dimensions of the three scaling variables. The dimension of $\Gamma$ are $[\Gamma] $ =  $M^\gamma L^{-3\gamma} T^{-1}$, where $\gamma$ can be any real number. The scaling variables $\rho / \rho_{\text{target}}(T)$ and $T/T_{\text{target}}(\rho)$ are both dimensionless meaning that the scaling of the reduced diffusion coefficients and the reduced viscosity is completely dimensionless. From a physical perspective, this is pleasing.

	
	\subsection{Structure}
	
	In the previous section, we showed that isochronal density scaling can collapse the reduced viscosity into a single curve. Cumene is a liquid for which power-law density scaling has been shown to work in a large pressure and temperature range \cite{Ransom2017,Wase2018_nature,Wase2018}, and isochronal density scaling reproduces the results quite well. The structural measures we compared for the first structure peak collapsed quite well, but not perfectly, as a function of $\Gamma$. It makes sense to examine the structural measurements again using isochronal density scaling to identify isochrones. Isochronal temperature scaling is still not possible for every state point, due to $T_{target}(\rho)$ corresponding to negative pressures for some state points.
	
\begin{figure}[H]
		\begin{subfigure}[t]{0.48\textwidth}
			
			\includegraphics[trim = 30mm 80mm 40mm 80mm, clip=true,width=0.98\textwidth]{Chapters/Figures/Gamma_qmax_scaled.pdf}
			\caption{Peak position in reduced units, $\tilde{q}=q\rho^{-\frac{1}{3}}$, plotted against $\Gamma$.  }
			\label{fig:Cumene_structure_IDS_a}
		\end{subfigure}\hfill
		\begin{subfigure}[t]{0.48\textwidth}
			
			\centering
			\includegraphics[trim = 30mm 80mm 40mm 80mm, clip=true,width=0.98\textwidth]{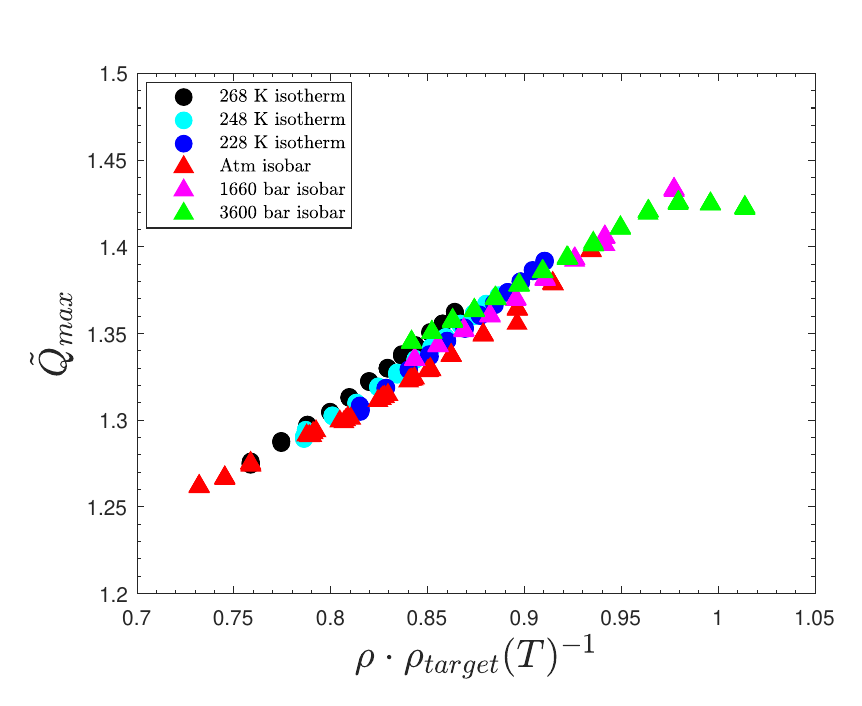}
			\caption{ Peak position in reduced units, $\tilde{q}=q\rho^{-\frac{1}{3}}$, plotted against $\rho / \rho_{\text{target}}(T)$.  }
			\label{fig:Cumene_structure_IDS_b}
		\end{subfigure}
		\begin{subfigure}[t]{0.48\textwidth}
			
			\centering
			\includegraphics[trim = 30mm 80mm 40mm 80mm, clip=true,width=0.98\textwidth]{Chapters/Figures/Gamma_skewness.pdf}
			\caption{Skewness plotted against $\Gamma$. }
			\label{fig:Cumene_structure_IDS_c}
		\end{subfigure}	\hfill
		\begin{subfigure}[t]{0.48\textwidth}
			
			\centering
			\includegraphics[trim = 30mm 80mm 40mm 80mm, clip=true,width=0.98\textwidth]{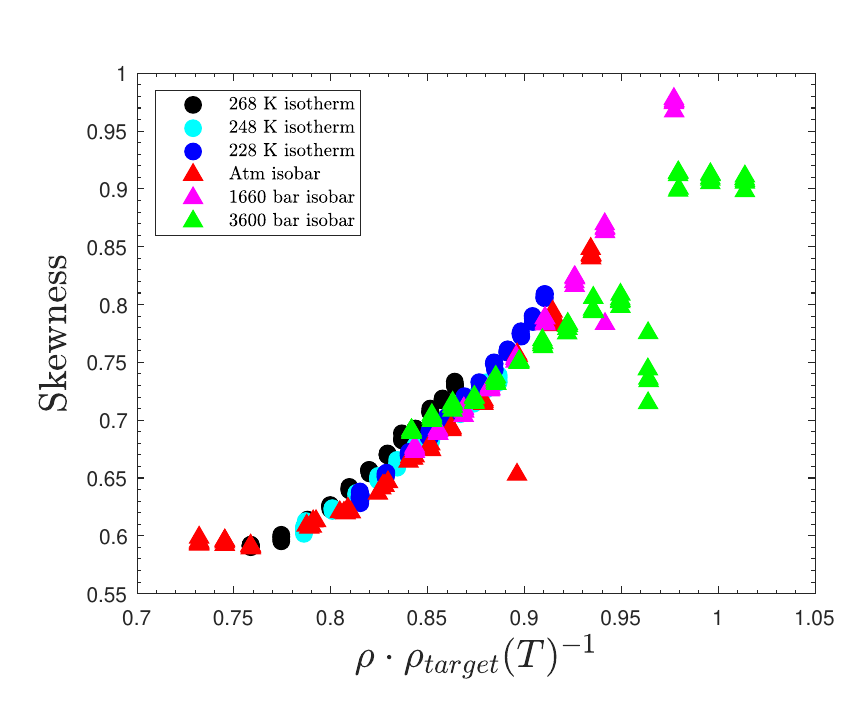}
			\caption{Skewness plotted against $\rho / \rho_{\text{target}}(T)$. }
			\label{fig:Cumene_structure_IDS_d}
		\end{subfigure}
		\begin{subfigure}[t]{0.48\textwidth}
			
			\centering
			\includegraphics[trim = 30mm 80mm 40mm 80mm, clip=true,width=0.98\textwidth]{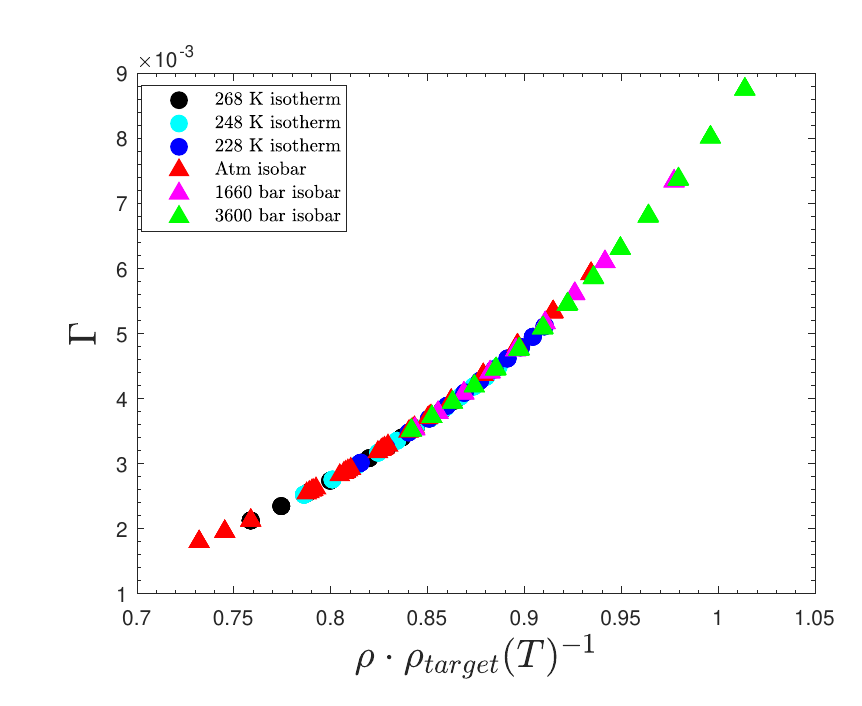}
			\caption{The value of the two scaling variables $\Gamma$ and $\frac{\rho}{\rho_{\text{target}}(T)}$  plotted against each other. }
			\label{fig:Cumene_structure_IDS_e}
		\end{subfigure} \hfill
		\begin{subfigure}[t]{0.48\textwidth}
			
			\centering
			\includegraphics[trim = 30mm 80mm 40mm 80mm, clip=true,width=0.98\textwidth]{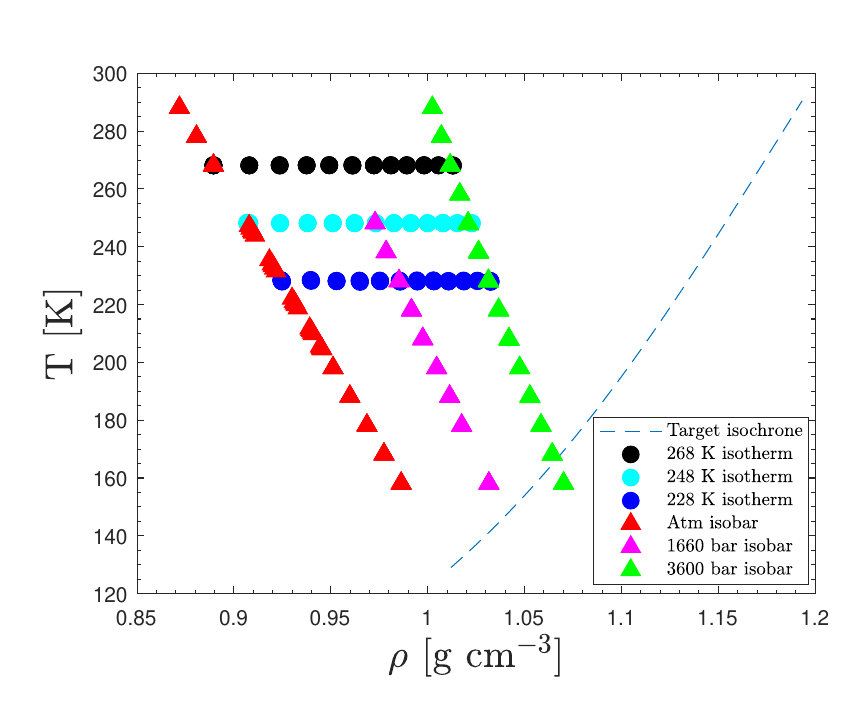}
			\caption{Each state point in the ($\rho$, T) phase diagram. }
			\label{fig:Cumene_structure_IDS_f}
		\end{subfigure}				
		\caption{Comparison between power-law density scaling and isochronal density scaling for the experimental data. Figures \ref{fig:Cumene_structure_IDS_a} and \ref{fig:Cumene_structure_IDS_b} show the reduced peak position, against $\Gamma$ and $\rho / \rho_{\text{target}}(T)$. Figures  \ref{fig:Cumene_structure_IDS_c} and \ref{fig:Cumene_structure_IDS_d} show the skewness of the first structure peak against $\Gamma$ and $\rho / \rho_{\text{target}}(T)$. Isochronal density scaling reproduces the same results as using power-law density scaling to identifying isochrones. In figure \ref{fig:Cumene_structure_IDS_e} $\Gamma$ and  $\rho / \rho_{\text{target}}(T)$ are plotted against each other, and they collapse into a single line. This mapping from $\frac{\rho}{\rho_{\text{target}}(T)}$ to $\Gamma$ is however, not a straight line. In figure \ref{fig:Cumene_structure_IDS_f} the position of each measurement in the $(\rho,T)$ phase diagram is show. 
		For several state points $T_{target}(\rho)$ would be at negative pressures. }
		\label{fig:Cumene_structure_IDS}
		
	\end{figure}

\begin{figure}[H]
		\centering
		\includegraphics[trim = 30mm 80mm 40mm 80mm, clip=true,width=0.6\textwidth]{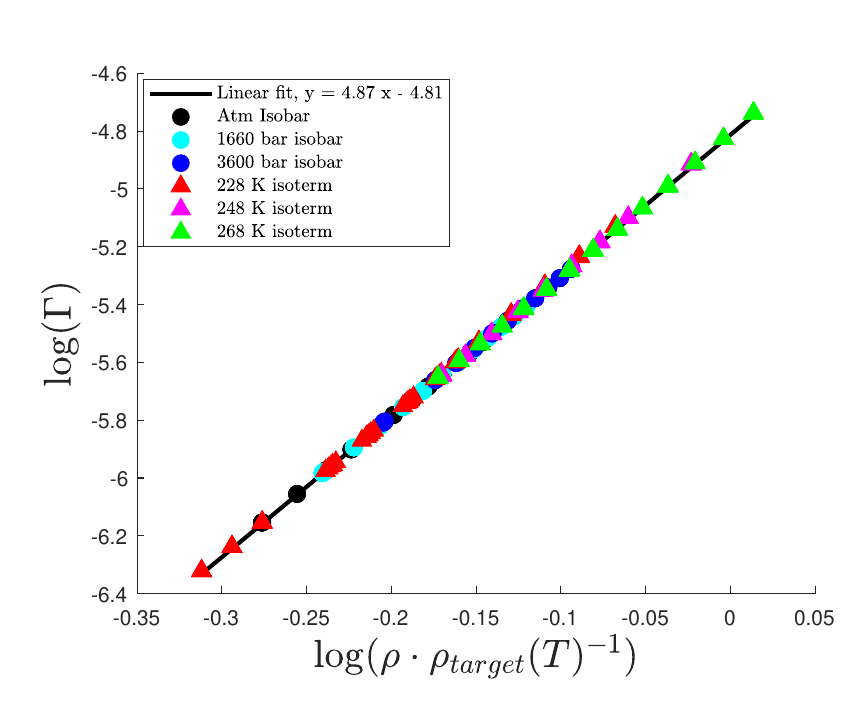}
		\caption{The logarithm of the two scaling variables $\Gamma$ and $\frac{\rho}{\rho_{\text{target}}(T)}$  plotted against each other at each state point. It can be fitted with a first-order polynomial, $\log(\Gamma$) = 4.87 $\log(\frac{\rho}{\rho_{\text{target}}(T)})$ -4.81. The slope of the power law is close to the value of $\gamma = 4.8$.}
		\label{fig:IDS_Gamma_log_log}
	\end{figure}

	In figure \ref{fig:Cumene_structure_IDS} we revisit the experimental results for the first peak of cumene. In chapter \ref{chapter:Cumene}, we observed that the FWHM is practically invariant when moving around the phase diagram; thus, we will compare the peak position, $q_{max}$, and the skewness of the peak. The target isochrone used is the same used for the dynamical data, namely the fit to the Andersson-Andersson equation from \Mycite{Ransom2017}. In figure \ref{fig:Cumene_structure_IDS_a} and \ref{fig:Cumene_structure_IDS_b} the peak position of the main structure peak is compared, while in figure \ref{fig:Cumene_structure_IDS_c} and \ref{fig:Cumene_structure_IDS_d} the skewness of the peak is compared. Both power-law density scaling and isochronal density scaling can collapse the data well. In figure \ref{fig:Cumene_structure_IDS_e} the $\Gamma$ is plotted against the scaling variable $\frac{\rho}{\rho_{\text{target}}(T)}$, and they collapse into a single line. This explains why both power-law density scaling and isochronal density collapse the data very well as there is a one-to-one mapping between the two scaling variables. The mapping between $\Gamma$ and $\frac{\rho}{\rho_{\text{target}}(T)}$ looks like as power-law relationship. In figure \ref{fig:IDS_Gamma_log_log} $\Gamma$ and $\frac{\rho}{\rho_{\text{target}}(T)}$ are plotted in a log-log plot, and the relationship is fitted with a first-order polynomial. The slope is 4.87, which is approximately $\gamma$.
	
	In section \ref{sec:IDS_intro} it was shown that for systems where $\gamma$ is independent of both density and temperature, both isochronal density scaling and isochronal temperature scaling are expected to scale the data. From the assumption that $\gamma$ is a material specific constant, it is possible to derive power-law density scaling in the isomorph theory framework. Power-law density scaling assumes that the shape of the isochrones is a power-law and for cumene it seems to be true. Isochronal density scaling assumes that the shape of the isochrone is independent of the density. That shape can be a power-law as is the case for cumene.
	
	
	\section{DC704}
	
	In section \ref{sec:IDS_cumene} isochronal density scaling was tested on the structure and dynamics of cumene. For cumene isochronal density scaling was able to reproduce the results of power-law density scaling. There was a one-to-one mapping between power-law density scaling and isochronal density scaling, consistent with the fact that the shape of the isochrones for cumene can be described by a power-law in the experimental range. In chapter \ref{chapter:Diamond} we tested DC704 for pseudo-isomorphs using power-law density scaling to identify the isochrones.	 As mention in the sample introduction for DC704, section \ref{sec:DC704_background}, \Mycite{Ransom_DC704} have observed a breakdown of power-law density scaling for DC704. In this section we will test isochronal density scaling and isochronal temperature scaling on literature data for the dynamics of DC704. Then use this new approach to find isochrones to revisit the structural measurements presented in chapter \ref{chapter:Diamond}.
	
	\subsection{Breakdown of power-law density scaling}
\Mycite{Ransom_DC704} measured the relaxation time in an unusually large range of temperatures and pressures. The temperature range was from 425.4 K to 218.6 K and the pressure ranged from ambient pressures up to $\approx$ 1 GPa. They found that in their large temperature and pressure range of the experiment, power-law density scaling breaks down for DC704. In section \ref{sec:IDS_intro} we showed that power-law density scaling can be derived from a constant $\gamma$. The authors argued that the breakdown of power-law density scaling for DC704 stems from the power-law exponent, $\gamma$, being state point dependent. The original description of density scaling\cite{Alba-Simionesco2002,Alba-Simionesco2004}, can be derived by assuming that $\gamma(\rho)$ is a function of density. This would be an obvious idea for the origin of the breakdown. We can probe the origin of the breakdown of power-law density scaling using isochronal temperature scaling and isochronal density scaling.
	
\begin{figure}[H]
		\begin{subfigure}[t]{0.48\textwidth}
			\centering
			\includegraphics[width=0.95\textwidth]{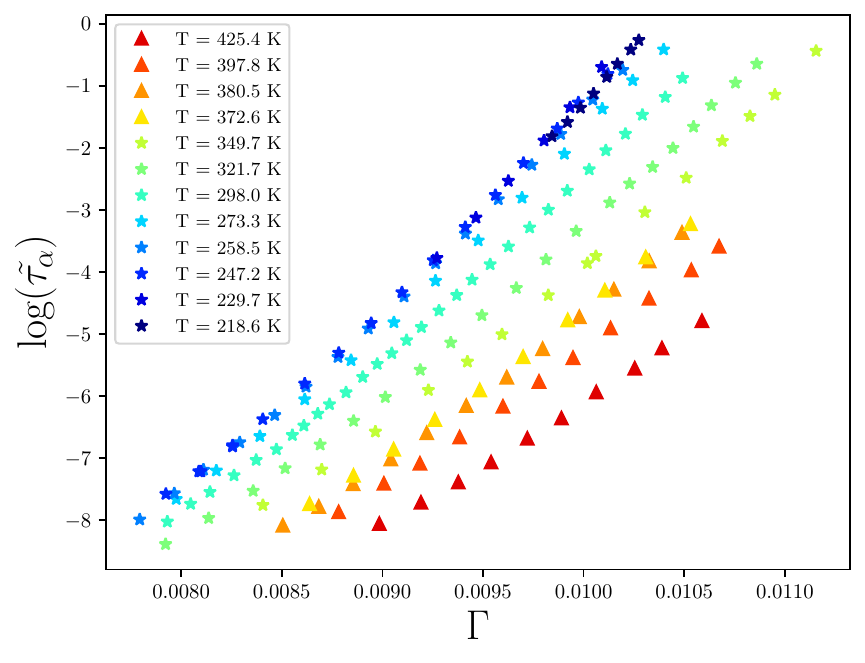}
			\caption{ $\gamma = 6.2$ }
			\label{fig:Ransom_gamma_62}
		\end{subfigure}\hfill
		\begin{subfigure}[t]{0.48\textwidth}
			\centering
			\includegraphics[width=0.95\textwidth]{Chapters/Figures/DC704_Ransom_gamma_513_notitle.pdf}
			\caption{$\gamma=5.13$}
			\label{fig:Ransom_gamma_513}
		\end{subfigure}
		\caption{Recreation of figure 3 and its insert in \Mycite{Ransom_DC704}. \Mycite{Ransom_DC704} demonstrated that power-law density scaling breakdown for DC704. The data cannot be collapsed using the literature value of $\gamma =6.2$  \cite{GundermannDitte2011Ptde}, figure \ref{fig:Ransom_gamma_62}, and even with a best fit value of $\gamma =5.13$ the relaxation times do not collapse, figure \ref{fig:Ransom_gamma_513}.}
	\end{figure}

In figures \ref{fig:Ransom_gamma_62} and \ref{fig:Ransom_gamma_513} the breakdown of density scaling is shown reproducing the data from \Mycite{Ransom_DC704}. Density scaling coefficients of $ \gamma=6.1$ and $\gamma = 6.2$ have been previously reported in ref.  \cite{AdrjanowiczKarolina2017PNDo,GundermannDitte2011Ptde}. In figure \ref{fig:Ransom_gamma_62} $\gamma = 6.2$ can collapse the low temperature data quite well, but the breakdown of power-law density scaling occurs at high temperature. The experimental temperature ranges in  ref.  \cite{AdrjanowiczKarolina2017PNDo,GundermannDitte2011Ptde} are similar to the low temperature measurements by \Mycite{Ransom_DC704}.

\subsection{Isochronal density scaling for DC704}
	
\begin{figure}[H]
		\centering
		\begin{subfigure}[t]{0.48\textwidth}
			
			\includegraphics[width=\textwidth]{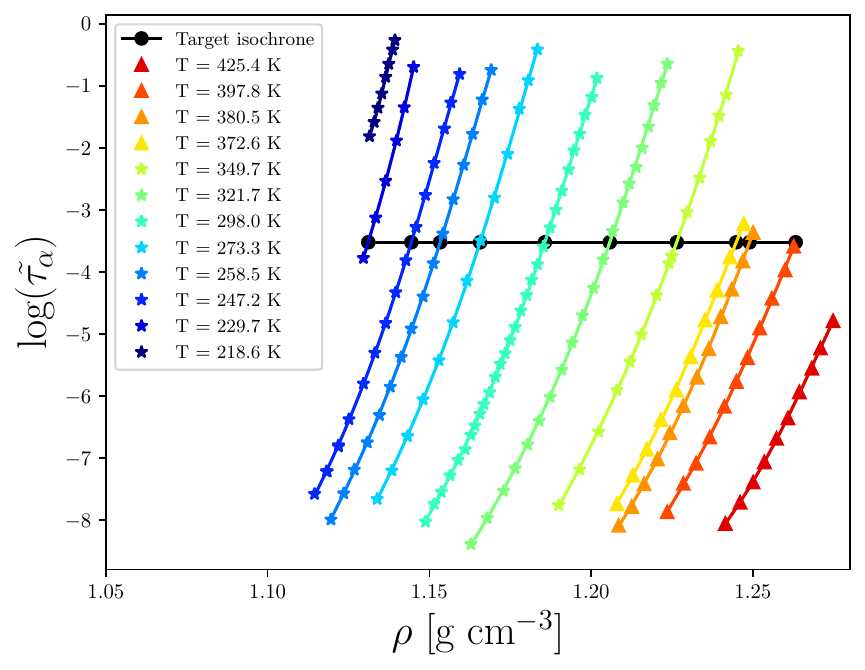}
			\caption{ $ \tilde{\tau}_{\text{target}}(\rho,T)= 3 \times 10^{-4}$}
			\label{fig:DC704_isochrone_show_a}
		\end{subfigure}\hfill
		\begin{subfigure}[t]{0.48\textwidth}
			
			\includegraphics[width=\textwidth]{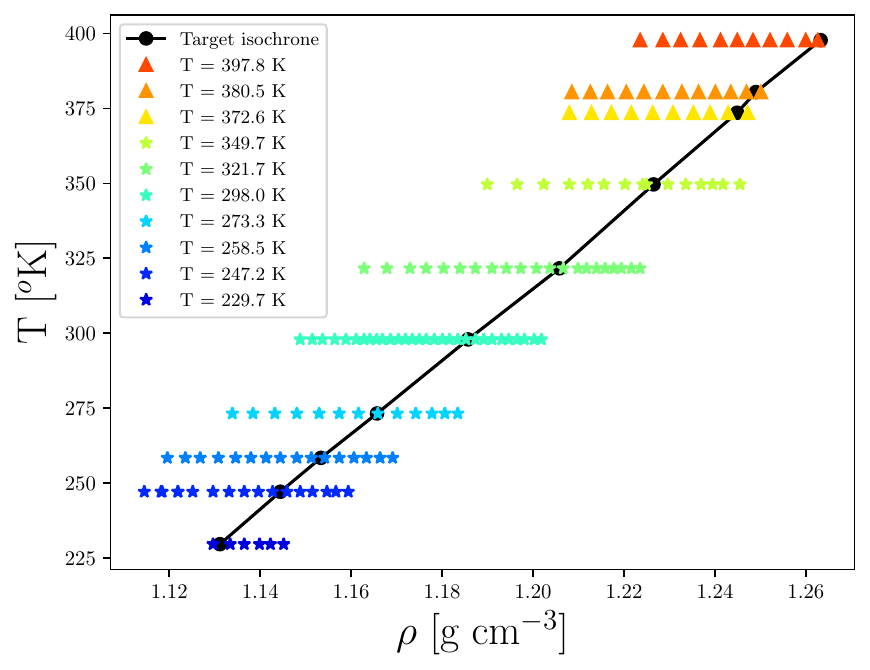}
			\caption{$ \tilde{\tau}_{\text{target}}(\rho,T)= 3 \times 10^{-4}$ }
			\label{fig:DC704_isochrone_show_b}
		\end{subfigure}	
		\caption{ The measurements of the relaxation time for DC704 originally published by \Mycite{Ransom_DC704}.
			In figure \ref{fig:DC704_isochrone_show_a}, the measured relaxation time, in reduced units $\tilde{\tau}_\alpha = \tau_\alpha T^{-\frac{1}{2}} \rho^{-\frac{1}{3}} $, along each isotherm was fitted with a third order polynomial to interpolate the density at the target isochrone. There is no overlap with the target isochrone and the  $ T= 425.4$ K  and $T = 218.6$ K isotherms. When we test isochronal density scaling, we exclude these measurement series. The black line is the target isochrone, $ \tilde{\tau}_{\text{target}}(\rho,T)$ that we scale with. In figure \ref{fig:DC704_isochrone_show_b}, the position of each measurement is shown in the $(\rho,T)$-phase diagram. The black line is the target isochrone used for isochronal density scaling. The $ T= 425.4$ K  and $T = 218.6$ K isotherms are excluded from the figure because there is no overlap with the target isochrone. }
		\label{fig:DC704_isochrone_show}
	\end{figure}

	In \Mycite{Ransom_DC704} the relaxation time is measured along isotherms, making it relatively easy to test if isochronal density scaling can collapse the data. We do not have to interpolated to calculate $\rho_{\text{target}}(T)$ since the relaxation time is measured along isotherms. In figure \ref{fig:DC704_isochrone_show_a} all the measured relaxation times from \Mycite{Ransom_DC704} are shown as a function of density. The density used in this section are from the equation of state made in \Mycite{Ransom_DC704}. This is also presented in section \ref{sec:DC704_background}. We selected a target relaxation time of $ \tilde{\tau}_{\text{target}}(\rho,T)= 3 \times 10^{-4}$, however the scaling should work for any choice of target isochrone. To obtain the exact $(\rho,T)$ values for the isochrone, each measurement series was fitted with a third order polynomial to interpolate the density at $ \tilde{\tau}_{\text{target}}$. In figure \ref{fig:DC704_isochrone_show_b} all measurements are shown in the $(\rho,T)$ phase diagram along with the target isochrone. 
	
\begin{figure}[H]
		\centering
		\includegraphics[width=0.75\textwidth]{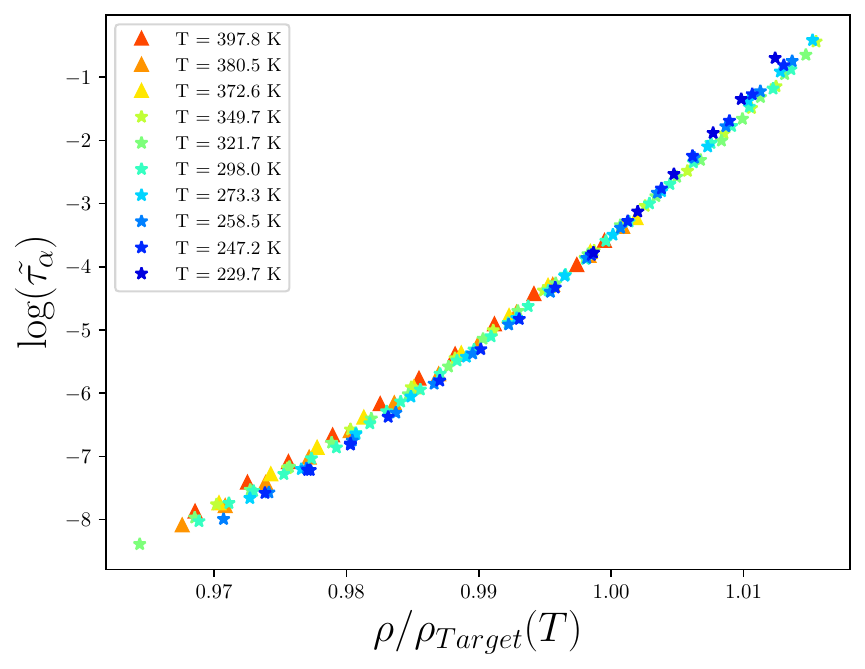}
		
		\caption{The reduced relaxation time plotted against $\rho / \rho_{\text{target}}(T)$. The relaxation time can be collapsed using isochronal density scaling, using the target isochrone $\tilde{\tau}_{target} = 3 \cdot 10^{-4}$.  The reduced relaxation times collapsed much better than with $\Gamma$, see fig. \ref{fig:Ransom_gamma_62} and \ref{fig:Ransom_gamma_513}. 
			Isochronal density scaling only use the data itself to obtain the data collapse.	The strength of isochronal density scaling is that the scaling is parameter free and makes no assumptions about the shape of the isochrones.  }
		\label{fig:DC704_Isochronal_density_scaling_collapse}
	\end{figure}

\begin{figure}[H]
		\centering
		\begin{subfigure}[t]{0.48\textwidth}
			\centering
			\includegraphics[width=0.98\textwidth]{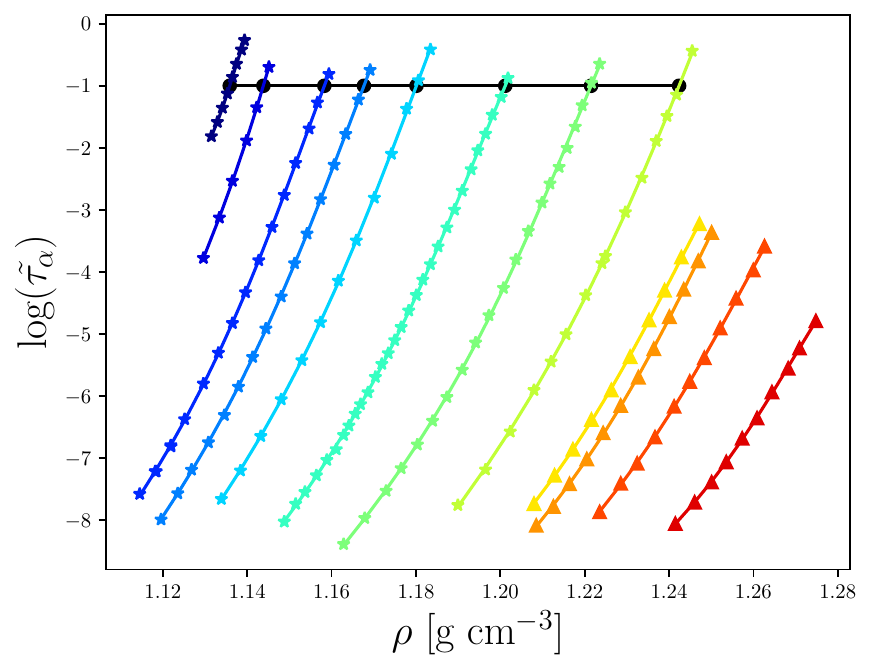}
			\caption{ The reduced relaxation time plotted against $\rho$. The target isochrone of $\tilde{\tau}_{target}$ = 0.1 is shown in black.}
		\end{subfigure}\hfill
		\begin{subfigure}[t]{0.48\textwidth}
			\centering
			\includegraphics[width=0.98\textwidth]{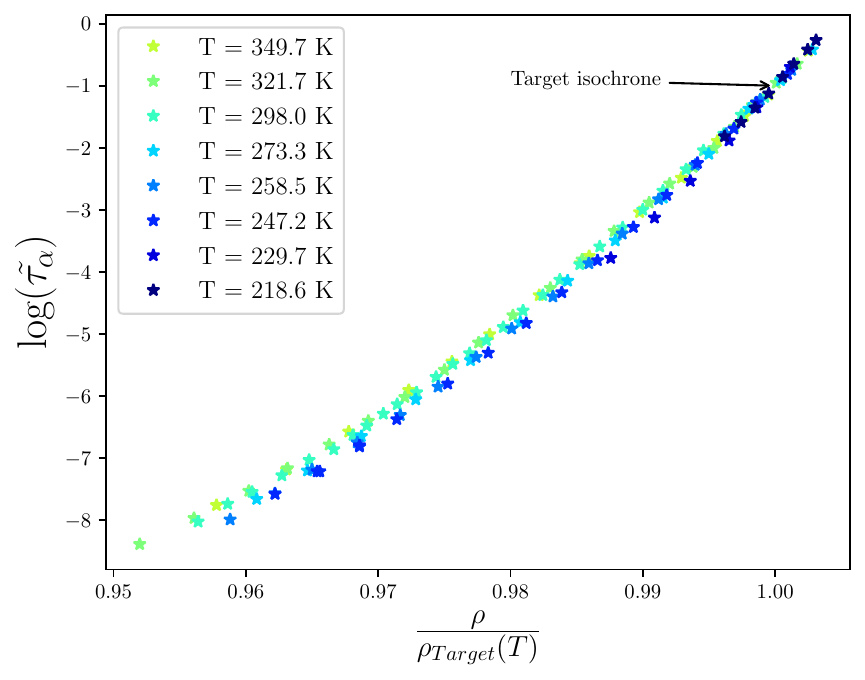}
			\caption{ The reduced relaxation time plotted against $\rho / \rho_{\text{target}}(T) $, for a target isochrone of $\tilde{\tau}_{target}$ = 0.1 . }
		\end{subfigure}
		\begin{subfigure}[t]{0.48\textwidth}
			\centering
			\includegraphics[width=0.98\textwidth]{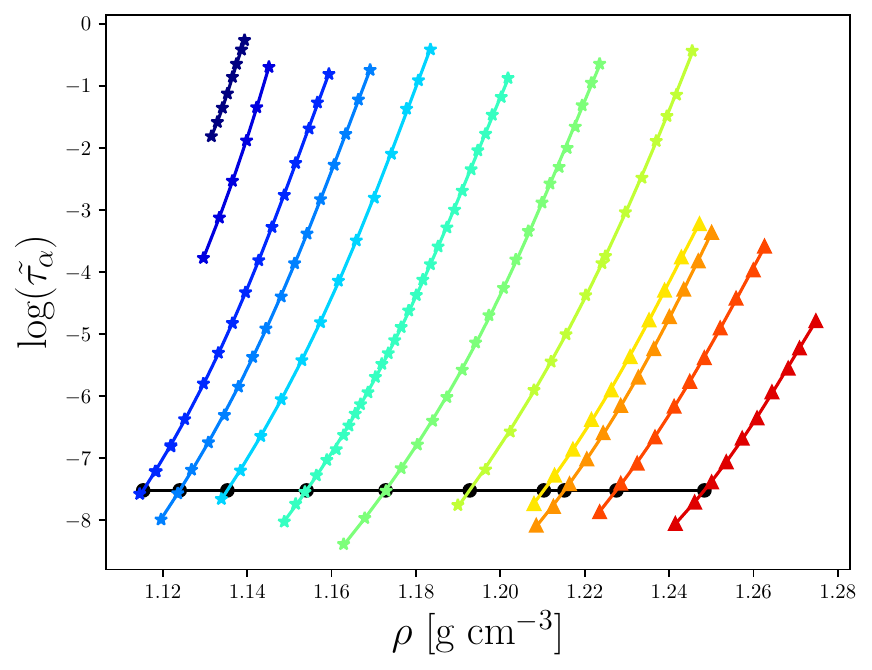}
			\caption{The reduced relaxation time plotted against $\rho$. The target isochrone of $\tilde{\tau}_{target}$ =  3 $\cdot 10^{-8}$ is shown in black.}
		\end{subfigure}\hfill
		\begin{subfigure}[t]{0.48\textwidth}
			\centering
			\includegraphics[width=0.98\textwidth]{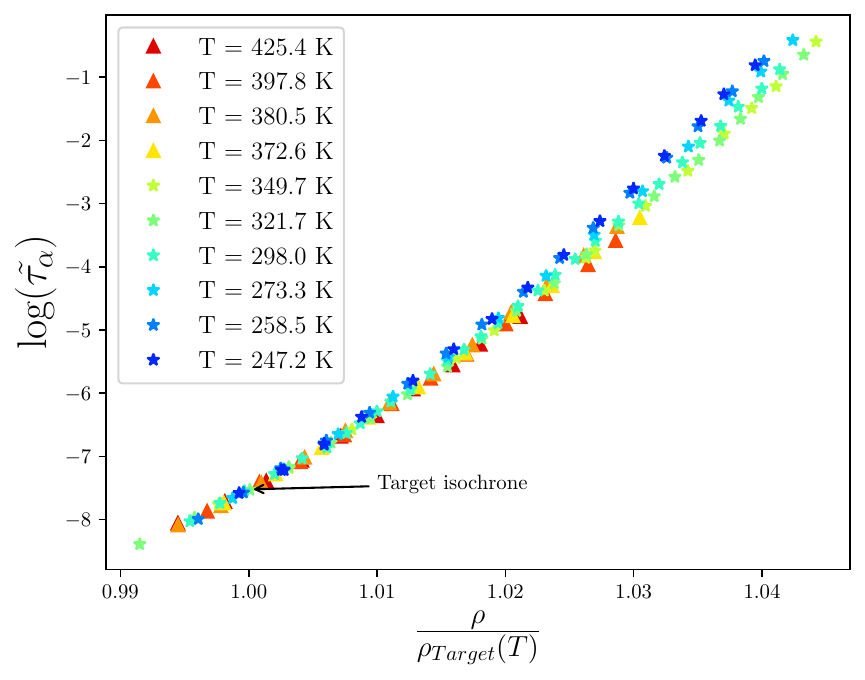}
			\caption{ The reduced relaxation time plotted against $\rho / \rho_{\text{target}}(T) $, for a target isochrone of $\tilde{\tau}_{target}$ = 3 $\cdot 10^{-8}$ }
		\end{subfigure}	
		
		\caption{In this figure the independence of the choice of target isochrone is tested for the DC704 data from \Mycite{Ransom_DC704}. In figures (a) and (b), an isochrone in the viscous regime is chosen, while in figures (c) and (d), a more fluid isochrone is chosen. On the left side, the target isochrone is plotted along with the data. On the right side the isochronal density scaling results are shown. Any measurement series not overlapping with the target isochrone is excluded. In general, isochronal density scaling works quite well regardless of the choice of target isochrone, however, it seems to work slightly better when choosing the target isochrone in the middle of the data.   }
		\label{fig:Isochronal_density_scaling_dif_isochrones}
	\end{figure}

	In figure \ref{fig:DC704_Isochronal_density_scaling_collapse} the reduced relaxation time is plotted against $\rho / \rho_{\text{target}}(T)$, for the target isochrone at $\tilde{\tau}_{target} = 3\cdot 10^{-4}$. The $ T= 425.4$ K  and $T = 218.6$ K isotherms are excluded because there is no overlap with the target isochrone. The measurements of relaxation time, collapse onto a single line using isochronal density scaling. It is worth noting that for the measured relaxation times to collapse we have only assumed that the shape of the isochrone is independent of density. The only requirement for the collapse is knowing $\rho_{\text{target}}(T)$ along a target isochrone, and $\rho$ at each state point, both of which can be measured directly. \Mycite{Ransom_DC704} have measure along isotherms so  $\rho_{\text{target}}(T)$ can found without interpolation. The collapse should be independent of the choice of the target isochrone. This is tested in figure \ref{fig:Isochronal_density_scaling_dif_isochrones}, where we use isochronal density scaling on two target isochrones at the extremes of the measured relaxation times. The scaling works very well regardless of the choice of target isochrones; however, the collapse seems to be best for an isochrone in the middle of the data set. For the derivation of isochronal density scaling, we assume that $\gamma$ is independent of density. When choosing the target isochrone in the middle of the measurements, this assumption is stronger. If $\gamma$ changes a little with density then, we are not changing the density enough to see an effect.
	
\begin{figure}[H]
		
		\begin{subfigure}[t]{0.48\textwidth}	
			\centering
			\includegraphics[width=0.98\textwidth]{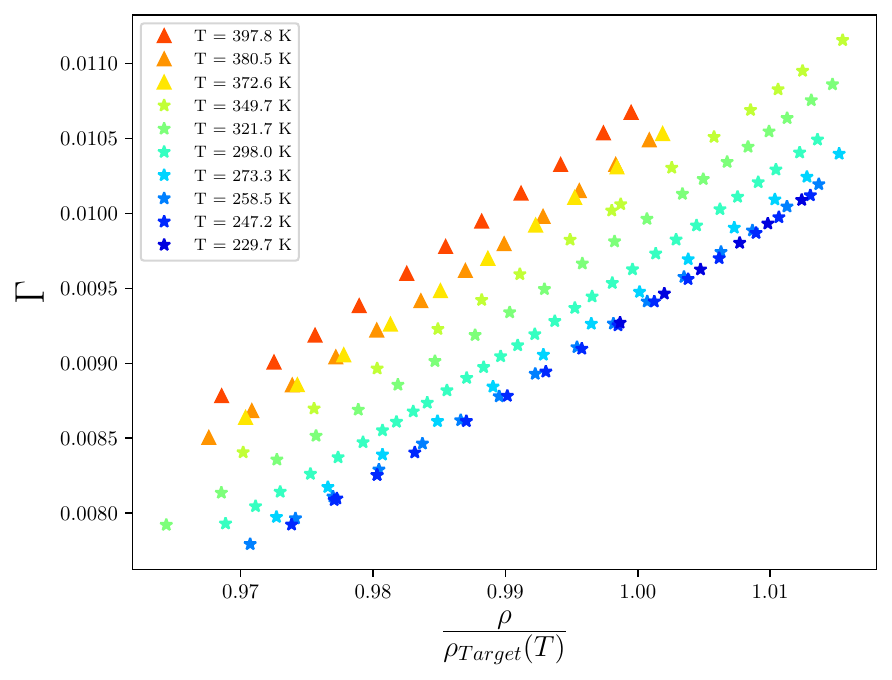}
			\caption{$\Gamma = \frac{\rho^{6.2}}{T}$ plotted against  $\rho / \rho_{\text{target}}(T) $ for every state point.}
		\end{subfigure}\hfill
		\begin{subfigure}[t]{0.48\textwidth}		
			\centering
			\includegraphics[width=0.98\textwidth]{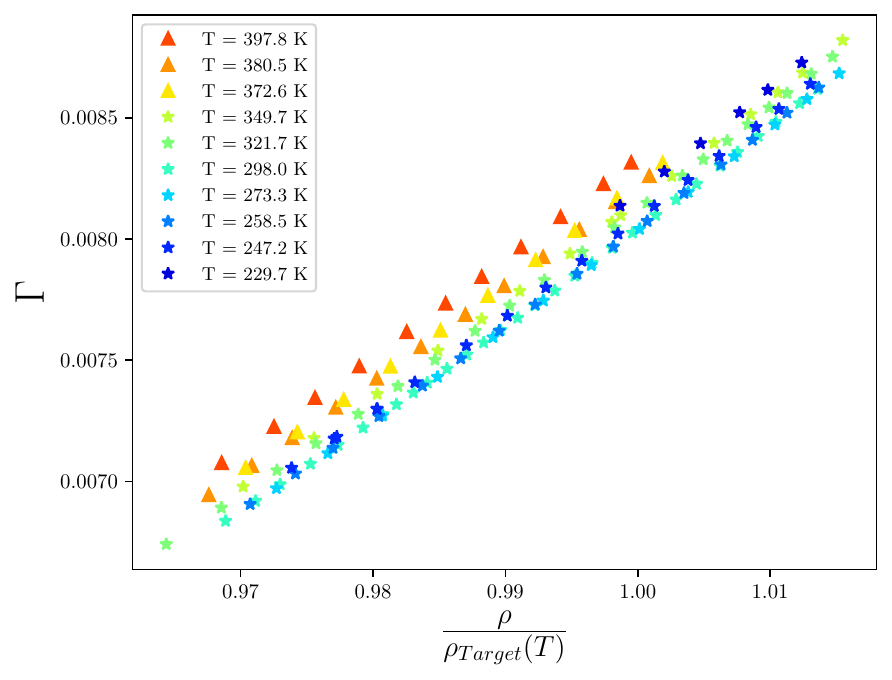}
			\caption{$\Gamma = \frac{\rho^{5.13}}{T}$ plotted against $\rho / \rho_{\text{target}}(T) $  for every state point.}
			
		\end{subfigure}
		\caption{Comparison between $\Gamma$ shown in figures \ref{fig:Ransom_gamma_62} and \ref{fig:Ransom_gamma_513}, and  $\rho / \rho_{\text{target}}(T) $ for DC704. }
		\label{fig:Isochrone_density_scaling_Gamma_compare}
	\end{figure}

	The previous figures have shown that isochronal density scaling can cause the dynamics of DC704 to collapse into a single curve, even when power-law density scaling with a constant $\gamma$ breaks down. In figure \ref{fig:Isochrone_density_scaling_Gamma_compare} the $\Gamma$ value for each state point is compared with the value of $\rho / \rho_{\text{target}}(T)$. Being able to scale the data using isochronal density scaling is by itself a very interesting result. Isochronal density scaling assumes that $\gamma$ is independent of the density, and only a function of temperature. The origin of the breakdown of power-law density is most likely that $\gamma$ is state point dependent. However assuming that $\gamma$ is independent of density gives a good collapse, this indicates that it is the temperature dependence of $\gamma$ that cause the breakdown of power-law density scaling. If it the temperature dependence of $\gamma$ causing the breakdown of power-law density scaling, then isochronal temperature scaling should not be able to collapse the data.
	

	\subsection{Testing isochronal temperature scaling for DC704}

\begin{figure}[H]
		
		\begin{subfigure}[t]{0.49\textwidth}	
			\centering
			\includegraphics[width=0.99\textwidth]{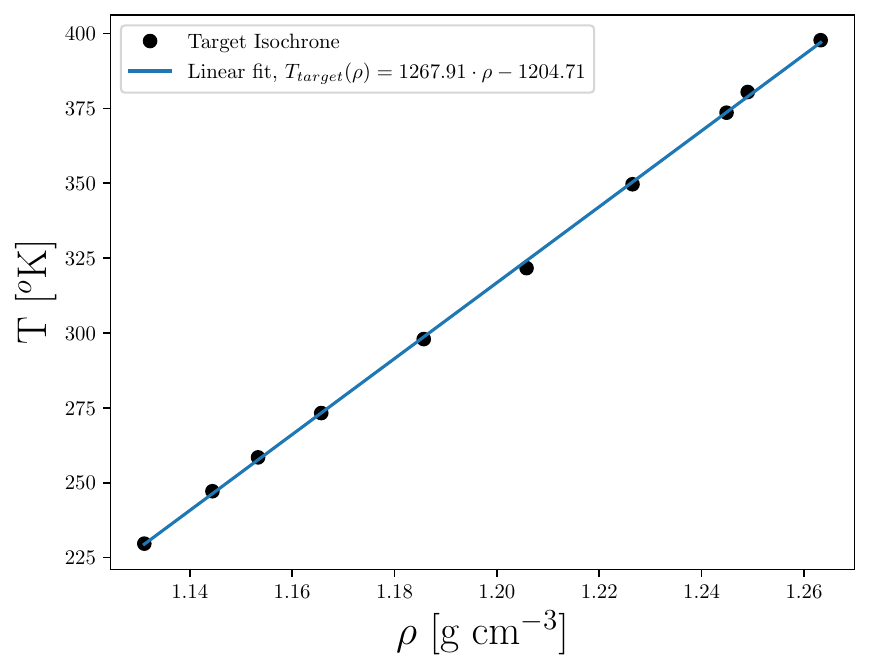}
			\caption{A linear fit to $T_{target}(\rho)$ for the state points along the target isochrone.}
			\label{fig:DC704_Isochronal_density_scaling_T_a}
		\end{subfigure}\hfill
		\begin{subfigure}[t]{0.49\textwidth}		
			\centering
			\includegraphics[width=0.99\textwidth]{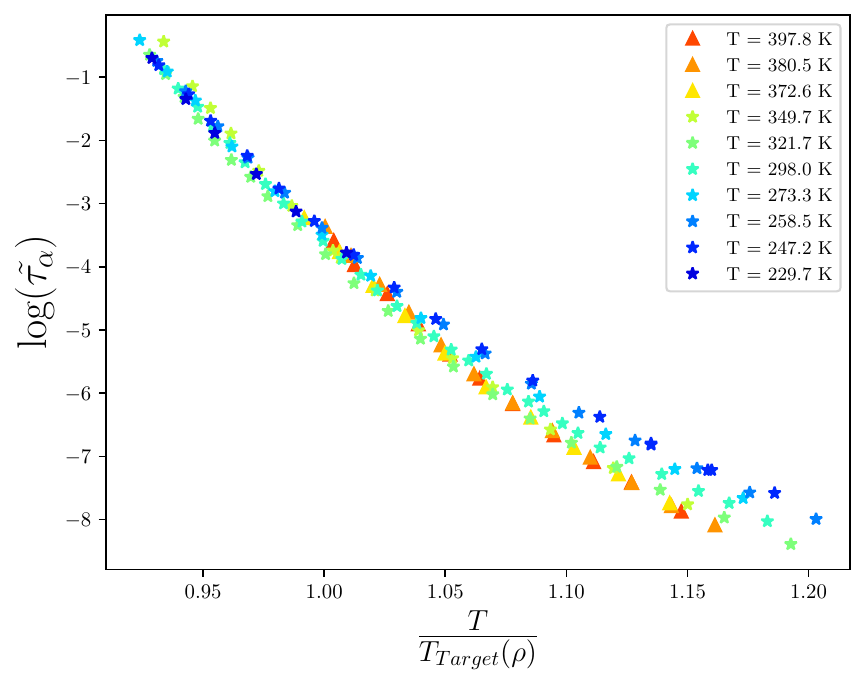}
			\caption{The reduced relaxation time plotted against $T / T_{\text{target}}(\rho)$, for the same data as in figure \ref{fig:DC704_Isochronal_density_scaling_collapse}}
			\label{fig:DC704_Isochronal_density_scaling_T_b}
		\end{subfigure}
		\caption{Test of isochronal temperature scaling, $T / T_{target}(\rho)$ for literature data of DC704\cite{Ransom_DC704}. The data are measured along isotherms. In order to find the target isochrone, we have interpolated between the measurements to find $T_{target}(\rho)$. We have fitted the isochrone using a straight line. The fit is shown in figure \ref{fig:DC704_Isochronal_density_scaling_T_a}. In figure \ref{fig:DC704_Isochronal_density_scaling_T_b} the reduced relaxation time is plotted against the scaling parameter $T / T_{target}(\rho)$. The data collapse using $\rho / \rho_{\text{target}}(T)$ performs better, see figure \ref{fig:DC704_Isochronal_density_scaling_collapse}. This indicates that it is the temperature dependence of $\gamma$ that cause the breakdown of power-law density scaling. It is also worth noting that at higher values of  $T / T_{target}(\rho)$, some data points were extrapolated onto the  $T_{target}(\rho)$ line, which might partially justify the worst collapse. This can be seen clearly in figure \ref{fig:DC704_isochrone_show_b}. }
		\label{fig:DC704_Isochronal_density_scaling_T}
	\end{figure}

We have introduced two new scalings, isochronal density scaling, $\rho  / \rho_{\text{target}}(T)$, and isochronal temperature scaling,  $T / T_{target}(\rho)$. So far we have focused on isochronal density scaling, but with the measurements of \Mycite{Ransom_DC704} we can test isochronal temperature scaling as well. This is shown in figure \ref{fig:DC704_Isochronal_density_scaling_T}. The \Mycite{Ransom_DC704} data were measured along isotherms; thus, to calculate $T_{target}(\rho)$ we have fitted linearly the state points along the target isochrone to interpolate $T_{target}(\rho)$ at a given density. As a first approach, the linear relationship works quite well, as shown in figure \ref{fig:DC704_Isochronal_density_scaling_T_a}, however, one should be careful when extrapolating outside the data range. 


In figure \ref{fig:DC704_Isochronal_density_scaling_T_b} the reduced relaxation time is plotted against $T / T_{target}(\rho)$. The data collapse at $T / T_{target}(\rho) = 1$, but this is also by design. At higher values of $T / T_{target}(\rho)$ there seem to be some scatter in the collapse. Isochronal density scaling collapses the data better than isochronal temperature scaling. This would indicate that it is the temperature dependence of $\gamma$ causing the breakdown of power-law densitys scaling. It is also worth noting that at higher values of  $T / T_{target}(\rho)$, some data points were extrapolated onto the  $T_{target}(\rho)$ line. This can be seen clearly in figure \ref{fig:DC704_isochrone_show_b}. 

Isochronal density scaling were able to collapse the data into a single curve, while isochronal temperature scaling were not able to reproduce a similar collapse. The breakdown of power-law density scaling, arises from the state point dependence of $\gamma$, and it is most likely the temperature dependence of $\gamma$ being the main contribution factor. 

	
For the \Mycite{Ransom_DC704} data we have tested both isochronal density scaling, and isochronal temperature scaling. There are a couple of practical advantages using isochronal density scaling compared to isochronal temperature scaling. First, the data are measured along isotherms making it possible to do isochronal density scaling without making any assumption about the functional form of the isochrone. Temperature is constant so $\rho_{\text{target}}(T)$, is simply the density at the target relaxation time along the measurement series. It is also possible to do the same for isochronal temperature scaling, if one measures along isochores \cite{Lorenzo2018,Lorenzo2019}. Experimentally it is easier to measure at constant temperature, than constant density. A second advantage of isochronal density scaling is that finding $\rho_{\text{target}}(T)$ is easier than finding $T_{\text{target}}(\rho)$, because $T_{\text{target}}(\rho)$ can correspond to nonphysical negative pressures. This is clear for the cumene data, section \ref{sec:IDS_cumene} where the target isochrone is the glass transition line. For many of the state points we have viscosity data for cumene, $T_{\text{target}}(\rho)$ would be at unphysical negative pressures.
	
For DC704 \Mycite{Ransom_DC704} showed that power-law density scaling breaks down when measured over a large temperature and pressure range. Power-law density scaling assumes that $\gamma$ is independent of both density and temperature. The quality of this assumption is extremely dependent on the experimental range of both temperatures and densities. The relevance of the experimental range is captured by comparing isochronal density scaling and isochronal temperature scaling, simply by looking at the values of $T / T_{target}(\rho)$ and $\rho / \rho_{target}(T)$. In comparison, the experimental range of temperature is larger than the experimental range of densities. The breakdown of power-law density scaling is in this case, not something that could be fixed by the original formulation of density scaling, $e(\rho) / T$. In the previous section, we have used $\Gamma$ to find isochrones; however, with the results presented in this section in mind, the structural results need to be reexamined.


	\subsection{Revisiting the structural measurements}
\begin{figure}[H]
		\centering
		\begin{subfigure}[t]{0.49\textwidth}
			\centering
			\includegraphics[width=0.95\textwidth]{Chapters/Figures/Gamma_vs_1st_qmax_scaled_bck.pdf}
			\caption{Reproduction of figure	\ref{fig:DC704_1st_fit_d}. The position of the first peak of DC704 plotted against $\Gamma$, with $\gamma=6.2$.  }
			\label{fig:DC704_ICS_structure_a}
		\end{subfigure}\hfill
		\begin{subfigure}[t]{0.49\textwidth}
			\centering
			\includegraphics[width=0.95\textwidth]{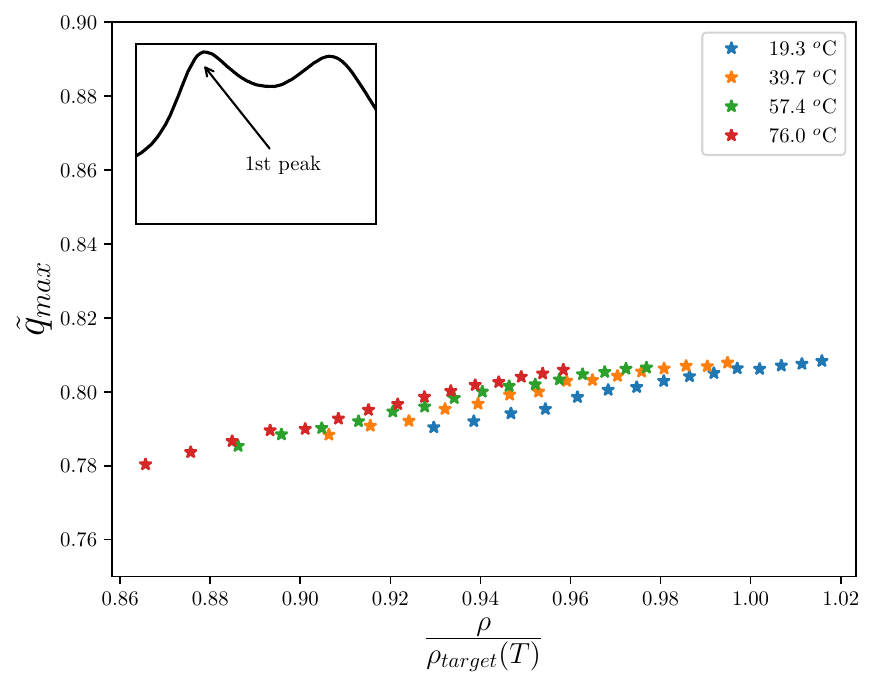}
			\caption{The position of the first peak of DC704 is plotted against $\rho / \rho_{\text{target}}(T)$, for the target isochrone at $\tilde{\tau}_{target} = 3 \cdot 10^{-4}$ }
			\label{fig:DC704_ICS_structure_b}
		\end{subfigure}
		\begin{subfigure}[t]{0.49\textwidth}
			\centering
			\includegraphics[width=0.95\textwidth]{Chapters/Figures/Gamma_vs_2nd_qmax_scaled_bck.pdf}
			\caption{Reproduction of figure \ref{fig:DC704_2nd_fit_d}. The position of the second peak of DC704 plotted against $\Gamma$, with $\gamma=6.2$. }
			\label{fig:DC704_ICS_structure_c}
		\end{subfigure}\hfill
		\begin{subfigure}[t]{0.49\textwidth}
			\centering
			\includegraphics[width=0.95\textwidth]{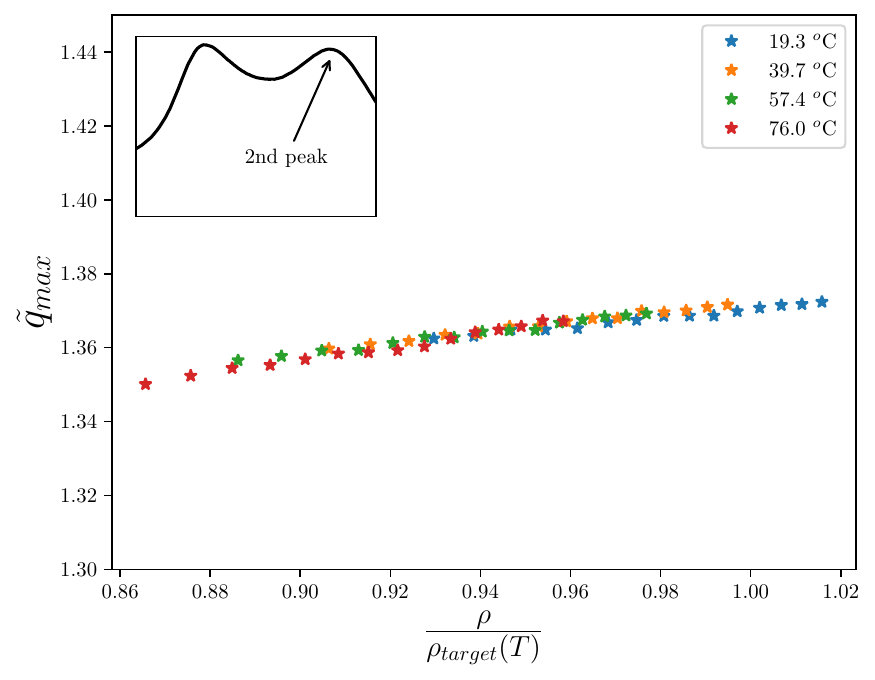}
			\caption{The position of the second peak of DC704 plotted against $\rho / \rho_{\text{target}}(T)$, for the target isochrone at $\tilde{\tau}_{target} = 3 \cdot 10^{-4}$ }
			\label{fig:DC704_ICS_structure_d}
		\end{subfigure}	
		\begin{subfigure}[t]{0.49\textwidth}
			\centering
			\includegraphics[width=0.95\textwidth]{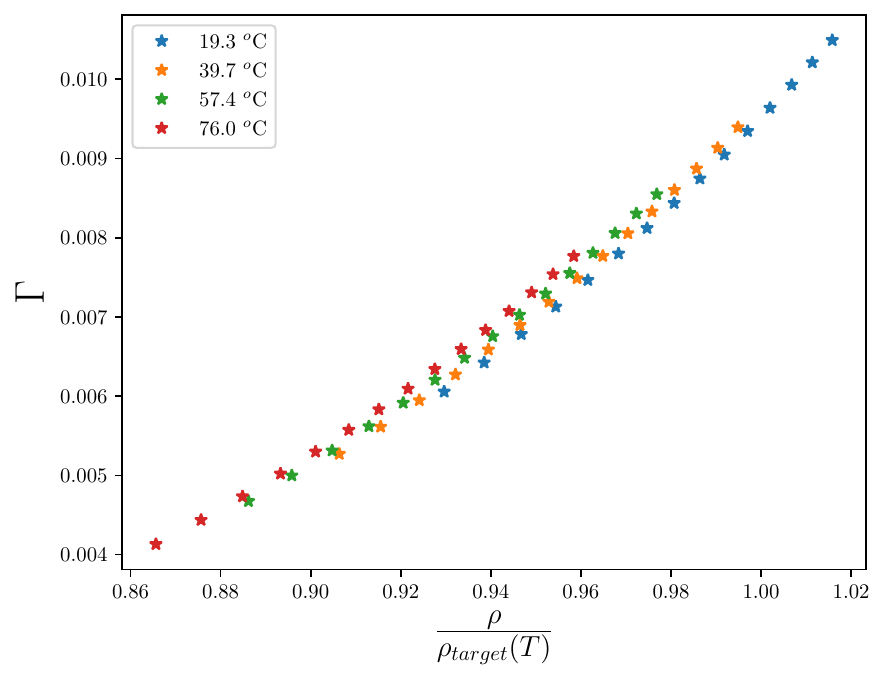}
			\caption{ $\Gamma$ plotted against $\rho / \rho_{\text{target}}(T)$. }
			\label{fig:DC704_ICS_structure_e}
		\end{subfigure}\hfill
		\begin{subfigure}[t]{0.49\textwidth}
			\centering
			\includegraphics[width=0.95\textwidth]{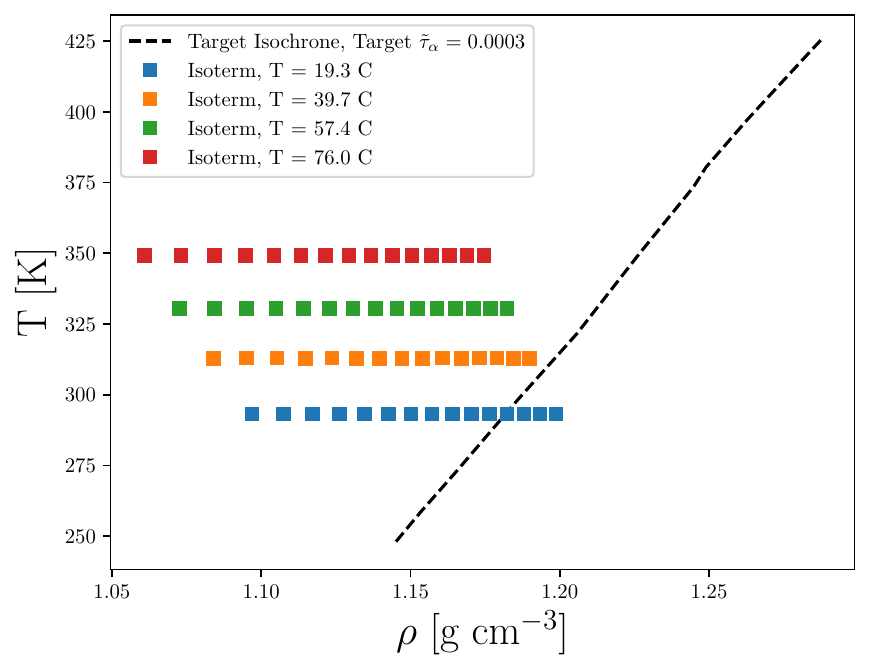}
			\caption{The measured state points in the $(\rho,T)$ phase diagram. The target isochrone used for the scaling is shoen as dashed black line. }
			\label{fig:DC704_ICS_structure_f}
		\end{subfigure}	
		
		\caption{Revisiting the structural measurements of DC704, originally presented in chapter \ref{chapter:Diamond}. Figure \ref{fig:DC704_ICS_structure_a} and \ref{fig:DC704_ICS_structure_c} shows the position of the first and second peak, using $\Gamma$ to identify isochrones. Figure \ref{fig:DC704_ICS_structure_b} and \ref{fig:DC704_ICS_structure_d} shows  the position of the first and second peak, using isochronal density scaling to identify isochrones. The second peak of DC704 collapse as a function of both $\Gamma$ and $\rho / \rho_{\text{target}}(T)$. Figure \ref{fig:DC704_ICS_structure_e} show the $\Gamma$ as a function of $\rho / \rho_{\text{target}}(T)$. In our experimental range there are only minor differences between $\Gamma$ and $\rho / \rho_{\text{target}}(T)$.
		The structural changes when moving from isochrone to isochrone are small for DC704, and any error arising from not being able to identify isochrones correctly is therefore negligible.  }
		\label{fig:DC704_ICS_structure}
	\end{figure}

	In figure \ref{fig:DC704_ICS_structure} the results obtained from fitting the two peaks of DC704 originally presented in chapter \ref{chapter:Diamond} are revisited.  In subfigure \ref{fig:DC704_ICS_structure_a} and  \ref{fig:DC704_ICS_structure_c} the peak positions of the first and second peak of DC704 are plotted against $\Gamma = \frac{\rho^{6.2}}{T}$. These are reproductions of figures \ref{fig:DC704_1st_fit_d} and \ref{fig:DC704_2nd_fit_d}. In figures \ref{fig:DC704_ICS_structure_b} and  \ref{fig:DC704_ICS_structure_d} the peak positions are plotted against 
	$\rho / \rho_{\text{target}}(T)$ for the target isochrone $\tilde{\tau}_{target} = 3 \cdot 10^{-4}$. The results obtained using $\Gamma$ reproduce well with isochronal density scaling. By eye, the collapse for the second peak position seems slightly better for $\rho / \rho_{\text{target}}(T)$, compared with $\Gamma$ but the differences are small. For the first peak the behavior seemed to be the opposite, the differences for each measurement series seemed worse for $\rho / \rho_{\text{target}}(T)$. In chapter \ref{chapter:Cumene} we discussed that the first peak of DC704 is like affected more by the scaling deviation from the intramolecular structure than the second peak. The scaling deviation arises when comparing measurement at different densities.	In figure \ref{fig:DC704_ICS_structure_e}, $\rho / \rho_{\text{target}}(T)$ is compared with $\Gamma$. The differences between $\Gamma$ and $\rho / \rho_{\text{target}}(T)$, while not as large as for the dynamical measurements (see figure	\ref{fig:Isochrone_density_scaling_Gamma_compare}), are still visible.
	Why do the results of the two approaches reproduce relatively well, when there are differences between $\Gamma$ and $\rho / \rho_{\text{target}}(T)$? When scaling out density's effect on the structure, the structural changes observed are small. The difference between $\Gamma$ and $\rho / \rho_{\text{target}}(T)$ does not play a significant role because that structural change when moving from isochrone to isochrone is small an so the error resulting from using $\Gamma$ to identify which isochrone each state point is on is barely notable. 
	
	In figure \ref{fig:DC704_ICS_structure_f} each state point is shown in the $(\rho,T)$ phase diagram. It is worth noting that  in the experimental range of the structure measurements the change in the relaxation time is quite large. The structural changes were small in comparison. The structural measurements were also relatively far from the target isochrone compared to the dynamical measurements. For the dynamics, there is some scatter when moving further away from the target isochrone, see figure 	\ref{fig:Isochronal_density_scaling_dif_isochrones}. However, any error in identifying isochrones does not really affect the structural collapse simply because the structure changes so little.
	
	\section{DPG}
	
\begin{figure}[H]
		\begin{subfigure}[t]{0.49\textwidth}
			\centering
			\includegraphics[width=0.99\textwidth]{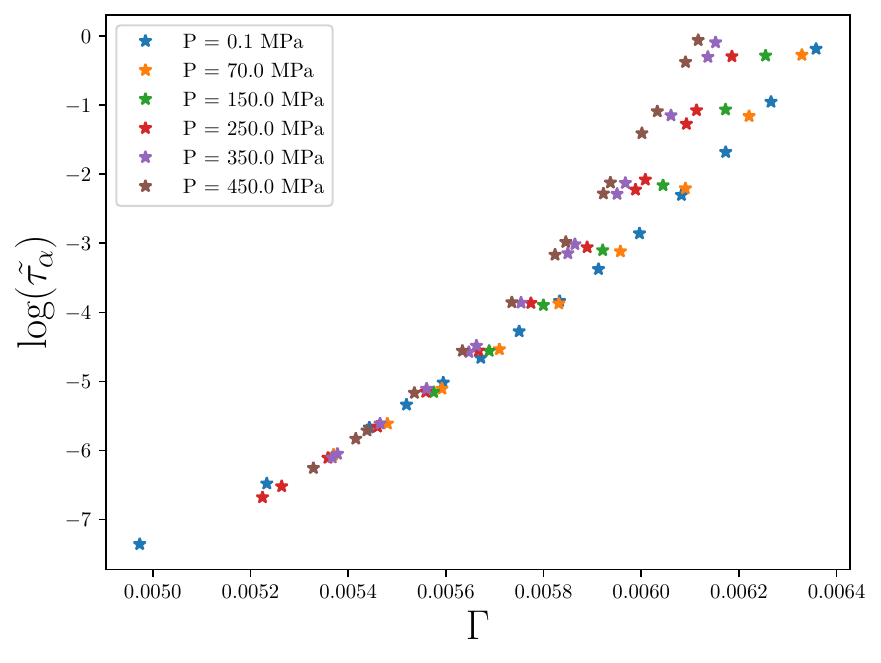}
			\caption{ Recreation of the inset of figure 2 in ref. \cite{ChatKatarzyna2019Tdsi}. The dimensionless relaxation time $\tilde{\tau}_\alpha = \tau_\alpha T^{-\frac{1}{2}} \rho^{-\frac{1}{3}} $ is plotted against $\Gamma$, for $\gamma =1.5$. }
			\label{fig:DPG_Density_Scaling_Collapse_15}
		\end{subfigure}\hfill
		\begin{subfigure}[t]{0.49\textwidth}
			\centering
			\includegraphics[width=0.99\textwidth]{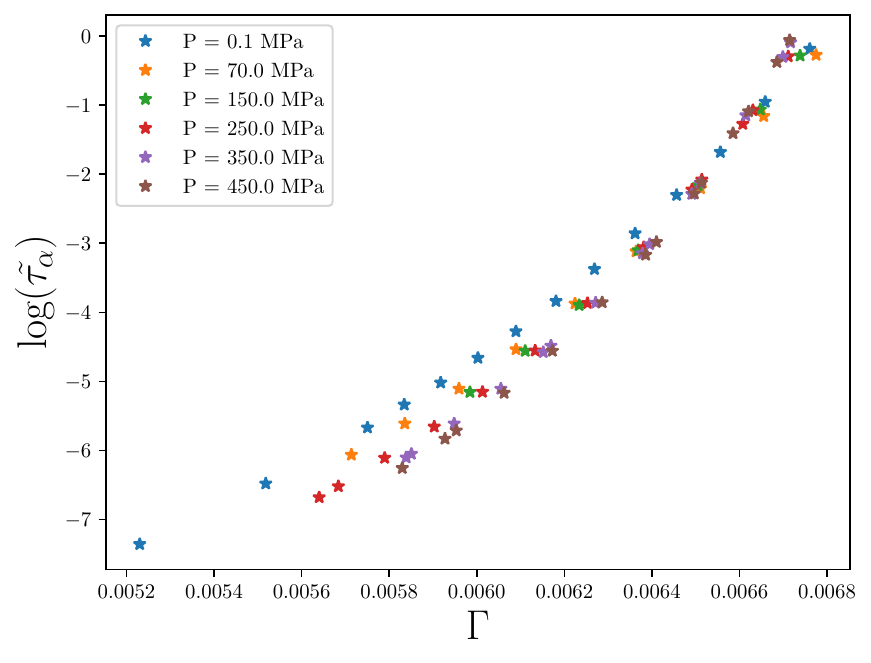}
			\caption{Recreation of figure 2 in ref.\cite{ChatKatarzyna2019Tdsi} The dimensionless relaxation time $\tilde{\tau}_\alpha = \tau_\alpha T^{-\frac{1}{2}} \rho^{-\frac{1}{3}} $ is plotted against $\Gamma$, for $\gamma =1.9$}
			\label{fig:DPG_Density_Scaling_Collapse_19}
		\end{subfigure}	
		
		\caption{Breakdown of power-law density scaling for DPG. Recreation of figure 2 in ref.\cite{ChatKatarzyna2019Tdsi}, showing the breakdown of power-law density scaling for DPG. }
		\label{fig:DPG_Density_Scaling_Collapse}
	\end{figure}

	In chapter \ref{chapter:Diamond}, DPG was tested for pseudo-isomorphs and we found that DPG does not have pseudo-isomorphs. The isomorph theory is not expected to work for hydrogen bonded liquids \cite{Ingebrigtsen2012_simple_liquid}, so it is unlikely that isochronal temperature scaling or isochronal density scaling would scale the structure and dynamics of DPG. As mentioned in section \ref{sec:Diamond_samples}, there has been evidence of a breakdown of power-law density scaling for DPG \cite{ChatKatarzyna2019Tdsi}. Having the results for DC704 in mind, revisiting the DPG data with isochronal density scaling seems anyway obvious. Figure \ref{fig:DPG_Density_Scaling_Collapse} shows the reduced relaxation times as a function of $\Gamma$. The power-law density scaling exponent $\gamma$ have previously been reported in litterature to be $\gamma = 1.5$ \cite{Wase2018} and $\gamma = 1.9$ \cite{ChatKatarzyna2019Tdsi}.  Figure \ref{fig:DPG_Density_Scaling_Collapse} is a recreation of figure 2 in \Mycite{ChatKatarzyna2019Tdsi} who shows that for a literature value of $\gamma =1.5$, it is not possible to collapse the measured relaxation times into a single curve. They found a best-fit value of $\gamma = 1.9$; however, even with this best-fit value the relaxations times did not collapse very well.
	
\begin{figure}[H]
		\begin{subfigure}[t]{0.48\textwidth}
			\centering
			\includegraphics[width=\textwidth]{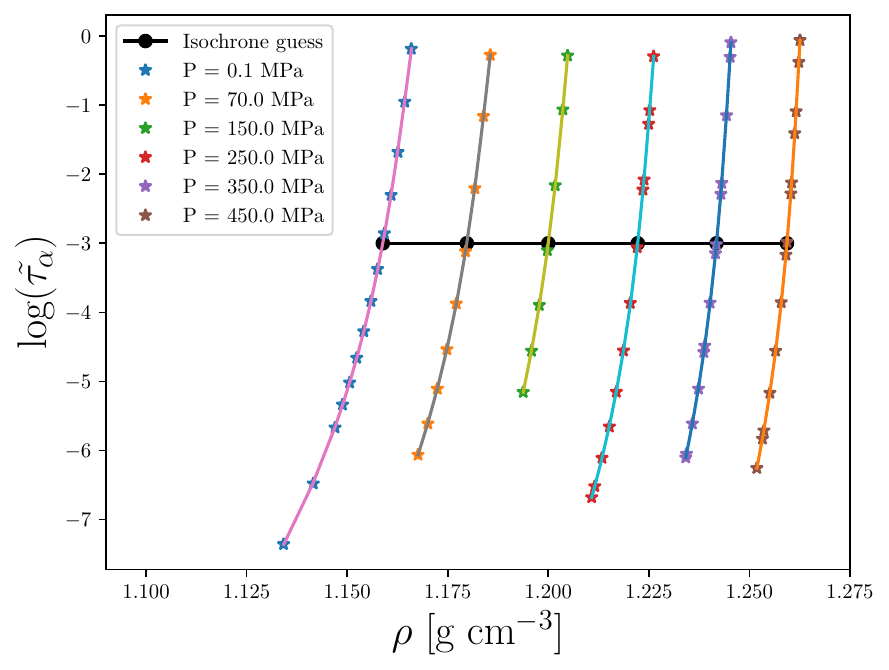}
			\caption{}
			\label{fig:DPG_target_isochrone_a}
		\end{subfigure}\hfill
		\begin{subfigure}[t]{0.48\textwidth}
			\centering
			\includegraphics[width=\textwidth]{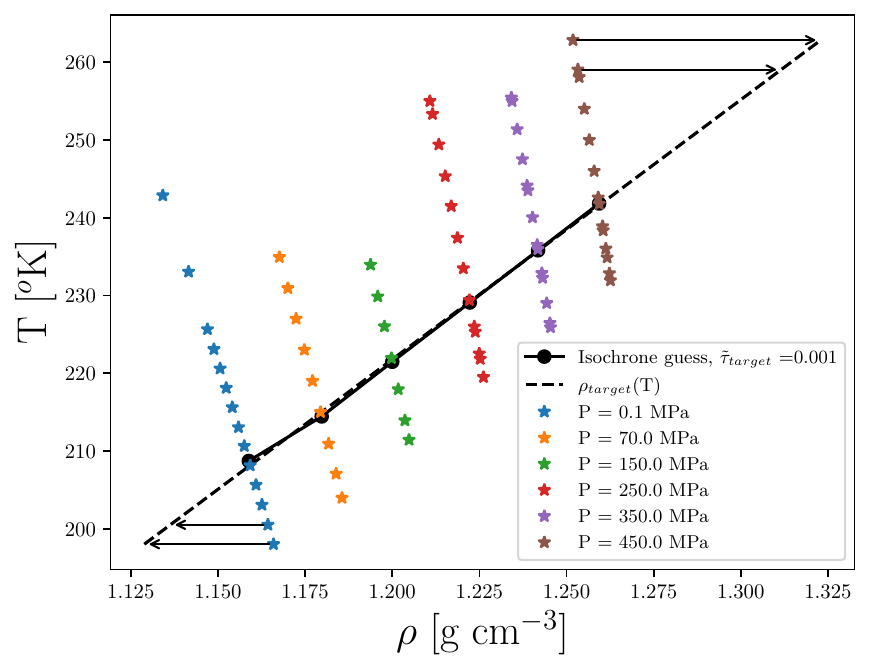}
			\caption{}
			\label{fig:DPG_target_isochrone_b}
		\end{subfigure}	
		
		\caption{Finding the target isochrone for isochronal density scaling. In figure \ref{fig:DPG_target_isochrone_a} the target isochrone is shown along with the reduced relaxation times from ref. \cite{ChatKatarzyna2019Tdsi}. Each isobar was fitted with a 3rd order polynomial to interpolate the density at the target relaxation time. In figure \ref{fig:DPG_target_isochrone_b} the target isochrone is plotted in the ($\rho,T$)-phase diagram, along with the extrapolated $\rho_{\text{target}}(T)$. The data from \Mycite{ChatKatarzyna2019Tdsi} are measured along isobars; thus it was not possible to find $\rho_{\text{target}}(T)$ for all state points without extrapolating the data. The arrows are a guide to eye for the extrapolation. For the ambient pressure measurements, $\rho_{\text{target}}(T)$ have to be extrapolated to nonphysical negative pressures.}
		\label{fig:DPG_target_isochrone}
	\end{figure}

	In figure \ref{fig:DPG_target_isochrone_a} the target isochrone, $\rho_{\text{target}}(T)$ used for isochronal density scaling is shown. The data measured by \Mycite{ChatKatarzyna2019Tdsi} are measured along isobars, and a consequence of this is that for most data points, we have to interpolate to get $\rho_{\text{target}}(T)$. For a few of the data points, some extrapolation is necessary to get $\rho_{\text{target}}(T)$. In figure \ref{fig:DPG_target_isochrone_b} all the measured state points are shown in the $(\rho,T)$ phase diagram. Examples of some the data points we need to extrapolate is shown in the figure.
	

\begin{figure}[H]
		\begin{subfigure}[t]{0.48\textwidth}
			\centering
			\includegraphics[width=0.99\textwidth]{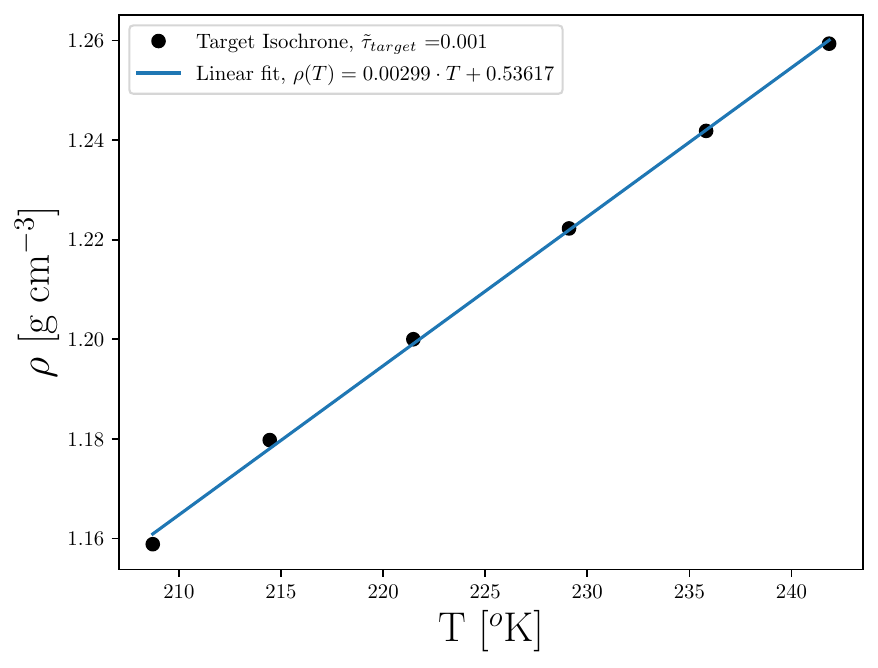}
			\caption{The density along the target isochrone as function of temperature. The relationship between density and temperature is fitted linearly, $\rho_{\text{target}}(T) = 0.00299 \cdot T + 0.53617$. This equation is used to interpolate and extrapolate $\rho_{\text{target}}(T)$. }
			\label{fig:DPG_IDS_collapse_a}
		\end{subfigure}\hfill
		\begin{subfigure}[t]{0.48\textwidth}
			\centering
			\includegraphics[width=0.99\textwidth]{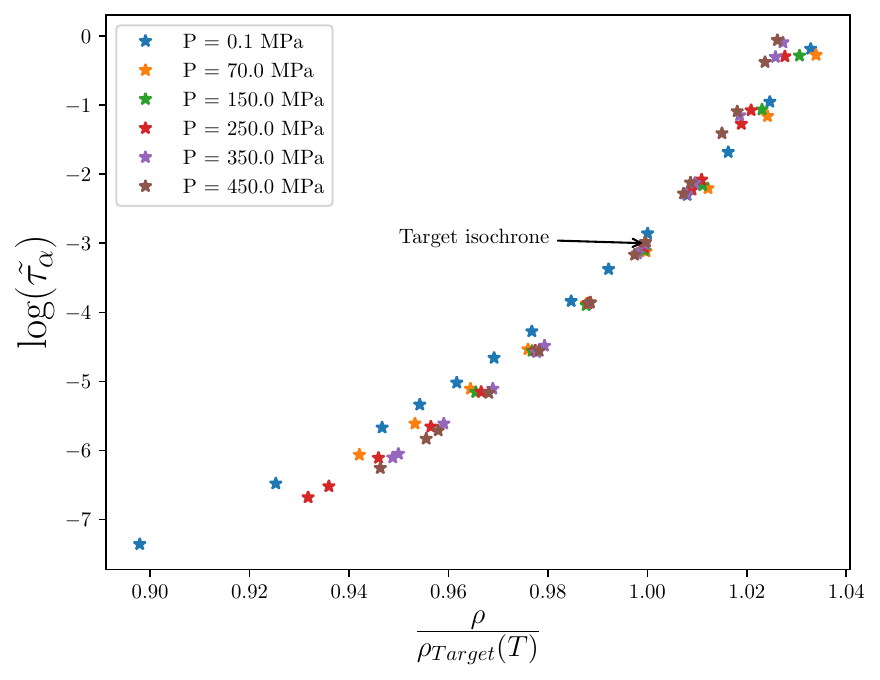}
			\caption{Isochronal density scaling for DPG. Although we do observe a slightly better collapse than with $\Gamma = \frac{\rho^\gamma}{T}$, using the best-fit value $\gamma =1.9$, the relaxation times do not collapse when plotted against $\rho / \rho_{\text{target}}(T)$. }
			\label{fig:DPG_IDS_collapse_b}
		\end{subfigure}	
		\begin{subfigure}[t]{0.48\textwidth}
			\centering
			\includegraphics[width=0.99\textwidth]{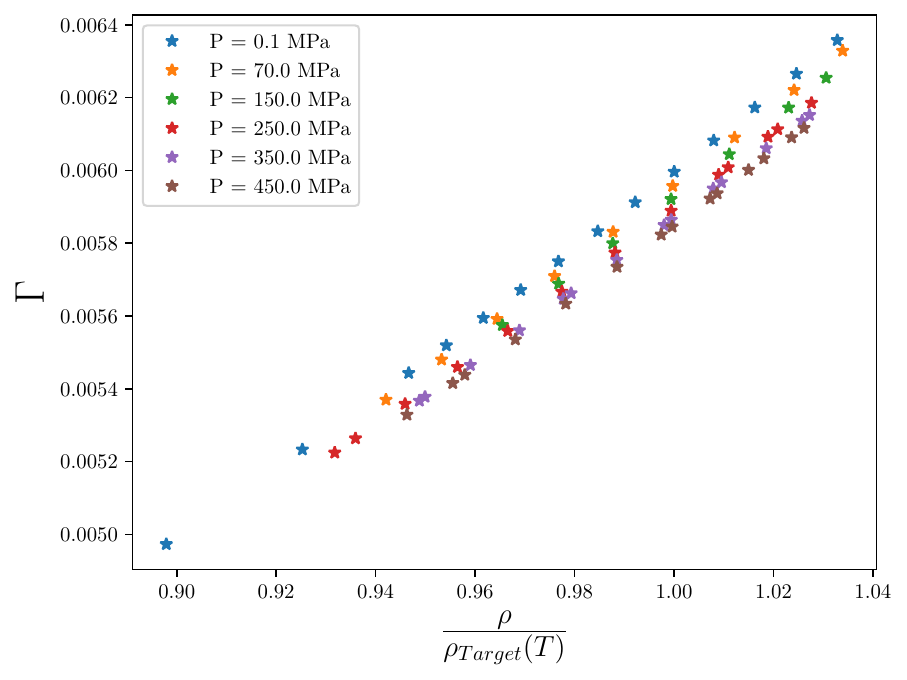}
			\caption{Comparison between   $\rho / \rho_{\text{target}}(T)$ and $\Gamma$, for the literature $\gamma$-value of 1.5 \cite{Wase2018}. }
			\label{fig:DPG_IDS_Gamma_a}
		\end{subfigure}\hfill
		\begin{subfigure}[t]{0.48\textwidth}
			\centering
			\includegraphics[width=0.99\textwidth]{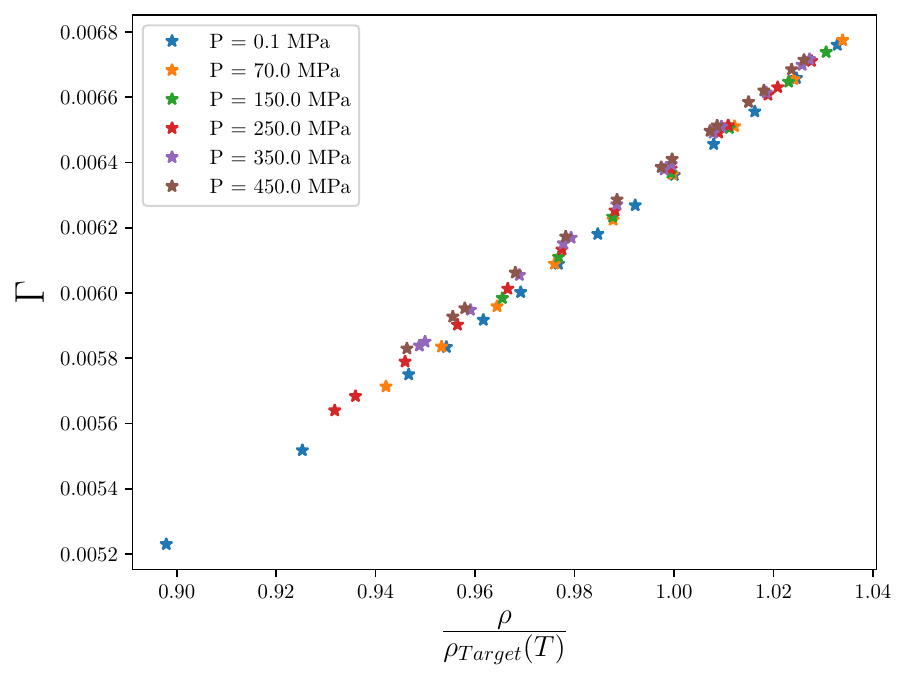}
			\caption{Comparison between   $\rho / \rho_{\text{target}}(T)$ and $\Gamma$, for the best fit $\gamma$-value of 1.9 from ref. \cite{ChatKatarzyna2019Tdsi}. }
			\label{fig:DPG_IDS_Gamma_b}
		\end{subfigure}	
		\caption{Testing isochronal density scaling for DPG. In figure \ref{fig:DPG_IDS_collapse_a}, $\rho_{\text{target}}(T)$ along the target isochrone is fitted with a straight line. The resulting fit is used for interpolating and extrapolating  $\rho_{\text{target}}(T)$. In figure \ref{fig:DPG_IDS_collapse_b} the dimensionless relaxation time is plotted against $\rho / \rho_{\text{target}}(T)$ and it does not collapse. This result is discussed in detail in the text.  Isomorph theory is not expected to work for hydrogen-bonded liquids \cite{Ingebrigtsen2012_simple_liquid}, and, as a consequence, isochronal density scaling is not expected to work either. }
		\label{fig:DPG_IDS_collapse}
	\end{figure}

	In figure \ref{fig:DPG_IDS_collapse_a} the target isochrone is fitted linearly to find $\rho_{\text{target}}(T)$ and in figure \ref{fig:DPG_IDS_collapse_b} the reduced relaxation time is plotted against $\rho / \rho_{\text{target}}(T)$. 
	Unlike DC704, the data do not collapse onto a single line for DPG. The data point collapse may be slightly better than with a best fit value $\gamma =1.9$; however, it is still not a good collapse. To obtain $\rho_{\text{target}}(T)$ we have to extrapolate for some state points. While extrapolation could give an error, this can not alone explain why the dynamics do not collapse.
	

\begin{figure}[H]
	\begin{subfigure}[t]{0.48\textwidth}
		\centering
		\includegraphics[width=0.99\textwidth]{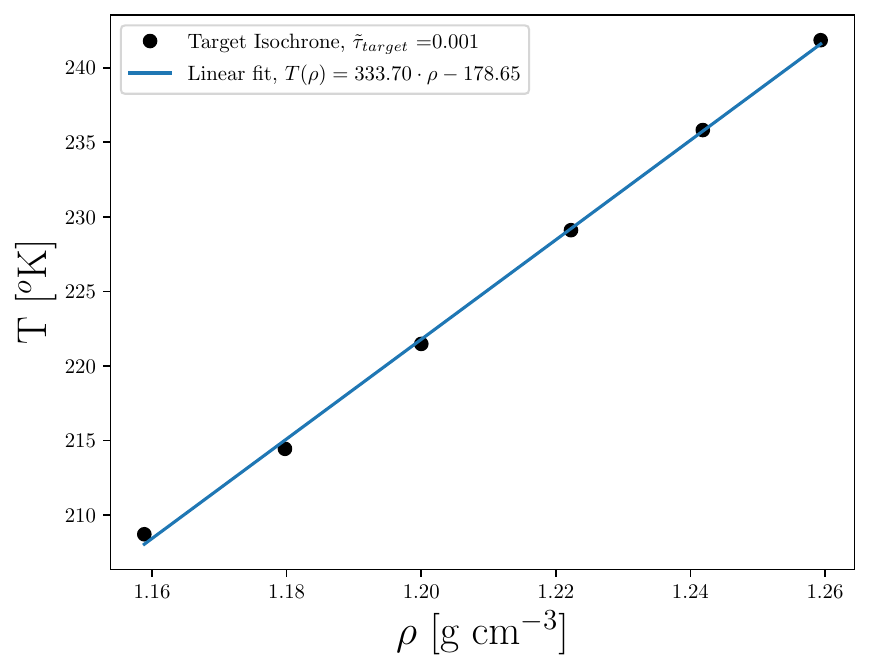}
		\caption{$T_{target}(\rho)$ is fitted. }
		\label{fig:DPG_ITS_a}
	\end{subfigure}\hfill
	\begin{subfigure}[t]{0.48\textwidth}
		\centering
		\includegraphics[width=0.99\textwidth]{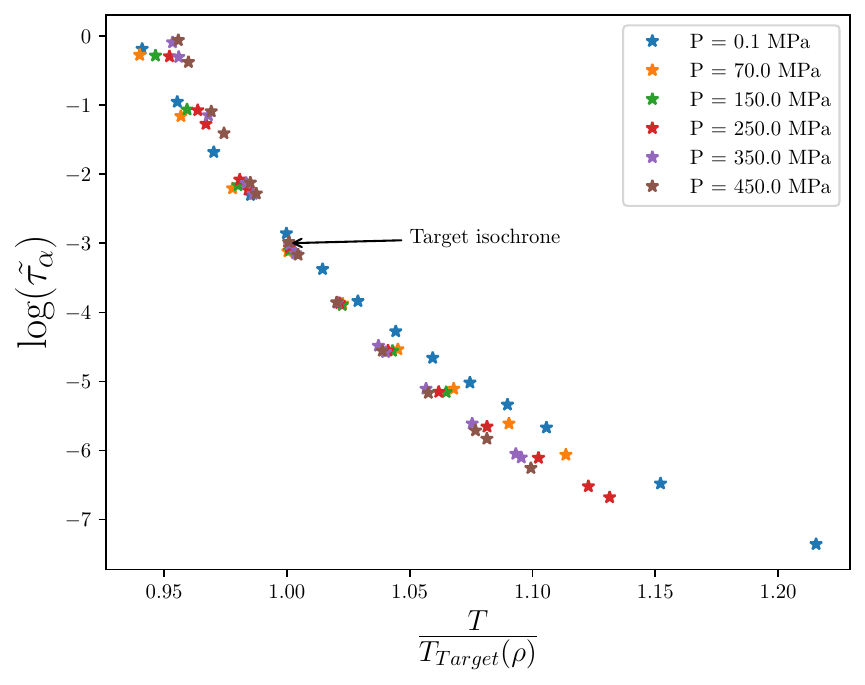}
		\caption{Isochronal temperature scaling for DPG. }
		\label{fig:DPG_ITS_b}
	\end{subfigure}	
	\caption{Isochronal temperature scaling for DPG. Figure \ref{fig:DPG_ITS_a} shows the fit to the target isochrone. It can be fitted linearly $T(\rho) = 333.70 \cdot \rho -178.65$. In figure \ref{fig:DPG_ITS_b} isochronal temperature scaling is tested on literature data for the relaxation time for DPG \cite{ChatKatarzyna2019Tdsi}. The origin of the breakdown of power-law density scaling is not the same as for DC704.  }
	\label{fig:DPG_ITS}
\end{figure}

In figure \ref{fig:DPG_ITS} isochronal temperature scaling is tested using the same isochrone as used to test isochronal density scaling. In figure \ref{fig:DPG_ITS_a} a fit to $T(\rho)$ along the target isochrone is shown. In figure \ref{fig:DPG_ITS_b} the reduced relaxation time is plotted against $T / T_{target}(\rho)$. Isochronal temperature scaling is not able to collapse the data.

	For DC704, we observed that the breakdown of power-law density scaling occurred not due to the density dependence of the scaling parameter $\gamma$, but because of the temperature dependence of $\gamma$. Given that for DPG isochronal density scaling cannot cause the dynamics to collapse onto a single line, the reason for the breakdown of density scaling must be different for DPG. The mechanism for the breakdown of power-law density scaling differs for hydrogen-bonded liquids and for van der Waals liquid like DC704  \cite{Casalini2011,Ransom_DC704}. With increasing pressure, hydrogen bonds disassociate and the longer-range structure in the liquid breaks down. For hydrogen-bonded liquids $\gamma$ tends to be lower than for van der Waals bonded liquids, and with increasing pressure the state point dependent $\gamma$ increase for hydrogen bonded liquids \cite{Roland2008,Ransom_DC704}. 
	
	In chapter \ref{chapter:Diamond} we measured the structure of DPG as a function of pressure. The pressure dependence of the breakdown of the long-range order is also visible in our structural measurements. The pre-peak in the measured intensity in figure \ref{fig:DPG_data} is associated with long-range ordering caused by hydrogen-bonds \cite{Bolle2020}. The intensity of the pre-peak decrease with increasing pressure. The pre-peak also decreased with increasing temperature. When we increase the pressure along an isotherm the density increases, while increasing the temperature along an isobar decrease the density. Increasing the density by temperature and by pressure has opposite effects on the prepeak associated with the long-range order caused by the hydrogen bonds. However, in both chapter \ref{chapter:Cumene} and  \ref{chapter:Diamond} we show that the first peak of $S(q)$ is not trivial to analyze and that the first peak of $S(q)$ effected by both intramolecular and intermolecular contributions. 
	
	
	In figure \ref{fig:DPG_IDS_collapse_b} he change of the relaxation time as a function of $\rho / \rho_{\text{target}}(T)$ is different for the ambient pressure measurements compared to the high-pressure measurement. This seems to harmonize well with the interpretation that the breakdown of density scaling comes from increasing pressure causing the long-range order due to hydrogen bonds to break down.  \Mycite{Roland2008} speculate that if, with increasing pressure, the hydrogen-bonded liquid stops to associate, there could exist a high-pressure range where density scaling would work. The same have been speculated to be true for isomorph theory \cite{Papini2011} and if that is the case then isochronal density scaling and isochronal temperature scaling would also work at high pressures.

	\subsection{Revisiting the structural measurements}
\begin{figure}[H]
		\begin{subfigure}[t]{0.48\textwidth}
			\centering
			\includegraphics[width=0.99\textwidth]{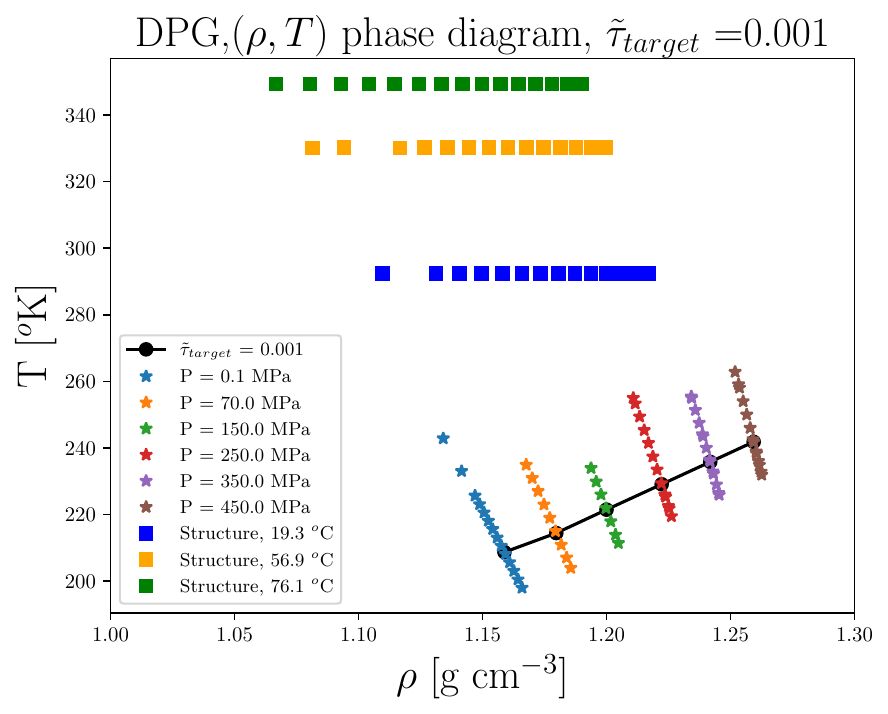}
			\caption{}
			\label{fig:DPG_structure_a}
		\end{subfigure}\hfill
		\begin{subfigure}[t]{0.48\textwidth}
			\centering
			\includegraphics[width=0.99\textwidth]{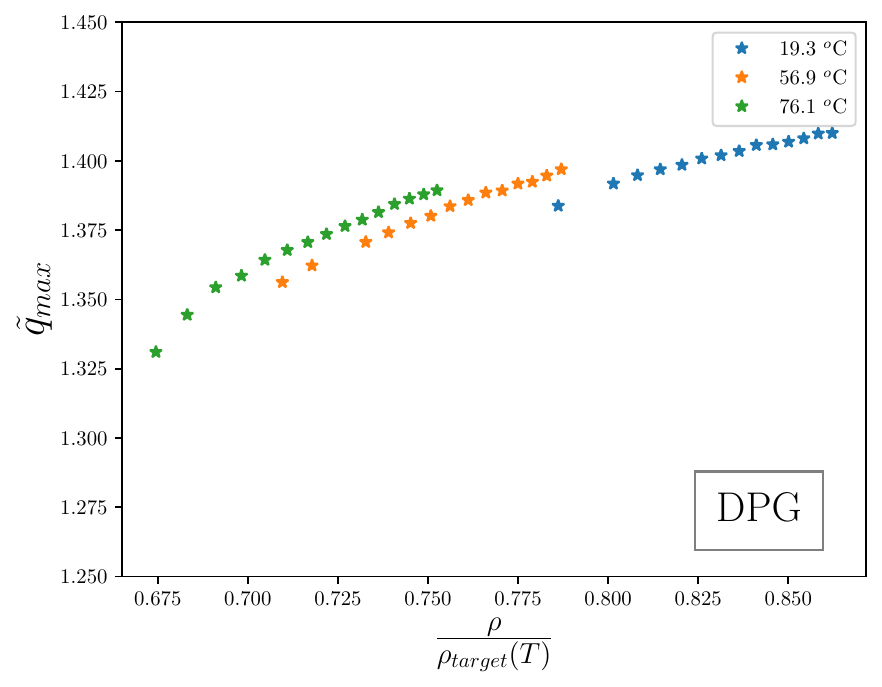}
			\caption{ }
			\label{fig:DPG_structure_b}
		\end{subfigure}	\hfill
		\centering
		\begin{subfigure}[b]{0.48\textwidth}
			\centering
			\includegraphics[width=0.99\textwidth]{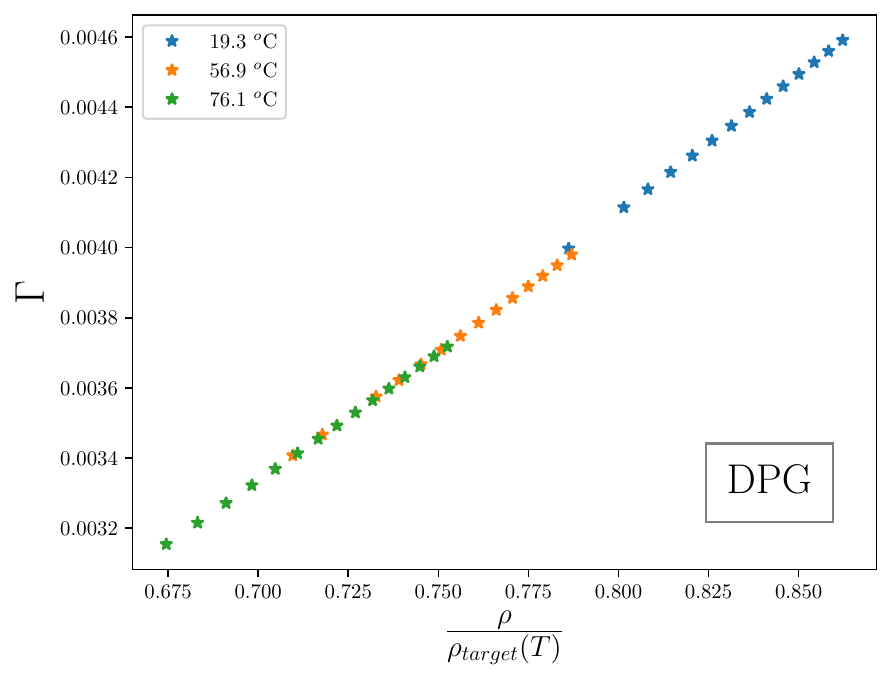}
			\caption{}
			\label{fig:DPG_structure_c}
		\end{subfigure}	
		\caption{Isochronal density scaling for the structure of DPG. In figure \ref{fig:DPG_structure_a} the position of dynamical measurements by \Mycite{ChatKatarzyna2019Tdsi} and the structural measurements from chapter \ref{chapter:Diamond} are plotted in the $(\rho,T)$ phase diagram. The target isochrone used for the scaling is also plotted. For isochronal density scaling we need substantial extrapolation to calculate $\rho_{\text{target}}(T)$. For isochronal temperature scaling we still need extrapolate $T_{\text{target}}(\rho)$, but the problem is smaller.   In figure \ref{fig:DPG_structure_b} $\tilde{q}_{max}$ data is plotted against $\rho / \rho_{\text{target}}$. $\rho_{\text{target}}$ used for scaling is an extrapolation of the fit found in \ref{fig:DPG_IDS_collapse_a}. In subfigure \ref{fig:DPG_structure_c} $\Gamma = \rho^{1.5}/T$ is plotted against  $\rho / \rho_{\text{target}}(T)$. There are almost a one-to-one mapping between $\Gamma$ and $\rho / \rho_{\text{target}}(T)$.
		The $\Gamma$ used in chapter \ref{chapter:Diamond}, seem to find the same candidates for isochrones as isochronal density scaling. To calculate $\rho_{\text{target}}(T)$ we have fitted the isochrone linearly, which is a power-law with power 1. In chapter \ref{chapter:Diamond} we used power-law density scaling, with $\gamma = 1.5$. This could be an explanation, for why power-law density scaling and isochronal density scaling find the same candidates for isochrones. }
		 
		\label{fig:DPG_structure}
	\end{figure}

	For the dynamics of DPG neither power-law density scaling nor isochronal density works. Power-law density scaling could not make the structure collapse, and it is unlikely this would change for isochronal density scaling. However this should be tested.	In figure \ref{fig:DPG_structure} we test if isochronal density scaling can collapse the position of the main peak for DPG. In figure \ref{fig:DPG_structure_a} the position of the measurements from chapter \ref{chapter:Diamond}, are   are plotted along with the measurement of dynamics from \Mycite{ChatKatarzyna2019Tdsi} in the $(\rho,T)$ phase diagram. The target isochrone is also plotted. The structural measurement from chapter \ref{chapter:IDS} and the dynamical data are relatively distant from each other. In order to calculate $\rho_{\text{target}}(T)$ we substantially have the extrapolate the fit. This can be seen in figure \ref{fig:DPG_structure_a}.
	
\begin{figure}[H]
		\begin{subfigure}[t]{0.48\textwidth}
			\centering
			\includegraphics[width=0.99\textwidth]{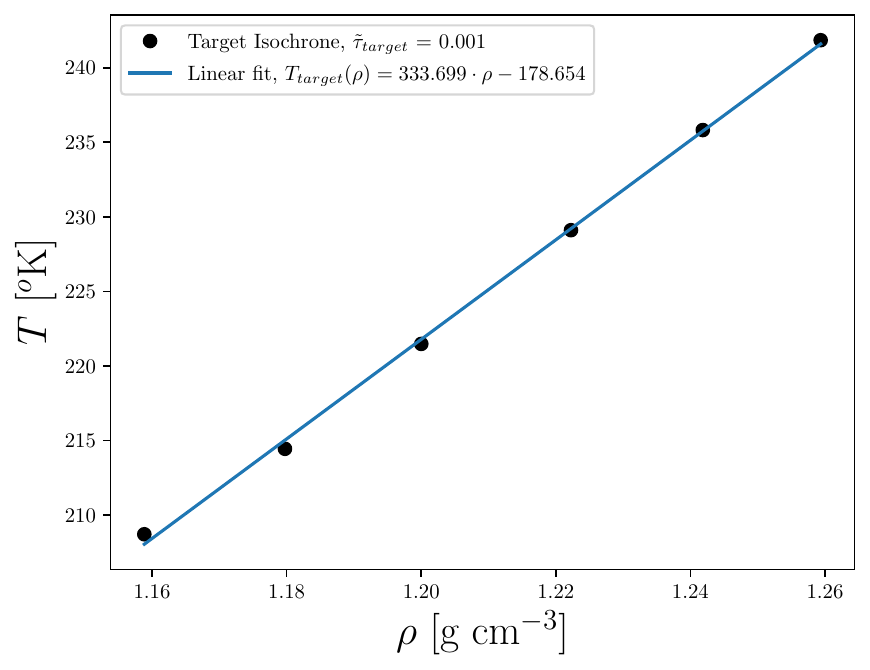}
			\caption{ }
			\label{fig:DPG_structure_T_a}
		\end{subfigure}\hfill
		\begin{subfigure}[t]{0.48\textwidth}
			\centering
			\includegraphics[width=0.99\textwidth]{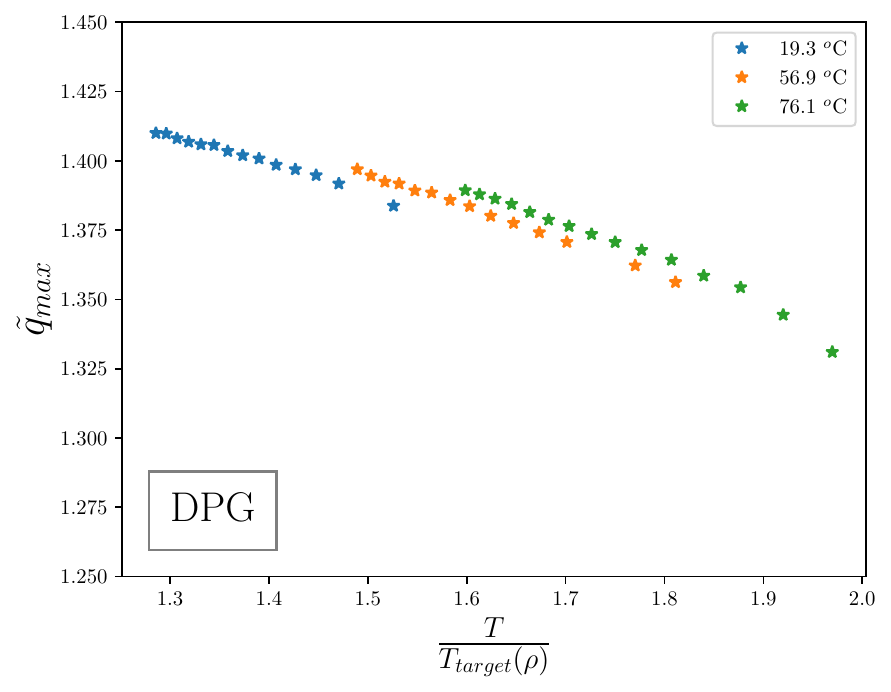}
			\caption{  }
			\label{fig:DPG_structure_T_b}
		\end{subfigure}	
		\centering
		\begin{subfigure}[t]{0.48\textwidth}
			\centering
			\includegraphics[width=0.99\textwidth]{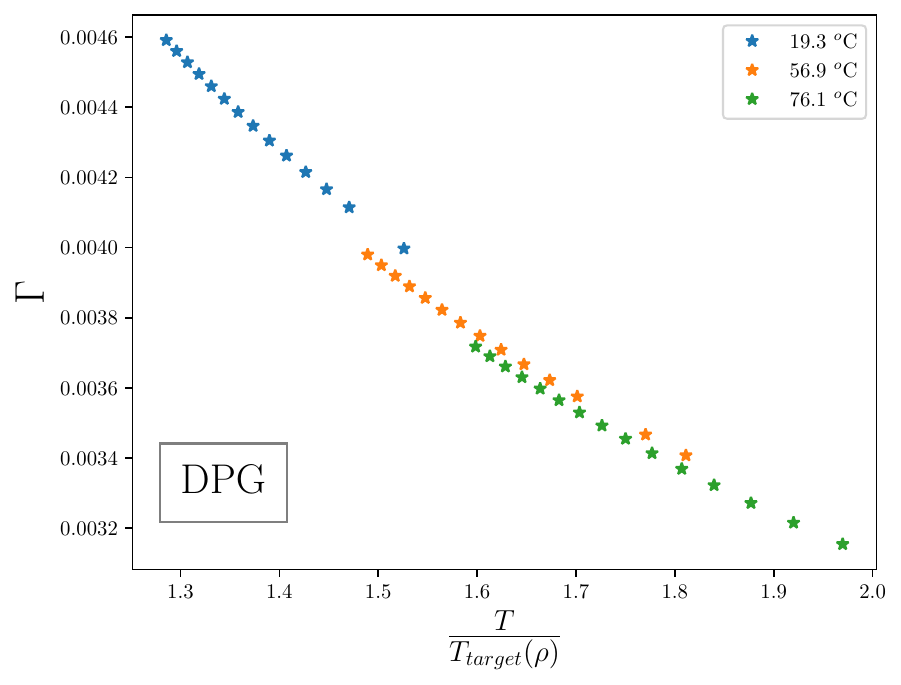}
			\caption{}
			\label{fig:DPG_structure_T_c}
		\end{subfigure}	\hfill
		\begin{subfigure}[t]{0.48\textwidth}
			\centering
			\includegraphics[width=0.99\textwidth]{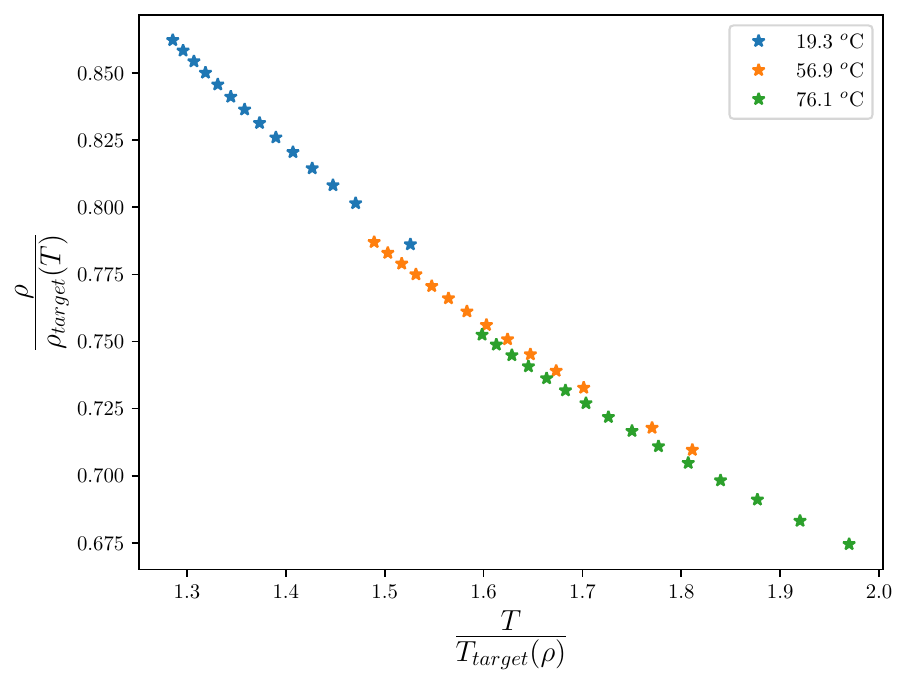}
			\caption{}
			\label{fig:DPG_structure_T_d}
		\end{subfigure}	
		
		\caption{Isochronal temperature scaling for the structure of DPG.
			In figure \ref{fig:DPG_structure_T_a} the target isochrone is fitted linearly: $T_{target}(\rho) = 333.699 \cdot \rho -178.654$. Unlike the results from figure \ref{fig:DPG_structure}, we do not have to extrapolate very far for $T_{target}(\rho)$, however we do have to extrapolate into negative pressures for the state points where $\rho < 1.16$ g cm$^{-3}$. In figure \ref{fig:DPG_structure_T_b} $\tilde{q}_{max}$ is plotted against $T / T_{target}(\rho)$. .
			In figure \ref{fig:DPG_structure_T_c} $T / T_{target}(\rho)$ is plotted against $\Gamma$ and in subfigure \ref{fig:DPG_structure_T_d} $T / T_{target}(\rho)$ is plotted against $\rho / \rho_{\text{target}}(T)$. In theory we should have a perfect collapse between the two, however this is not the case. To calculate $\rho / \rho_{\text{target}}(T)$ we extrapolate quite far from the fitted region. This is likely the cause of the difference between the two.}
		\label{fig:DPG_structure_T}
	\end{figure}

	From figure \ref{fig:DPG_structure_a} it is clear that for the structural measurements,  we can calculate $T_{target}(\rho)$ without to much extrapolation. The only issue here is that for some of the state points, we have to extrapolate the fit to negative pressures. The resulting scaling with $T / T_{target}(\rho)$ is shown in figure \ref{fig:DPG_structure_T}. 
	The reason for choosing DPG as the sample originally was to provide a counter example, and it has fulfilled its role quite well
	Regardless of using isochronal temperature scaling and isochronal density scaling, we cannot collapse the structural measurement presented in chapter \ref{chapter:Diamond}. The dynamical measurements from \Mycite{ChatKatarzyna2019Tdsi} do not collapse either using isochronal temperature scaling and isochronal density scaling. $\gamma$ for DPG seem to be dependentz of both temperature and density, in the experimental range, 
	
	
\section{Discussion and conclusions}
	
	
	In this chapter we have tested isochronal density scaling and isochronal temperature scaling on dynamical and structural measurements on three different glass-formers, Cumene, DC704 and DPG. In the introduction of the chapter we named three overall goals for this study, first to verify the existence of pseudo-isomorphs by revisited the results from chapters \ref{chapter:Cumene} and \ref{chapter:Diamond}, using this new scaling to identify isochrones. Second, to test isochronal density scaling and isochronal temperature scaling on literature data for the dynamics of different glass-formers. Third, use isochronal density scaling and isochronal temperature scaling to study the breakdown of power-law density scaling. Isochronal density scaling and isochronal temperature scaling assume different dependencies for the slope of the isochrones, and give information about the state point dependence of $\gamma$.
	
	The first goal was to verify that the isochrones used to find pseudo-isomorphs in chapters \ref{chapter:Cumene} and \ref{chapter:Diamond} are actually isochrones. Luckily, this seem to be the case. For cumene the shape of the isochrone is a power-law, and power-law density scaling works very well. Isochronal density scaling find the same candidates for isochrones that power-law density scaling does. For DC704, \Mycite{Ransom_DC704} showed that power-law density scaling breaks-down when considering measurements in a large temperature and pressure range. The experimental range for the structure experiment in chapter \ref{chapter:Diamond} is small enough for power-law density scaling to be a good approximation. While there are small changes between $\Gamma$ and $\rho / \rho_{\text{target}}(T)$, this is not enough to affect the collapse of the second peak of $S(q)$. For a hydrogen bonded liquid like DPG isomorph theory is not expected to work \cite{Ingebrigtsen2012_simple_liquid}, so isochronal density scaling and isochronal temperature scaling is not expected to work either. DPG is included in this thesis as a counterexample, and it has fulfilled its role quite well.
	
So what is the origin of the breakdown of power-law density scaling for DPG? For DPG \Mycite{ChatKatarzyna2019Tdsi} showed that power-law density scaling breaks down in a pressure range of up to 450 MPa. Isochronal density scaling and isochronal temperature scaling cannot collapse the dynamical data, unlike what we observed for DC704. Hydrogen-bonded liquids like DPG have strong directional bonds, and the cause of the breakdown of power-law density scaling differs for DPG. Hydrogen-bonded liquids often have long-range structures due to clustering. In a structure measurement this is often seen as a pre-peak or shoulder at lower q-values \cite{Bolle2020}. Increasing pressure will cause hydrogen bonds to break down, which is also seen in the scattered intensities measured in our experiment. Because of the hydrogen bonds any type of scaling derived from isomorph theory will break down. Since hydrogen bonds can be suppressed by applying pressure, it has been speculated that there exists a high-pressure regime where isomorph theory would be true \cite{Papini2011,Roland2008}. There could be a high-pressure regime, where both isochronal density scaling and isochronal temperature scaling could work.
	
For the van der Waals liquids Cumene and DC704, isochronal density scaling was able to collapse the dynamical data into a single curve. For Cumene we were not able to test isochronal temperature scaling, but for DC704 we did. For DC704 \Mycite{Ransom_DC704} measured the relaxation time in a large temperature and pressure range and showed that power-law density scaling breaks down. Using their data, we showed that isochronal density scaling can collapse the relaxation times into a single curve only using $\rho_{\text{target}}(T)$ along a target isochrone. This is shown in figure \ref{fig:DC704_Isochronal_density_scaling_collapse}. The dynamical measurements were measured along isotherms, so we do not need to make any assumptions about the functional form of the isochrones. Isochronal density scaling becomes parameter-free and makes no assumptions about the shape of the target isochrone.  

For DC704 we saw that isochronal density scaling worked better than isochronal temperature scaling. Why is that? In table \ref{tab:Scaling_guide}, we have compare the different types of scaling and when they are expected to work, depending on the behavior of $\gamma(\rho,T)$. The definition of $\gamma(\rho,T)$ is defined in equation \ref{eq:gamma_def_2}.

\begin{equation*}
	\gamma(\rho,T) = \left. \frac{\partial \ln(T)}{\partial \ln(\rho)}\right|_{\tau_\alpha}
\end{equation*}

\begin{table}[H]
	\centering
	\begin{tabular}{|l|l|l|l|l|}
		\hline
		& Power-law density scaling &  Density scaling & Isochro. $\rho$ scaling & Isochro. T scaling \\ \hline
		$\gamma$ = constant         & Yes                       & Yes             & Yes                     & Yes                \\ \hline
		$\gamma(\rho)$   & No                        & Yes             & No                      & Yes                \\ \hline
		$\gamma(T)$     & No                        & No              & Yes                     & No                 \\ \hline
		$\gamma(\rho,T)$ & No                        & No              & No                      & No                 \\ \hline
	\end{tabular}
	\caption{Expected scaling depending on the behavior of $\gamma$ within the experimental range. Density scaling, $e(\rho) / T$, and isochronal temperature scaling is derived from the same assumptions of $\gamma$. It is worth stressing that the $\gamma$ used in this table is defined as in equation \ref{eq:gamma_def_2}, i.e. the slope of the isochrones in a double logarithmic $\rho,T$-phase diagram.  }
	\label{tab:Scaling_guide}
\end{table}
	
The origin of power-law density scaling, density scaling, isochronal density scaling and isochrone temperature scaling, arises from assumption made about the slopes of the isochrones in a double logarithmic $\rho,T$-phase diagram. The validity of these assumptions is dependent on the experimental range and the liquid studied. If the experimental range is small enough, $\gamma(\rho,T) = \gamma$ is a good assumption. What constitutes a "small enough" experimental range is sample dependent. The measurements presented by \Mycite{Ransom_DC704} was measured in a very large temperature and pressure range. If we look at the state points of the measurements in the $(\rho,T)$-phase diagram, figure \ref{fig:DC704_isochrone_show_b}, the density range is not that big compared to the temperature range. The origin of the break-down of power-law density scaling for DC704 is the \textit{temperature} dependence of $\gamma$, not the density dependence. 

Isochronal temperature scaling and isochronal density scaling can be used to test the assumptions made. In this way this can be used as a 'quick and dirty' approach to test the density and temperature dependence of $\gamma$ in the experimental range. For DC704, we see isochronal density scaling collapsing the data better than  isochronal temperature scaling. This indicates that in the experimental range the temperature dependence of $\gamma$ is more important than the density dependence. From the figure presented it is hard to quantify in what experimental range it is a good assumption that $\gamma$ is independent of density and/or temperature. This would be interesting for further research.

Another interesting research question that arises from isochronal temperature scaling is the consequences for fragility. The target isochrone used for the scaling, can be any reduced units isochrone. The glass transition is approximately an isochrone. If $\gamma$ is independent of temperature, the relaxation time would scale as function of $T / T_g(\rho)$. As a consequence the isochoric fragility should be independent the density. This was also shown for the original description of density scaling \cite{TarjusG2003Ddat}, which is based on the same assumptions as isochronal temperature scaling. However as seen in figure \ref{fig:DC704_Isochronal_density_scaling_T_b}, isochronal temperature scaling seems to work less well for DC704, and as a consequence DC704 will not have a constant isochoric fragility. 

In conclusion we have tested isochronal density scaling on experimental data of the dynamics of three different glass-formers along with MD-simulation data from the united atom model of cumene. We have also tested isochronal temperature scaling on DC704, DPG, and on the MD model. The candidates for isochrones for Cumene and DC704 that were found by power-law density scaling seem to be isochrones. We can still conclude that cumene and DC704 have pseudo-isomorphs. 

For the liquids tested isochronal density scaling seem to work better than isochronal temperature scaling, but this is dependent on the experimental range. Surprisingly, isochronal density scaling could collapse measurements of the relaxation time for DC704 in a large temperature and pressure range where power-law density scaling breaks down. The analysis indicates that it is the temperature dependence of $\gamma$ that cause the breakdown of power-law density scaling, and not the density dependence of $\gamma$.

	\chapter{Concluding Discussion} \label{chapter:discussion}

In this thesis, we have shown the existence of pseudo-isomorphs for two different van der Waal's glass-formers, which means that there exist lines in the phase diagram where both the dynamics and the intermolecular structure are invariant when presented in reduced units. This is the first experimental for the existence of pseudo-isomorphs. With the existence pseudo-isomorph in real molecular liquids, there exist glass-formers where the intermolecular structural changes and the dynamical changes follow each other in lockstep when moving around the phase diagram. The isomorph theory does not makes any predictions about how the structure or dynamics would change when moving between pseudo-isomorphs only that both structure and dynamics are invariant along the pseudo-isomorph. The results of this work would be of interest for several fields of active research.

\subsubsection{Isomorph theory}

\Mycite{Ingebrigtsen2012_simple_liquid} categorizes glass-formers into catagories based on the type of bonds in the liquids. In this picture, liquids that are metallic bonded and (only) van der waals bonded are expected to be R-simple liquids, while liquids with covalent bonds, hydrogen bonds, and ionic bonds are expected not to be R-simple. A liquid can of course have several bond types presented, for example a hydrogen-bonded liquid like DPG, has van der Waals interactions along with hydrogen bonds.In this thesis, we have tested two liquids with van der Waals bonds for pseudo-isomorphs, cumene in chapter \ref{chapter:Cumene} and DC704 in chapter \ref{chapter:Diamond}, and both of them have pseudo-isomorphs. In chapter \ref{chapter:Diamond} also showed that DPG, a hydrogen-bonded liquid, do not have pseudo-isomorphs. What kind of glass-formers are likely to have pseudo-isomorphs? Most likely, molecular liquids with only van der Waals bonds and metallic liquids. For metallic liquids, the problem of intramolecular bonds causing the correlation between the instantaneous virial and instantaneous potential energy to breakdown is not an issue.  In simulations of three different compositions of the CuZr system, all three compositions have been shown to have isomorphs \cite{Friedeheim2021}. By definition a liquid with isomorphs also has pseudo-isomorphs.  In united-atom simulations of the room temperature ionic Pyr14TFSI, it was shown that it has isodynes \cite{Knudsen2024,PeterPHD}, that is lines where several different types of dynamics are invariant, but the structure in reduced units was not invariant. Dependent on the type of bonds, it is possible for different liquids to have, invariance of both dynamics and structure, invariance of only dynamics, or neither structure and dynamics.

In this thesis, the role of the scaling deviation arising from the intramolecular structure not scaling with the reduced units has been discussed many times. In section \ref{sec:strucral changes} the changes seen in $S(\tilde{q})$ when moving along isochores and isochrones are discussed.  For systems with pseudo-isomorphs, we see that along the isochrones, the intermolecular structure is invariant, but there are still changes in $S(\tilde{q})$ from the scaling deviation.  Along the isochore, the measured change in $S(\tilde{q})$ is purely intermolecular, arising from the structural changes originating from something else than density changes. This can also be used for experimental measurements for $S(\tilde{q})$ to liquids for pseudo-isomorphs. The intermolecular structural change between two different pseudo-isomorphs should be invariant, and it is possible to isolate the intermolecular structural change by subtracting two isochoric state points. This is written in the following equation:

\begin{equation}\label{eq:Mohammed_Zidan}
	S^{(1)}\left(\tilde{q}\right) - S^{(2)}\left(\tilde{q}\right) \approx 	S^{(1)}_{inter}\left(\rho^{-\frac{1}{3}}q\right) - S^{(2)}_{inter}\left(\rho^{-\frac{1}{3}}q\right)
\end{equation}

It should be possible to calculate equation \ref{eq:Mohammed_Zidan}, for several different isochores for the same two pseudo-isomorphs. This quantity can be calculated for experimental data, without any knowledge of the intramolecular and intermolecular contributions to  $S(\tilde{q})$. 

 In the discussion of chapter \ref{chapter:Cumene} we discuss the possible origin of the structural changes from moving between pseudo-isomorphs. There have been several studies of the liquid structure of similar aromatic molecules \cite{Headen2010,Headen2019,Falkowska2016}. \Mycite{Headen2010} compare the local structures of benzene and toluene. They found that benzene appeared more structured, compared to toluene. The presence of the methyl group on toluene affects the ability of the molecules to pack. It would be interesting to study the structure of cumene with more complex structural measures than the static structure factor and pair distribution function. This is already possible with the simulations presented in chapter \ref{chapter:Cumene}. A neutron scattering experiment as a function of pressure and temperature analyzed with isotopic substitution and EPSR would be an interesting supplement to the results of this thesis. This would give insight into how the cumene prefers to pack and if the structural motifs change when moving between pseudo-isomorphs.

With the existence of pseudo-isomorphs, there exists a group of glass-formers where both the structure and dynamics are invariant along the same lines in the phase diagram. The obvious question to follow is the structure and dynamics of a liquid connected? The mode-coupling theory attempts to predict the relaxation dynamics of glass-forming materials directly from the static, time-independent structure of the liquid \cite{Janssen2018}. It is promising for the mode-coupling theory that there exist pseudo-isomorphs in molecular liquids. In section \ref{sec:strucral changes} we discuss the structural changes to $S(\tilde{q})$ when moving around the phase diagram. It is only the intermolecular contributions to the structure factor that are invariant along an isochrone. The intramolecular structure does not scale with the reduced unit set, $\tilde{q} = q \rho^{-\frac{1}{3}}$, resulting in a scaling deviation in the total $S(\tilde{q})$. Because of this, the total $S(\tilde{q})$ is not invariant along an isochrone, only the intermolecular contribution. For the mode-coupling theory the static structure factor $S(q)$ is the input to calculate the dynamics \cite{Janssen2018}, so this scaling deviation would most likely be an issue when trying to extent the mode-coupling theory into molecular liquids.

\subsubsection{Connecting structure and fragility}

The overall goal of this thesis is to investigate the connection between the dynamics and structure of molecular glass formers, by testing different scaling laws for both. In the introduction of the thesis we motivated for this goal by discussing the possible structural origin of fragility. The existence of pseudo-isomorphs is very interesting in the context of fragility. In subsection \ref{sec:fragility} a few empirical correlations between structure and fragility were introduced. A short recap of the correlation is presented here.	\Mycite{Voylov2016} correlated the change in the FWHM of the first peak in S(q) with the fragility of several different glass-forming liquids, including metallic glass-formers, oxid glasses, polymers, and molecular glass-formers. The authors correlate the relative change between the FWHM in the liquid and the glass with the fragility.
	
			\begin{equation}  \label{eq:Voylov_dis}
		\frac{FWHM_{liquid} - FWHM_{glass}}{FWHM_{liquid}}
	\end{equation}

	\Mycite{Ryu2020} correlate the medium-range order and the fragility for several different metallic glass-formers, using either experimental data or simulations. \Mycite{Ryu2019} is an earlier paper by the same group arguing for the origin.  They found a correlation between fragility and the structural measure: 
\begin{equation}\label{eq:Ryu_dis}
	\frac{\xi_s(T_g)}{a} = \frac{4}{3\pi} \frac{q_{max}(T_g)}{FWHM(T_g)}
\end{equation} 

where $\xi_s(T_g)$ is the structural correlation length evaluated at $T_g$ and $a$ is the average nearest-neighbor distance. Assuming that the peak shape of the first peak is a Lorentzian, this quantity can be calculated by fitting the first peak. The study of \Mycite{Voylov2016} is one of the only studies trying to connect between structure and fragility that also includes molecular glass-formers.

When trying to connect the two for molecular liquids, there is a relatively simple separation into two different sub question; If there is a connection between structure and fragility, then what is the right structural measure to compare with fragility? and what is the right measure of fragility to compare with structure? 

First, the fragility is a measure of how the relaxation time deviates from the Arrhenius behavior when moving in the phase diagram. The existence of pseudo-isomorphs shows that there exists a class of liquid where both the relaxation time and the intermolecular structure are invariant in reduced units along the same lines in the phase diagram. Isomorph theory only makes the prediction that structure and dynamics should be invariant along the same lines, it does not make any prediction for how either should change when moving in the phase diagram. However, it is clear that the reduced intermolecular structure and the reduced relaxation time are in lockstep for pseudo-isomorphic liquids when one moves around in the phase diagram. If the origin for the deviation from Arrhenius behavior is from structural changes, then pseudo-isomorphic liquids are interesting candidates to study because the reduced intermolecular structure and the reduced relaxation time are invariant along the same lines in the phase diagram. 
	
The change in the relaxation time and structure when moving around the phase diagram is different along isotherms and isochores.  In section \ref{sec:strucral changes}, we discuss the structural changes seen in $S(\tilde{q})$ happens when moving along isotherms, isochores and isochrones. The role of the scaling deviation arising from the intramolecular contributions to the static structure factor was clear and visible in the first peak of $S(\tilde{q})$. The correlation of \Mycite{Ryu2020} was tested on simulation data of metallic glass-formers, and the literature data of different types of glass-formers, originally published by \Mycite{Voylov2016}. For metallic glass-formers, the structural unit is an atom, so there is no intramolecular contribution to the structure. The structural measure of \Mycite{Ryu2020}, equation \ref{eq:Ryu_dis}, would for molecular liquids be affected by the scaling deviation from the intramolecular contributions to $S(\tilde{q})$. The structural measure of \Mycite{Voylov2016}, takes the difference in FWHM between two state points. As shown in section \ref{sec:strucral changes} the difference in $S(q)$ represents a purely intermolecular change; however, if the difference between the two state points is not taken along an isochore, it is not possible to use the reduced units. Another approach could be to find a correlation between the fragility and the real-space structure, where the intramolecular contribution to the structure is easier to control. However, the structural changes moving between isochrones are less visible in real space compared to reciprocal space. This can be seen by  comparing figure \ref{fig:cumene_gr_isochrone_compare} and figure \ref{fig:Cumene_MD_sq_compare}. Finding a structural measure that captures these delicate changes could be difficult.

In chapter \ref{chapter:IDS} we tested isochronal density scaling and isochronal temperature on measurements of both structure and dynamics for Cumene, DPG, and DC704. Isochronal density scaling and isochronal temperature scaling are interesting in the context of fragility.  It has been shown that for liquids with density scaling, the isochoric fragility is density independent \cite{TarjusG2003Ddat}. In section \ref{sec:IDS_intro} we have shown that liquids with density scaling also have isochronal temperature scaling. It is easily shown that liquids with isochronal temperature scaling also have constant isochoric fragility. In chapter \ref{chapter:IDS} we saw that for the van der Waals liquid DC704 and the hydrogen-bonded liquid DPG, that isochronal temperature scaling was not able to collapse the data completely. As a consequence, the isochoric fragility is not constant for the two liquids. For DC704, isochronal density scaling could create a better collapse of the data than isochronal temperature scaling. Perhaps it would make sense to calculate a 'isothermal fragility'

\begin{equation}
m_T = \left.\frac{\text{d} \log_{10}(\tau_{\alpha})}{\text{d} \rho_g/\rho}\right|_{\rho=\rho_g, T = \text{constant}}
\end{equation}

While this expression would be difficult to measure experimentally, this fragility could be independent of temperature for DC704 and Cumene. However, the isothermal fragility would not be invariant for the DPG data. 

Isochronal density scaling and isochronal temperature scaling are both derived by making assumptions about the density and temperature dependence of $\gamma(\rho,T)$. When deriving isochronal temperature scaling, we assumed that $\gamma$ is only a function of the density. This means that if that assumption is true, the isochores are also lines of constant $\gamma$. When deriving isochronal density scaling, we assumed that $\gamma$ is only a function of temperature. In this case, the isotherms are also lines of constant $\gamma$. In both cases, the fragility along these iso-$\gamma$ lines is invariant. For a system like DPG, it would be possible to calculate a line of constant $\gamma(\rho,T)$. The role of $\gamma(\rho,T)$ in fragility should be explored in further work.

So how could this discussion be relevant for trying to connect structure and fragility? We expect metallic glass-formers and van der Waals molecular liquids to have pseudo-isomorphs. For these liquids, we expect the structure and dynamics to follow in lockstep when moving around the phase diagram. If one measures the structure along an isochore, the change in the structure measured in reduced units is purely intermolecular. This seems to me to be an interesting structural quantity to compare with fragility. If $\gamma(\rho,T)$ can be assumed to be independent of the temperature, the isochoric fragility is independent of the chosen isochore. This assumption is of course dependent on the experimental range. In such an experiment, the structural change would be assumed to be purely intermolecular and the isochoric fragility would be independent of the chosen isochore. This could be a way to compare the purely intermolecular structural change and the change in the relaxation time.

\subsubsection{Conclusion}
	
The fundamental prediction of isomorph theory states that \textit{if two state points, ($\rho_1$,$T_1$) and ($\rho_2$,$T_2$) are isomorphic then they have same excess entropy, structure and dynamics in reduced units} \cite{Dyre2014}. Pseudo-isomorphs are a translation of isomorphs into more realistic systems, where the requirement of constant excess entropy is relaxed.
 In this thesis, we have shown the existence of pseudo-isomorphs in molecular liquids. This is the first experimental evidence of pseudo-isomorphs existing beyond computer simulations. In chapter \ref{chapter:Cumene}, we show that the van der Waals bonded liquid cumene has pseudo-isomorphs. We measured the structure in a relatively large temperature and pressure range and found that the measured structure seem to collapse along isochrones. The high-pressure experiment was supplemented with a united atom MD simulation, showing that the intermolecular structure collapsed along isochrones when presented in reduced units. The intramolecular structure caused a scaling deviation when the structure is presented in reduced units. In chapter \ref{chapter:Diamond} the search for pseudo-isomorphs are extended to several other glass-formers. We show that the van der Waals bonded-liquid DC704 also has pseudo-isomorphs. We also show that the hydrogen-bonded liquid DPG does not have pseudo-isomorphs. In chapter \ref{chapter:IDS} we introduce the two new types of scaling laws isochronal density scaling and isochronal temperature scaling. We discuss how isochronal density scaling and isochronal temperature scaling can be used to find the origin of the reported breakdowns of power-law density scaling. Using isochronal density scaling, we can collapse the relaxation time measurements of \Mycite{Ransom_DC704}. Since isochronal density scaling can collapse the relaxation time measurements of \Mycite{Ransom_DC704}, it is likely that the temperature dependence of $\gamma$ causing the breakdown of power-law density scaling for DC704. 
	

\newpage
\printbibliography
\appendix
\clearpage

\chapter{Appendix} 

\section{Equation of state}\label{app:EOS}

In this study, knowing the density at a given state point is important. Firstly, In the reduced unit set we use to plot the data, the reduced length is calculated via the density, $\tilde{l} =l \rho^{\frac{1}{3}}$. Secondly, we wish also to compare the structure along isochores with isochrones. We need accurate data of the density for both. The most common way obtain the density of a liquid as a function of pressure and temperature is via the Tait equation. Technically, what we call the Tait equation should be named Tait-Tammann equation or the modified tait equation, to reflect that it is an extension of the orginal Tait equation \cite{Dymond1988}. In the literature however this often refered to just the Tait equation or the Tait equation of state (Tait EoS) \cite{Ransom2017,Ditte_PHD,Burk2021}. The Tait equation is derived by modeling the derivative in the isothermal bulk modulus:

\begin{equation}
	K_T =  - V_0 \left.  \frac{\partial P}{\partial V}\right|_T  \label{eq:Pogacar}
\end{equation}

Where the inverse of the derivative is modeled by

\begin{equation}
	\left. \frac{\partial P}{\partial V}\right|_T = \frac{B(T) +P}{C} \label{eq:Pogacar_1}
\end{equation}

where $B(T)$ and $C$ are parameters used for the fit. $B(T)$ is almost always expressed as a function of temperature, while $C$ is sometimes a constant independent of temperature, but in other cases, it is also assumed dependent on temperature as well. In the thesis we will present five different equations of states from the literature. For Cumene $C$  depends on temperature, while for the equations of state presented in chapter \ref{chapter:Diamond}, $C$ is independent of temperature. Integrating the expression in equation \ref{eq:Pogacar} with \ref{eq:Pogacar_1} inserted gives:

\begin{equation}
	V(T,P) =  V_0(T) \left[1- C\cdot \ln\left(\frac{B(T)+P}{B(T)+P_0}\right)\right] \label{eq:EoS_v1}
\end{equation}

Using the identity $\frac{1}{V} =\rho$, the Tait equation can be expressed as a function of density instead of volume

\begin{equation}
	\rho(T,P) = \rho_0(T) \left[1- C\cdot \ln\left(\frac{B(T)+P}{B(T)+P_0}\right)\right]^{-1} \label{eq:EoS_v2}
\end{equation}

In this thesis, when the equation of state is presented, it will be presented as in the original source. The equation of states presented, are all representation of the original equation as shown in equation \ref{eq:EoS_v1} or \ref{eq:EoS_v2}. A common representation of equation \ref{eq:EoS_v2} is letting $P_0 = 0$, then equation \ref{eq:EoS_v2}

\begin{equation}
	\rho(T,P) = \rho_0(T) \left[1- C\cdot \ln\left(1 + \frac{P}{B(T)}\right)\right]^{-1}
\end{equation}

This is the case for DC704, 5PPE, and DPG presented in chapter \ref{chapter:Diamond}. $B(T)$ is often defined by either a polynomial \cite{Ransom2017,Burk2021} or by an exponential \cite{Ditte_PHD,CasaliniRiccardo2003Dαai}. The temperature dependence of the ambient pressure density, $\rho_{0}(T)$, are also described in different ways. In the literature, an exponential \cite{Ditte_PHD,GundermannDitte2011Ptde}, an linear expression \cite{Ransom2017,Wase2018} and a polynomial \cite{Burk2021} have been used.


\end{document}